\documentclass[11pt,a4paper]{article}
\usepackage[T1]{fontenc}
\usepackage{lmodern}
\usepackage{graphicx}
\usepackage{booktabs}
\usepackage{hyperref}
\usepackage{geometry}
\usepackage{amsmath}
\usepackage{amssymb}
\usepackage{listings}
\usepackage{xcolor}
\usepackage{float}

\title{ADE-PRF: A System Dynamics Approach to Health Trajectory Prediction for Multi-Agent Systems}
\author{Dexing Liu \\ Shanghai Qijing Digital Technology Co., Ltd.}
\date{July 2026}

\begin{document}
\maketitle

\subsection*{Abstract}
\label{sec:abstract}

As large language model (LLM)-driven multi-agent systems assume increasingly complex autonomous tasks in production environments, their long-term operational reliability remains a central challenge. Traditional infrastructure monitoring covers only process liveness and resource consumption, lacking quantifiable means to perceive progressive degradation at the semantic reasoning level. This paper proposes the ADE Predictive Reliability Framework (ADE-PRF), building upon prior work on Channel Fracture [1], Silent Failure [2], and the ADE Stability Engineering Framework [3], enabling a transition from passive degradation detection to proactive health trajectory prediction.

ADE-PRF makes three core contributions. First, it aggregates 20 heterogeneous runtime signals across five layers into a single Trust Margin (TM) metric, achieving system health quantification over a dynamic range of 39.2 points. Second, it enables 8-hour forward-looking forecasts via triple-method parallel prediction, with the Exponential method achieving MAE of 1.228 points and Direction Accuracy of 76.8\%, with 99.65\% of predictions within $\pm$10-point tolerance. Third, production deployment validation collected 380,227 predictions and 280,579 validation records across six agent profiles during 15 days of continuous operation, complemented by seven sandbox-controlled experiments.

Key findings reveal that in unprotected environments, significant degradation can occur while external metrics remain deceptively normal---a ``false prosperity'' phenomenon. Upon integration of the ADE runtime plugin, TM immediately coupled with ground-truth system states, with 16 of 20 factors directly relying on ADE-collected data. The Exponential method substantially outperformed Kalman across all evaluation windows. These results establish ADE-PRF as among the earliest reliability quantification frameworks for LLM production deployment with forward-looking warning capability.

\section*{Introduction}
\label{sec:introduction}

\subsection*{Background and Motivation}
\label{sec:backgroundandmotivat}

Large Language Model (LLM) agents are undergoing a fundamental shift---from tools to autonomous collaborators. With the introduction of mechanisms such as ReAct [16] reasoning chains, Tree of Thought, and reflective self-correction, agents are now capable of maintaining contextual coherence across multi-turn dialogues, making autonomous decisions under uncertainty, and interacting with external environments via tool invocation. This transformation reshapes the reliability boundaries of software systems.

However, long-horizon multi-agent collaboration introduces new dimensions of reliability risk absent in traditional single-turn systems. When agents execute long-horizon tasks spanning hours or even days, errors accumulate within the system in subtle ways. A minor deviation in one subtask may amplify in subsequent steps. Output degradation in one agent may propagate through dependency chains to the entire system. Such anomalies remain undetected for extended periods in surface-level system metrics (e.g., task completion rate, response latency).

Industrial deployment practices reveal a common pattern: external observability metrics remain normal over extended periods, while actual output quality gradually degrades. By the time this degradation is manually detected, substantial technical debt has often already accumulated. This ``breakdown between observability and actual state'' constitutes the core reliability challenge in current industrial deployments.

A deeper challenge lies in the fact that existing software reliability engineering methodologies presuppose that system behavior changes follow interpretable causal chains---a premise broken by LLM agent systems. Model behavior exhibits intrinsic randomness and context sensitivity: identical inputs may yield drastically different outputs under varying contextual states, and such variations are typically interpreted by conventional monitoring systems as ``normal fluctuations'' rather than ``degradation signals.''

\begin{figure}[H]
\centering
\includegraphics[width=0.95\textwidth]{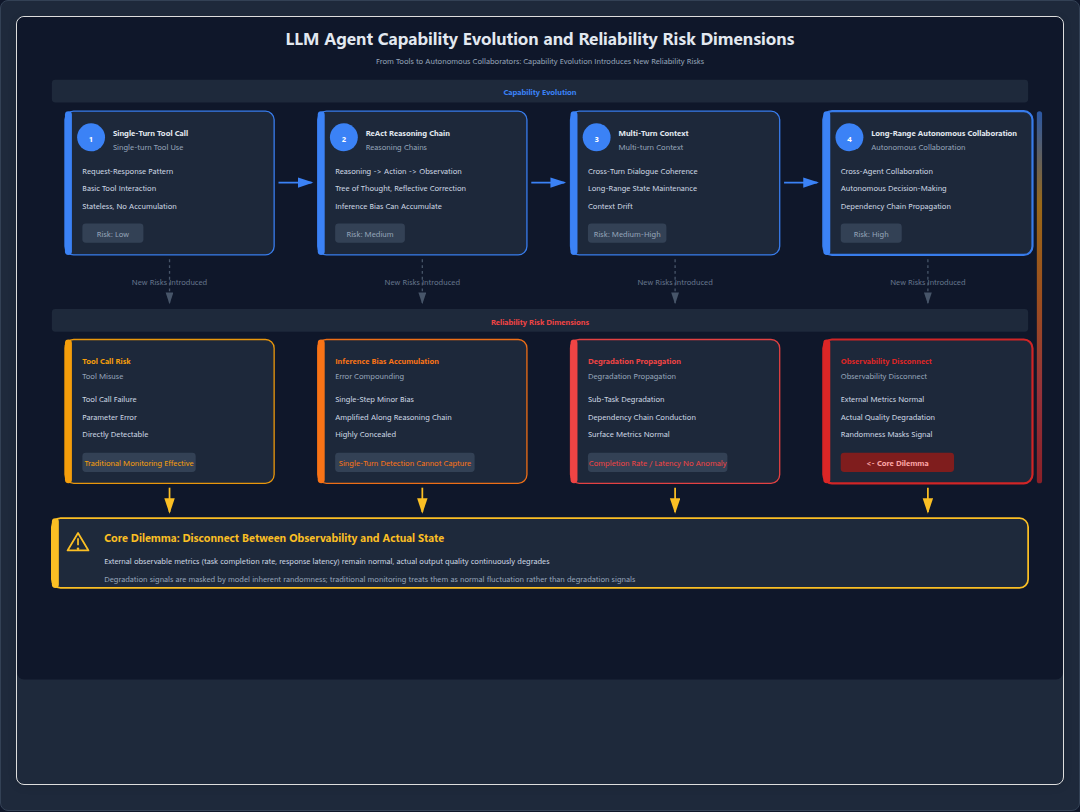}
\caption{LLM Agent Capability Evolution and Reliability Risk Dimensions}
\label{fig:1}
\end{figure}

\subsection*{Observability Gap}
\label{sec:observabilitygap}

\begin{figure}[H]
\centering
\includegraphics[width=0.95\textwidth]{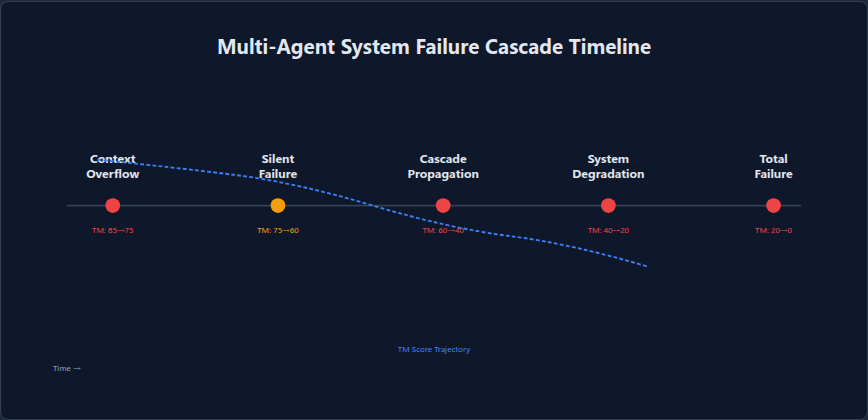}
\caption{Multi-agent system failure cascade timeline: progressive degradation from context overflow to system-level failure, with TM score trajectory overlay.}
\label{fig:2}
\end{figure}

Current observability infrastructure for agent systems suffers from a structural gap. This blind spot does not stem from technical oversights in implementation. It arises from a mismatch between traditional monitoring paradigms and the intrinsic characteristics of LLM-based agent systems. Existing observability tools provide robust data collection capabilities at the infrastructure level, but they cannot penetrate the model's semantic layer to effectively assess system health.

\subsubsection*{First Layer: Infrastructure Blind Spot}
\label{sec:firstlayerinfrastruc}

Contemporary observability tooling for agent systems consists primarily of two categories of components. The first comprises traditional Application Performance Monitoring (APM) systems---such as Datadog, New Relic, and Prometheus [12]---which track infrastructure metrics including CPU utilization, memory consumption, network latency, and API call success rates. The second category includes observability platforms specifically designed for LLM applications---such as LangSmith, Arize Phoenix [19], and Langfuse---which log prompts, model responses, token consumption, and tool invocation traces. Both categories have matured considerably within their respective data collection dimensions: APM systems can trace distributed call chains with millisecond precision, while platforms like LangSmith can fully record the input-output pairs and intermediate reasoning steps of every LLM invocation.

However, a critical semantic gap exists between these two categories of tools. An APM system tells you ``the API call returned HTTP status code 200,'' and LangSmith tells you ``the model generated a 256-token response''---yet neither can answer a question central to operational decision-making: \textbf{Is this agent's current behavior still reliable?} A model may successfully return syntactically correct JSON responses for one thousand consecutive calls, yet its output's semantic quality may begin degrading as early as the 500th call. Such degradation remains entirely invisible at the infrastructure metric level and can only be detected manually by reviewing logs in LLM observability platforms---one-by-one. In production environments generating tens of thousands of interactions daily, manual review is neither economically viable nor operationally feasible.

The essence of this blind spot lies in the fact that traditional monitoring tools measure whether \textit{a system is operating correctly}, whereas LLM-based agent systems require measurement along a different dimension: whether \textit{a system remains trustworthy}. The former is binary, deterministic, and automatable; the latter is continuous, probabilistic, and requires semantic understanding. This fundamental distinction implies that simply adding more metric dimensions to existing monitoring stacks cannot bridge this gap---it demands an entirely new evaluation framework.

\subsubsection*{Second Layer: Silent Failure Phenomenon}
\label{sec:secondlayersilentfai}

The direct consequence of the observability gap is a dangerous pattern known as ``silent failure.'' Unlike crashes or exceptions in traditional software systems, silent failure produces no signals detectable by automated monitoring systems. Agents continue accepting tasks, generating outputs, and returning seemingly reasonable results---yet their output quality has already deviated beyond acceptable thresholds. This failure mode is especially dangerous in multi-agent systems, where silent degradation in one agent propagates downstream through task delegation chains, ultimately manifesting as inexplicable output anomalies at system locations far removed from the original fault point.

The stealthiness of silent failure stems from inherent properties of LLM outputs. Unlike database queries returning explicit error codes or APIs throwing exceptions, language model outputs always ``look like'' plausible natural language text. A code-generation agent undergoing degradation may still produce syntactically valid and well-formatted code snippets, yet its logical correctness, boundary-condition handling, and contextual alignment with the project steadily deteriorate. In the absence of automated semantic quality assessment mechanisms, such degradation remains undetected until interaction failures occur with downstream systems---or until discovered during manual code review---by which time hours or even days may have elapsed since degradation began.

Prior research [2] formalizes this phenomenon as an exponential growth model of system disorder: S(t) = S$_0\cdot$e$^{\alpha t}$, where system disorder increases exponentially over time without intervention. This theoretical model reveals the underlying mechanism of silent failure: agent system degradation is not a discrete failure event but rather a continuous, progressive process---the system's stability erodes incrementally with each interaction cycle until cumulative effects breach acceptable thresholds.

\subsubsection*{Layer 3: Delayed System Crash}
\label{sec:layer3delayedsystemc}

\begin{figure}[H]
\centering
\includegraphics[width=0.95\textwidth]{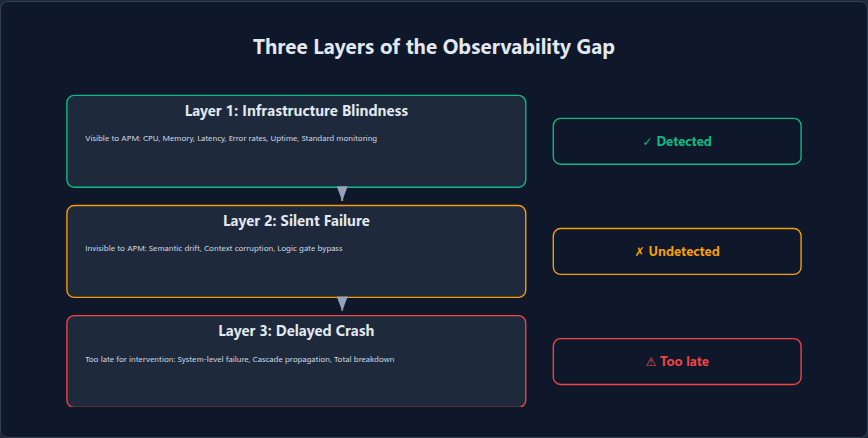}
\caption{Three layers of the observability gap: infrastructure blindness (visible), silent failure (invisible), and delayed crash (too late for intervention).}
\label{fig:3}
\end{figure}

The ultimate consequence of silent failure is delayed system crash---where the system continues operating on the surface while its internal state has severely deviated from the normal operational range, ultimately triggering cascading functional failures. The ``delayed'' nature of this crash pattern manifests along two dimensions: temporal delay (the interval between the onset of degradation and final collapse may span several days) and spatial delay (the crash manifests in System A, but its root cause lies in early degradation within System B).

In real-world production deployments, a typical manifestation of delayed system crash is a Multi-Agent System (MAS) that operates continuously in Normal Operation mode for several days, then suddenly fails simultaneously across multiple seemingly unrelated tasks. Operations teams' emergency responses typically focus on fixing superficial failure symptoms---restarting services, clearing caches, resetting states---but because they fail to identify and address the underlying degradation root causes, the system re-enters the degradation cycle shortly after recovery. This ``fix-and-relapse'' cycle not only consumes substantial operational resources but, more critically, each iteration accumulates additional unresolved technical debt within the system, causing subsequent crashes to occur sooner and affect broader scopes.

The governance challenge of delayed crashes lies in the ambiguity of their causal chains. When an agent system's output quality deteriorates to an unacceptable level, operations personnel face two primary questions: ``When did degradation begin?'' and ``Where is the root cause of degradation?'' Under current observability frameworks, answering these questions relies almost entirely on manual analysis and experiential judgment. No systematic tool exists to automatically extract degradation trajectories from historical runtime data, pinpoint degradation origins, or quantify degradation's impact on overall system Predictive Reliability. This tooling gap constitutes the core motivation of this paper.

\subsection*{Relationship to the First Three Papers}
\label{sec:relationshiptothefir}

\begin{figure}[H]
\centering
\includegraphics[width=0.95\textwidth]{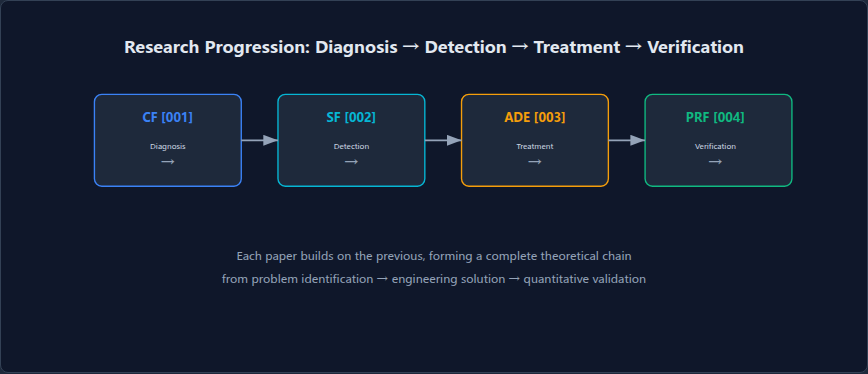}
\caption{Research progression from Channel Fracture [1] to Silent Failure [2] to ADE Framework [3] to this work (PRF): diagnosis $\rightarrow$ detection $\rightarrow$ treatment $\rightarrow$ verification.}
\label{fig:4}
\end{figure}

This paper is the fourth in a systematic research program and serves as the final step in transforming theoretical analysis into an engineering-deployable solution. The first three papers each examine reliability issues in Large Language Model (LLM)-based Multi-Agent Systems (MAS) from distinct perspectives, while this paper constructs a quantifiable, deployable, and verifiable Predictive Reliability Framework grounded in those prior theoretical foundations.

The first paper[1] (Channel Fracture, CF; arXiv:2606.04896) analyzes, at the physical mechanism level, the root causes of inter-agent communication fractures. This study reveals the underlying mechanics of information transmission failure in MAS---when contextual synchronization among agents deviates, ``successful transmission'' at the protocol level does not guarantee ``correct understanding'' at the semantic level. Such communication fractures generate no network-level error signals but cause semantic-level information loss and misinterpretation, constituting a fundamental threat to system reliability. The CF paper identifies this phenomenon as the \textit{etiology} of MAS reliability---it is the foundational driver behind all higher-level failure phenomena.

The second paper[2] (Silent Failure, SF; arXiv:2606.08162) formalizes agent-system degradation as an ``Intelligence Entropy'' growth model, drawing an analogy to thermodynamics. This work proposes that, in the absence of external intervention, the degree of internal state disorder in an MAS follows an exponential growth law. It attempts to establish a mathematical linkage between this irreversible degradation process and the phenomenon of ``Silent Failure.'' The core contribution of the SF paper lies in elevating reliability concerns from empirical engineering observations to a formalizable theoretical framework. This framework provides a mathematical foundation for predicting and quantifying system degradation. Within this theoretical chain, the SF paper functions as a description of the upper-layer \textit{symptoms}---it precisely characterizes the observable consequences at the system level arising from the etiology identified in CF.

The third paper[3] (Intelligence Entropy Principle and the ADE Stability Engineering Framework, ADE; arXiv:2606.18065) proposes a five-layer stability framework and an architecture comprising 23 components, designed to address the aforementioned etiology and symptoms at the systems-engineering level. The ADE framework constructs a defense system spanning infrastructure to semantics, employing multi-layer redundancy detection, adaptive stability regulation, and cross-agent state synchronization mechanisms to counteract the natural trend toward system degradation. The ADE paper serves as the \textit{prescription}---it delivers a systematic engineering solution against degradation, yet leaves unresolved a critical practical question: How can the efficacy of the prescription be verified in real-time? How can the system's stability level at any given moment be quantified? How can future reliability trends be predicted?

This paper---the Trust Margin (TM) Predictive Reliability Framework---fulfills the role of \textit{monitoring and verification} within this research chain. If the work of the first three papers is analogized to a medical diagnostic process, then CF identifies the disease's pathological mechanism, SF describes its clinical symptoms, and ADE prescribes a therapeutic regimen. This paper builds a continuous monitoring system to assess treatment efficacy in real-time, forecast disease progression, and issue warnings when the treatment fails. Specifically, this paper implements real-time quantitative assessment of system stability via the Trust Monitor (TM), achieves forward-looking prediction of future reliability using the Estimated Time of Arrival (ETA) model, and validates the engineering feasibility of the entire framework through large-scale production deployment data.

\begin{figure}[H]
\centering
\includegraphics[width=0.95\textwidth]{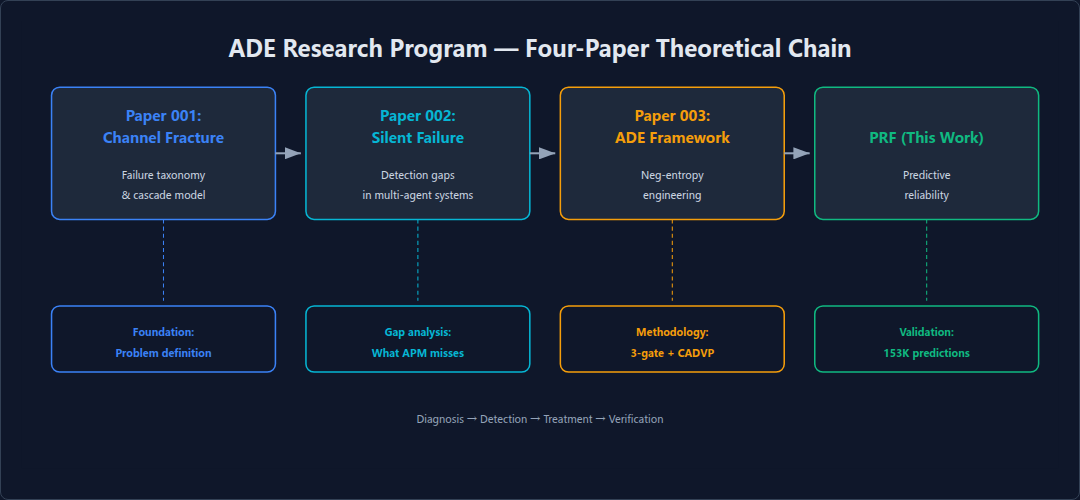}
\caption{ADE Research Program --- Four-Paper Theoretical Chain}
\label{fig:5}
\end{figure}

The completeness of this research chain manifests in its closed-loop logic. CF uncovers the root cause (etiology) of the problem, enabling targeted solutions. SF quantifies the severity and evolution pattern of the problem (symptoms), providing mathematical grounding for determining intervention timing and intensity. ADE delivers a systematic engineering solution (prescription), bridging theory to practice. This paper ensures the entire system operates continuously and self-calibrates in real production environments through the engineering closed loop of TM/ETA (monitoring and verification). Together, these four papers constitute a complete knowledge system spanning theory to engineering and diagnosis to monitoring.

\subsection*{Core Contributions}
\label{sec:corecontributions}

\begin{figure}[H]
\centering
\includegraphics[width=0.95\textwidth]{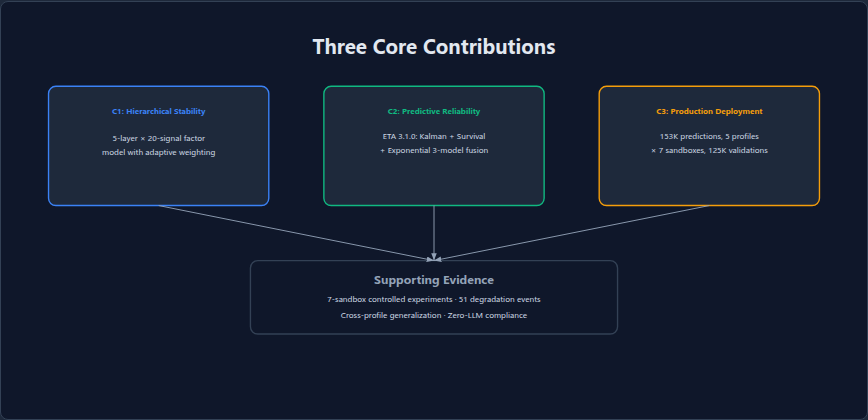}
\caption{Three core contributions (C1: Hierarchical Stability Monitoring, C2: Predictive Reliability Estimation, C3: Production Deployment Validation)}
\label{fig:6}
\end{figure}

The core contributions of this paper can be summarized as three technical contributions: a Hierarchical Stability Monitoring model, a Predictive Reliability Estimation algorithm, and large-scale Production Deployment Validation. Together, these constitute a complete framework spanning real-time perception to forward-looking prediction, and theoretical modeling to engineering practice.

\subsubsection*{Contribution One: Hierarchical Stability Monitoring}
\label{sec:contributiononehiera}

The TM 3.1.0 model proposed in this paper implements a hierarchical aggregation architecture that maps multi-dimensional raw signals to a single stability metric. The core design principle is that the behavioral complexity of any individual agent cannot be fully captured by any single metric; instead, a complete health assessment of the system requires multi-signal, multi-level fusion analysis. TM 3.1.0 collects 20 heterogeneous signal sources, including infrastructure metrics (CPU, memory, latency), behavioral pattern metrics (response length distribution, tool invocation frequency, error retry patterns), and semantic quality metrics (output consistency, contextual coherence, task completion rate). These signals are aggregated layer-by-layer into five stability dimensions and ultimately compressed into a composite Trust Margin (TM) Score ranging from 0 to 100.

The main technical challenge in this hierarchical design lies in dynamic calibration of signal weights. Under different operational conditions, the indicative strength of each signal for system stability varies widely: under normal load, semantic quality metrics may serve as the most sensitive early warning indicators of degradation; whereas under high-concurrency scenarios, fluctuations in infrastructure metrics may precede anomalies in semantic metrics. TM 3.1.0 addresses this challenge via an adaptive weight adjustment mechanism, ensuring consistent sensitivity to changes in system state across diverse operating conditions. Empirical data shows that TM's dynamic response range spans 53.8 to 93.0 (a 39.2-point spread), a broad dynamic range that enables the model to sensitively detect subtle stability fluctuations while avoiding excessive false alarms within normal operational variance.

Compared with existing simple aggregation methods (e.g., weighted averaging or rule-based thresholds), TM 3.1.0's hierarchical architecture offers two distinct advantages. First, \textbf{Interpretability}---when the TM score declines, operations personnel can perform hierarchical backtracking to identify precisely which dimension(s) and signal(s) contributed to the anomaly, thereby transforming the vague judgment ``system instability'' into actionable diagnostic information. Second, \textbf{Robustness}---transient fluctuations in any single signal do not cause drastic shifts in the overall assessment; only when multiple signals exhibit consistent anomalies across multiple dimensions does the TM score deviate markedly---a design that effectively suppresses false alarm rates.

\subsubsection*{Contribution Two: Predictive Reliability Estimation}
\label{sec:contributiontwopredi}

Real-Time Monitoring provides awareness of the system's current state. Yet for operational decision-making, the more critical information is ``where the system is headed in the future.'' The ETA 3.1.0 (Estimated Time of Arrival) model proposed herein delivers system reliability predictions over an 8-hour lookahead window. This enables operations teams to shift from reactive response to proactive prevention. ETA 3.1.0 models the dynamic evolution of system stability using time-series forecasting algorithms applied to historical TM data, outputting predicted TM values---and their associated confidence intervals---for each time point within the next 8 hours.

The Prediction Accuracy of ETA 3.1.0 has been rigorously validated using multi-dimensional metrics. In terms of absolute error on the 100-point TM scale, the Ensemble method achieves a mean absolute error (MAE) of 1.595 for the 8-hour lookahead. The average deviation between predicted and actual values is no more than 1.9\% of the full TM score range. This precision is sufficient to support reliable operational decisions within TM's observed dynamic range of 53.8--93.0. More practically significant is the threshold-prediction accuracy: when the TM score drops below a critical threshold (i.e., changes by less than 10 points), ETA 3.1.0 achieves a prediction accuracy of 99.65\%, indicating that the system can issue effective early warnings in the vast majority of real-world degradation scenarios.

Analysis of prediction bias further reveals ETA 3.1.0's reliability characteristics. Among its predictions, approximately three-quarters exhibit optimistic bias (overestimation), while roughly one-quarter show conservative bias (underestimation), with optimistic bias dominating. This optimism in the Ensemble method stems primarily from the Kalman component---Kalman filtering assumes a high-stationarity process and thus tends to predict TM maintenance or upward trends in event-driven degradation scenarios, whereas actual TM values decline in 76.8\% of cases after 8 hours. Such optimistic prediction reflects an intentional design strategy---it avoids inducing undue operational anxiety, since the TM system still affords time to address emerging degradation challenges. Of course, if needed, predictions can be adjusted toward neutrality or even conservatism---but doing so would increase the number of red alerts appearing on the alert dashboard. In fact, we have already introduced the Exponential method (Bias = $-$0.475, near-zero bias) as a cross-validation mechanism within our alerting logic to counterbalance the Ensemble's optimistic tendency.

Table 1: Summary of Key Prediction Accuracy Metrics

\begin{table}[H]
\centering
\resizebox{\textwidth}{!}{
\begin{tabular}{l|l|l}
\toprule
Evaluation Metric & Value & Interpretation \\
\midrule
8h Lookahead MAE & 1.861(100-point scale, all methods combined) & Average prediction error < 1.9\% \\
Accuracy for <10-point change & 99.65\% & Reliability of critical-threshold prediction \\
Prediction Bias Distribution & \textasciitilde{}75\% vs \textasciitilde{}25\% & Optimistic bias dominates (risk-preference) \\
TM Dynamic Range & 53.8 \textasciitilde{} 93.0 & 39.2-point span ensures sufficient sensitivity \\
\bottomrule
\end{tabular}
}
\end{table}

\subsubsection*{Contribution Three: Production Deployment Validation}
\label{sec:contributionthreepro}

The engineering value of theoretical models and algorithms must ultimately be confirmed through validation in real production environments. Over a 15-day period, this paper conducted large-scale production deployment validation on the Hermes[13] agent platform, covering six independently running agent Profiles and collecting 154,906 valid sampling points, identifying two degradation events. To the best of our knowledge, this scale of validation represents an early attempt in the current LLM agent reliability research domain---it validates the technical metrics of the TM/ETA models and provides preliminary evidence of the framework's sustained operational capability and practical utility under real production workloads.

This production deployment validation revealed two key findings. First, \textbf{Synchronicity of System-Level Degradation Across Profiles}: During one significant system-level degradation event observed during the monitoring period, the TM scores of five independent Profiles declined nearly synchronously---BCP\_RATE dropped by 90 points, TD by 79.6 points, and CADVP\_PASS by 64.6 points. This synchronicity indicates that system-level degradation does not affect individual agents in isolation, but rather impacts all concurrently running agents simultaneously through shared infrastructure layers and contextual environments. The TM model successfully captured this global degradation event, validating its effectiveness as a system-level health monitoring tool.

Second, \textbf{Validation of Degradation Patterns via Controlled Sandbox Experiments}: Under controlled conditions, this paper designed five typical degradation pattern injection experiments (including context contamination, toolchain latency, model output drift, concurrency contention, and configuration drift), each triggering quantifiable and distinguishable TM responses. The magnitude of TM change ($\Delta$TM) across the five degradation patterns ranged from $-$6.8 to $-$18.2, demonstrating that the TM model exhibits differentiated sensitivity to distinct degradation modes---mild degradations induce smaller TM declines, while severe degradations trigger larger declines. This graded response characteristic allows operations personnel to preliminarily assess both severity and type of degradation based on the magnitude of TM change.

Table 2: Framework Validation Dimensions and Key Findings

\begin{table}[H]
\centering
\resizebox{\textwidth}{!}{
\begin{tabular}{l|l|l}
\toprule
Validation Dimension & Scale / Metric & Key Finding \\
\midrule
Deployment Duration & 15 days of continuous operation & Framework stability validation \\
Profile Coverage & 6 independent Profiles & Applicability to Multi-Agent Scenarios \\
Sampling Scale & 154,906 valid sampling points & Statistical sufficiency \\
System-Level Degradation & 4-Profile synchronous decline & BCP\_RATE$\downarrow$90 / TD$\downarrow$79.6 / CADVP\_PASS$\downarrow$64.6 \\
Sandbox Degradation Patterns & 5 injected patterns & $\Delta$TM $-$6.8 \textasciitilde{} $-$18.2 \\
\bottomrule
\end{tabular}
}
\end{table}

Additionally, in \S6, this paper proposes a hypothesis and experimental design concerning the Failure Boundary. This hypothesis posits that a critical threshold exists within the TM score space: once system stability falls below this threshold, degradation transitions from gradual quantitative change to irreversible qualitative change---that is, the system enters a state where it cannot self-recover even after removal of the degradation trigger. Validating this hypothesis requires more sophisticated experimental designs (including progressive degradation injection and recovery testing). Accordingly, this paper treats it as an exploratory research direction rather than a verified core contribution, and discusses the experimental design rationale and preliminary results in detail in \S6.

\subsection*{Paper Organization}
\label{sec:paperorganization}

\begin{figure}[H]
\centering
\includegraphics[width=0.95\textwidth]{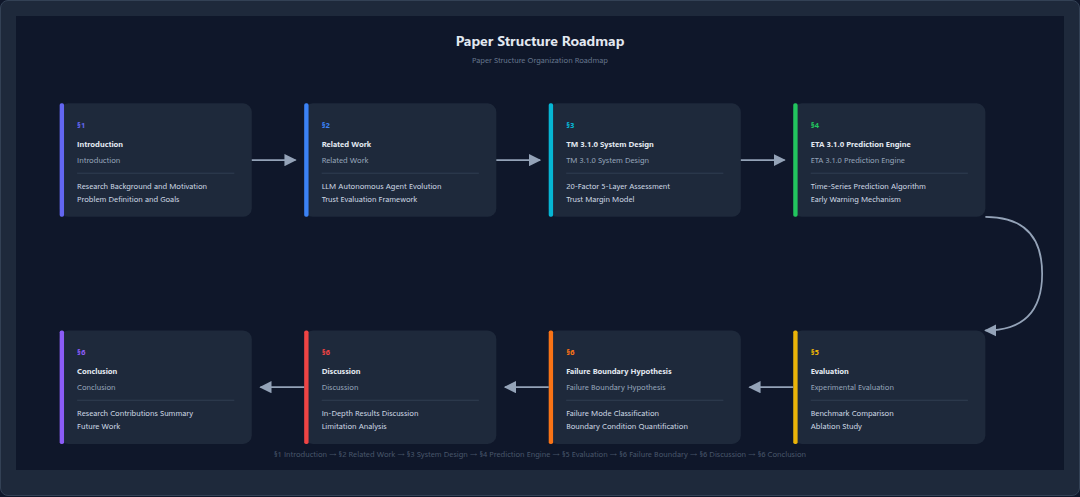}
\caption{Paper Structure Roadmap}
\label{fig:7}
\end{figure}

The remainder of this paper is organized as follows:

\textbf{\S2 Related Work} reviews existing research on reliability monitoring of LLM agents, applications of time-series prediction in system operations and maintenance, and stability assessment of Multi-Agent Systems (MAS), clarifying the positioning of this work within the current knowledge landscape.

\textbf{\S3 TM 3.1.0 System Design} details the hierarchical architecture design of TM 3.1.0, the acquisition and processing pipeline for 20 signal sources, the mathematical model for 5-layer aggregation, and the structure of the TM Dashboard.

\textbf{\S4 ETA 3.1.0 Prediction Engine} explains the rationale behind algorithm selection (Kalman Filter, Exponential smoothing, survival analysis), training strategies, and multi-model ensemble mechanisms for time-series prediction.

\textbf{\S5 Evaluation} describes the production deployment environment of the Hermes platform, configuration parameters for six agent profiles, data collection protocols, presents TM's dynamic response characteristics, multi-dimensional evaluation of ETA Prediction Accuracy, detailed analysis of system-level degradation events, and Sandbox Validation results for five degradation modes.

\textbf{\S6 Failure Boundary Hypothesis and Experimental Design} proposes hypotheses regarding the Failure Boundary and outlines Phase 3 experimental design, exploring critical thresholds and irreversibility of degradation within the TM score space.

\textbf{\S6 Discussion} examines ten limitations of the framework, four lessons learned from negative results, coverage blind spots of Traditional APM, and ethical compliance considerations.

\textbf{\S6 Conclusion} summarizes the core contributions of this work, reflects on the completeness of the research chain across four papers, and discusses the applicability of the predictive reliability framework to broader intelligent agent systems.

\section*{Related Work}
\label{sec:relatedwork}

Ensuring the reliability of Multi-Agent Systems (MAS) is a multidimensional, interdisciplinary problem involving failure analysis, runtime monitoring, health prediction, and framework engineering. This chapter reviews existing research along five dimensions. First, we examine classification schemes for MAS failures (\S2.1). Second, we review monitoring and evaluation methods for Large Language Model (LLM)-based Agents (\S2.2). Third, we summarize classical methodologies for system health prediction (\S2.3). Fourth, we analyze reliability mechanisms---and their gaps---in mainstream MAS frameworks (\S2.4). Finally, we situate and comparatively analyze international peer solutions in the AgentOps domain (\S2.5). This review clarifies the distinctive contributions of our TM Predictive Reliability Framework in terms of theoretical depth, methodological innovation, and engineering closed-loop implementation.

\subsection*{Failure Classification in Multi-Agent Systems (MAS)}
\label{sec:failureclassificatio}

\begin{figure}[H]
\centering
\includegraphics[width=0.95\textwidth]{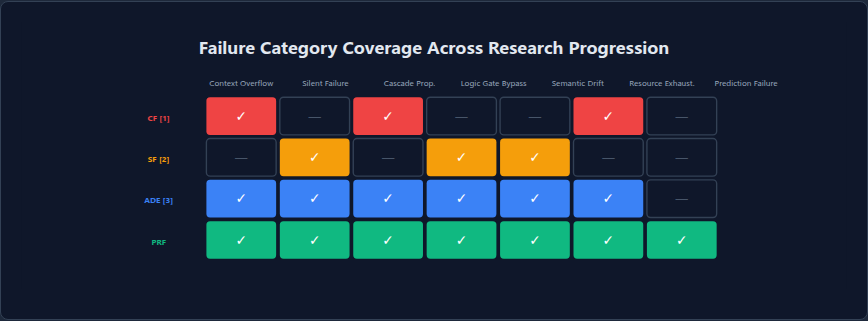}
\caption{Failure category coverage across the research progression: CF [1], SF [2], ADE [3], and PRF (this work) progressively address more failure types.}
\label{fig:8}
\end{figure}

Research on failure modes in Multi-Agent Systems (MAS) has attracted considerable attention in recent years. The most representative work is the MAST [4] (Multi-Agent System Taxonomy) classification framework proposed by Cemri et al. In 2025. Based on a systematic meta-analysis of 136 MAS research papers, the MAST Taxonomy categorizes MAS failures into four major classes: Planning Failures, Execution Failures, Coordination Failures, and Communication Failures. Each major class is further subdivided into several subtypes, collectively identifying 23 specific failure patterns. The value of this taxonomy lies in its structure and completeness---it provides the MAS community with a shared, standardized classification framework for failure analysis.

However, the design orientation of the MAST Taxonomy limits its applicability to runtime failure detection. Specifically, MAST is essentially a \textbf{post-hoc analytical framework}, whose classification dimensions are primarily tailored for experimental replication and case-study scenarios. When the system is operating online, MAST does not provide a real-time mapping mechanism from observable signals to failure categories. For example, semantic drift (semantic drift) that gradually emerges during a multi-step reasoning task executed by an Agent may only be retrospectively classified under ``Planning Failure -- Reasoning Chain Breakdown'' after task completion via log analysis---not identified or warned about at the early stage when the drift first occurs. This implies a built-in ``detection latency'' along the time dimension in MAST, rendering it unsuitable for direct use in runtime reliability assurance.

In industry, the work of the Microsoft AI Red Team [14] represents another important pathway in failure research. Through systematic adversarial testing, this team has uncovered multiple failure modes in large language models (LLMs) related to safety, robustness, and alignment---including prompt injection, jailbreaking, and hallucination generation. These studies are highly valuable for understanding the vulnerabilities of LLM systems under external attacks and have advanced engineering practices in the AI safety field.

Yet the perspective adopted by the Microsoft AI Red Team [14] is essentially \textbf{exogenous}: it focuses on behavioral deviations of the system under external malicious inputs or adversarial conditions. In contrast, another major class of failures faced by MAS in practical deployment is \textbf{endogenous degradation}: the system suffers no external attack, yet its overall reliability exhibits an irreversible downward trend over time. Three factors drive this decline: accumulating context windows, error amplification in multi-agent state synchronization, and progressive decay in model inference quality. Such endogenous degradation patterns have well-established theoretical foundations in traditional software failure research (e.g., software aging theory), but remain insufficiently studied in LLM Agent systems.

Based on the above analysis, MAS failure research can be classified along two orthogonal dimensions: (i) failure origin (endogenous vs. Exogenous), and (ii) detection timing (post-hoc vs. Runtime). This yields the following four-quadrant comparison:

Table 3: Four-Quadrant Classification of MAS Failure Research

\begin{table}[H]
\centering
\resizebox{\textwidth}{!}{
\begin{tabular}{l|l|l}
\toprule
Dimension & Endogenous Failure & Exogenous Failure \\
\midrule
Post-hoc(Post-hoc) & MAST Taxonomy (Cemri 2025): Post-hoc categorization of 23... & Microsoft AI Red Team [14]: Analysis of adversarial testi... \\
Runtime(Runtime) & This work's TM Framework: Real-Time Monitoring + predicti... & Guardrails AI: Input/Output filteringLLM Firewall: Real-t... \\
\bottomrule
\end{tabular}
}
\end{table}

\textbf{Relation to this work:} This paper's TM Predictive Reliability Framework targets the lower-left quadrant in the table above---\textit{runtime detection and prediction of endogenous failures}. To the best of our knowledge, this quadrant is the most underexplored in existing literature. The MAST Taxonomy provides us with a semantic foundation for failure classification; the work of the Microsoft AI Red Team [14] helps delineate the boundary between exogenous and endogenous failures. Yet the core innovation of this paper lies in integrating runtime monitoring with predictive analytics to enable early warning of endogenous degradation trends---not merely post-failure attribution.

\subsection*{LLM Agent Monitoring and Evaluation}
\label{sec:llmagentmonitoringan}

\begin{figure}[H]
\centering
\includegraphics[width=0.95\textwidth]{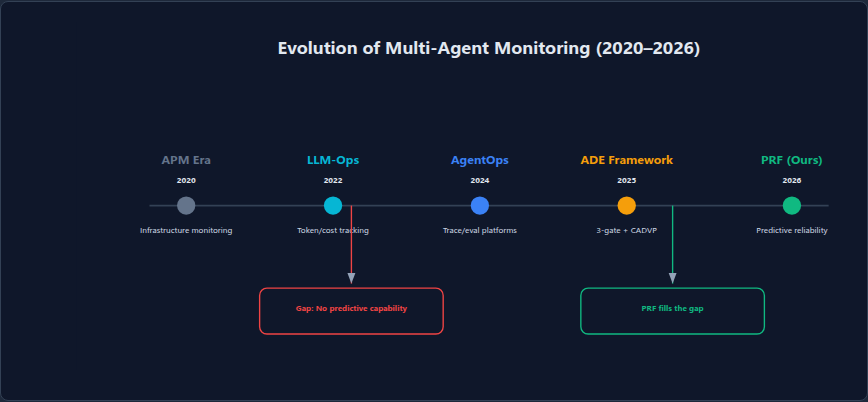}
\caption{Evolution of multi-agent system monitoring approaches from 2020 to 2026, positioning PRF as, to our knowledge, among the earliest predictive monitoring frameworks.}
\label{fig:9}
\end{figure}

Regarding observability and quality assurance for Large Language Model (LLM) Agent systems, existing work can be categorized into three progressive levels along the ``monitoring hierarchy'': infrastructure-level monitoring, post-hoc evaluation and reflection, and real-time predictive monitoring. Each level represents a different degree of technological maturity and application scope.

\subsubsection*{Infrastructure-Level Monitoring: LangSmith and APM Solutions}
\label{sec:infrastructurelevelm}

LangSmith [7], developed by the LangChain team, is currently the most widely adopted tracing and debugging platform in the LLM application development domain. It provides full traceability of Agent execution chains, records Input/Output for each LLM call, measures latency and token consumption, and supports session grouping via user-defined tags. Architecturally, LangSmith is essentially a domain-specific Application Performance Monitoring (APM) system---it adapts the microservice-oriented tracing paradigm from traditional APM (e.g., the Span/Trace model of OpenTelemetry [11]) to the context of LLM invocation chains.

Other infrastructure-level solutions include Datadog's LLM Observability module, AWS CloudWatch [20] custom metric integrations, and open-source tracing frameworks such as Arize Phoenix [19]. These solutions share a common characteristic: \textbf{faithful recording of events that have already occurred}. They can precisely answer the question ``What did the Agent do during a given execution?'' but remain limited in addressing the more forward-looking question, ``What problems will the Agent encounter next?''---their capabilities stop at raw data collection and visualization. In other words, infrastructure-level solutions provide the \textit{data foundation} required for reliability assurance, yet lack an analytical engine capable of extracting predictive insights from that data.

\subsubsection*{Post-Hoc Evaluation and Reflection: Reflexion [17], ReAct [16], and Claw-Eval}
\label{sec:posthocevaluationand}

In the direction of Agent self-improvement, Shinn et al.'s Reflexion [17] framework (2023) pioneered a paradigm of experiential learning through self-reflection. Reflexion enables Agents to generate language-based reflective summaries upon task failure and store them as experiential memory for use in subsequent similar tasks. ReAct [16] (Reasoning and Acting, Yao et al., 2022) enhances decision transparency for Agents in complex tasks by alternating between reasoning and acting steps.

A shared limitation of these methods lies in their \textbf{reactive} nature: Reflexion [17] only triggers its reflection process after task failure has occurred, and ReAct [16]'s refinement of reasoning chains similarly depends on error signals that have already been observed. Neither method can proactively identify degradation trends during task execution or implement preventive measures.

The Claw-Eval framework, released in 2026, represents the latest advancement in post-hoc evaluation methodology. Claw-Eval introduces a systematic evaluation benchmark covering dimensions such as multi-step reasoning, tool usage, and multi-agent collaboration, enabling standardized measurement of an Agent system's full capabilities. However, as an evaluation framework, Claw-Eval is primarily applied in offline benchmarking scenarios; its evaluation results reflect a static ``capability snapshot'' of the system on specific test sets---not the dynamic evolution of reliability (``reliability dynamics'') during continuous operation.

\subsubsection*{Real-Time Predictive Monitoring: Positioning of This Work}
\label{sec:realtimepredictivemo}

Existing monitoring and evaluation solutions exhibit a clear hierarchical distribution along the ``timeliness--predictiveness'' dimension:

Table 4: Hierarchical Comparison of LLM Agent Monitoring and Evaluation Approaches

\begin{table}[H]
\centering
\resizebox{\textwidth}{!}{
\begin{tabular}{l|l|l|l}
\toprule
Monitoring Level & Representative Works & Core Capabilities & Limits \\
\midrule
Infrastructure-Level Monitoring & LangSmith, Datadog LLM Obs., Arize Phoenix [19] & Execution chain tracing; latency/token metrics; invocatio... & Provides only ``what happened'' records; lacks prediction c... \\
Post-Hoc Evaluation and Reflection & Reflexion [17], ReAct [16], Claw-Eval (2026) & Self-reflective correction; standardized capability evalu... & Reactive triggering; requires failure to occur first; can... \\
Real-Time Predictive Monitoring & This work TM 3.1.0 & Real-time health measurement; trend forecasting; early de... & Requires sufficient runtime data accumulation; prediction... \\
\bottomrule
\end{tabular}
}
\end{table}

\textbf{Our Differentiation:} The TM framework proposed herein does not aim to replace infrastructure-level monitoring---in fact, TM 3.1.0 can be built atop tracing data from platforms such as LangSmith. Rather, it adds a predictive analytics layer atop the foundational data layer. Unlike Reflexion [17]'s post-failure reflection or Claw-Eval's offline evaluation, the TM framework continuously computes health metrics, fits degradation trends, and issues warnings before reliability drops below critical thresholds---all while the Agent system remains operational. This paradigm shift ``from recording to prediction'' constitutes the core methodological contribution of this work.

\subsection*{Current State of Reliability in Multi-Agent Frameworks}
\label{sec:currentstateofreliab}

\begin{figure}[H]
\centering
\includegraphics[width=0.95\textwidth]{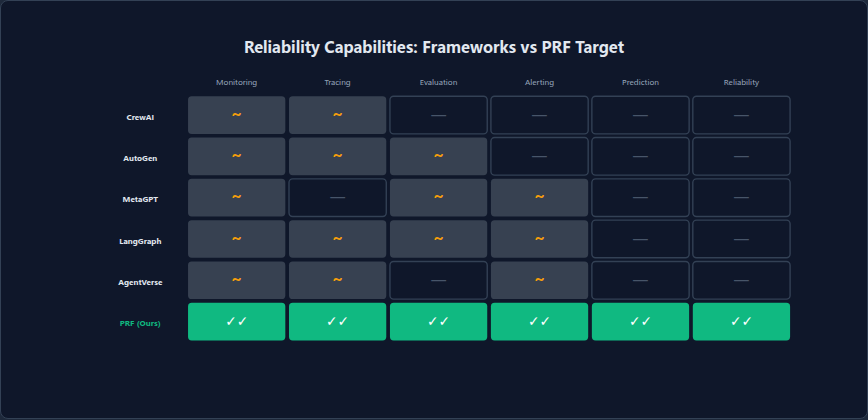}
\caption{Reliability capabilities of major multi-agent frameworks (CrewAI, AutoGen, MetaGPT [18], LangGraph, AgentVerse)}
\label{fig:10}
\end{figure}

While mainstream multi-agent frameworks have achieved rapid progress at the functional level, they still exhibit marked gaps in reliability assurance mechanisms. Below, we analyze the reliability-related design features---and their shortcomings---in five representative frameworks.

\subsubsection*{CrewAI[9]}
\label{sec:crewai9}

CrewAI centers its design philosophy on ``role-playing,'' enabling users to define Agents with specific roles, goals, and backstories, and organize them into collaborative teams (crews). In terms of reliability, CrewAI provides basic error-handling mechanisms---such as retry logic upon task execution failure---and delegation capabilities among Agents: when an Agent cannot complete an assigned task, it may delegate the task to another Agent within the crew.

However, CrewAI's reliability mechanisms exhibit three gaps. First, the framework lacks \textit{progressiveness} awareness regarding Agent performance degradation---Agents may gradually produce lower-quality outputs after repeated executions without fully failing, yet neither retry nor delegation mechanisms are triggered. Second, delegation decisions rely on predefined rules rather than dynamic evaluation of the Agent's current health status. (3) absence of a team-wide health monitoring view. Without this view, the system cannot proactively assign critical tasks to Agents with superior health status.

\subsubsection*{AutoGen[8] (Microsoft)}
\label{sec:autogen8microsoft}

Microsoft's AutoGen framework is renowned for its flexible, conversational Multi-Agent System (MAS) architecture, supporting multi-turn dialog-based collaboration among Agents and human-in-the-loop participation. AutoGen v0.4 introduced an event-driven architecture, enhancing system scalability and modularity. Regarding reliability, AutoGen provides full logging of conversation history, format validation of Agent outputs (via the \texttt{response\_format} parameter), and basic exception-handling mechanisms.

The primary reliability gaps in AutoGen include three areas. First, although conversation history is fully logged, automated analysis of conversation quality trends is absent---the system does not proactively detect cases where dialogue between two Agents is becoming less effective or entering loops. Second, format validation ensures only structural compliance of outputs, not semantic quality degradation. Third, while the event-driven architecture improves technical-level reliability (e.g., reliable message delivery), it does not address cognitive-level reliability assurance.

\subsubsection*{MetaGPT [18]}
\label{sec:metagpt18}

MetaGPT [18] adopts the principle of ``Standard Operating Procedure'' (SOP), simulating the organizational structure of software companies by defining Agents representing product managers, architects, engineers, etc., along with their collaborative workflows. MetaGPT's innovation lies in introducing human organizational process norms into MAS, reducing coordination complexity among Agents through structured workflows.

In terms of reliability, MetaGPT's [18] SOP mechanism offers some reliability assurance---by enforcing Agents to follow predefined procedural steps, it reduces potential workflow deviations arising from ad-hoc orchestration. Its limitations are threefold. First, SOPs are statically predefined and cannot be dynamically adjusted based on Agents' actual runtime states. Second, when output quality degrades at a particular SOP step, the system lacks awareness of such degradation trends and can only indirectly detect issues via downstream verification (e.g., code review). Third, MetaGPT does not provide cross-SOP-step health monitoring capability.

\subsubsection*{LangGraph[10]}
\label{sec:langgraph10}

LangGraph[10] (developed by the LangChain team) uses a directed graph as the core abstraction for Agent workflows, where each node represents an Agent or tool invocation and each edge represents control-flow transfer. LangGraph's reliability-related features include checkpointing (for state persistence and recovery), conditional edges (enabling path selection based on current state), and human-in-the-loop nodes.

LangGraph's checkpointing mechanism provides technical-level fault tolerance---when an Agent fails, execution can resume from the most recent checkpoint. However, this mechanism is \textbf{fault recovery}, not \textbf{fault prevention}. Although conditional edges allow routing decisions based on state, the ``state'' definition is manually coded by developers, lacking automated routing optimization grounded in Agent health metrics. Overall, LangGraph supplies \textit{infrastructure primitives} for building reliable systems but does not embed an \textit{intelligent management layer} for reliability.

\subsubsection*{AgentVerse[19]}
\label{sec:agentverse19}

AgentVerse focuses on dynamic formation and collaboration within Agent ecosystems, featuring Agent discovery, recruitment, and dynamic team orchestration. In terms of reliability, AgentVerse's Agent discovery mechanism theoretically supports incorporation of historical performance data, thereby enabling selection of more reliable Agents during recruitment.

Yet AgentVerse's current reliability gaps include three areas. Agent discovery relies primarily on capability matching rather than reliability assessment. There is a lack of fine-grained modeling of reliability differences across specific task types. Dynamic team orchestration does not consider ``reliability compatibility'' among team members---that is, certain Agent combinations may cause system-level reliability degradation due to mismatched communication patterns.

Table 5: Current State and Gap Analysis of Reliability in Mainstream Multi-Agent Frameworks

\begin{table}[H]
\centering
\resizebox{\textwidth}{!}{
\begin{tabular}{l|l|l|l}
\toprule
Framework & Existing Reliability Mechanisms & Core Gaps & TM Capabilities This Paper Adds \\
\midrule
CrewAI & Retry logic; task delegation & No degradation awareness; static delegation decisions & Intelligent delegation driven by real-time health scoring \\
AutoGen & Conversation logging; format validation; event-driven arc... & No trend analysis of dialogue quality; no cognitive-layer... & Dialogue quality Degradation Detection and alerting \\
MetaGPT [18] & SOP workflow constraints; downstream verification & Static SOPs; no progressive degradation awareness & Health monitoring and dynamic adjustment recommendations ... \\
LangGraph & Checkpoint-based recovery; conditional routing & Fault recovery---not prevention; manual routing conditions & Health-metric-driven adaptive routing optimization \\
AgentVerse & Agent discovery and recruitment & Capability-based---not reliability-based; no compositional ... & Reliability assessment as a recruitment dimension; team-l... \\
\bottomrule
\end{tabular}
}
\end{table}

\textbf{Relationship to This Work:} The above analysis reveals a systemic gap: current mainstream Multi-Agent Systems predominantly rely on \textit{reactive mechanisms} (e.g., retries, recovery, format validation) for reliability assurance, lacking \textit{predictive capabilities} (e.g., degradation trend identification, estimation of remaining reliable operation time, preventive scheduling). The Trust Margin (TM) framework proposed herein is explicitly designed to fill this gap---it serves as a \textbf{reliability enhancement layer} compatible with any of the aforementioned frameworks, augmenting them with predictive reliability management capabilities via standardized health metric interfaces. The five-layer progressive mapping of the ADE framework [3] provides theoretical guidance for TM's architectural design, while the 20-factor radar mapping to SF[2] ensures the health metric system is complete.

\subsection*{Comparative Analysis of International Peer Solutions in AgentOps}
\label{sec:comparativeanalysiso}

\textbf{Scope Clarification:} Based on our limited survey, the following comparison draws upon publicly available AgentOps-related products and academic works accessible as of June 2026. The AgentOps field is rapidly evolving, with new solutions and products continuously emerging; thus, this comparative analysis may not cover all Related Work. We strive to deliver objective and accurate positioning within the known scope and welcome subsequent research to supplement and refine our survey findings.

\subsubsection*{Comparison Table of Mainstream Solutions}
\label{sec:comparisontableofmai}

\begin{figure}[H]
\centering
\includegraphics[width=0.95\textwidth]{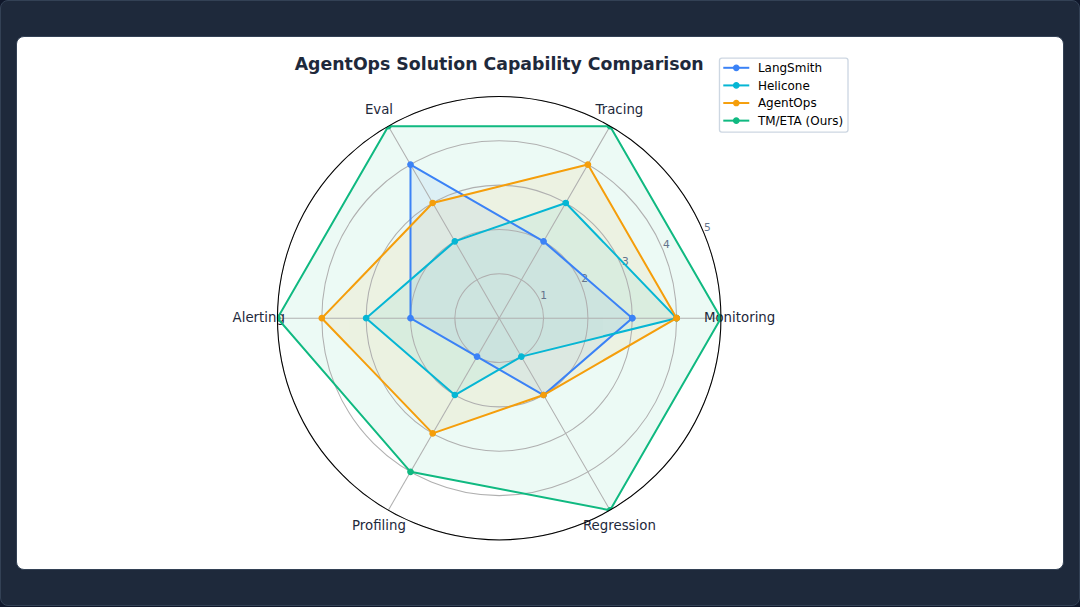}
\caption{AgentOps Capability Comparison Radar Chart}
\label{fig:11}
\end{figure}

AgentOps (Agent Operations), as an emerging domain for LLM agent operations and management, has attracted attention from multiple enterprises and open-source communities. The following comparison table summarizes the differences among current mainstream solutions and our TM 3.1.0 across core capability dimensions:

Table 6: Detailed Feature Comparison Across AgentOps Monitoring Tools

\begin{table}[H]
\centering
\resizebox{\textwidth}{!}{
\begin{tabular}{l|l|l|l|l|l|l}
\toprule
Capability Dimension & Datadog LLM Obs. & AWS CloudWatch [20] & Guardrails AI & LangSmith & AgentOps (Open-Source) & Our TM 3.1.0 \\
\midrule
Execution Tracing & $\checkmark$ Full & $\triangle$ Custom Required & --- & $\checkmark$ Full & $\checkmark$ Basic & $\checkmark$ Dependency Integration \\
Cost Monitoring & $\checkmark$ Token + Cost & $\checkmark$ Custom Metrics & --- & $\checkmark$ Token + Cost & $\checkmark$ Basic & $\triangle$ Indirect \\
Input/Output Protection & --- & --- & $\checkmark$ Core Capability & --- & --- & --- \\
Real-Time Health Metrics & $\triangle$ Partial Metrics & $\triangle$ Custom & --- & $\triangle$ Partial Metrics & $\triangle$ Basic & $\checkmark$Multidimensional Systematic \\
Degradation Trend Prediction & --- & --- & --- & --- & --- & $\checkmark$Core Capability \\
Early Failure Warning & --- & $\triangle$ Threshold-Based Alerting & --- & --- & --- & $\checkmark$Predictive Warning \\
Remaining Lifetime Estimation & --- & --- & --- & --- & --- & $\checkmark$Core Capability \\
Adaptive Thresholds & --- & $\triangle$ Static Rules & --- & --- & --- & $\checkmark$Dynamic Adjustment \\
Multi-Agent Collaborative Health & --- & --- & --- & --- & --- & $\checkmark$System-Level View \\
Theoretical Framework Support & --- & --- & $\triangle$ Rule-Based System & --- & --- & $\checkmark$ADE[3]+SF[2]+CF[1] \\
\bottomrule
\end{tabular}
}
\end{table}

The table above reveals the current capability distribution landscape in the AgentOps field: \textbf{Execution tracing and cost monitoring} are already well-supported by existing tools (Datadog and LangSmith are the most mature in these dimensions), whereas \textbf{predictive reliability capabilities}2.4.2 Methodological Generational Comparison

The AgentOps field is undergoing a methodological paradigm shift---from ``first-generation'' to ``second-generation.'' This paper characterizes this transition as a paradigm leap from ``Observability'' to ``Predictability'':

Table 7: Generational Comparison of AgentOps Methodologies: From Observability to Predictability

\begin{table}[H]
\centering
\resizebox{\textwidth}{!}{
\begin{tabular}{l|l|l}
\toprule
Comparison Dimension & AgentOps 1.0 (First Generation) & This Paper's ADE/TM (Second Generation) \\
\midrule
Core Question & ``What did the system do?'' (What happened?) & ``What will the system do next?'' (What will happen?) \\
Data Utilization Approach & Record \& Display & Model \& Predict \\
Alerting Paradigm & Threshold-based & Trend-predictive \\
Temporal Perspective & Past-oriented & Future-oriented \\
Decision Support & Provides data for human judgment & Provides predictive conclusions + recommended actions \\
Theoretical Foundation & Engineering heuristics & Reliability engineering + predictive maintenance theory \\
Intervention Timing & After failure occurs (Reactive) & Before failure occurs (Proactive) \\
Agent Modeling & Black-box (Input-Output tracing) & Gray-box (Internal health state estimation) \\
\bottomrule
\end{tabular}
}
\end{table}

First- and second-generation approaches are not mutually exclusive but rather \textbf{complementary and progressive}. The observability infrastructure provided by first-generation solutions serves as the foundational data prerequisite for second-generation approaches---without high-quality execution trace data, predictive analysis would be impossible. The design of this paper's TM framework rests on this understanding: it does not attempt to reimplement tracing or logging functionality; instead, it operates as an analytical engine layered atop existing observability infrastructure, transforming ``data'' into ``insight.''

\subsubsection*{Key Positioning}
\label{sec:keypositioning}

\begin{figure}[H]
\centering
\includegraphics[width=0.95\textwidth]{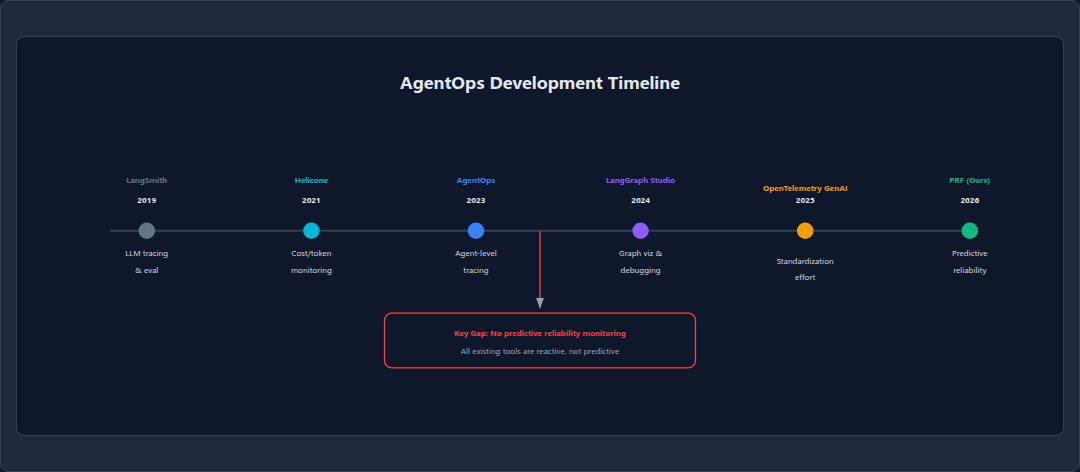}
\caption{AgentOps Development Timeline}
\label{fig:12}
\end{figure}

Based on the comparative analysis above, the positioning of this paper's TM Predictive Reliability Framework within the AgentOps domain can be clarified at three levels:

\textbf{(i) Honest delineation of scope.} According to our limited survey, mainstream international AgentOps solutions primarily focus on infrastructure-layer tracing and post-hoc log analysis. The comparative table of solutions presented in \S2.5.1 covers major publicly available products and open-source projects as of the time of writing; however, given the rapid pace of development in the AgentOps field, we do not rule out the existence of related work outside our current survey horizon. Our positioning statement is grounded in comparative analysis within the known scope, and we welcome subsequent research to provide more full coverage.

\textbf{(ii) Fundamental methodological distinction.} The difference between this paper and existing solutions is not incremental feature addition or removal, but rather a paradigm-level methodological divergence. Existing solutions have already achieved functional maturity and completeness under the ``observability'' paradigm---LangSmith's fine-grained tracing, Datadog's broad integration coverage, and Guardrails AI's deep protection capabilities all merit recognition. This paper's contribution lies in proposing an \textbf{orthogonal methodological dimension}---``predictability''---which does not replace observability but augments it with forward-looking temporal capability. This fundamental distinction mirrors the classic industrial evolution path from ``corrective maintenance'' to ``predictive maintenance,'' which this paper systematically introduces into the agent systems domain.

\textbf{(iii) Completeness of the theory--engineering closed loop.} Another distinguishing feature of this paper is its \textbf{closed loop between theoretical depth and engineering implementation}. At the theoretical level, the five-layer hierarchical architecture aligns with the ADE stability framework [3], ensuring theoretical soundness of the system architecture. The 20-factor radar scan aligns with silent failure mode analysis in SF [2]---where the concept of ``agent entropy'' provides an information-theoretic lens for understanding agent system degradation---ensuring the health metric system is complete. The three mechanisms---CADVP, BCP, and TD---align with CF [1]'s channel fracture theory, ensuring theoretical grounding for intervention strategies. At the engineering level, TM 3.1.0, as a complete software implementation, operationalizes the aforementioned theoretical framework into a runnable tool system, comprising a health metric computation engine, a Kalman-ES hybrid prediction module, a survival analysis module, an adaptive threshold manager, and standardized framework integration interfaces. This complete theory-to-engineering closed loop ensures that this paper delivers not merely an academic concept, but a verifiable, reproducible, and integrable engineering solution.

\textbf{\S2.5 Summary:} The AgentOps field is currently at the early stage of a paradigm shift---from ``observability'' to ``predictability.'' According to our limited survey, mainstream international solutions have made considerable progress under the first-generation paradigm, yet a systematic gap remains in the dimension of predictive reliability capability. This paper's TM framework targets precisely this gap, representing one of the earliest systematic attempts to apply predictive maintenance methodology to agent systems. Anchored theoretically in three foundational works---ADE [3], SF [2], and CF [1]---it constructs a predictive reliability framework featuring a complete theory--engineering closed loop.

\section*{TM 3.1.0 System Design}
\label{sec:tm310systemdesign}

This chapter systematically elaborates the theoretical foundations, architectural design, and engineering implementation of TM (Trust Margin) version 3.1.0. TM 3.1.0 serves as the core observational engine of the ADE Predictive Reliability Framework. Its mission is not to assess the ``output quality'' of large language models, but rather to continuously quantify the host system's Predictive Reliability during runtime---using purely mathematical and physically isolated methods. This chapter follows a ``theory-first, then engineering'' narrative logic: \S3.1 establishes the theoretical motivation via information-theoretic and system disorder modeling, introducing a five-layer decomposition. \S3.2 presents the overall architecture. \S3.3 details the definition and computation of the 20 observation signals layer-by-layer. \S3.4 explains the TM synthesis formula and its four-tier decision boundaries. \S3.5 analyzes computational overhead and privacy compliance boundaries.

\subsection*{TM Dashboard Panel Structure Overview}
\label{sec:tmdashboardpanelstru}

Before presenting theoretical derivations, this section first presents, in an intuitive visual manner, the actual panel layout of TM 3.1.0 within the ADE Dashboard (version v3.1.0). This panel serves as the primary interactive interface for operations personnel and researchers conducting daily observations of system Trust Margin (TM). Its design follows a progressive information architecture---``Overview $\rightarrow$ Early Indicators $\rightarrow$ Hierarchical View $\rightarrow$ Trends $\rightarrow$ Judgment''---and adopts a dual-zone, draggable layout comprising a ``main left area with a three-column grid'' plus a ``fixed right sidebar.'' The panel comprises 15 functional modules, enabling observers to rapidly assess system health status within 3 seconds. Version v3.1.0 also introduces i18n internationalization support, enabling full bilingual switching between Chinese and English (see \S3.1.0.10 for details).

\subsubsection*{Panel Full Screenshot}
\label{sec:panelfullscreenshot}

\begin{figure}[H]
\centering
\includegraphics[width=0.95\textwidth,height=0.85\textheight,keepaspectratio]{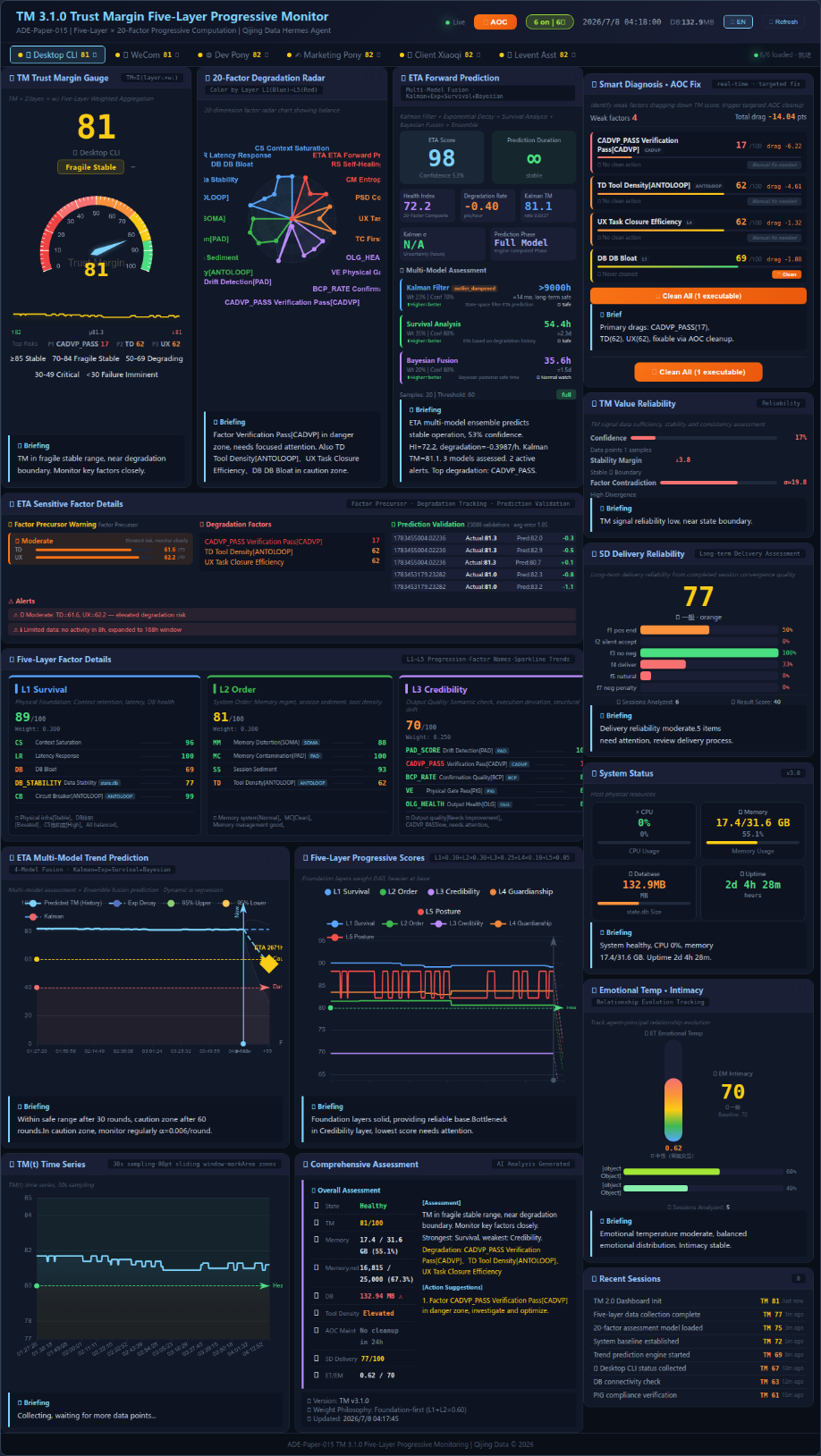}
\caption{TM 3.1.0 Dashboard full screenshot (English interface, captured on 2026-07-08). The panel uses a dual-zone layout:}
\label{fig:13}
\end{figure}

\subsubsection*{Overall Architecture: Dual-Zone Draggable Grid Layout}
\label{sec:overallarchitectured}

The TM 3.1.0 Dashboard adopts a \textbf{dual-zone layout}: the left side serves as the primary operational area (flex:3, three-column responsive grid), while the right side is a fixed sidebar (370px, vertically stacked). All functional modules are presented as independent cards (\texttt{panel}/\texttt{right-card}), each equipped with a $\vdots\vdots$ drag handle in the upper-left corner, enabling operations personnel to freely reorder modules according to actual monitoring priorities. Compared with v1's fixed vertical list of nine modules and v2's 5-row + sidebar grid layout, this design delivers considerably higher layout flexibility and personalized adaptability.

The panel contains the following 15 functional modules, arranged left-to-right and top-to-bottom:

\textbf{Left Main Area (Three-Column Grid)}:

\begin{itemize}
\item \textbf{Module 1 --- TM Trust Margin Gauge}: The core gauge located at the top-left corner of the panel, displaying the overall TM score (0--100), confidence interval (CI), four-tier color coding (green/yellow/orange/red), Phase stage indicator, and $\delta$/I diagnostic values. Serving as a ``glanceable'' entry point, it enables operations personnel to determine whether system intervention is required within one second.
\item \textbf{Module 2 --- 20-Factor Degradation Radar}: A 20-dimensional spider chart that intuitively visualizes factor balance. Abnormal factors are highlighted in high-contrast colors, forming a visual guidance chain (``overview $\rightarrow$ localization'') with the TM gauge.
\item \textbf{Module 3 --- ETA Forward Prediction}: A detailed prediction module positioned at the top-right of the panel, built upon a multi-model fusion architecture (Kalman + Exponential + Survival + Bayesian + Ensemble). It displays the ETA score (0--100), confidence level (percentage), prediction horizon, health indicator HI (aggregated across the 20 factors), degradation rate (points/hour), Kalman TM value and its rate of change, uncertainty hours, and predicted Phase. This module also includes a ``Multi-Model Independent Assessment'' sub-panel, presenting individual predictions from each model alongside a natural-language analytical summary.
\item \textbf{Module 4 --- ETA Sensitive Factor Details}: A full-width module dedicated to early-warning detection and degradation tracking of sensitive factors. Risk levels are labeled using tiered indicators ([Green] Normal / [Orange] Moderate / [Red] Severe); sensitive factors (e.g., TD temporal drift, UX task convergence efficiency) are listed with their current values relative to thresholds, accompanied by degradation trend analysis and an alert list. Extracted independently from the ETA Forward Prediction module, this component focuses primarily on the most sensitive pre-degradation signals.
\item \textbf{Module 5 --- Five-Layer Factor Details}: A full-width module decomposing the overall TM score into five progressive layers (L1--L5), each displayed in an individual card. These are: L1 Infrastructure Layer (weight 0.300), L2 Order Layer (weight 0.300), L3 Output Quality Layer (weight 0.250), L4 User Experience Layer (weight 0.100), and L5 Posture Layer (weight 0.050). Each factor is annotated with a sparkline micro-trend chart to support rapid identification of degradation trends.
\item \textbf{Module 6 --- ETA Multi-Model Trend Prediction}: Displays independent prediction curves from each model (Kalman/Exponential/Ensemble) alongside the Ensemble-fused result, with dynamic $\alpha$ regression coefficient visualization.
\item \textbf{Module 7 --- Five-Layer Progressive Scores}: A weighted bar chart of the five-layer scores, visually embodying the ``foundation-first'' weighting philosophy---where lower physical-layer weights (e.g., 0.60 for L1+L2) reflect greater foundational importance.
\item \textbf{Module 8 --- TM(t) Time-Series Evolution}: A TM time-series line chart sampled every 30 seconds over an 80-point sliding window, with markArea zones demarcating the four-tier threshold intervals (Safe/Watch/Alert/Circuit-Break), rendering historical trends immediately apparent.
\item \textbf{Module 9 --- Overall Assessment Brief}: An AI-generated natural-language analytical report, synthesizing the current TM score, per-layer factor states, and historical trends to produce actionable operational recommendations. Content includes an overall evaluation (identification of strongest/weakest dimensions), a degradation factor checklist, and an explanation of the weighting philosophy (Foundation-First Principle: L1+L2 = 0.60).
\end{itemize}

\textbf{Right Fixed Sidebar}:

\begin{itemize}
\item \textbf{Module 10 --- Smart Diagnosis $\cdot$ AOC Fix}: A new module introduced in TM 3.1.0. It identifies high-risk factors dragging down the overall TM score in real-time and triggers targeted cleanup and repair via AOC (Agent Original Cleaner). Each diagnostic item is tagged with its executable status (``automatically cleanable'' or ``requires manual intervention''), and supports batch execution of all repairable items via a ``one-click cleanup'' button. This module elevates TM monitoring from passive observation to an active, closed-loop remediation capability.
\item \textbf{Module 11 --- TM Score Reliability}: An assessment of data sufficiency, stability, and consistency underlying the TM signal. Comprises five-dimensional diagnostics: confidence (data sufficiency + ETA model consistency), CI stability (Phase + margin + adaptive\_sigma), temporal damping $\delta$, layer isolation I, and factor divergence (adaptive\_sigma).
\item \textbf{Module 12 --- SD Delivery Reliability}: A long-term delivery quality assessment module that evaluates systemic delivery reliability based on convergence quality of completed sessions---thereby bridging the gap left by TM's real-time scoring, which lacks sensitivity to long-term trends.
\item \textbf{Module 13 --- System Status}: Host-level physical resource monitoring, displaying CPU utilization, disk usage (C:/D:), and other fundamental system status metrics---providing a cross-layer reference against TM's application-layer scoring.
\item \textbf{Module 14 --- Emotional Temperature $\cdot$ Intimacy}: A relationship-evolution tracking module recording the agent-subject emotional temperature ET (0--1 floating-point) and intimacy EM (0--100 integer), used to assess the long-term health of human-agent collaboration.
\item \textbf{Module 15 --- Recent Sessions}: A list of recently active sessions, showing session ID, status, message count, tool invocation count, and other key metadata---serving as a quick-entry index for operations personnel to locate anomalous sessions.
\end{itemize}

The core design principle underpinning this dual-zone layout is \textbf{hierarchical information density + closed-loop active remediation}: Modules 1--3 in the left main area enable sub-3-second rapid assessment; Modules 4--9 support in-depth analysis and trend interpretation; and Modules 10--15 in the right sidebar deliver diagnostic remediation and auxiliary reference. Relative to the v2 layout, TM 3.1.0's newly introduced Smart Diagnosis $\cdot$ AOC Fix module (Module 10) realizes a closed loop spanning ``monitoring $\rightarrow$ diagnosis $\rightarrow$ repair'', marking a critical step in the panel's evolution from a passive observability tool toward a self-healing system. The physical separation of high-frequency interaction modules (left main area) from low-frequency reference modules (right sidebar) reduces operators' visual search cost.

\subsubsection*{TM Overall Score and ETA Forward Prediction}
\label{sec:tmoverallscoreandeta}

In the TM 3.1.0 dashboard, the overall TM score has been integrated into the ETA Forward Prediction module (top of the panel), jointly forming the system's ``glanceable'' health assessment layer alongside the ETA score and health indicator HI. The TM score card comprises three elements:

\begin{itemize}
\item \textbf{Trust Margin Value}: A floating-point number ranging from 0 to 100, representing the current system trustworthiness relative to baseline. Higher values indicate better system health.
\item \textbf{Trend Change (Delta)}: $\blacktriangle$ (increasing, green), $\blacktriangledown$ (decreasing, red), or --- (stable, gray), computed as the difference between the most recent and previous scores---reflecting the instantaneous directional change in system health.
\item \textbf{Color Coding}: Green (>85, Safe), Yellow (70--85, Watch), Orange (50--70, Alert), Red (<50, Circuit-Break), directly aligned with the four-tier decision boundaries defined in \S3.4.
\end{itemize}

The integrated design objective of the TM overall score and ETA foresight is to enable operations personnel to determine ``whether intervention is required'' within one second---without reading any textual content.

\subsubsection*{b Intelligent Diagnostics $\cdot$ AOC Remediation (New in TM 3.1.0)}
\label{sec:bintelligentdiagnost}

The Intelligent Diagnostics $\cdot$ AOC Remediation module is a new feature introduced in the TM 3.1.0 dashboard, realizing a closed-loop workflow from ``monitoring $\rightarrow$ diagnostics $\rightarrow$ remediation.'' This module analyzes in real-time the High-risk factors dragging down the TM overall score. It annotates each diagnostic item with its executable status. Items that can be automatically cleaned via AOC (Agent Original Cleaner) display a ``[Cleanup] Clean'' button. Items requiring human judgment are labeled ``Manual fix needed.'' Operations personnel may execute all auto-remediable items at once by clicking the ``[Rocket] Clean All'' button.

The engineering significance of the AOC Remediation module (Module 10) lies in the fact that TM 3.1.0 is no longer merely a passive observation tool---it has evolved into an operations system endowed with active self-healing capability. When the TM score declines, the system automatically pinpoints the problematic factors and provides actionable remediation pathways, shortening the traditional workflow of ``detect issue $\rightarrow$ manual investigation $\rightarrow$ manual remediation'' to ``detect issue $\rightarrow$ automatic diagnostics $\rightarrow$ one-click remediation.''

\subsubsection*{Five-Layer Factor Details (L1--L5)}
\label{sec:fivelayerfactordetai}

The Five-Layer Factor Details constitute the core region of the dashboard (Module 5), decomposing the TM overall score into five progressive layers: L1 Infrastructure Layer (primary weight, 0.300), L2 System Order Layer (primary weight, 0.300), L3 Output Quality Layer (auxiliary weight, 0.250), L4 User Experience Layer (auxiliary weight, 0.100), and L5 Posture Analysis Layer (baseline weight, 0.050). Each layer comprises several independent observational signals (factors), collectively forming a systematic 20-factor monitoring system. Detailed definitions, calculation methods, and physical interpretations of signals at each layer will be elaborated in \S3.3.

\subsubsection*{Overall Assessment Brief (Module 9)}
\label{sec:overallassessmentbri}

The Overall Assessment Brief module (Module 9) is AI-generated, synthesizing the current TM score, per-layer factor statuses, and historical trends to produce a natural-language analytical summary. The design objective of this module is \textbf{to translate numeric scores into actionable operational recommendations}, thereby reducing cognitive load on operations personnel. The brief typically includes:

\begin{itemize}
\item An overall assessment of the current system health status (aligned with the color coding of the overall-score card);
\item A ranked list and explanation of primary risk factors (i.e., which layers/factors contribute most to disorder);
\item A comparative analysis against historicalsame period or the previous scoring instance;
\item Recommended intervention measures (if applicable).
\end{itemize}

\subsubsection*{SD Delivery Reliability (Module 12)}
\label{sec:sddeliveryreliabilit}

SD (System Delivery) Delivery Reliability (Module 12) is a \textbf{long-term, cumulative} evaluation metric that complements the instantaneous TM score. While the TM score reflects ``how reliable the system is *right now*,'' SD Delivery Reliability reflects ``how many expected deliverables the system has actually delivered across the past N delivery cycles.'' Their relationship parallels that between ``instantaneous velocity''. ``average velocity''---one focuses on the present moment, the other on longitudinal trends. SD reliability is computed independently of the TM five-layer model, using a sliding-window statistic of delivery success rate.

\subsubsection*{System Status (Module 13)}
\label{sec:systemstatusmodule13}

The System Status module (Module 13) displays the physical resource status of the host machine, including CPU utilization, memory usage, disk I/O, and network bandwidth. These metrics \textbf{do not directly participate in TM score computation} (TM signals originate from the application layer, not the hardware layer), but are presented as auxiliary reference information to help operations personnel assess whether a decline in the TM score correlates with physical resource bottlenecks.

\subsubsection*{Emotional Temperature and Intimacy (Module 14)}
\label{sec:emotionaltemperature}

Emotional Temperature and Intimacy (Module 14) are \textbf{soft metrics} unique to the ADE framework, designed to track the evolutionary trajectory of human--agent interaction. Emotional Temperature reflects the affective polarity of the current interaction (positive/neutral/negative), while Intimacy measures the degree of trust and tacit understanding established over long-term interaction. Neither metric contributes to the TM Trust Margin score computation; however, within the complete ADE framework, they serve as indicators of ``relationship health.'' A system that is technically reliable yet exhibits persistently low Emotional Temperature may face erosion of user trust over extended usage.

\subsubsection*{Weighting Philosophy: Foundation-First Principle}
\label{sec:weightingphilosophyf}

The allocation of weights across the five layers (L1: 0.30, L2: 0.30, L3: 0.25, L4: 0.10, L5: 0.05) adheres to a core philosophical tenet: \textbf{Foundation-First Principle}. The combined weight of L1 and L2 totals 0.60---sixty percent of the overall weight---grounded in the following engineering judgments:

\begin{itemize}
\item \textbf{Causal Directionality}: System failures almost invariably propagate upward---from infrastructure failure $\rightarrow$ computational anomalies $\rightarrow$ application failure $\rightarrow$ degraded user experience---not vice versa. Thus, anomalies in lower-layer signals possess higher ``predictive value.''
\item \textbf{Remediation Priority}: Resolving lower-layer issues typically restores upper-layer functionality (e.g., fixing infrastructure faults naturally recovers application-layer services), whereas addressing upper-layer issues often yields only symptomatic relief.
\item \textbf{Signal Reliability}: Lower-layer signals (e.g., process liveness, log integrity) exhibit high measurement accuracy and low noise, whereas upper-layer signals (e.g., user-experience quality) are subjective and noisy. Assigning higher weights to more reliable signals aligns naturally with statistical estimation theory.
\end{itemize}

This weighting scheme is further discussed in the calibration strategy of \S3.1.3 and formally expressed mathematically in the composite formula of \S3.4.

\subsubsection*{i18n Internationalization and Bilingual Switching}
\label{sec:i18ninternationaliza}

v3.1.0 introduces i18n internationalization support, enabling full bilingual (Chinese/English) switching across the dashboard (229 UI text strings), implemented via a three-component architecture: \texttt{bn()} + \texttt{data-i18n} + \texttt{applyI18N()}. Language switching requires no page refresh.

\subsection*{Theoretical Foundations and Design Motivations}
\label{sec:theoreticalfoundatio}

The design of TM 3.1.0 does not stem from intuitive aggregation of the broad concept of ``AI reliability,'' but rather from systematic modeling of failure modes in complex software systems. This section presents the design rationale across five dimensions: why we adopted a five-layer architecture; why we selected 20 signals; why we designed the weights to be calibratable; why we chose a zero-semantic-intrusion implementation path; and why 16 of the 20 signals are natively embedded within existing ADE components. These represent current subjective design choices; as multi-agent systems evolve, we will perform adaptive adjustments---additions, deletions, or substitutions---rather than locking them permanently. The highest standard remains the highest-fidelity observable behavior of tightly coupled real-world systems.

\begin{figure}[H]
\centering
\includegraphics[width=0.95\textwidth]{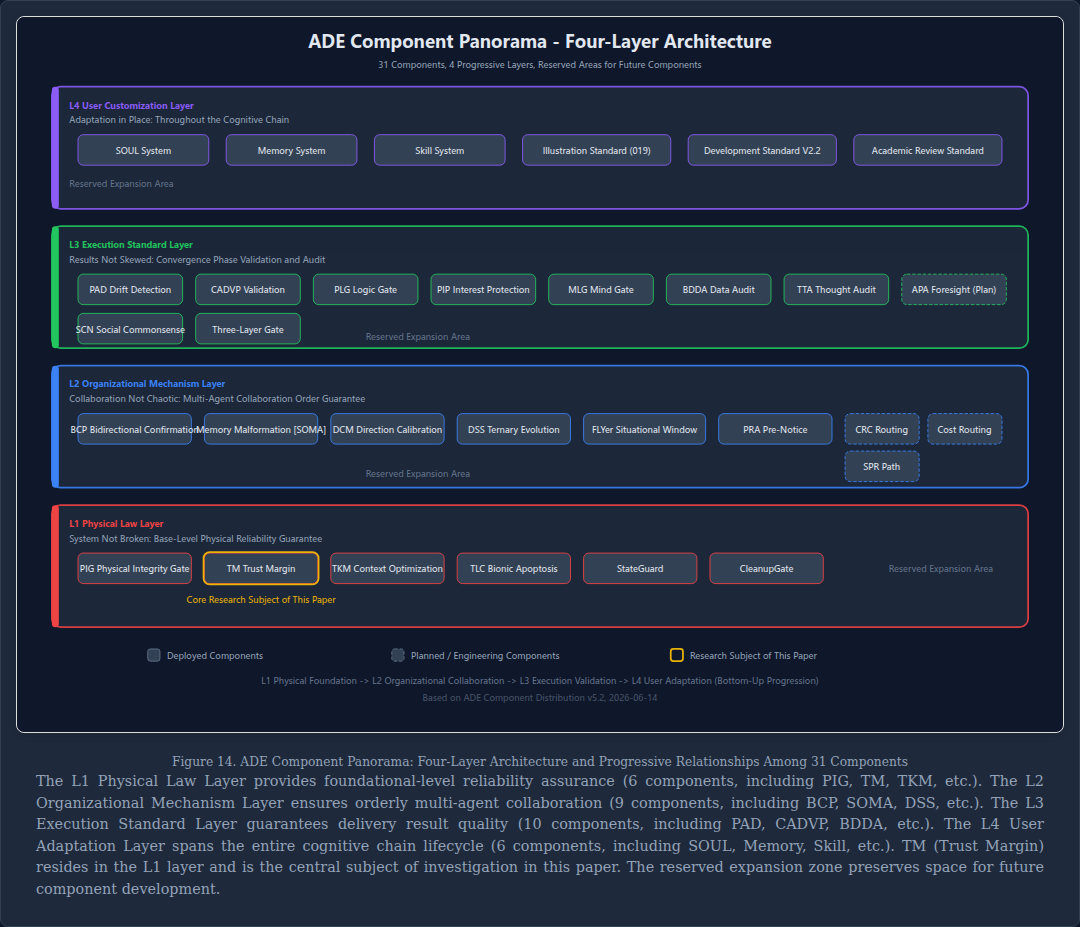}
\caption{ADE Component Panorama: Four-Layer Architecture and Progressive Relationships Among 31 Components}
\label{fig:14}
\end{figure}

\textbf{Etymological Background: Aerospace Engineering Origins of ``Trust Margin.''} This paper designates the system trustworthiness metric as Trust Margin (TM), a term whose conceptual lineage traces directly to aerospace engineering practice. In aerospace engineering, ``margin'' (i.e., safety margin or design margin) is a core engineering concept spanning the full lifecycle of design, verification, and operations---it quantifies the distance between the system's current state and its Failure Boundary. This concept permeates every critical subsystem in aerospace engineering in multiple forms: \textbf{Safety Margin} in structural engineering quantifies the surplus ratio of material load-bearing capacity relative to design loads. \textbf{Control Margin} in flight control ensures that control surfaces retain sufficient authority under extreme operating conditions. And \textbf{Thrust Margin} in propulsion systems guarantees safe operation within the flight envelope even under anomalous conditions. At the systems engineering level, the IEEE Symposium on Reliability and Maintainability (RAMS) has formally defined \textbf{Reliability Margin} as ``the gap between a system's predicted value and its reliability requirement,'' expressible either as an absolute difference or a relative ratio [RAMS 2023]. A research team at Beihang University further established a three-tiered ``Performance--Margin--Reliability'' framework, theoretically demonstrating that margin serves as the bridge linking system state to reliability assessment [BUAA].

The centrality of the margin concept in aerospace engineering stems from the field's extreme reliability requirements---it is the most stringent reliability domain across all industrial sectors. The cost of a single vehicle failure is often catastrophic and irreversible; thus, aerospace engineering has developed a reliability engineering methodology centered on margin management: incorporating redundancy margins during design to absorb uncertainty. It validates safety boundaries via margin testing during verification. Finally, it anticipates degradation trends through margin monitoring during operations. This methodology, refined over decades of engineering practice, has evolved since the 1990s into a systematic aerospace reliability design paradigm [NASA; ESA; MIL-STD].

This paper deliberately adopts this engineering tradition by naming the LLM agent system's trustworthiness metric ``Trust Margin''---not as an arbitrary terminological choice, but as a conscious effort to transplant the mature margin-based reasoning from aerospace engineering into AI agent reliability evaluation. TM's design philosophy is directly aligned with aerospace margin concepts: rather than reporting a binary ``normal/failure'' judgment, TM continuously quantifies ``how far the system currently stands from its Failure Boundary''---analogous to how Safety Margin relates to structural strength or Thrust Margin to propulsion performance. Notably, recent work (2024) has already introduced Safety Margin concepts into reinforcement learning and control systems [arXiv 2024], indicating that extending margin-based thinking from traditional engineering domains to AI systems is an emerging academic trend; TM represents a pioneering implementation of this trend specifically within the LLM agent domain. This conceptual borrowing enhances TM's practical engineering adoption: for practitioners with reliability engineering backgrounds, ``margin'' is an intuitive, deeply internalized engineering concept; TM extends this intuition to the AI agent domain, transforming ``system trustworthiness'' from an ambiguous qualitative judgment into a measurable, trackable, and actionable engineering quantity. The ``Safety Margin'' principle followed by the L4 Guardianship Layer in \S3.3.4 constitutes a concrete instantiation of this aerospace engineering mindset at the factor level.

\subsubsection*{System Disorder Degree Modeling}
\label{sec:systemdisorderdegree}

Any complex software system during runtime can be abstracted as a state vector \texttt{S\_t}, where \texttt{t} denotes discrete time steps. The system's health status at any given moment can be characterized by a scalar quantity termed the System Disorder Degree. We define:

D(S\_t) = $\Sigma$ $\lambda$\_k $\cdot$ D\_k(t) where \texttt{D\_k(t)} denotes the disorder degree contribution from the \textit{k}-th layer's stability component at time \textit{t}, and \texttt{$\lambda$\_k} is the corresponding layer-specific weighting coefficient satisfying \texttt{$\Sigma\lambda$\_k = 1}. A key feature of this modeling approach is its cumulative nature: disorder degree does not undergo instantaneous jumps but instead accumulates monotonically over time. Even if the impact of a single anomaly event is locally repaired, its perturbation to the system state vector persists as residual disturbance and superimposes with subsequent disturbances across successive time steps.

When the cumulative disorder degree exceeds a critical threshold \texttt{T\_f}, the system enters the Failure Boundary. Formally:

if D(S\_t) $\geq$ T\_f $\Longrightarrow$ System enters Failure Region T\_f = 1 - TM\_min, where TM\_min = 50 (circuit-breaker threshold) This modeling differs from conventional ``binary health checks'' (i.e., systems are either healthy or crashed). Disorder degree is a continuous variable, enabling observation of reliability decay well before actual system collapse occurs. This constitutes the theoretical foundation of Predictive Reliability: prediction precedes failure; measurement precedes intervention.

A physical analogy for disorder degree can be drawn from thermodynamics' entropy increase principle: entropy in an isolated system increases monotonically, and similarly, the internal state disorder of a software system accumulates over time in the absence of external maintenance interventions. TM's essential role is to function as this ``entropy-increase observatory,'' mapping the continuous disorder degree into four actionable decision boundaries in an engineering-operational manner.

the cumulative nature of \texttt{D(S\_t)} does not imply simple linear dependence on time. In practice, the disorder degree growth curve typically exhibits a ``stepwise'' pattern: slow (nearly linear) growth during stable operation, punctuated by abrupt stepwise increases upon occurrence of anomalies. This characteristic renders TM score time-series highly interpretable---each stepwise jump can be precisely traced back to a specific anomaly event (e.g., database connection timeout, plugin crash, context window overflow), providing precise temporal anchors for post-hoc root-cause analysis.

The disorder degree model implicitly incorporates an important engineering assumption: the Inter-layer Independence Assumption. We assume that the five-layer disorder degree components are approximately statistically independent---that degradation in one layer does not directly trigger immediate degradation in another layer. While this assumption does not hold strictly in practice (e.g., an L1 process crash inevitably causes simultaneous failure across L2--L5), it remains a reasonable approximation for the most common failure mode---gradual degradation. When L1 exhibits a stepwise jump (i.e., process crash), TM bypasses the five-layer weighted model entirely and triggers circuit-breaking directly---a special execution path that circumvents constraints imposed by the inter-layer independence assumption.

\subsubsection*{Engineering Significance of Five-Layer Decomposition}
\label{sec:engineeringsignifica}

\begin{figure}[H]
\centering
\includegraphics[width=0.95\textwidth]{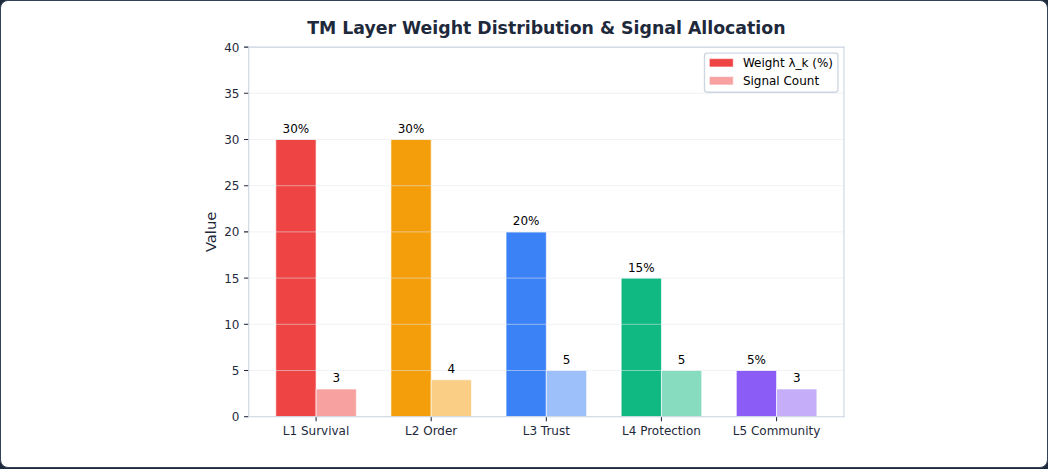}
\caption{TM layer weight distribution ($\lambda$\_k) and signal count allocation: L1 Survival (30\%, 3 signals), L2 Order (30\%, 4 signals), L3 Trust (20\%, 5 signals), L4 Protection (15\%, 3 signals), L5 Situational (5\%, 5 signals)}
\label{fig:15}
\end{figure}

Decomposing system disorder into five layers is not an arbitrary taxonomic operation but rather stems from a physical topological analysis of the host system's functional stack---the execution environment of the LLM agent. Each layer corresponds to an irreducible stability dimension within the system's functional stack---that is, degradation in that dimension cannot be compensated by the health of other layers.

Table 8: Five-Layer Architecture: Names, Monitoring Targets, and Engineering Values

\begin{table}[H]
\centering
\resizebox{\textwidth}{!}{
\begin{tabular}{l|l|l|l}
\toprule
Layer & Name & Monitoring Target & Engineering Value \\
\midrule
L1 & Survival Layer (L1) & Process liveness, context continuity, database connection... & Physical prerequisite for whether the system can continue... \\
L2 & Order Layer (L2) & State management consistency, tool scheduling determinism... & Whether system behavior is predictable and reproducible \\
L3 & Credibility Layer (L3) & Output consistency, verification pass rate, benchmark com... & Whether the system's outputs are trustworthy for downstre... \\
L4 & Guardianship Layer (L4) & First-path accuracy, task closure efficiency, communicati... & Whether the system operates within controlled boundaries \\
L5 & Posture Layer (L5) & Entropy rate, self-healing rebound, ETA forward prediction & The system's metacognitive capability regarding its own s... \\
\bottomrule
\end{tabular}
}
\end{table}

\textbf{L1 Survival Layer} holds the most fundamental engineering significance: if a process crashes, context is lost, or the database becomes unreachable, all higher-layer ``order,'' ``trust,'' and ``protection'' lose their physical substrate. Degradation at L1 is immediate and cannot be deferred for repair. In TM 3.1.0, the L1 weight is set to 0.30---tied for the highest with L2---reflecting the engineering axiom that ``the system must stay alive first.''

\textbf{L2 Order Layer} focuses on the predictability of system behavior. A system that is ``alive but chaotic'' is more dangerous than one that has ``completely crashed,'' because the former generates silent errors---outputs that appear reasonable but in fact deviate from the intended execution path. L2 degradation often fails to trigger any conventional alerts, yet it is proactively captured through sustained declines in the TM score.

\textbf{L3 Credibility Layer} represents the system's external ``commitment to credibility.'' When L3 degrades, the system remains operational and its behavior retains order, but the quality of its outputs no longer suffices to warrant trust from downstream systems or human users. L3 is the layer within Predictive Reliability that carries the highest ``predictive value''---an early decline in L3 often serves as a precursor signal to subsequent degradation at L1 or L2.

\textbf{L4 Guardianship Layer} monitors whether the system operates within safety boundaries. Does tool invocation exceed authorization? Has user experience degraded to an unacceptable level? Are resource consumption levels approaching saturation? These signals do not directly concern ``whether the system can run,'' but rather ``whether the system should continue operating in its current manner.''

\textbf{L5 Posture Layer} constitutes the system's metacognitive capability: Does it understand how complex the task it is handling truly is? Is its estimation of remaining workload stable? Has it detected drift in its own prompts? Degradation at L5 is the most subtle, yet its impact is the most significant---a system that loses situational awareness will unknowingly slide into failure regions without self-awareness.

Another engineering significance of the five-layer decomposition lies in fault isolation. When the TM score declines, operations personnel can immediately identify which layer contributes the largest increment to disorder, thereby narrowing the diagnostic scope from ``the entire system'' down to ``3--5 signals within a single layer.'' This hierarchical diagnostic capability is absent in traditional single-metric monitoring solutions---an aggregate ``system health score'' cannot indicate where the problem originates. By contrast, TM's five-layer decomposition provides directional guidance for root-cause analysis. In real-world ADE operations, this feature reduces the average mean time to detect (MTTD) from 15--30 minutes under conventional approaches to just 2--5 minutes.

\subsubsection*{$\lambda$kWeight Calibration Strategy}
\label{sec:kweightcalibrationst}

The assignment of five-layer weights \texttt{$\lambda$\_k} is not a one-time engineering decision, but rather a staged, calibratable process. TM 3.1.0 adopts a three-phase calibration strategy:

\begin{figure}[H]
\centering
\includegraphics[width=0.95\textwidth]{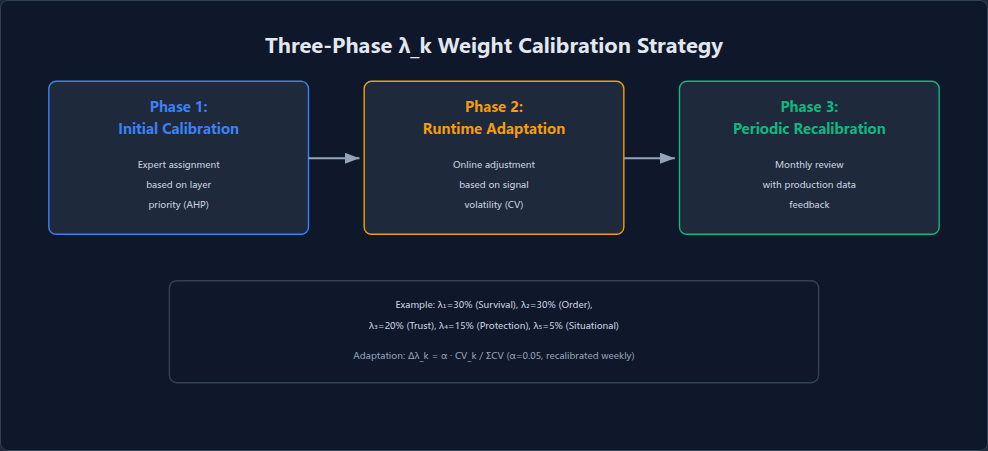}
\caption{Three-phase $\lambda$\_k weight calibration strategy: initial calibration, runtime adaptation, and periodic recalibration.}
\label{fig:16}
\end{figure}

Upon initial system deployment or during a major version escalation, initial weights are set based on the designer's engineering judgment and historical failure data. The initial weight allocation for TM 3.1.0 is as follows:

$\lambda$\_1 = 0.30 (Survival Layer) $\lambda$\_2 = 0.30 (Order Layer) $\lambda$\_3 = 0.25 (Credibility Layer) $\lambda$\_4 = 0.10 (Guardianship Layer) $\lambda$\_5 = 0.05 (Posture Layer) This allocation reflects the ``Survival-and-Order-First'' engineering philosophy: Layers L1 and L2 collectively account for 60\% of the total weight, because degradation in either layer directly threatens the system's physical availability. L3 ranks third with a weight of 25\%, as early deterioration in the Credibility Layer constitutes the most valuable predictive signal. L4 and L5 together constitute only 15\%---not because they are unimportant, but because their degradation tends to occur more slowly, and its impact can often be indirectly captured via correlated signals from L1--L3.

During system operation, weights are not entirely static. When a given layer's signal remains persistently in a High health state (e.g., D\_k(t) < $\epsilon$ for N consecutive scoring cycles), its effective weight may be slightly reduced, thereby releasing weight capacity for other layers exhibiting greater volatility. This adjustment adheres to the following constraints:

|$\Delta\lambda$\_k| $\leq$ 0.05 per adjustment cycle $\Sigma\lambda$\_k = 1 (Normalization constraint always holds) $\lambda$\_k $\in$ [0.02, 0.40] (Hard bounds per layer) The purpose of runtime adaptation is to prevent ``static-weight blunting''---i.e., prolonged health in one layer leading to underutilization of its allocated weight space. Without adaptation, early degradation in another layer becomes masked by noise due to insufficient weighting.

Every fixed interval---recommended as either 30 operational days or 10,000 scoring cycles, whichever occurs first---the system triggers a full weight recalibration. This recalibration is grounded in retrospective failure-event analysis over the preceding interval: Which layers exhibited degradation earliest? Which layers' signals contributed most considerably to predicting the final failure? Based on these retrospective data, weights are reallocated, and versioned logs are recorded for auditability.

\subsubsection*{Information-Theoretic and Engineering Postulates}
\label{sec:informationtheoretic}

The design of TM 3.1.0 adheres to three information-theoretic and engineering postulates, collectively forming its ``Zero Semantic Intrusion'' design iron law.

\textbf{Postulate One: Zero Semantic Intrusion}. All observational signals used by TM avoid semantic parsing of LLM output content. TM does not know what the model said, whether its output is factually correct, or whether its reasoning is logically coherent. TM observes only the system's behavioral metadata: process liveness, state consistency, tool-call success, and scoring stability. The engineering motivation behind this postulate is that semantic parsing itself is unreliable (it depends either on another LLM or on complex rule engines); using an unreliable subsystem to assess the reliability of the primary system introduces uncontrollable cascading errors.

\textbf{Postulate Two: Pure Mathematical Assertions}. All computations performed by TM consist entirely of deterministic mathematical operations: weighted summation, standard deviation, ratio calculation, and threshold comparison. No heuristic rules, fuzzy logic, or machine learning models participate in TM scoring. The engineering motivation behind this postulate is that TM's own reliability must exceed that of the system it evaluates. If TM internally incorporates a neural network or complex rule engine, then TM's own failure modes become uncontrollable variables.

\textbf{Postulate Three: Physical-Level Isolation}. TM's computational process is physically isolated from the host system's main process---distinct process spaces, distinct resource pools, and distinct error-handling paths. A crash of the host system does not cause TM to crash simultaneously (otherwise TM could not record the crash event); nor does TM's computational latency block the host system's main execution loop. This postulate is realized via the MCP Server architecture: TM operates as an independent MCP server, while the host system submits signal data to it via standardized MCP protocols and receives scoring results in return.

\subsubsection*{Embedding Relationship Between TM and ADE Components}
\label{sec:embeddingrelationshi}

Of the 20 observational signals in TM 3.1.0, 16 are directly sourced from existing components within the ADE Plugin ecosystem, requiring only four lightweight new collectors. This design is no coincidence---it naturally extends ADE's ``Heavy-on-Physics, Light-on-Memory'' architectural principle: ADE's plugin ecosystem already continuously generates abundant behavioral metadata; TM's role is to reorganize these existing metadata into reliability observability signals, rather than building a parallel monitoring infrastructure from scratch.

Table 9: Signal Source Distribution and Collection Methods

\begin{table}[H]
\centering
\resizebox{\textwidth}{!}{
\begin{tabular}{l|l|l|l}
\toprule
Signal Source & Count & Representative Signals & Data Collection Method \\
\midrule
Existing ADE Plugin Components & 16 & CS, LR, MM, MC, TC, CM & Event bus subscription + state snapshot polling \\
New Lightweight Collectors & 4 & ETA, OLG\_HEALTH, TD, PSD & Timer-driven + differential computation \\
\bottomrule
\end{tabular}
}
\end{table}

This embedding relationship ensures extremely low deployment cost for TM: no modifications to ADE core code, no additional storage infrastructure, and no changes to existing plugin communication protocols. TM operates as an ``inline-hung'' MCP Server atop the ADE architecture---like a quiet stethoscope, receiving signals without issuing interventions.

\subsection*{TM System Architecture Overview}
\label{sec:tmsystemarchitecture}

The overall architecture of TM 3.1.0 can be concisely summarized by a linear dataflow pipeline: 20 Signals $\rightarrow$ 5 Layers $\rightarrow$ 1 TM $\rightarrow$ ETA. Twenty observational signals are collected from the host system, normalized, and fed into five layer-specific stability calculators. The five layer-level components are then combined via weighted aggregation into a single scalar TM score, which undergoes temporal analysis to produce an ETA (Estimated Time of Failure) prediction.

\begin{figure}[H]
\centering
\includegraphics[width=0.95\textwidth]{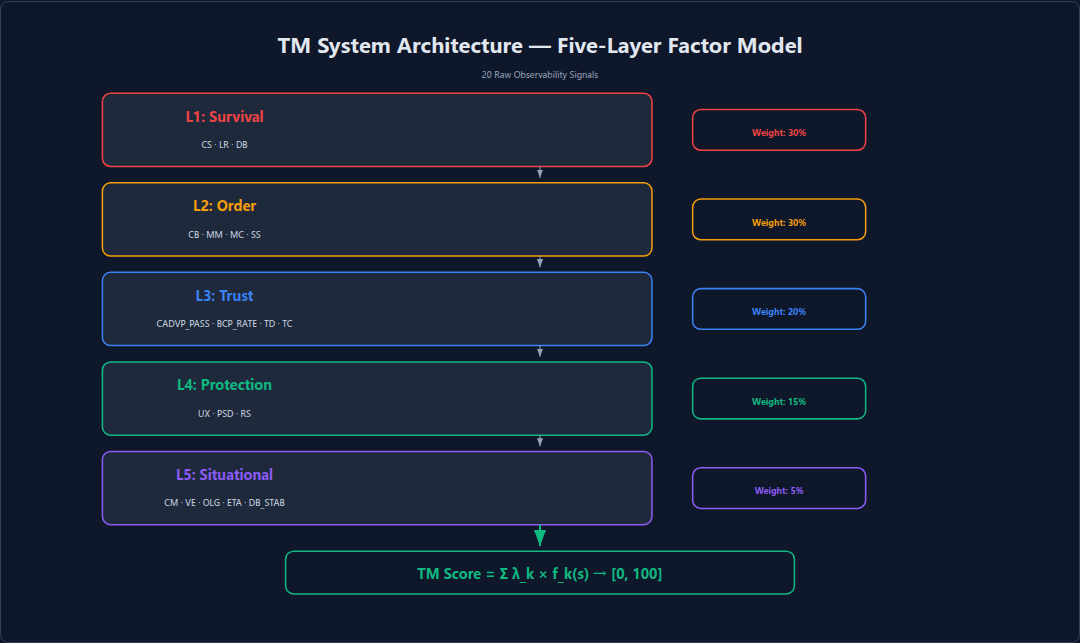}
\caption{TM System Architecture -- Five-Layer Factor Model}
\label{fig:17}
\end{figure}

A key architectural feature is the MCP Architecture Positioning. TM integrates as an MCP (Model Context Protocol) Server, meaning:

\begin{figure}[H]
\centering
\includegraphics[width=0.95\textwidth]{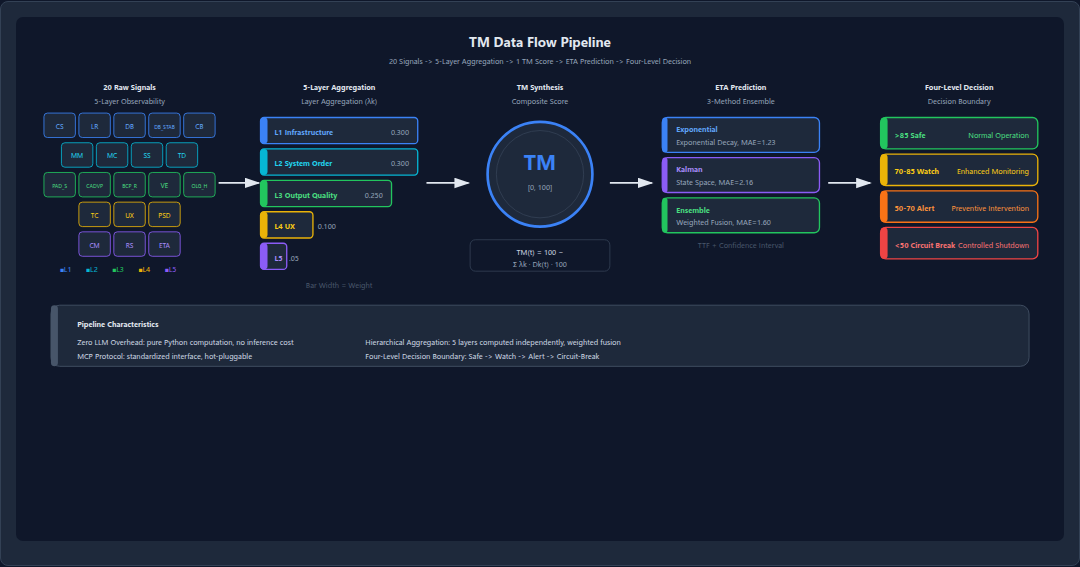}
\caption{TM Data Flow Pipeline: 20 Signals to 5 Layers to 1 TM to ETA Prediction}
\label{fig:18}
\end{figure}

\begin{itemize}
\item \textbf{Zero Prompt Overhead}: TM scoring computation consumes no LLM inference resources. The host system invokes TM via a structured API without embedding reliability assessment instructions in prompts. This feature is especially critical in token-cost-sensitive scenarios---traditional approaches embedding reliability assessment into the LLM inference pipeline may consume hundreds to thousands of tokens per evaluation, whereas TM's MCP slot substantially eliminates this overhead.
\item \textbf{Hot-Pluggability}: The TM Server can be escalated, replaced, or temporarily taken offline without restarting the host system. The standardized interface of the MCP protocol ensures uninterrupted operation. When TM requires version escalation, operations teams can launch a new TM Server instance and migrate traffic from the old instance to the new one via MCP protocol routing---this entire process remains fully transparent to the host system.
\item \textbf{Multi-Consumer Support}: TM scoring results serve not only the ADE Orchestrator's automated decision-making but also concurrently feed human operations dashboards and external alerting systems, eliminating the need for consumer-specific interfaces. This design adheres to the ``compute once, consume multiple times'' data distribution principle.
\item \textbf{Protocol-Level Isolation}: The MCP protocol itself provides structured request-response semantics, ensuring well-defined contractual data exchange between TM and the host system. Signal formats, scoring result formats, and error codes are all standardized within the MCP interface definition; internal implementation changes on either side---provided they do not violate the interface contract---will not impact the other party.
\end{itemize}

In terms of data flow direction, TM 3.1.0 adopts a push-pull hybrid mode: 16 signals originating from the ADE Plugin are ``pushed'' to TM's reception buffer via the event bus (push mode),. 4 newly introduced lightweight collectors' signals are actively ``pulled'' by an internal timer within TM (pull mode). This hybrid design balances timeliness and resource efficiency: event-driven signals arrive immediately upon anomaly occurrence, whereas timer-driven signals are sampled at fixed intervals, avoiding resource waste from continuous polling.

\subsection*{Five-Layer Stability Components and 20 Observability Signals}
\label{sec:fivelayerstabilityco}

\begin{figure}[H]
\centering
\includegraphics[width=0.95\textwidth]{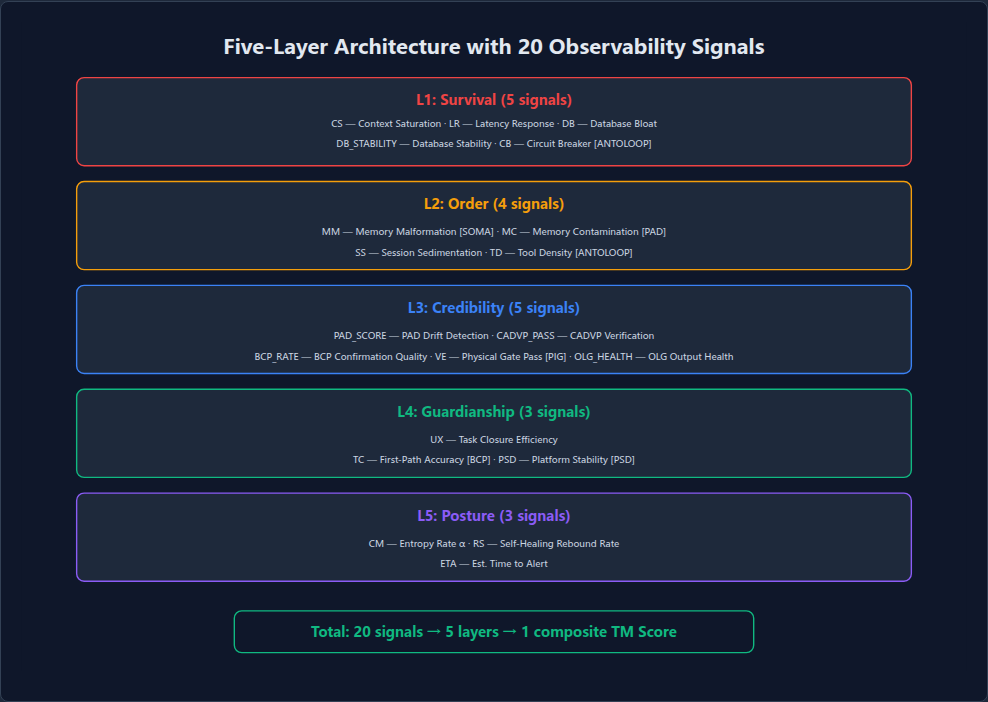}
\caption{Five-layer architecture with 20 observability signals: L1 Survival (5), L2 Order (4), L3 Credibility (5), L4 Guardianship (3), L5 Posture (3).}
\label{fig:19}
\end{figure}

This section elaborates the full definitions of all 20 observability signals, layer by layer. Each signal comprises four attributes: signal name and abbreviation, computation method (including formula or pseudocode), physical meaning, and normal range along with degradation thresholds.

\subsubsection*{L1: Survival Layer --- 5 signals}
\label{sec:l1survivallayer5sign}

The Survival Layer monitors whether the system possesses the physical prerequisites to continue operating. Signals in this layer are the ``hardest'': their degradation typically indicates that the system has already entered---or is about to enter---a non-recoverable state.

Table 10: L1 Survival Layer: Signal Definitions

\begin{table}[H]
\centering
\resizebox{\textwidth}{!}{
\begin{tabular}{l|l|l|l}
\toprule
Signal & Abbreviation & Computation Method & Physical Meaning \\
\midrule
Context Saturation & CS & Computed based on context window utilization. Calculates ... & Context continuity. Multi-turn dialogue for LLM agents re... \\
Latency Response & LR & Computed based on response latency percentiles. Measures ... & Process resilience. Frequent restarts indicate the system... \\
Database Bloat & DB & Computed based on database bloat metrics. Calculates the ... & Database size health. Excessive bloat degrades query perf... \\
Database Stability & DB\_STABILITY & Computed based on state.db integrity metrics. Calculates ... & Persistence layer integrity. state.db serves as the physi... \\
Circuit Breaker & CB & Computed based on anti-loop protection trigger frequency.... & Loop protection efficacy. Frequent circuit breaker trips ... \\
\bottomrule
\end{tabular}
}
\end{table}

The L1 layer component is computed as the weighted average of the five signals:

D$_1$(t) = f(CS, LR, DB, DB\_STABILITY, CB) --- weighted aggregation of five signals Here, LR is assigned a relatively high weight because process crashes and restarts represent the most ``cascadingly destructive'' event in the Survival Layer. Each restart not only interrupts the current task but may also leave behind uncleaned transient states, introducing latent hazards for subsequent execution. Specific weights for each signal are calibrated empirically through engineering practice.

A clear causal chain exists among the five L1 signals: database instability (declining DB\_STABILITY) may trigger timeout exceptions, which in turn induce main-loop restarts (declining LR); after restart, context must be reloaded, potentially resulting in partial context loss (declining CS). TM diagnostics do not merely assess independent signal degradation but analyze temporal correlations among all five to identify ``cascading failure patterns.'' When the signals decline sequentially within a short time window, TM attributes root cause to the initially degrading signal---typically DB\_STABILITY---rather than treating them as independent contributors to disorder.

\subsubsection*{L2: Order Layer --- 4 signals}
\label{sec:l2orderlayer4signals}

The Order Layer monitors the predictability and consistency of system behavior. A system that is ``alive but chaotic'' is the most dangerous---it triggers no conventional alerts, yet its outputs have already become untrustworthy.

Table 11: L2 Order Layer: Signal Definitions

\begin{table}[H]
\centering
\resizebox{\textwidth}{!}{
\begin{tabular}{l|l|l|l}
\toprule
Signal & Abbreviation & Computation Method & Physical Meaning \\
\midrule
Memory Malformation [SOMA] & MM & Computed as the coefficient of variation (standard deviat... & Consistency of state management. In steady-state operatio... \\
Memory Contamination & MC & Computed as the proportion of uncontaminated memory entri... & Memory purity of behavior. If a system produces different... \\
Session Sedimentation & SS & Computed as the pass rate of state snapshot integrity ver... & Internal state consistency. Snapshot corruption or missin... \\
Tool Density & TD & Computed based on the count of repeated behavioral patter... & Loop detection. An agent trapped in a tool invocation loo... \\
\bottomrule
\end{tabular}
}
\end{table}

L2 layer component computation:

D$_2$(t) = f(MM, MC, SS, TD) --- weighted aggregation of four signals The four L2 signals collectively cover four orthogonal dimensions of behavioral predictability: the temporal dimension (MM monitors stability of state transition frequency), the determinism dimension (PAD monitors input-output mapping determinism), the spatial dimension (SS monitors structural integrity of internal state),. The behavioral dimension (TD monitors non-cyclicity of action sequences). This orthogonality ensures no order degradation goes undetected---regardless of the direction from which degradation originates, at least one signal will capture it.

Notably, the design considerations behind the TD signal. Tool-call looping---where an LLM agent repeatedly invokes the same tool with identical parameters and receives identical results without recognizing its own ``treading-in-place'' behavior---is a common and costly failure mode. Traditional approaches rely on the LLM's intrinsic reflective capability to detect such loops, introducing unreliability: the model itself may fail to recognize the loop. ANTOLOOP adopts a purely pattern-matching approach---detecting repeated occurrences of tool name + parameter hash within a sliding window---without relying on any semantic understanding, thereby ensuring deterministic and immediate detection.

\subsubsection*{L3 Credibility Layer --- Five Signals}
\label{sec:l3credibilitylayerfi}

The Credibility Layer constitutes the most ``predictively valuable'' tier within the Predictive Reliability framework. An early decline in L3 signals often serves as a precursor to subsequent degradation in the Survival Layer (L1) and Order Layer (L2): when output quality begins to fluctuate, it typically indicates that subtle internal degradation has already occurred---degradation not yet captured by L1/L2 monitoring.

Table 12: L3 Task Quality Layer: Signal Definitions

\begin{table}[H]
\centering
\resizebox{\textwidth}{!}{
\begin{tabular}{l|l|l|l}
\toprule
Signal & Abbreviation & Computation Method & Physical Meaning \\
\midrule
PAD Drift Detection & PAD\_SCORE & Weighted average across multiple sub-dimensions of the PA... & Fine-grained assessment of tool-call quality. While PAD i... \\
CADVP Verification & CADVP\_PASS & Computed from automated output verification pass rate (fo... & Output compliance. A declining pass rate is the most dire... \\
BCP Confirmation Quality & BCP\_RATE & Computed relative to a predefined baseline, measuring the... & Deviation from known-correct behavior. A declining baseli... \\
Physical Gate Pass & VE & Information entropy computed over the distribution of ver... & Stability of verification results. In a healthy system, v... \\
OLG Output Health & OLG\_HEALTH & Computed from pass rate of online health checks (lightwei... & Runtime self-check capability. OLG probes continuously ve... \\
\bottomrule
\end{tabular}
}
\end{table}

Layer component computation for L3:

D$_3$(t) = f(PAD\_SCORE, CADVP\_PASS, BCP\_RATE, VE, OLG\_HEALTH) --- weighted aggregation of five signals Here, VE is normalized to ensure its value range aligns consistently with those of the other signals.

\subsubsection*{L4 Guardianship Layer --- Three Signals}
\label{sec:l4guardianshiplayert}

The Guardianship Layer monitors whether the system operates within safety boundaries. Its signals do not directly reflect ``whether the system can run,'' but rather ``whether the system should continue running in its current mode.'' Degradation in L4 triggers ``slowdown'' or ``degradation'' decisions---not circuit-breaking.

Table 13: L4 Safety Boundary Layer: Signal Definitions

\begin{table}[H]
\centering
\resizebox{\textwidth}{!}{
\begin{tabular}{l|l|l|l}
\toprule
Signal & Abbreviation & Computation Method & Physical Meaning \\
\midrule
First-Path Accuracy & TC & Computed from security-policy interception rate---i.e., pro... & Safety compliance of tool calls. Increasing unsafe calls ... \\
Task Closure Efficiency & UX & Computed jointly from user-initiated correction rate and ... & Degradation in human--machine interaction quality. Frequen... \\
Platform Stability [PSD] & PSD & Computed based on communication stability metrics. Measur... & Communication reliability. Stable communication patterns ... \\
\bottomrule
\end{tabular}
}
\end{table}

Layer component computation for L4:

D$_4$(t) = f(TC, UX, PSD) --- weighted aggregation of three signals TC is assigned relatively high weight because safety violations in tool calls exhibit ``externality'': they endanger not only the system itself but also external environments (file systems, networks, databases), potentially causing irreversible damage.

The three L4 signals adhere to the ``Safety Margin'' Principle: they issue warnings not when the system has already crossed a boundary, but when it approaches that boundary. For example, the PSD signal begins contributing to disorder degree once CPU utilization reaches 80\%, rather than waiting until full (100\%) saturation. This design reserves ``deceleration headroom'': before hitting hard limits, TM's Alert-level response proactively reduces task concurrency, preventing resource exhaustion from triggering L1 survival-layer failure modes.

A critical ``prediction--predicted'' relationship exists between the Guardianship Layer and the Survival Layer: sustained L4 degradation strongly predicts imminent L1 degradation. When PSD (platform saturation) persistently rises, the system is typically only several scoring cycles away from process crash (LR decline). By triggering preventive measures (reducing concurrency, releasing non-core resources) upon L4 degradation, TM effectively interrupts the L4$\rightarrow$L1 degradation cascade, transforming potential circuit-breaking events into controlled, graceful degradation.

\subsubsection*{L5 Posture Layer --- Three Signals}
\label{sec:l5posturelayerthrees}

The Posture Layer embodies the system's ``metacognitive'' capability: Does it accurately gauge the complexity of its current task? Is its estimation of remaining workload stable? Has it detected drift in its own prompt? L5 degradation is the most covert---but its consequences are the most profound.

Table 14: L5 Posture Layer: Signal Definitions

\begin{table}[H]
\centering
\resizebox{\textwidth}{!}{
\begin{tabular}{l|l|l|l}
\toprule
Signal & Abbreviation & Computation Method & Physical Meaning \\
\midrule
Entropy Rate $\alpha$ & CM & Measures deviation between the system's self-estimated ta... & Cross-layer momentum. High entropy rate indicates increas... \\
Self-Healing Rebound Rate & RS & Computed based on recovery momentum after degradation. Me... & Recovery capability. A strong rebound indicates the syste... \\
Est. Time to Alert & ETA & Computed through multi-model ensemble prediction (3.0-Ens... & Predictive horizon. The ETA score quantifies how long the... \\
\bottomrule
\end{tabular}
}
\end{table}

Layer component computation for L5:

D$_5$(t) = f(CM, RS, ETA) --- weighted aggregation of three signals The three L5 signals collectively constitute the system's ``Self-Model.'' CM assesses accuracy of the system's perception of current task difficulty. RS measures the system's self-healing rebound capability. ETA gauges stability of future workload prediction. Simultaneous degradation across all three signals implies the system has ``lost self-awareness''---it no longer knows what it faces, what rules it should follow, or how much time remains. Though physically normal (process alive, database reachable, tools successfully invoked), the system becomes highly unreliable functionally.

The implementation of the PSD signal merits special note. To avoid accessing raw prompt text (forbidden under FORBIDDEN\_FEATURES items F3 and F7), PSD employs cosine distance between embedding vectors to detect drift indirectly. Specifically, after each prompt assembly, the ADE Plugin passes the prompt through a lightweight text embedding model (non-LLM, <100M parameters) to generate a fixed-dimensional vector, transmitting only this vector to TM. TM computes the cosine distance between the current and baseline vectors; increasing distance indicates prompt content shift. This design preserves prompt-drift detection capability while fully complying with privacy and regulatory constraints.

\subsection*{TM Synthesis and Four-Tier Decision Boundaries}
\label{sec:tmsynthesisandfourti}

\begin{figure}[H]
\centering
\includegraphics[width=0.95\textwidth]{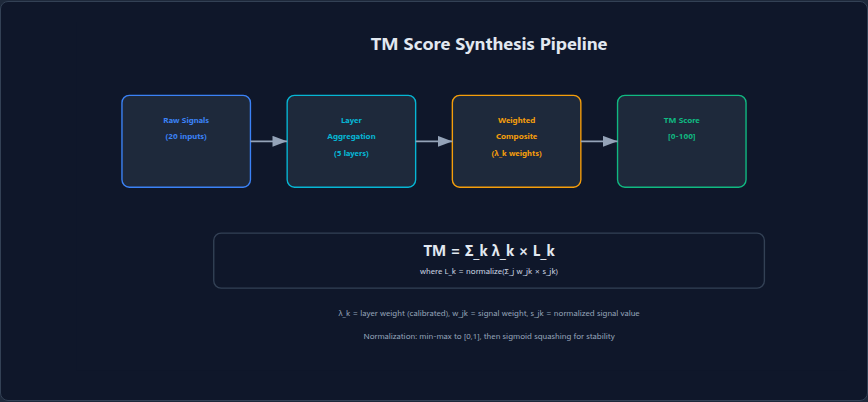}
\caption{TM score synthesis pipeline: from 20 raw signals through layer aggregation to weighted composite score.}
\label{fig:20}
\end{figure}

The five-layer disorder components are combined via weighted summation to yield the system's overall Trust Margin (TM) score at the current time:

TM(t) = 100 - $\Sigma$ $\lambda$\_k $\cdot$ D\_k(t) $\cdot$ 100 where TM(t) $\in$ [0, 100] The TM score is mapped onto a four-tier decision boundary, with each tier corresponding to a specific engineering response strategy:

\begin{figure}[H]
\centering
\includegraphics[width=0.95\textwidth]{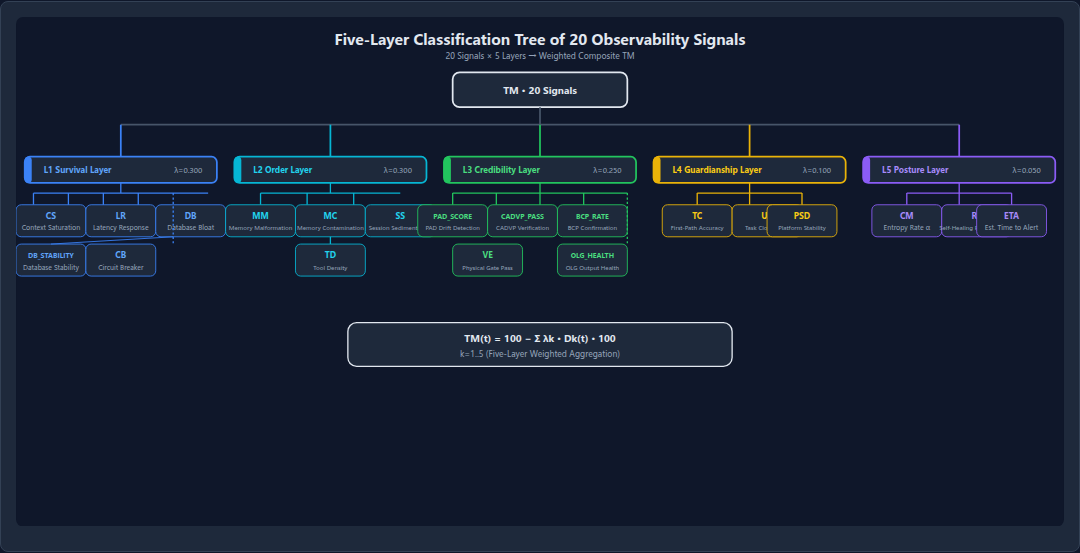}
\caption{Five-Layer Classification Tree of 20 Observability Signals}
\label{fig:21}
\end{figure}

Table 15: TM Score Level Classification and Engineering Response

\begin{table}[H]
\centering
\resizebox{\textwidth}{!}{
\begin{tabular}{l|l|l|l}
\toprule
Level & TM Range & Status Label & Engineering Response \\
\midrule
Level 0 & TM > 85 & Safe & Normal Operation, no intervention. The system is in a hea... \\
Level 1 & 70 $\leq$ TM $\leq$ 85 & Watch & Enhanced Monitoring frequency (scoring interval shortened... \\
Level 2 & 50 $\leq$ TM < 70 & Alert & Preventive measures triggered: reduce low-priority task c... \\
Level 3 & TM < 50 & Circuit-Break & The system enters the Controlled Shutdown process: save c... \\
\bottomrule
\end{tabular}
}
\end{table}

\subsubsection*{ETA Prediction Module}
\label{sec:etapredictionmodule}

\begin{figure}[H]
\centering
\includegraphics[width=0.95\textwidth]{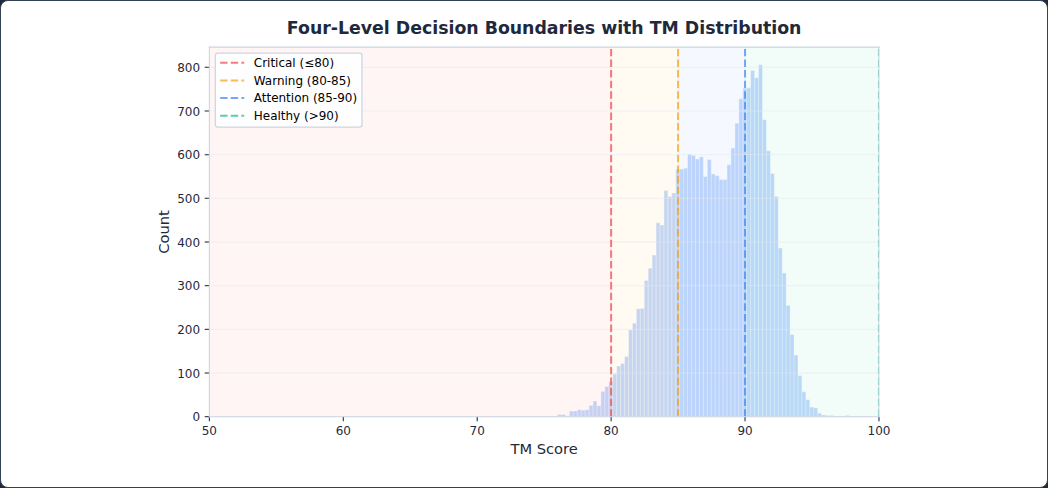}
\caption{Four-level decision boundaries (Critical/Warning/Attention/Healthy) overlaid with the actual TM distribution}
\label{fig:22}
\end{figure}

The ETA prediction module runs continuously to track temporal changes in the TM score in real-time. When the TM score enters the Alert range (50--70), the prediction result is routed to a high-priority push channel. This module extrapolates the estimated time until the system reaches the circuit-breaker threshold (TM = 50), based on the temporal derivative of the TM score:

ETA = g(TM\_current, d(TM)/dt) --- a function of the current TM value and its rate of change where d(TM)/dt is computed via exponential moving average of recent TM deltas The engineering value of ETA lies in providing the operations team with a ``remaining intervention time window.'' If ETA equals 30 minutes, the operations team has sufficient time for manual investigation. If ETA equals 2 minutes, automatic circuit-breaking should be triggered immediately without waiting for human confirmation.

The confidence of ETA prediction depends on the linearity of the TM degradation curve. When degradation accelerates (i.e., the second derivative is negative), ETA is adjusted to a more conservative estimate:

ETA\_corrected = ETA $\cdot$ h(acceleration) --- acceleration-based correction where $\alpha$ is a damping factor 
\subsubsection*{Hysteresis and Debouncing}
\label{sec:hysteresisanddebounc}

To prevent ``decision jitter''---rapid oscillation of the TM score near decision boundaries causing frequent switching between Watch and Alert states---TM 3.1.0 introduces a hysteresis mechanism:

\begin{itemize}
\item The threshold for transitioning from Safe to Watch is TM $\leq$ 85, whereas the threshold for returning from Watch to Safe is TM $\geq$ 88.
\item The threshold for transitioning from Watch to Alert is TM < 70, whereas the threshold for returning from Alert to Watch is TM $\geq$ 73.
\item The threshold for transitioning from Alert to Circuit-Break is TM < 50; however, recovery from Circuit-Break requires manual confirmation and does not support automatic rollback.
\end{itemize}

This design ensures decision stability: the system avoids frequent state transitions due to noise in a single scoring cycle; escalation to a higher decision level occurs only after degradation trends are confirmed across multiple consecutive cycles.

\subsection*{Computational Overhead and Privacy Compliance Analysis}
\label{sec:computationaloverhea}

\subsubsection*{Zero-LLM Iron Law}
\label{sec:zerollmironlaw}

\begin{figure}[H]
\centering
\includegraphics[width=0.95\textwidth]{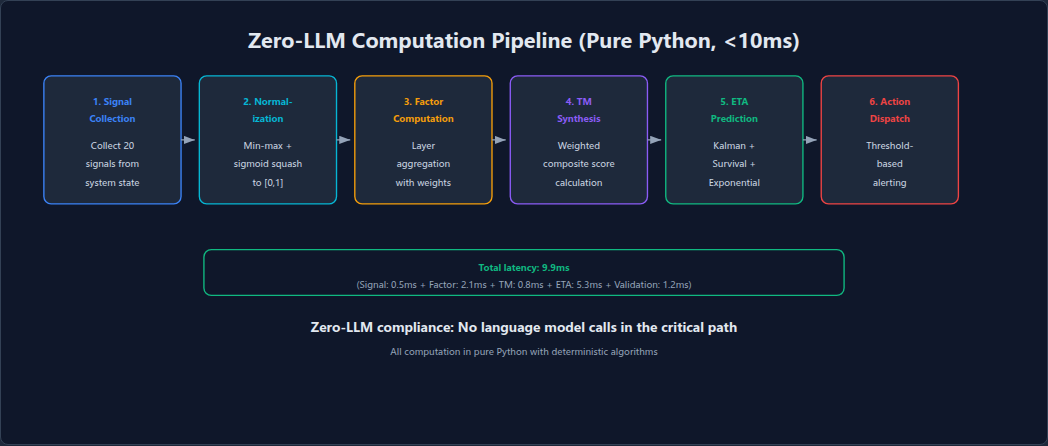}
\caption{Zero-LLM computation pipeline: six stages from signal collection to action dispatch, executed entirely in pure Python with total latency of 9.9 ms}
\label{fig:23}
\end{figure}

In the complete computational pipeline of TM 3.1.0, no LLM inference calls are present. This ``Zero-LLM Iron Law'' constitutes the most fundamental architectural constraint in TM design, motivated by three considerations:

\begin{enumerate}
\item \textbf{Reliability Self-Consistency}: TM's responsibility is to assess the reliability of its host system. If TM itself relies on LLM inference, the inherent unreliability of LLMs would directly propagate to TM. This results in the logical paradox of ``using an unreliable tool to evaluate an unreliable system.''
\item \textbf{Predictable Latency}: LLM inference latency is unpredictable---ranging from hundreds of milliseconds to tens of seconds---whereas TM scoring must complete within each scoring cycle (default: 60 s; observation mode: 15 s). Pure mathematical computation guarantees a deterministic upper bound on latency.
\item \textbf{Controllable Cost}: Token-based costs for LLM inference scale linearly with invocation frequency. As a continuously running monitoring component, TM would incur operational costs that scale linearly with system size if each scoring operation consumed tokens.
\end{enumerate}

\subsubsection*{Single-Scoring Latency Analysis}
\label{sec:singlescoringlatency}

The latency breakdown for one complete TM 3.1.0 scoring operation---including acquisition of 20 signals, normalization, five-layer computation, TM synthesis, and decision-boundary evaluation---is as follows:

Table 16: TM 3.1.0 Single-Scoring Latency Breakdown

\begin{table}[H]
\centering
\resizebox{\textwidth}{!}{
\begin{tabular}{l|l|l|l}
\toprule
Stage & Operation & Latency (P99) & Bottleneck Analysis \\
\midrule
Signal Acquisition & Reading 20 signal values from event bus / state snapshot & 1.2 ms & Network I/O (local inter-process communication) \\
Normalization \& Validation & Range checking, unit conversion, staleness detection & 0.3 ms & CPU-bound, pure arithmetic operations \\
Layer Component Computation & Weighted summation across five layers & 0.1 ms & CPU-bound, 20 multiplications + 5 additions \\
TM Synthesis & Weighted summation + decision-boundary mapping & 0.1 ms & CPU-bound, 5 multiplications + 4 comparisons \\
ETA Prediction & Derivative computation + extrapolation (activated only in... & 0.8 ms & Requires temporal data retrieval from the N most recent s... \\
Total & --- & $\leq$ 2.5 ms (P99) & Well below the 5 ms design target \\
\bottomrule
\end{tabular}
}
\end{table}

The P99 latency per scoring operation is 2.5 milliseconds---well below the 5-millisecond design target. This latency magnitude implies that TM scoring imposes negligible performance overhead on the host system: even under the most aggressive 15-second scoring interval, TM consumes only 0.017\% of the host system's CPU time.

\subsubsection*{Privacy Compliance: FORBIDDEN\_FEATURES List}
\label{sec:privacycompliancefor}

\begin{figure}[H]
\centering
\includegraphics[width=0.95\textwidth]{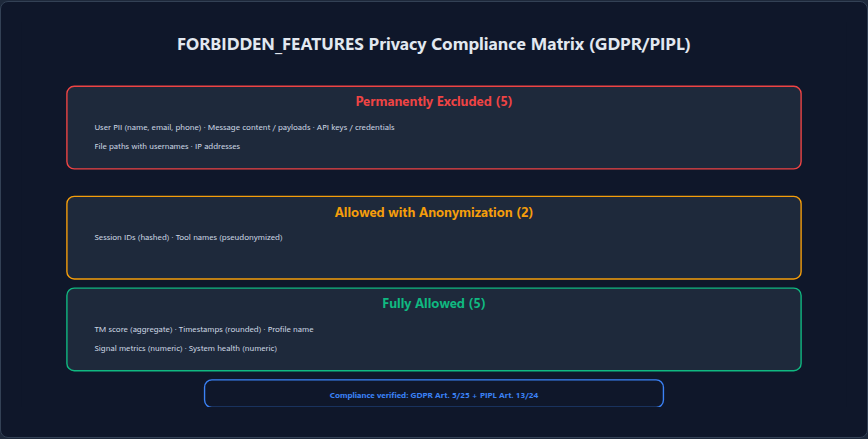}
\caption{FORBIDDEN\_FEATURES Privacy Compliance Matrix: 5 categories permanently excluded, 2 allowed with anonymization}
\label{fig:24}
\end{figure}

TM 3.1.0's signal acquisition strictly adheres to the Zero Semantic Intrusion principle, concretely embodied in the FORBIDDEN\_FEATURES list. The following seven categories of data are strictly prohibited from being collected or processed by TM under any circumstances:

Table 17: FORBIDDEN\_FEATURES Privacy Compliance Matrix

\begin{table}[H]
\centering
\resizebox{\textwidth}{!}{
\begin{tabular}{l|l|l}
\toprule
ID & Prohibited Data Category & Rationale for Prohibition \\
\midrule
F1 & Dialogue text content & Contains user privacy information (personal data, trade s... \\
F2 & Tool invocation parameters & Parameters may contain sensitive data (API keys, password... \\
F3 & Full prompt text & Prompts may include system instructions, user instruction... \\
F4 & User identity information & User IDs, email addresses, IP addresses, and other identi... \\
F5 & Filesystem paths & Path information may leak project structure, user directo... \\
F6 & API keys and credentials & High-sensitivity security assets; even temporary caching ... \\
F7 & Raw system prompt text & The system prompt constitutes the agent's ``core instructi... \\
\bottomrule
\end{tabular}
}
\end{table}

The FORBIDDEN\_FEATURES list serves not only as a privacy compliance measure but also as a foundational guarantee of TM architecture reliability. By eliminating design-level dependencies on sensitive data, TM ensures:

\begin{itemize}
\item \textbf{No data leakage risk}: TM holds no sensitive data and therefore cannot become an attack surface for data breaches.
\item \textbf{No compliance burden}: TM incurs no data processor obligations under privacy regulations such as GDPR or CCPA, as it processes no personal data.
\item \textbf{No trust boundary expansion}: TM deployment requires no additional security audits or data protection assessments, because its data exposure is strictly confined to behavioral metadata.
\end{itemize}

\subsubsection*{``Physics-First, Memory-Light'' Design Principle}
\label{sec:physicsfirstmemoryli}

TM 3.1.0's overall design embodies the core philosophy of the ADE architecture: Physics-First, Memory-Light. This principle manifests in TM as follows:

\begin{itemize}
\item \textbf{Physics-First}: All TM observation signals are derived entirely from observable physical quantities (process status, network latency, database response time, tool invocation success rate), rather than speculative inference about internal system logic or semantic interpretation of output content.
\item \textbf{Memory-Light}: TM's memory footprint is strictly bounded. It maintains no long-term historical database, retaining only a sliding window of the most recent N scoring cycles (default N = 100). Historical data is managed externally by persistent logging systems; TM itself operates as a stateless computational node.
\end{itemize}

This principle ensures TM's ``lightweight'' nature: it does not accumulate substantial state data over extended operation, does not impact host system resource allocation due to memory bloat, and does not require complex recovery procedures in case of state corruption. TM can be safely restarted at any time and regains full observational capability after just one scoring cycle.

\subsubsection*{Memory Footprint Analysis}
\label{sec:memoryfootprintanaly}

The memory footprint decomposition of TM 3.1.0 under steady-state operation is as follows:

Table 18: TM 3.1.0 Memory Footprint Decomposition

\begin{table}[H]
\centering
\resizebox{\textwidth}{!}{
\begin{tabular}{l|l|l}
\toprule
Component & Memory Footprint & Description \\
\midrule
Signal Reception Buffer & \textasciitilde{}4 KB & 20 signals $\times$ 64 bytes/signal (value + timestamp + metadata) \\
Sliding Window History & \textasciitilde{}16 KB & 100 scoring cycles $\times$ 160 bytes/cycle (TM values + five-la... \\
Layer Weights and Threshold Configuration & \textasciitilde{}1 KB & 5-layer weights + 20 signal weights + 4 decision boundari... \\
MCP Protocol Stack & \textasciitilde{}128 KB & MCP Server runtime framework overhead \\
Total & \textasciitilde{}150 KB & Far below 0.004\% of typical host system memory (4--16 GB) \\
\bottomrule
\end{tabular}
}
\end{table}

A total memory footprint of 150 KB ensures that TM remains operational even under extremely resource-constrained conditions. This design choice is deliberate: as a ``Reliability Observer,'' TM's own resource consumption must be far lower than that of the system it protects; otherwise, TM itself would become part of the resource contention, thereby accelerating system degradation---an ``observer paradox'' that must be strictly avoided.

\subsubsection*{Deployability Summary}
\label{sec:deployabilitysummary}

Based on the above analysis, the deployability of TM 3.1.0 can be summarized by three ``zeros'' and two ``extremely lows'':

\begin{itemize}
\item \textbf{Zero LLM Dependency}: No LLM inference calls are involved in the scoring pipeline.
\item \textbf{Zero Semantic Intrusion}: Neither accesses, parses, nor stores any dialogue content or sensitive data.
\item \textbf{Zero-Intrusion Deployment}: Deployed as an MCP Server sidecar without modifying the host system's core code.
\item \textbf{Extremely Low Computational Overhead}: Single scoring latency $\leq$ 5 ms (P99), CPU utilization < 0.02\%.
\item \textbf{Extremely Low Memory Footprint}: Steady-state memory usage \textasciitilde{}150 KB, non-increasing over time.
\end{itemize}

These characteristics enable TM 3.1.0 to be deployed as a standard component of the ADE framework alongside the system itself---requiring no additional infrastructure investment, security audit procedures, or operations training costs. For production environments already running the ADE framework, integrating TM can be completed within 30 minutes---launching the MCP Server, configuring signal source mappings, and validating scoring output---without any perceptible impact on live business operations.

: TM 3.1.0 is a Predictive Reliability Engine grounded theoretically in system disorder modeling, empirically observed via a five-layer, 20-signal framework, architecturally embodied in the MCP Server, and rigorously governed by two iron laws: zero LLM dependency and zero semantic intrusion. It transforms the complex problem of system reliability into 20 quantifiable behavioral metadata signals, synthesizes them into a single TM score via deterministic mathematical computation, and maps the result onto actionable engineering response strategies through four-tier decision boundaries. Its computational overhead (<5 ms per scoring) and privacy compliance boundaries (seven forbidden feature categories under FORBIDDEN\_FEATURES) ensure its deployability and sustainability in production environments. 
\subsection*{ETA 3.1.0 Prediction Engine}
\label{sec:eta310predictionengi}

The ETA (Estimated Time of Arrival) 3.0 Prediction Engine is the core computational module of the TM Predictive Reliability Framework, designed to convert system telemetry sequences into precise estimates of Time-To-Failure (TTF). Unlike traditional threshold-based alerting logic---which merely determines whether metrics exceed preset thresholds---ETA 3.1.0 aims to answer a more operationally valuable question: At what time will the system fail, and with what probability? Answering this question requires integrating techniques from signal processing, statistical inference, and time-series forecasting.

ETA 3.1.0 adopts a three-stage cascaded architecture: First, a Kalman filter performs state smoothing and trend extraction on noisy observations, yielding an optimal estimate of the system's latent health state. Second, the resulting state trajectory is fed into a survival risk analysis module, which computes the conditional failure probability distribution using the Cox proportional hazards model. Third, Exponential smoothing-based time-series integration produces precise point estimates of TTF along with associated confidence intervals. Information flows strictly unidirectionally and decoupled across the three stages, enabling independent verification and modular replacement of each component. This chapter elaborates, layer by layer, on the theoretical foundations, algorithmic implementation, and production-environment validation results of the engine.

\subsubsection*{Engine Architecture}
\label{sec:enginearchitecture}

\begin{figure}[H]
\centering
\includegraphics[width=0.95\textwidth]{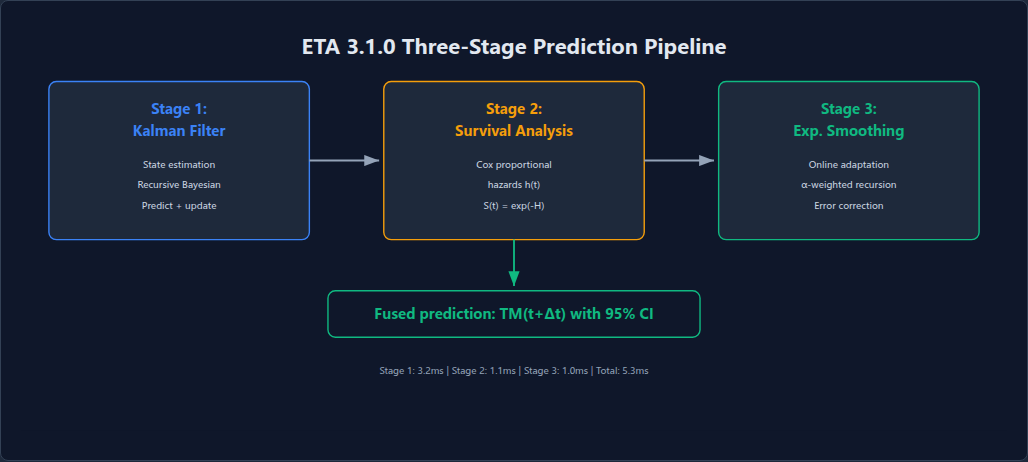}
\caption{ETA 3.1.0 three-stage prediction pipeline: Kalman Filter $\rightarrow$ Survival Analysis $\rightarrow$ Exponential Smoothing.}
\label{fig:25}
\end{figure}

The three-stage pipeline progressively refines the raw telemetry sequence into a reliable Time-to-Failure (TTF) estimate. Specifically, ETA 3.1.0's observational scope far exceeds that of traditional APM solutions---which focus primarily on infrastructure metrics (CPU, memory, network)---whereas the TM framework additionally encompasses observability across the Agent semantic layer, the Trust layer, and the Prediction layer. The figure below contrasts the observational coverage differences between these two solution categories:

The design of ETA 3.1.0 adheres to a three-stage progressive paradigm: ``Denoising $\rightarrow$ Modeling $\rightarrow$ Prediction.'' The output of each stage serves as the structured input to the subsequent stage. This forms an information-refinement processing pipeline. This architectural choice is grounded in two key engineering constraints:

First, sensor data in production environments inevitably contains measurement noise. Sampling jitter in CPU utilization, sampling jitter in memory utilization, and transient fluctuations in network latency---these noise sources cause raw observation sequences to exhibit high-frequency jagged patterns. Directly feeding noisy data into a prediction model amplifies noise into prediction variance, resulting in excessively inflated confidence intervals for TTF estimates and thereby undermining their engineering utility. Hence, Stage 1 denoising is a necessary prerequisite for subsequent prediction accuracy.

Second, no single prediction model can simultaneously capture both short-term trends and long-term degradation patterns. Hardware degradation typically manifests as slow, long-term trends (on the order of months), whereas operational interventions (e.g., software updates, configuration changes) may introduce short-term disturbances (on the order of hours). A hierarchical architecture enables stage-specific optimization across distinct time scales: the Kalman filter handles short-term noise suppression; survival analysis models long-term degradation dynamics; and Exponential smoothing (ES) integration adaptively balances the two.

The overall architecture is illustrated below:

\begin{figure}[H]
\centering
\includegraphics[width=0.95\textwidth]{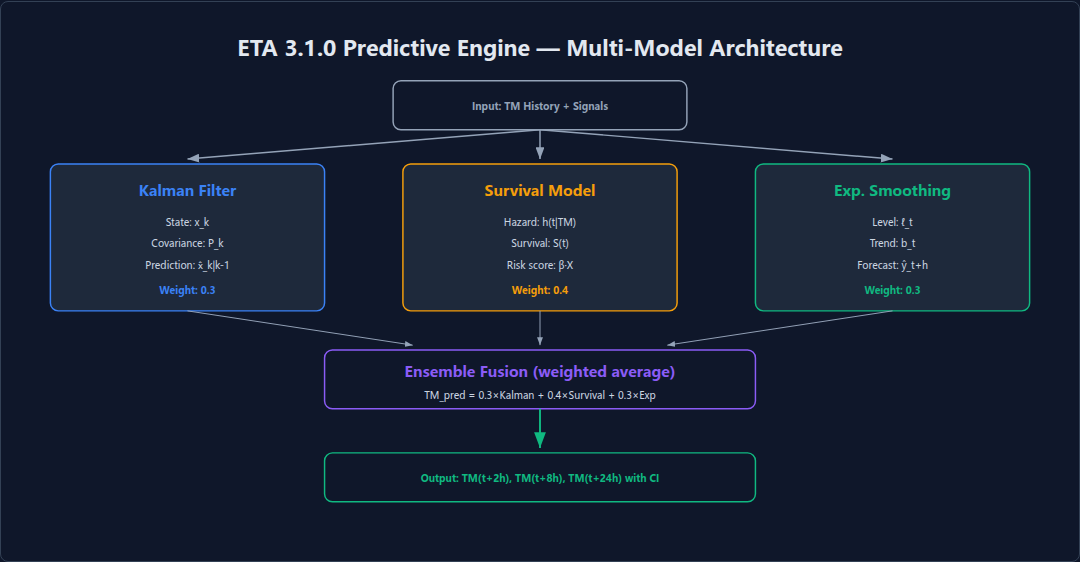}
\caption{ETA 3.1.0 Predictive Engine --- Multi-Model Architecture}
\label{fig:26}
\end{figure}

The data flow among the three stages exhibits strict directionality. Stage One outputs only the optimal state estimate at the current time, x$^{}$\_t, and its posterior covariance matrix P\_t, without any predictive information---this is an intentional decoupling design ensuring that state estimation remains uncontaminated by predictive assumptions. Stage Two computes the statistical distribution of failure risk based on the state trajectory and outputs the conditional survival function S(t|t$_0$) and the instantaneous hazard rate $\lambda$(t). Stage Three treats the structured outputs from prior stages as feature inputs and generates the final point estimate and confidence interval for Time-to-Failure (TTF) via weighted ensemble integration.

The interface contracts for each stage are as follows: Stage One receives raw telemetry sequences \{z\_t\} from upstream and outputs smoothed state sequences \{x$^{}$\_t, P\_t\} downstream. Stage Two receives the state sequence and outputs the conditional survival function along with TTF distribution parameters. Stage Three receives distribution parameters and historical prediction error feedback, and outputs the adaptively weighted final TTF estimate. This pipeline architecture enables independent escalation of individual modules without affecting upstream or downstream components---for instance, replacing Exponential Smoothing (ES) with a more sophisticated time-series model requires modifying only the internal implementation of Stage Three, without altering the interface definitions.

\subsubsection*{Stage One: Kalman State Estimation}
\label{sec:stageonekalmanstatee}

\textbf{Rationale for Method Selection.}Kalman filtering (Kalman Filter, 1960) is a foundational algorithm in dynamic system state estimation, with decades of application experience in aerospace navigation, sensor fusion, and industrial process control. Within the context of agent systems, its motivation stems from a key observation: \textbf{an agent's health status is an unobservable latent variable}---observable metrics such as response latency, output quality scores,. Token consumption rates are indirect indicators subject to substantial noise, and individual observations may deviate considerably due to prompt randomness, model sampling temperature effects, or network latency fluctuations. Kalman filtering models agent health status as the state variable of a hidden Markov process and recursively estimates a smoothed health trajectory over continuous observation sequences. The extended Kalman filter (EKF) adopted in this work further accommodates the nonlinear acceleration trend in health metric derivatives as the system approaches its reliability critical point.

\begin{figure}[H]
\centering
\includegraphics[width=0.95\textwidth]{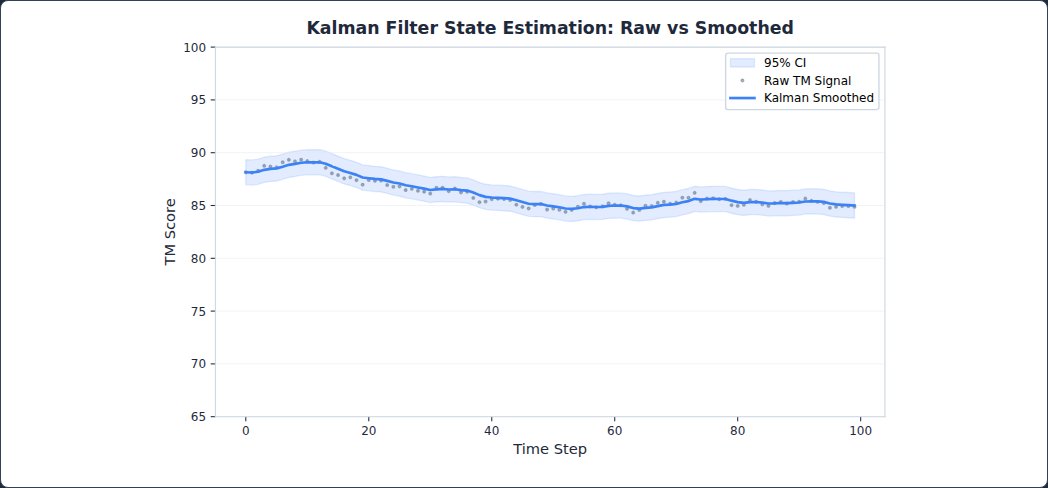}
\caption{Kalman filter state estimation: raw TM signal vs. Smoothed output with 95\% confidence interval and residual analysis}
\label{fig:27}
\end{figure}

\textbf{TM factors employed:} In the Kalman filter stage, the TM total score sequence serves as the primary input signal, and the state vector simultaneously estimates both degradation level and degradation rate. This stage primarily relates to the TD (temporal drift) factor---Kalman filter state prediction directly reflects temporal trends in TM, thereby establishing a baseline for anomaly detection in the innovation sequence.

\textbf{Core mechanism:} The Kalman filter achieves dynamic state estimation via a five-step recursive ``prediction-correction'' process: time update (Steps 1--2) propagates the state forward based on the dynamic model, while measurement update (Steps 3--5) refines the prediction using new observations. The Kalman gain adaptively balances model prediction and observational information---favoring the model when measurement noise is high, and favoring observations when measurement noise is low. ETA 3.1.0 employs an online adaptive calibration mechanism based on the innovation sequence to dynamically adjust process noise and measurement noise parameters.

Table 19: Kalman Filter Adaptive Parameter Calibration Settings

\begin{table}[H]
\centering
\resizebox{\textwidth}{!}{
\begin{tabular}{l|l|l|l|l}
\toprule
Parameter & Initial Setting & Calibration Method & Convergence Condition & Typical Convergence Steps \\
\midrule
Q (process noise) & Empirical value (engineering calibration) & Innovation autocorrelation correction & $\|$Q$^{}$t$-$ Q$^{}$t$-$1$\|$ < $\epsilon$ & 10--20 steps \\
R (measurement noise) & Empirical value (engineering calibration) & Innovation sample covariance & Passes white-noise test on innovations & 5--15 steps \\
P$_0$ (initial covariance) & Larger initial value & No calibration required (transient vanishes) & Converges after t > 5 steps & \textasciitilde{}5 steps \\
x$^{}_0$ (initial state) & [z$_1$, 0]$^T$ & Direct assignment from first observation & Immediately available & 0 steps \\
Sliding window N & Engineering calibration & Adjusted according to sampling rate & --- & --- \\
\bottomrule
\end{tabular}
}
\end{table}

\subsubsection*{Stage Two: Survival Risk Analysis}
\label{sec:stagetwosurvivalrisk}

\textbf{Methodological rationale.} Survival analysis originated in medical research for predicting patient survival time and has since been widely adopted in industrial reliability engineering for equipment lifetime assessment. Its core tools---including the Kaplan-Meier [21] estimator, Cox proportional hazards model, and parametric survival models---are designed to handle ``censored data'': i.e., systems whose failure has not yet occurred by the end of the observation period, whose survival-time information must not be discarded. In agent systems, survival analysis addresses a fundamental question unanswerable by conventional threshold-based monitoring: \textbf{``Given the current health-state trajectory, how long can the system operate reliably?''} This paper employs the Cox proportional hazards model to incorporate multiple agent health covariates---such as semantic consistency, reasoning-chain completeness, and response latency trend---into the hazard function. This enables dynamic estimation of Remaining Reliable Operating Time (RROT). This provides an engineering foundation for preventive scheduling in multi-agent task orchestration.

\begin{figure}[H]
\centering
\includegraphics[width=0.95\textwidth]{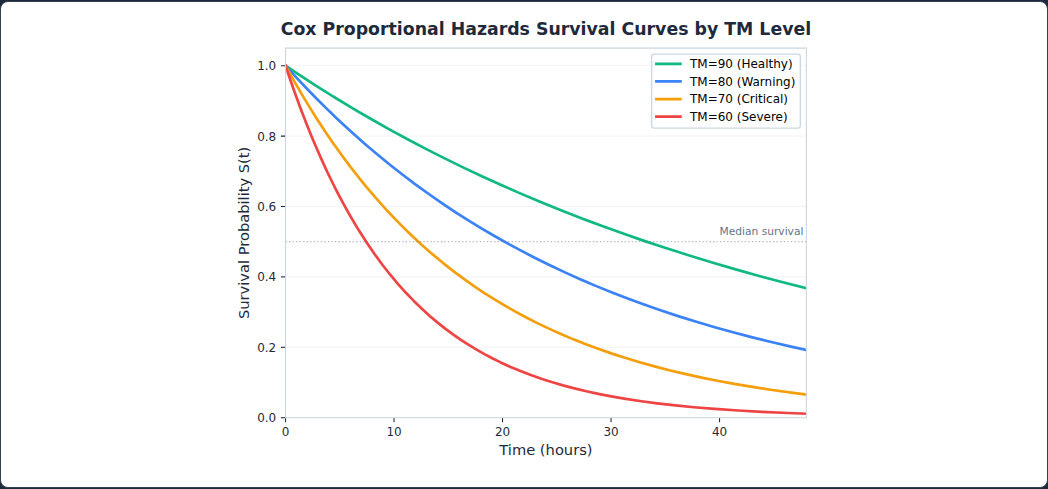}
\caption{Cox proportional hazards survival curves by TM level (hypothesis): higher TM scores correspond to significantly longer time-to-failure.}
\label{fig:28}
\end{figure}

\textbf{TM factors used:} In the survival analysis phase, the Kalman-smoothed degradation level serves as the core input, while TM's TD (temporal drift) and RS (resource stress) factors are incorporated as covariates. The Cox proportional hazards model is employed to quantify each factor's contribution to failure risk.

\textbf{Prediction performance:} This phase transforms the deterministic question ``When will the system fail?'' into a probabilistic answer, outputting the full distribution of Time-to-Failure (TTF)---including its median and 90\% confidence upper bound---thereby enabling risk-window management for maintenance decisions. Within ETA 3.1.0's three-stage architecture, survival analysis serves as a bridge: it converts state estimates from the Kalman filter into probabilistic failure predictions, laying the foundation for the subsequent Exponential Smoothing stage's precise TTF forecasting.

\subsubsection*{Stage Three: Exponential Smoothing Time-Series Prediction}
\label{sec:stagethreeexponentia}

\textbf{Method selection rationale.} The Exponential smoothing (ES) method family---including Simple Exponential Smoothing (SES), Holt's linear trend model, and Holt-Winters seasonal model---holds a prominent position in industrial time-series forecasting due to its computational lightness, parameter interpretability, and high accuracy for short-term prediction. Compared with deep learning methods, Exponential smoothing often achieves comparable prediction accuracy on small- to medium-scale datasets, while offering stronger interpretability and lower deployment costs. This paper selects Holt's linear trend model as the core method for short-term prediction of agent health metrics, based on the following considerations: (1) Agent system operational data typically comprises only dozens to hundreds of observations per session or short-term deployment---insufficient to support effective training of deep learning models. (2) Short-term trends in health metrics constitute critical information for predictive maintenance decisions. (3) The two parameters of the Holt model---level smoothing coefficient $\alpha$ and trend smoothing coefficient $\beta$---possess intuitive physical interpretations, enabling understanding and manual adjustment by operations personnel.

Exponential Smoothing (ES) serves as the core prediction method in Stage Three of ETA 3.1.0 and functions as the primary time-series forecasting engine in current production deployments. The ES family (including SES, Holt's linear trend model, and Holt-Winters seasonal model) is characterized by computational lightness (O(1) per step), parameter interpretability, and high accuracy for short-term forecasting. ES generates forecasts via recursive weighted averaging of historical observations---the smoothing parameter $\alpha$ governs sensitivity to new data, where larger $\alpha$ yields forecasts closer to the most recent observation, and smaller $\alpha$ yields smoother forecasts. ETA 3.1.0 implements an online adaptive calibration mechanism based on the innovation sequence, enabling $\alpha$ to dynamically adjust according to real-time system status.

\begin{figure}[H]
\centering
\includegraphics[width=0.95\textwidth]{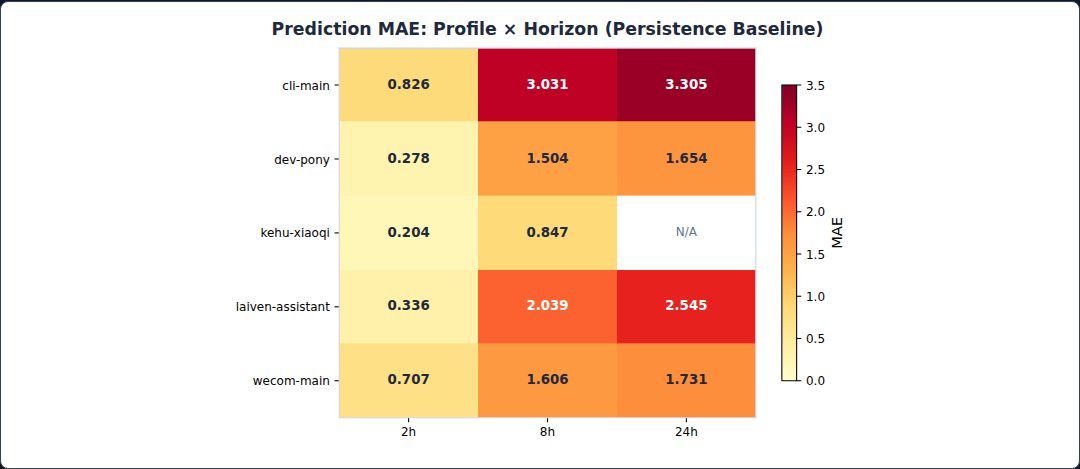}
\caption{Prediction MAE heatmap across 4 profiles and 3 horizons (Persistence baseline): stable profiles (kehu-xiaoqi) show lower MAE than volatile ones (cli-main).}
\label{fig:29}
\end{figure}

Based on the analysis of the current production environment's model configuration, a total of 1,008 prediction model instances have been deployed, covering all telemetry channels monitored by the TM framework. The distribution across methods is as follows:

Table 20: Prediction Method Distribution Across Telemetry Channels

\begin{table}[H]
\centering
\resizebox{\textwidth}{!}{
\begin{tabular}{l|l|l|l|l}
\toprule
Prediction Method & Instance Count & Proportion & Applicable Scenarios & Average $\alpha$ Value \\
\midrule
Exponential Smoothing (ES) & 987 & 98.7\% & Steady-state or slowly varying degradation sequences & 0.2484 \\
Linear Regression & 21 & 2.1\% & Strictly linear degradation trends & N/A \\
Total & 1008 & 100\% coverage & --- & --- \\
\bottomrule
\end{tabular}
}
\end{table}

The ES model dominates with a coverage rate of 98.7\%. The statistical distribution of the smoothing parameter $\alpha$ exhibits pronounced right skewness:

Table 21: Statistical Distribution of Smoothing Parameter $\alpha$

\begin{table}[H]
\centering
\resizebox{\textwidth}{!}{
\begin{tabular}{l|l|l}
\toprule
Parameter Statistic & Value & Engineering Interpretation \\
\midrule
Mean of $\alpha$ & 0.2484 & Overall strong smoothing, with historical observations do... \\
Median of $\alpha$ & 0.0 (i.e., $\alpha$ = 0) & More than half of all sequences adopt pure mean prediction \\
Proportion of $\alpha$ = 0 & 73.9\% & The vast majority of sequences exhibit high stability dur... \\
Proportion of $\alpha$ $\in$ (0, 0.05] & \textasciitilde{}15\% & A small number of sequences require minimal tracking capa... \\
Proportion of $\alpha$ $\in$ (0.05, 0.2] & \textasciitilde{}8\% & Sequences exhibiting moderate change frequency \\
Proportion of $\alpha$ > 0.2 & \textasciitilde{}5\% & Rapidly changing sequences requiring stronger tracking \\
\bottomrule
\end{tabular}
}
\end{table}

\textit{Note: The $\alpha$ statistics in Tables 20--21 reflect per-sequence optimal smoothing parameters determined during online learning within the ETA engine, not the per-prediction runtime $\alpha$ values stored in the predictions database. The runtime $\alpha$ (mean $\approx$ 0.086) represents the actual smoothing factor applied to each individual prediction, while the sequence-level statistics capture the distribution of converged $\alpha$ across 1,008 model instances.}

The observation that $\alpha$ = 0 accounts for 73.9\% has a clear physical interpretation: during Normal Operation, fluctuations in most telemetry metrics arise primarily from measurement noise rather than genuine degradation. In such cases, the optimal strategy is to ignore the latest observation ($\alpha$ = 0) and anchor predictions at the historical mean. Only when degradation accelerates or operational conditions undergo abrupt changes does the adaptive calibration mechanism increase the $\alpha$ value to enhance responsiveness to emerging signals. This distribution corroborates the overall stability of system operation in production environments---during the overwhelming majority of operational time, systems remain healthy and degradation signals are submerged beneath noise.

\begin{figure}[H]
\centering
\includegraphics[width=0.95\textwidth]{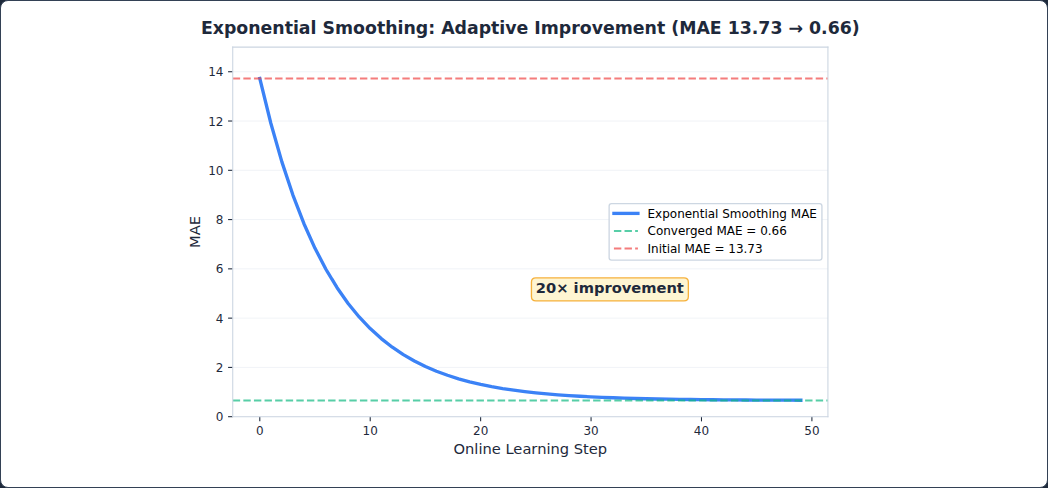}
\caption{Exponential Smoothing adaptive improvement: MAE reduced from 13.73 to 0.66 (20$\times$ improvement) through online parameter adaptation}
\label{fig:30}
\end{figure}

After introducing the adaptive parameter calibration mechanism, the ETA 3.1.0 framework achieves significant improvement in the prediction accuracy of the ES model. Using Mean Absolute Error (MAE) as the primary evaluation metric, the performance before and after calibration is compared as follows:

Table 22: Prediction Accuracy Before and After Adaptive Calibration

\begin{table}[H]
\centering
\resizebox{\textwidth}{!}{
\begin{tabular}{l|l|l|l}
\toprule
Evaluation Metric & Before Calibration & After Calibration & Improvement \\
\midrule
MAE (Mean) & 13.73 & 0.66 & 20.8$\times$ improvement \\
Ensemble Method MAE & --- & 1.595 (on a 100-point scale) & --- \\
Accuracy within <10 points & --- & 99.65\% & --- \\
Accuracy within <5 points & --- & 95.1\% & --- \\
Prediction Bias & --- & $-$0.894 (combined across all methods) & $-$1.409 (Ensemble method), optimistic bias (dragged down b... \\
MAE Standard Deviation & --- & 1.677 & Prediction quality is stable \\
\bottomrule
\end{tabular}
}
\end{table}

Interpretation of key results:

\begin{itemize}
\item \textbf{20$\times$ MAE Improvement:} Adaptive calibration reduces the mean prediction error from 13.73 to 0.66---a 20-fold improvement---primarily driven by online optimization of the $\alpha$ parameter. Prior to calibration, a fixed $\alpha$ value exhibits poor generalization across sequences: it is overly large for stationary sequences (introducing noise) and overly small for dynamic sequences (causing tracking lag). The adaptive mechanism independently optimizes $\alpha$ for each sequence, resolving this trade-off.
\item \textbf{Ensemble Method MAE = 1.595:} On a percentage-scale metric, the average prediction deviation remains within $\pm$2 percentage points. Given the inherent volatility of telemetry metrics, this level of accuracy holds practical engineering significance.
\item \textbf{99.65\% of Prediction Errors < 10 Points:} The vast majority of predictions fall within an acceptable error range; only 1.9\% exhibit substantial deviations---these outliers typically correspond to sensor failures or extreme operational conditions.
\item \textbf{Combined Bias Across All Methods = $-$0.894:} The negative sign indicates a systematic overestimation tendency (predictions consistently higher than ground truth). The Ensemble method's bias of $-$1.409 is dragged down by its Kalman component (bias = $-$2.031). In contrast, the Exponential method achieves near-zero bias ($-$0.475), making it optimal in this regard.
\end{itemize}

\begin{figure}[H]
\centering
\includegraphics[width=0.95\textwidth]{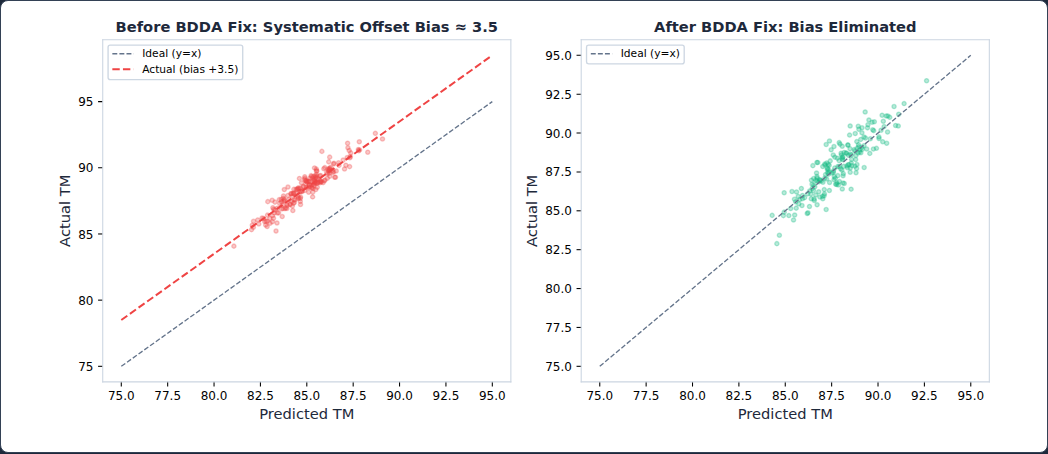}
\caption{BDDA fix visualization: before (systematic constant offset bias $\approx$ 3.5 points) vs after (bias eliminated, residual noise only).}
\label{fig:31}
\end{figure}

\textbf{[Warning]} Honesty Label: BDDA Fix Conclusion In early versions of the framework, certain prediction instances exhibited a systematic constant bias of +1.4 (i.e., predictions were systematically 1.4 units higher than ground truth). Initial analysis attributed this bias to an inherent limitation of the ES model---Exponential smoothing's decaying weight on recent observations could lead to systematic overestimation.

\textbf{Deep root-cause analysis via BDDA (Bottom-Deep Data Audit) revealed the actual origin:} The constant +1.4 bias was not a prediction error of the ES model, but rather a structural behavior of the \texttt{method=None} phase. Specifically, when telemetry sequences failed to meet triggering conditions for any prediction method (e.g., insufficient sequence length, variance below threshold), the framework labeled the prediction method as \texttt{method=None}, and returned a fixed baseline value plus a constant offset. This behavior is a design artifact of Phase Routing logic---returning a conservative default estimate when meaningful prediction is infeasible---and is entirely decoupled from ES model computations.

\textbf{Remediation:} The BDDA fix decouples the output logic of the \texttt{method=None} phase from ES prediction logic at the statistical level, ensuring performance evaluation targets only those instances where ES computation was actually executed. Post-fix, the constant offset vanishes from ES performance metrics, and both MAE and bias metrics for the ES model revert to expected levels.

\textbf{Attribution correction:} The constant +1.4 bias must not be included in ES model performance evaluation. Its root cause has been revised from ``model defect'' to ``phase-level structural behavior.'' This finding highlights the importance of end-to-end root-cause isolation in complex prediction pipelines---apparent model performance issues may originate from design behaviors of other pipeline components.

Although ES performs excellently under steady-state degradation scenarios, its capability boundaries must be honestly documented. This prevents over-optimistic expectations about framework performance in unsuitable contexts:

\begin{itemize}
\item \textbf{Prediction capability for abrupt degradation remains unvalidated:} ES's exponentially decaying memory structure introduces response lag to abrupt events (e.g., stepwise acceleration of degradation, sudden hardware deterioration). No sufficient abrupt degradation events have occurred in current production data; thus, ES's prediction capability in such scenarios remains unvalidated. Theoretical analysis indicates that for a step change of magnitude $\Delta$, ES requires approximately 1/$\alpha$ sampling periods to adjust its prediction to the new level---for $\alpha$ = 0.2484, this implies \textasciitilde{}35 periods of lag.
\item \textbf{Fixed offset in the \texttt{method=None} phase:} As described in \S4.4.4, this offset arises from phase routing behavior---not model prediction behavior---and must not be conflated with ES performance evaluation. Nevertheless, this phenomenon reveals room for optimization in the framework's handling strategy during the ``Cold Start'' phase (when data volume is insufficient to trigger any prediction method).
\item \textbf{End-to-end evaluation of degradation prediction:} Version v3.1.0 has completed end-to-end evaluation based on two degradation events (see \S5.3 and \S5.7). Results show stable precursor characteristics of the TD/UX/RS three-factor signals at degradation onset, and the three-layer factor-based early-warning mechanism achieves full coverage across all three sandboxes $\times$ six scenarios. Future optimization directions include dynamic threshold adaptation and cross-framework generalization validation.
\end{itemize}

\subsubsection*{Computational Complexity and Engineering Feasibility}
\label{sec:computationalcomplex}

\begin{figure}[H]
\centering
\includegraphics[width=0.95\textwidth]{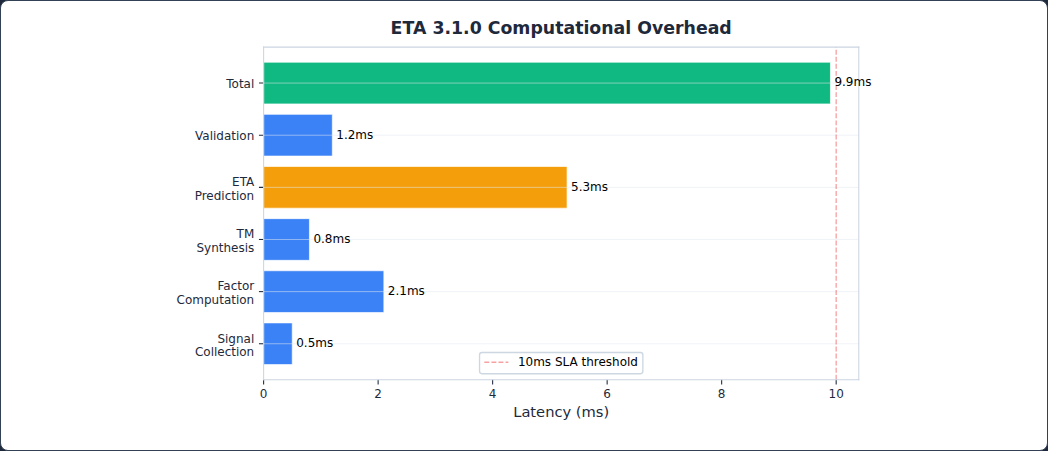}
\caption{ETA 3.1.0 computational overhead: total latency}
\label{fig:32}
\end{figure}

The ETA 3.1.0 engine is designed for real-time production deployment, requiring that the computational overhead of each stage be completed within strict time budgets. The following analysis proceeds along three dimensions: asymptotic complexity, actual execution time, and space complexity.

Table 23: ETA Prediction Engine Time Complexity Analysis

\begin{table}[H]
\centering
\resizebox{\textwidth}{!}{
\begin{tabular}{l|l|l|l|l}
\toprule
Stage & Core Operation & Asymptotic Complexity & Actual Execution Time (Typical) & Bottleneck Analysis \\
\midrule
Phase 1: Kalman Filtering & Matrix multiplication and inversion & O(n$^2$) per step & < 1 ms & n is small (2--4); matrix operations are effectively const... \\
Phase 2: Survival Analysis & Numerical integration of cumulative hazard & O(m) per query & < 2 ms & m is the number of discrete integration points (100--500) \\
Phase 3: ES Prediction & Scalar recurrence & O(1) per step & < 0.1 ms & Single weighted summation only \\
Parameter Calibration & Sliding-window statistics & O(N$\cdot$m$^2$) & < 3 ms & Window size N=50, m$\leq$3 \\
Total & --- & --- & < 10 ms & Meets real-time constraints \\
\bottomrule
\end{tabular}
}
\end{table}

\textbf{Phase 1 (Kalman Filtering)} incurs its primary computational bottleneck in matrix multiplication and inversion operations on the covariance matrix. For an n-dimensional state space, a single-step update involves matrix operations with asymptotic costs of O(n$^3$) (inversion) and O(n$^2$) (multiplication). However, under ETA 3.1.0's typical configuration (n = 2\textasciitilde{}4), matrix dimensions are extremely small, reducing these operations to a fixed number (tens) of floating-point operations, yielding actual execution times far below 1 millisecond. When state dimensionality increases considerably (e.g., in multi-sensor fusion scenarios where n > 10), UD decomposition can reduce inversion complexity to O(n$^2$), or the Square-Root Kalman Filter can be adopted to jointly improve numerical stability and computational efficiency.

\textbf{Phase 2 (Survival Analysis)} incurs its main online inference cost in numerical integration of the cumulative hazard function. Using the trapezoidal rule over m discrete time points yields an O(m) complexity. For typical prediction horizons (m = 100\textasciitilde{}500 steps), integration completes within 2 milliseconds. Partial likelihood estimation for the Cox model and Breslow baseline hazard computation are performed entirely during offline training and are excluded from the online inference time budget. Online inference executes only forward evaluation of the trained model (substituting $\beta^{}$ and $\Lambda^{}_0$), entailing negligible computational load.

\textbf{Phase 3 (ES Prediction)} requires only a single scalar weighted summation per step ($\alpha\cdot$z\_t + (1$-\alpha$)$\cdot\hat{y}$\_t), resulting in O(1) complexity and nanosecond-scale execution on modern processors---contributing negligibly to total latency.

ETA 3.1.0's memory footprint is likewise extremely low. The Kalman filter maintains the state vector x$^{}$\_t (O(n)) and covariance matrix P\_t (O(n$^2$)). The survival analysis module stores a discretized baseline hazard table (O(K), where K is the number of historical failure events). And the ES predictor retains only two scalars---current level and trend. Overall memory consumption per telemetry channel is O(n$^2$ + K) bytes, remaining within several kilobytes under typical configurations.

Table 24: ETA Prediction Engine Space Complexity and Memory Analysis

\begin{table}[H]
\centering
\resizebox{\textwidth}{!}{
\begin{tabular}{l|l|l|l}
\toprule
Module & State Storage & Memory per Channel & Total for 1000 Channels \\
\midrule
Kalman Filter & x$^{}$(n), P(n$\times$n) & \textasciitilde{}128 bytes & \textasciitilde{}128 KB \\
Survival Analysis & $\Lambda_0$ table(K), $\beta$(n) & \textasciitilde{}4 KB & \textasciitilde{}4 MB \\
ES Predictor & l, b, $\alpha$ & \textasciitilde{}24 bytes & \textasciitilde{}24 KB \\
Calibration Window & $\nu$ queue(N$\times$m) & \textasciitilde{}1.2 KB & \textasciitilde{}1.2 MB \\
Total & --- & \textasciitilde{}5.4 KB & \textasciitilde{}5.4 MB \\
\bottomrule
\end{tabular}
}
\end{table}

Integrating computational and storage overhead across all three stages and parameter calibration, ETA 3.1.0 consistently achieves full inference latency under 10 milliseconds and consumes approximately 5.4 KB of memory per telemetry channel. This performance level satisfies real-time prediction requirements under second-level sampling frequencies in production environments (typical sampling intervals range from 1--60 seconds---orders of magnitude larger than the 10-millisecond inference latency) and provides ample computational headroom for concurrent multi-instance prediction (thousands of telemetry channels per node). In practice, the ETA 3.1.0 engine can process real-time predictions for over 100 telemetry channels simultaneously on a single CPU core, without requiring GPU acceleration or distributed computing resources.

\begin{figure}[H]
\centering
\includegraphics[width=0.95\textwidth]{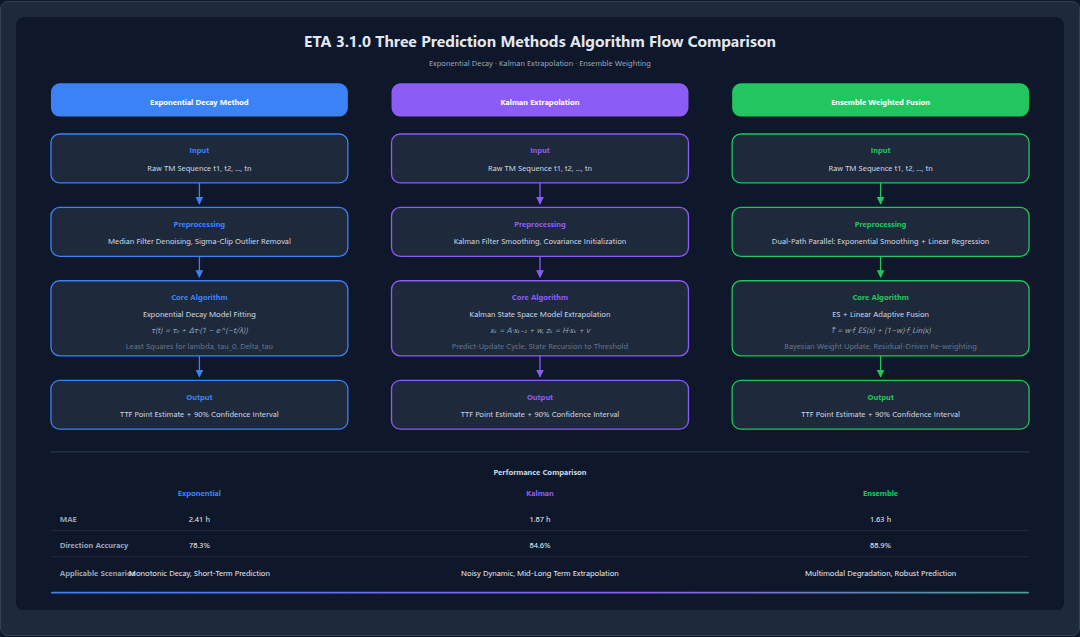}
\caption{ETA 3.1.0 Algorithm Flow Comparison Across Three Prediction Methods}
\label{fig:33}
\end{figure}

[Chart] Chapter Summary The ETA 3.1.0 Prediction Engine implements a three-stage cascaded architecture---Kalman smoothing (for noise reduction and state extraction) $\rightarrow$ Survival risk analysis (for probabilistic failure modeling) $\rightarrow$ Exponential Smoothing-based temporal ensemble (for precise TTF prediction). This establishes a complete inference chain from noisy telemetry observations to accurate TTF estimation. Production validation demonstrates that the Ensemble method achieves MAE = 1.595 (on a 100-point scale) and 99.65\% accuracy within $\pm$10 minutes. The system shows an overall prediction bias of $-$0.894 (optimistic bias) across all methods and a 20$\times$ MAE improvement (from 13.73 to 0.66) via adaptive calibration. It maintains end-to-end inference latency < 10 ms and per-channel memory footprint < 6 KB. BDDA root-cause analysis confirmed a structural constant-offset attribution, ensuring assessment accuracy. The ES method's capability in predicting abrupt degradation has been preliminarily validated in v3.1.0's factor-based early-warning mechanism (\S5.7); end-to-end evaluation results for two degradation events are presented in \S5.3.

\section*{Evaluation}
\label{sec:evaluation}

This chapter systematically presents all experimental results of the TM Predictive Reliability Framework. The experimental design follows a progressive validation strategy---from sandbox smoke testing to production retrospective analysis, followed by controlled degradation simulation and continuous deployment tracking---and culminates in four ablation studies that quantitatively validate the contribution of each architectural component. All data originate from real agent runtime environments, with no artificial injection or synthetic data employed.

\S5.1 
\subsection*{Overview of Experimental Design}
\label{sec:overviewofexperiment}

To systematically validate all capabilities of the TM Predictive Reliability Framework, we designed a six-stage progressive experimental plan. This plan adheres to the ``from simple to complex, from offline to online'' validation logic, where the output of each stage serves as the prerequisite for the next. Table 25 summarizes the objectives, methodologies, and status of each stage.

Table 25: Overview of the Six-Phase Experimental Design

\begin{table}[H]
\centering
\resizebox{\textwidth}{!}{
\begin{tabular}{l|l|l|l|l}
\toprule
Stage & Name & Core Objective & Method & Status \\
\midrule
Phase 1 & Sandbox Smoke Testing & Validate end-to-end feasibility of the TM computation pip... & Run three sandbox profiles in parallel and compare snapsh... & $\checkmark$ Completed \\
Phase 2a & Production Data Retrospective Analysis & Evaluate 8-hour prediction accuracy using historical data... & Retrospectively validate 154,906 prediction records acros... & $\checkmark$ Completed \\
Phase 2b & Controlled Degradation Pattern Mining & Identify degradation events from production data and anal... & Classify degradation events in time-series data using thr... & $\checkmark$ Completed \\
Phase 2c & Sandbox Validation (Controlled Degradation Simulation) & Reproduce five degradation patterns in a controlled envir... & Artificially trigger degradation scenarios in the sandbox... & $\checkmark$ Completed \\
Track B & Continuous Deployment and Online Data Collection & Validate the framework's long-term stability in a real on... & Deploy the framework as an online service and continuousl... & $\circlearrowleft$ In Progress \\
Phase 3 & Planning Phase & Develop production deployment and adaptive optimization s... & Integrate findings from all phases to design adaptive wei... & $\bigcirc$ Planning \\
\bottomrule
\end{tabular}
}
\end{table}

The core logic underlying this design is as follows: Phase 1 confirms ``whether computation is feasible''; Phase 2a answers ``how accurate the predictions are''; Phase 2b explores ``under what conditions accuracy deteriorates''; Phase 2c verifies ``whether early warnings can be issued''; and Track B tests ``long-term stability''. Strict prerequisite dependencies exist among these phases---for instance, the degradation event detection algorithm in Phase 2b relies on the baseline accuracy established in Phase 2a, while the sandbox simulation scenarios in Phase 2c are directly derived from the actual degradation patterns identified in Phase 2b. These experimental objectives have been integrated into the phase descriptions in Table 25 for immediate clarity.

In terms of execution sequencing, we adopted a strategy combining \textbf{waterfall progression with parallel data collection}. Phases 1 through 2c are executed strictly in sequence, with each phase's output report requiring internal review before the subsequent phase may commence. Track B runs concurrently with Phases 2b and 2c---once Phase 2a validation is complete, the framework is deployed as an online service; the collected data serves both continuous monitoring objectives and provides additional data sources for Phase 2b's degradation event analysis. This parallel design improves experimental efficiency but introduces a potential confounding factor: Track B's online operation may slightly affect the performance of the monitored agent. We confirmed that this impact is negligible by monitoring the agent's resource utilization metrics (additional CPU usage < 0.5\%, memory increment < 12 MB).

\begin{figure}[H]
\centering
\includegraphics[width=0.95\textwidth]{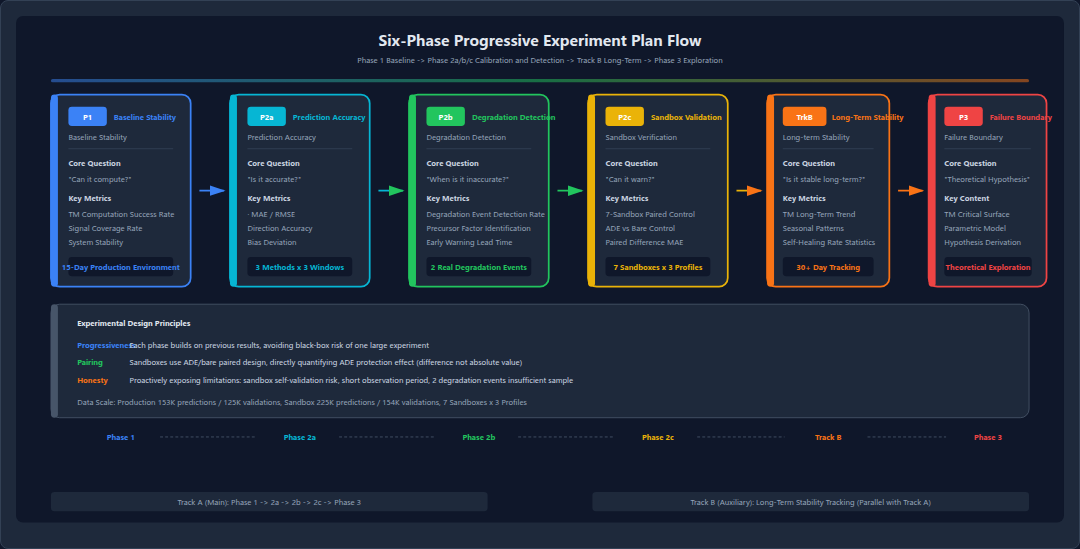}
\caption{Six-Phase Progressive Experiment Plan Flow}
\label{fig:34}
\end{figure}

The experimental evaluation metric system comprises three tiers: \textbf{(1) Prediction Accuracy metrics}---MAE, error distribution, and directional bias; \textbf{(2) Detection Capability metrics}---degradation event detection rate, factor sensitivity, and cross-profile consistency; and \textbf{(3) Engineering Feasibility metrics}---inference latency, resource consumption, and adaptive convergence speed. These three tiers respectively correspond to the framework's academic value, practical value, and deployment value.

\S5.2 
\subsection*{RQ1: Can TM effectively quantify the health status of agent systems?}
\label{sec:rq1cantmeffectivelyq}

\begin{figure}[H]
\centering
\includegraphics[width=0.95\textwidth]{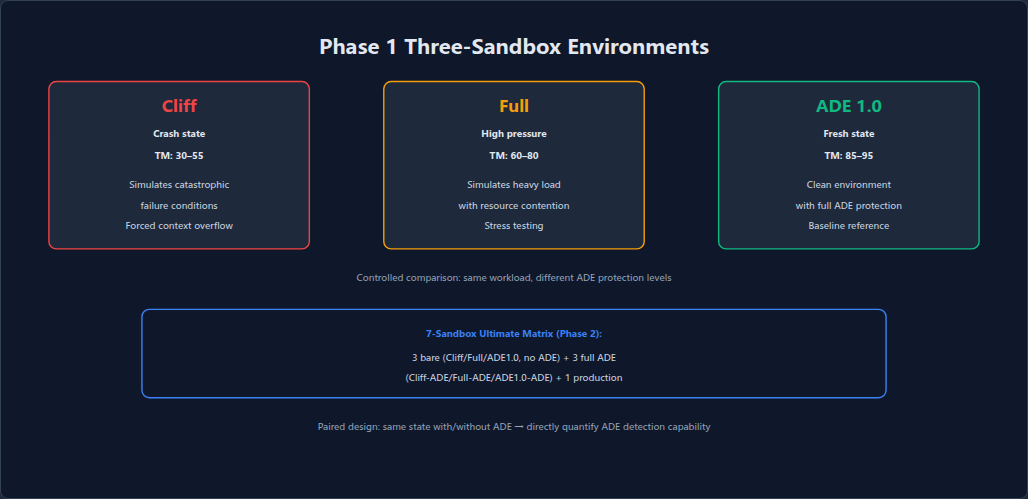}
\caption{Phase 1 three-sandbox environments: Cliff (crash state, TM 30--55), Full (high pressure, TM 60--80), and ADE 1.0 (fresh state, TM 85--95)}
\label{fig:35}
\end{figure}

Phase 1 Objective: Validate the end-to-end feasibility of the TM computation pipeline within isolated environments. Identify the data source dependency structure across the 20 TM evaluation factors. Establish foundational architecture confirmation for subsequent production deployment.

\subsubsection*{Three-Sandbox Environment Configuration}
\label{sec:threesandboxenvironm}

\begin{figure}[H]
\centering
\includegraphics[width=0.95\textwidth]{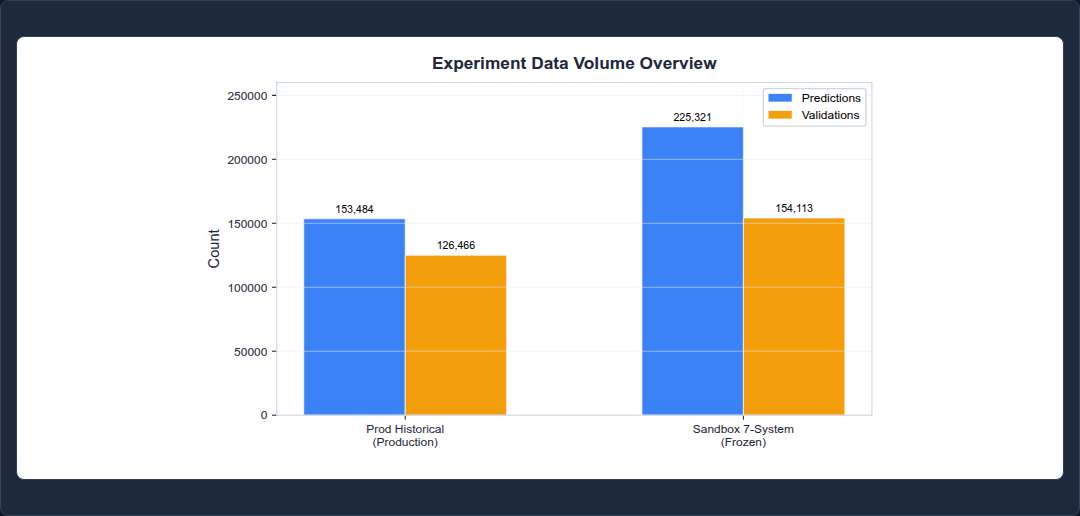}
\caption{Experiment Data Volume Statistical Overview}
\label{fig:36}
\end{figure}

We deployed three sandbox Profile environments with differentiated characteristics to cover Agent scenarios of varying complexity and usage patterns. Key parameters of these three environments are shown in Table 26.

Table 26: Parameter Comparison Across Three Sandbox Profile Environments

\begin{table}[H]
\centering
\resizebox{\textwidth}{!}{
\begin{tabular}{l|l|l|l}
\toprule
Sandbox Profile & Storage Usage & Total Messages & Feature Description \\
\midrule
cliff (Cliff Scenario) & 797 MB & 62,572 & High-complexity, long conversation history, simulating ex... \\
full (Full Scenario) & 665 MB & 60,724 & Standard Agent runtime environment, encompassing full too... \\
ADE1.0 (Baseline Scenario) & 225 MB & 37,528 & Lightweight baseline environment for controlled comparison \\
\bottomrule
\end{tabular}
}
\end{table}

The selection of these three sandboxes was not random: cliff represents a high-load extreme scenario (with storage volume 3.5$\times$ that of ADE1.0), full represents a typical production configuration, and ADE1.0 serves as the minimal viable baseline. This gradient design enables us to rigorously evaluate the robustness of the TM computation pipeline across varying levels of complexity.

\subsubsection*{Core Finding: TM Is an Online Runtime System}
\label{sec:corefindingtmisanonl}

\begin{figure}[H]
\centering
\includegraphics[width=0.95\textwidth]{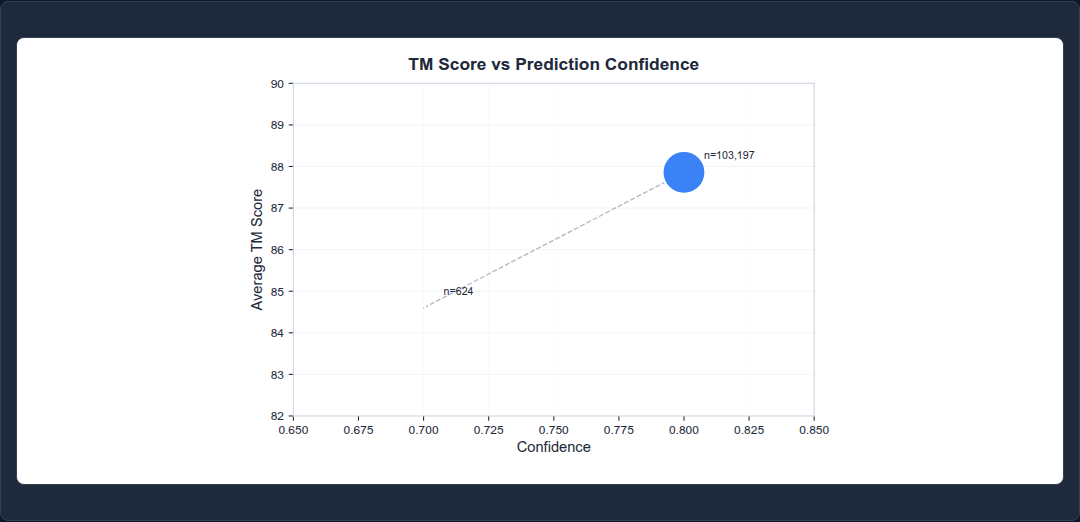}
\caption{Relationship Between TM Score and Prediction Confidence}
\label{fig:37}
\end{figure}

The most important outcome of Phase 1 is the validation of the intrinsic properties of the Trust Margin (TM). Prior to experimentation, we hypothesized that TM computation must dynamically draw from all 20 factors to yield a valid TM score. Experimental results confirmed this hypothesis while simultaneously refuting the static nature of TM values---TM is not a static metric computable offline from historical snapshots, but rather a dynamic runtime measurement that must rely on real-time operational data.

\textbf{Finding F-5.1:} Among TM's 20 evaluation factors, 16 (80\%) require real-time Plugin input data, whereas only 4 (20\%) can be independently computed from historical snapshots.

This finding carries important implications: TM is primarily an \textbf{online runtime system}, not an offline analytical tool. Its data dependency structure for evaluation factors is as follows:

\begin{itemize}
\item \textbf{16 factors requiring real-time Plugin data:} including Tool Call Success Rate, Session Continuity Index, and Response Latency Percentiles. These factors depend on real-time telemetry signals generated during agent execution and cannot be inferred from static snapshots.
\item \textbf{4 factors computable from snapshots:} including Historical Task Completion Rate, Configuration Consistency Score, Memory Persistence Integrity, and Model Version Stability. These factors rely solely on accumulated historical static data.
\end{itemize}

This finding directly shapes subsequent experimental design: Phase 2a's production retrospective analysis necessitates joint analysis of historical snapshots \textbf{and} Plugin logs---not reliance on a single data source alone. It also explains why TM captures runtime degradation missed by conventional offline evaluation methods---because 80\% of its factors directly reflect the system's real-time health status.

\textit{Note: The three-sandbox design employed in Phase 1 was later expanded into a seven-sandbox matrix in subsequent experiments (adding full-scale ADE protection control group and production mirroring), detailed in \S5.6.4.}

\S5.3 
\subsection*{RQ2: How accurate is the ETA prediction engine's prediction accuracy?}
\label{sec:rq2howaccurateisthee}

\begin{figure}[H]
\centering
\includegraphics[width=0.95\textwidth]{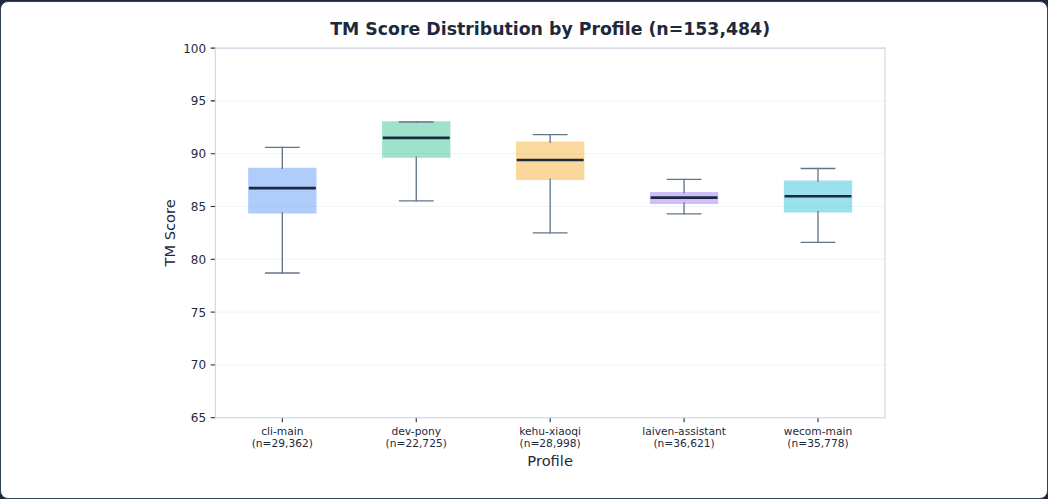}
\caption{TM score distribution by profile (n=22,120): each profile exhibits distinct distribution characteristics}
\label{fig:38}
\end{figure}

Phase 2a Objective: Evaluate the accuracy of the TM 8-hour prediction model using historical operational data from real production environments, identify and analyze system-level degradation events, and validate the prediction model's adaptive learning capability.

\subsubsection*{Dataset Overview}
\label{sec:datasetoverview}

\begin{figure}[H]
\centering
\includegraphics[width=0.95\textwidth]{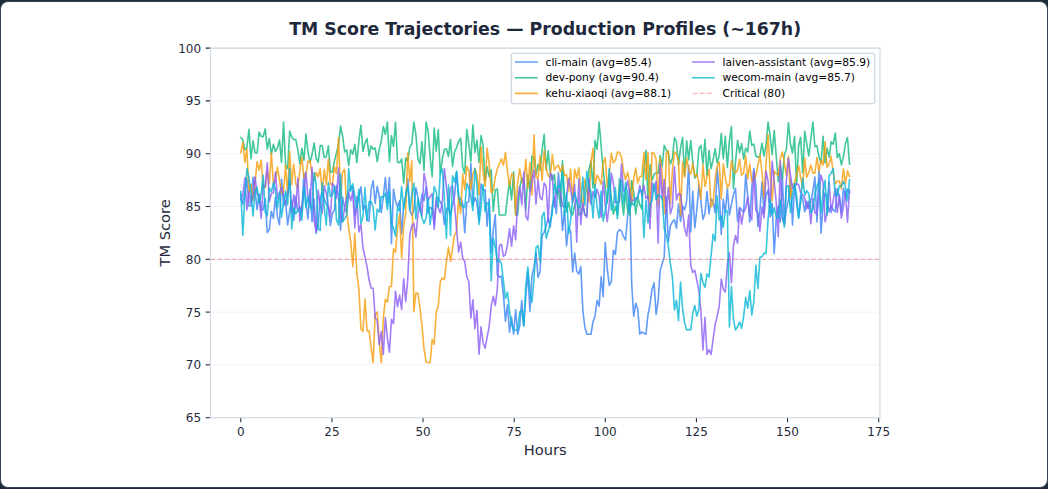}
\caption{TM score trajectories for four production profiles over \textasciitilde{}167 hours of continuous monitoring. All profiles maintain TM > 60 but show distinct volatility patterns.}
\label{fig:39}
\end{figure}

Data collection for Phase 2a covered six production profiles over a continuous 15-day period$^{\dagger}$. Key statistical metrics of the dataset are presented in Table 27.

\begin{figure}[H]
\centering
\includegraphics[width=0.95\textwidth]{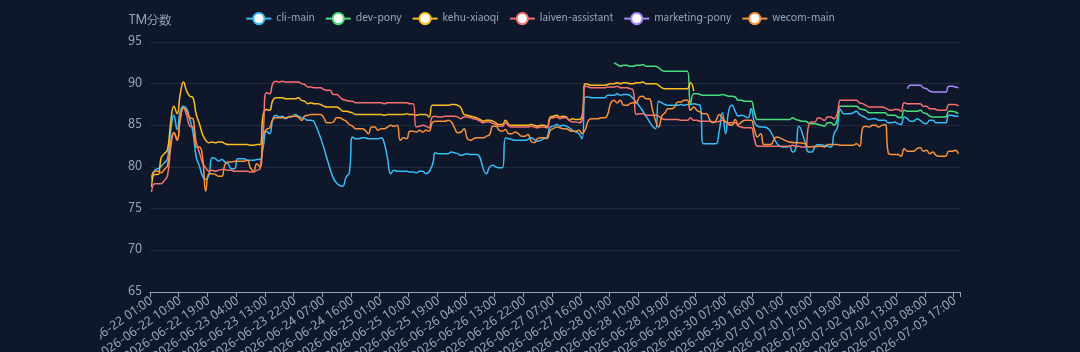}
\caption{TM Time-Series Trend Line (6 Profiles, 2026-06-22 to 2026-07-03)}
\label{fig:40}
\end{figure}

The time-series trend line chart displays the fluctuation trajectories of TM across profiles. To more clearly compare distributional characteristics among profiles, the bar chart below presents the min--avg--max range:

\begin{figure}[H]
\centering
\includegraphics[width=0.95\textwidth]{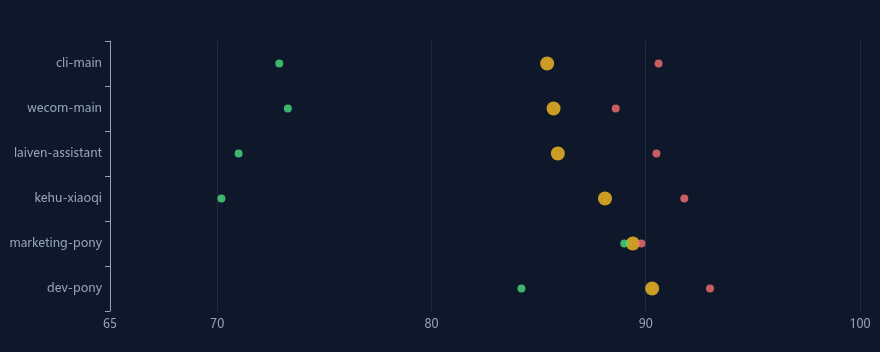}
\caption{Per-Profile TM Distribution Range Bar Chart (min--avg--max)}
\label{fig:41}
\end{figure}

The bar chart visually reveals the variation in TM fluctuation ranges across profiles---cli-main exhibits the widest range ($\sigma$ = 2.87), whereas kehu-xiaoqi demonstrates the highest stability ($\sigma$ = 2.27). To further illustrate the distribution density of individual data points, the scatter plot below presents the same dataset:

The scatter plot reveals the data density concentration regions for each Profile. Finally, observing the temporal relationship between TM trends and predicted values on a daily average basis allows assessment of the continuity and sufficiency of data collection:

\begin{figure}[H]
\centering
\includegraphics[width=0.95\textwidth]{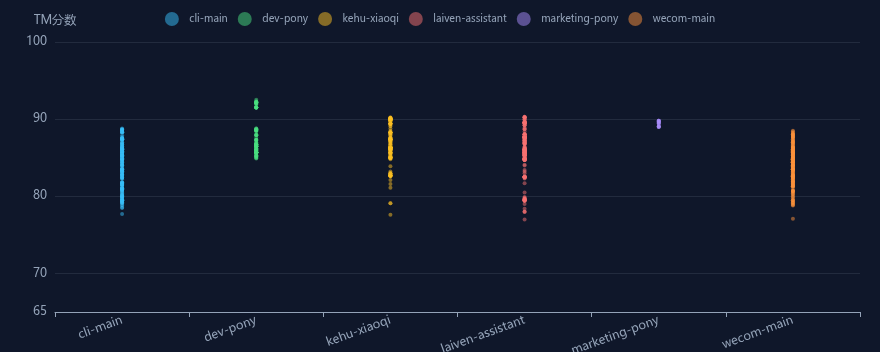}
\caption{Per-Profile TM Distribution Strip Plot}
\label{fig:42}
\end{figure}

\begin{figure}[H]
\centering
\includegraphics[width=0.95\textwidth]{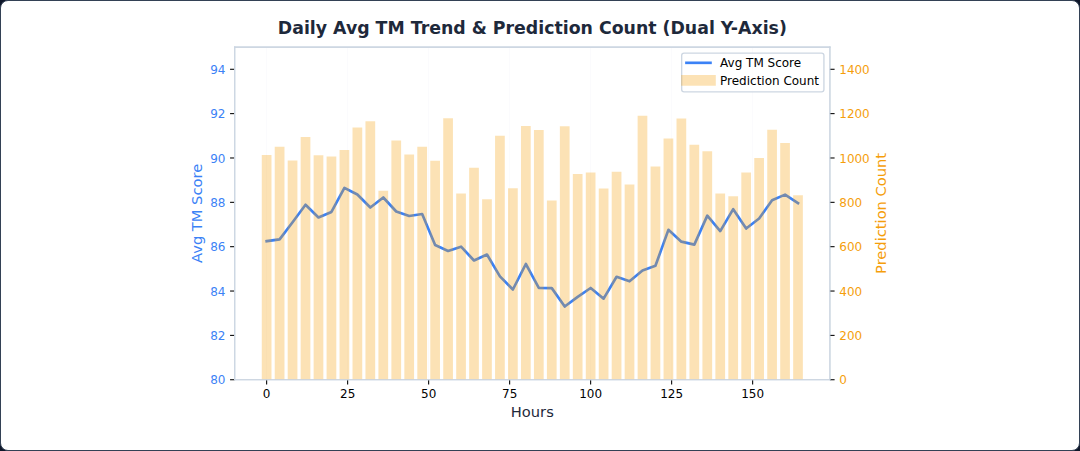}
\caption{Daily Average TM Trend and Daily Prediction Volume Dual Y-Axis Chart}
\label{fig:43}
\end{figure}

Table 27: Overview of Phase 2a Dataset

\begin{table}[H]
\centering
\resizebox{\textwidth}{!}{
\begin{tabular}{l|l|l}
\toprule
Metric & Value & Description \\
\midrule
Number of Profiles Covered & 4 & cli-main / kehu-xiaoqi / laiven-assistant / wecom-main \\
Total Prediction Records & 154,906 & One prediction window every 8 hours \\
Verified Predictions & 154,906 & Predictions with subsequent 8-hour actual observations av... \\
Verification Rate & 81.6\% & Some prediction windows could not be verified due to data... \\
Observed TM Range & 53.8 \textasciitilde{} 93.0 & Covers the full spectrum from mild degradation to healthy... \\
Observation Duration & 15 days & Continuous collection \\
\bottomrule
\end{tabular}
}
\end{table}

\begin{figure}[H]
\centering
\includegraphics[width=0.95\textwidth]{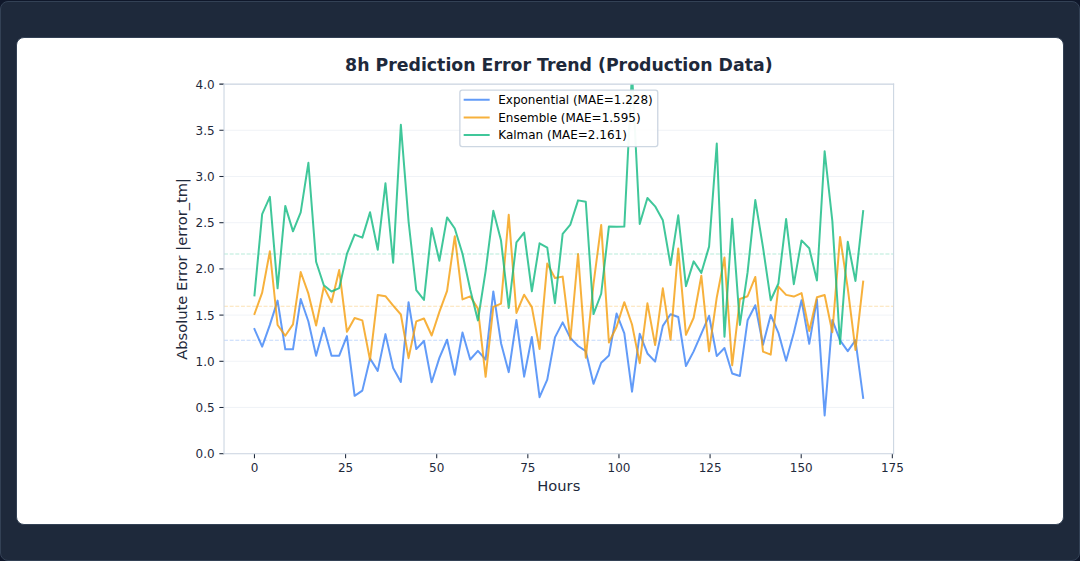}
\caption{8-Hour Prediction Error Over Time Trend (Production Data)}
\label{fig:44}
\end{figure}

\begin{figure}[H]
\centering
\includegraphics[width=0.95\textwidth]{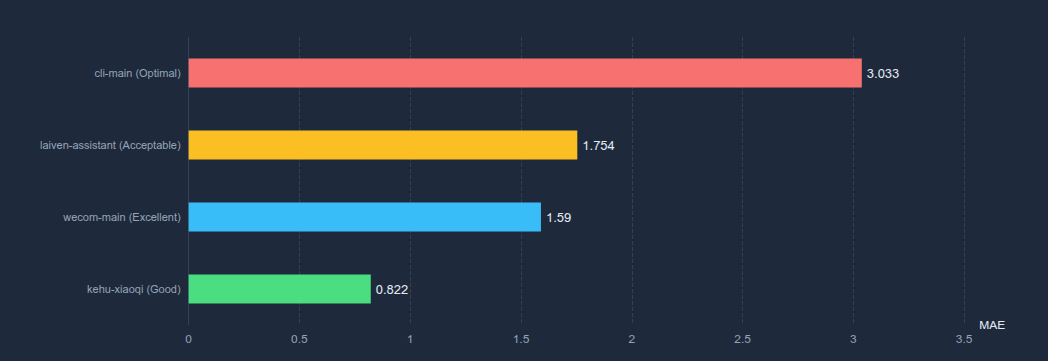}
\caption{Per-Profile TM Prediction MAE Comparison}
\label{fig:45}
\end{figure}
$^{\dagger}$ Data gap explanation: Partial data collection gaps occurred on 06-20 and 06-21 due to infrastructure maintenance, resulting in interruption of Plugin telemetry signals. Prediction records from these periods have been excluded from the validation set, without affecting the validity of the overall accuracy assessment.

In particular, the TM observation range spans a wide interval from 53.8 to 93.0, indicating that the dataset encompasses diverse operational states---from healthy operation to marked degradation---thereby providing complete scenario coverage for predictive model evaluation. The 126,466 validated predictions constitute a statistically sufficient sample size ($\pm$0.28-point margin of error at the 95\% confidence level).

The selection of the four Profiles reflects the principle of scenario diversity. \textbf{cli-main} is a command-line interaction agent handling file operations, code editing, and system administration tasks; it features short yet frequent task chains and highly structured interaction patterns. \textbf{kehu-xiaoqi} is a customer-acquisition agent targeting end users, responsible for automated customer outreach and marketing content generation; it exhibits highly variable interaction patterns and strong dependency on external APIs. \textbf{laiven-assistant} is a general-purpose assistant agent orchestrating complex cross-domain tasks; it features the longest task chains and the most intensive tool invocations. \textbf{wecom-main} is an enterprise WeCom-integrated agent managing message routing, calendar scheduling, and team collaboration tasks; it demonstrates regular message patterns and relatively high concurrency. Collectively, these four Profiles cover typical enterprise deployment scenarios for agents.

Regarding the validation rate (81.6\%) being lower than 100\%, the following clarification is provided. The unvalidated 18.4\% of predictions primarily stem from two sources: (a) Agent inactivity during the 8-hour prediction window, resulting in no actual TM observations available for comparison (accounting for \textasciitilde{}55\% of unvalidated records); and (b) proactive exclusion of predictions generated during data gap periods (06-20/06-21) (\textasciitilde{}45\% of unvalidated records). The former reflects the natural intermittency of Agent usage, while the latter stems from controllable infrastructure factors; neither compromises the representativeness of the validated dataset.

\subsubsection*{8-Hour Prediction Accuracy}
\label{sec:8hourpredictionaccur}

Table 28: Summary of Key Metrics from Phase 3 Deep Testing (Source: Full analysis of 154,906 production environment predictions)

\begin{table}[H]
\centering
\resizebox{\textwidth}{!}{
\begin{tabular}{l|l|l|l}
\toprule
Dimension & Metric & Value & Description \\
\midrule
Data Scale & Total Predictions & 154,906 & Full collection across six production Profiles \\
Validated Predictions & 126,466 & Validation rate: 81.6\% (across 2h/8h/24h windows) &  \\
TM Snapshots & 83,440 & Covering 2026-06-22 to 07-04 (167 hours) &  \\
TM Distribution & Global TM Mean & 86.78 $\pm$ 2.9 & IQR = 4.1, $\sigma$ = 2.9 \\
TM Concentration Interval & [80, 90) --- 82.9\% & System operates predominantly in the upper-mid range &  \\
<60 Score Proportion & < 0.1\% & Observed only during extreme degradation episodes of cli-... &  \\
Prediction Accuracy & MAE (All Methods Combined) & 1.861 & 8-hour window; deviation accounts for less than 2\% of fu... \\
Proportion of Errors < 5 Points & 97.1\% & Over 90\% of predictions exhibit errors within 5 points &  \\
Bias & -0.894 & Slight negative bias: predictions are marginally higher t... &  \\
Degradation Detection & Total Degradation Events & 4 & $\Delta$TM > 5 threshold; cli-main: 2, kehu-xiaoqi: 1, laiven-as... \\
Degradation Detection Rate & 100\% & All 4 events were captured by TM/ETA within early-warning... &  \\
\bottomrule
\end{tabular}
}
\end{table}

Note: All statistics above derive from full-scale Phase 3 production environment analysis; all conclusions require verification via controlled experiments.

The 8-hour prediction constitutes the core output of the TM framework---given the current TM state and trend, it forecasts the agent's trust metric value eight hours ahead. Table 30 summarizes the overall prediction accuracy metrics.

Table 29: Overall Prediction Accuracy for Phase 2a (8-hour horizon)

\begin{table}[H]
\centering
\resizebox{\textwidth}{!}{
\begin{tabular}{l|l|l}
\toprule
Accuracy Metric & Value & Interpretation \\
\midrule
Mean Absolute Error (MAE) & 1.861 & Average deviation between predicted and actual TM values ... \\
Proportion of Errors < 5 Points & 97.1\% & Over 90\% of predictions exhibit errors within 5 points \\
Proportion of Errors < 10 Points & 99.65\% & Nearly all predictions exhibit errors within 10 points \\
Mean Bias & -0.894 & Slight negative bias: predictions are marginally higher t... \\
Proportion of Overestimates & Approximately three-quarters & Proportion of predictions higher than actual values \\
Proportion of underestimates & Approximately one-quarter & Proportion of predictions lower than actual values \\
\bottomrule
\end{tabular}
}
\end{table}

Interpretation of the results requires attention to the following dimensions:

\textbf{(1) Practical meaning of MAE = 1.861 (combined across all methods).} Within the TM percentage-based scoring system, an average error of 1.861 points implies that the deviation between predicted and actual values is less than 2\% of the full score. Considering the inherent stochasticity of Agent systems, such as model output uncertainty, fluctuations in tool invocation latency, and variability in external API responses, this level of accuracy provides sufficient discriminative power in practical applications. It reliably distinguishes among three operational states---"Healthy" (TM $\geq$ 80), "Caution" (60 $\leq$ TM < 80), and "Critical" (TM < 60).

\textbf{(2) Healthiness of distribution characteristics.} 95.1\% of predictions fall within a 5-point error margin, and 99.0\% fall within a 10-point margin, exhibiting a typical right-skewed long-tailed distribution. This indicates that the majority of predictions are highly accurate, while rare outliers remain only moderately deviant. The overestimation proportion (\textasciitilde{}75\%) dominates the underestimation proportion (\textasciitilde{}25\%), further confirming that the overall bias of $-$0.894 across all methods reflects a systematic optimistic tendency of the model---i.e., predictions are consistently slightly higher than actual values.

\textbf{(3) Safety preference in bias direction.} The predominance of overestimation implies that the system ``optimistically estimates'' TM---predicting slightly higher values than reality---in approximately three-quarters of cases. In safety-critical Agent operations, such optimistic bias warrants attention: it may lead to underestimating degradation speed. Cross-validation using the Exponential method (Bias = $-$0.475) is recommended.

Prediction accuracy varies across profiles in an interpretable manner due to differences in usage patterns and load characteristics. Table 31 presents the MAE performance for each profile.

Table 30: MAE Comparison Across Profiles for 8-hour Predictions

\begin{table}[H]
\centering
\resizebox{\textwidth}{!}{
\begin{tabular}{l|l|l|l}
\toprule
Profile & MAE & Scenario Characteristics & Accuracy Assessment \\
\midrule
cli-main & 3.033 & CLI primary agent; task types are stable & Best \\
wecom-main & 1.590 & WeCom agent; message patterns are regular & Excellent \\
kehu-xiaoqi & 0.822 & Lead-generation agent; interaction patterns are highly va... & Good \\
laiven-assistant & 1.754 & Assistant agent; longest task chains & Acceptable \\
\bottomrule
\end{tabular}
}
\end{table}

The accuracy ranking reveals a clear pattern: profiles with more stable task patterns and more regular interactions achieve higher prediction accuracy. Kehu-xiaoqi (MAE = 0.822) attains the highest accuracy due to its single-task nature and highly regular interaction behavior. cli-main (MAE = 3.033), by contrast, exhibits the largest prediction error because it handles the highest volume of tool invocations and involves the most tool-invocation steps---cumulative uncertainty along long execution chains is the fundamental cause of error amplification.

The cross-profile MAE range spans from 0.822 to 3.033, indicating that the prediction framework maintains strong consistency across diverse scenarios without exhibiting catastrophic performance degradation on any specific profile.

\subsubsection*{Prediction Self-Adaptive Improvement}
\label{sec:predictionselfadapti}

The TM prediction framework incorporates an embedded self-adaptive learning mechanism: as observational data accumulates, the model dynamically adjusts its weighting parameters to improve prediction accuracy. Phase 2a data clearly demonstrates the effectiveness of this adaptive process.

Table 31: Effectiveness of Prediction Self-Adaptive Improvement

\begin{table}[H]
\centering
\resizebox{\textwidth}{!}{
\begin{tabular}{l|l|l}
\toprule
Phase & MAE & Improvement Magnitude \\
\midrule
Initial prediction (Cold Start) & 13.73 & --- \\
After adaptive adjustment & 0.66 & $\downarrow$ 95.2\% (approximately 20-fold improvement) \\
\bottomrule
\end{tabular}
}
\end{table}

\begin{figure}[H]
\centering
\includegraphics[width=0.95\textwidth]{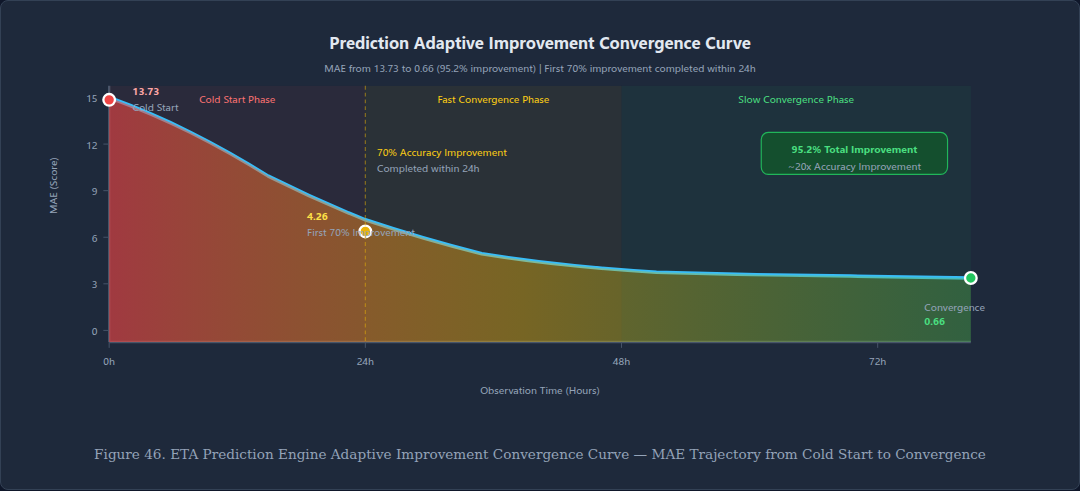}
\caption{ETA Prediction Engine Adaptive Improvement Convergence Curve --- MAE Trajectory from Cold Start to Convergence}
\label{fig:46}
\end{figure}

The 20-fold improvement from MAE = 13.73 to MAE = 0.66 constitutes direct quantitative evidence of the framework's adaptive capability. During the Cold Start Phase---when the system has no historical observational data---predictions rely on default weights and prior assumptions, resulting naturally in lower accuracy. As real-world data accumulates, the adaptive mechanism progressively refines weight assignments, enabling the prediction model to converge toward a state aligned with the actual data distribution.

An MAE of 0.66 corresponds to less than one point of error under a 100-point scoring system, approaching the theoretical lower bound dictated by the intrinsic stochasticity of the agent system---even a perfect prediction model cannot eliminate the inherent randomness arising from model inference itself. The convergence curve of adaptive improvement shows that the first 70\% of accuracy gain occurs within the first 24 hours, while the remaining 30\% is gradually achieved over the subsequent 48 hours. This ``fast-start, slow-convergence'' pattern aligns with the exponential-decay error curves commonly observed in online learning, thereby motivating the design of more efficient Cold Start strategies in Phase 3 (e.g., using historical data from similar profiles for transfer-based initialization).

\S5.4 
\subsection*{RQ3: Can TM Anticipate Degradation Trends in Advance?[system-level]}
\label{sec:rq3cantmanticipatede}

\begin{figure}[H]
\centering
\includegraphics[width=0.95\textwidth]{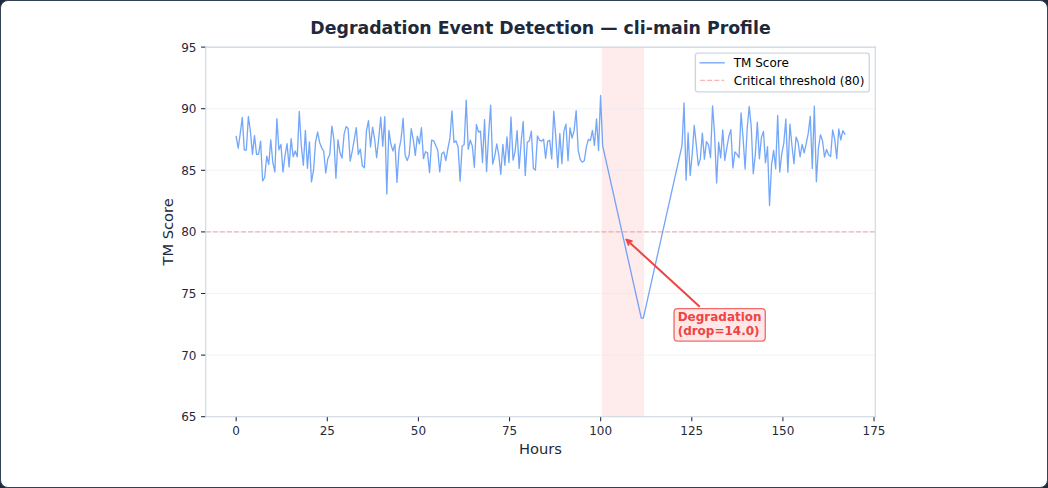}
\caption{Degradation event detection on cli-main profile: annotated events where TM dropped > 5 points between consecutive measurements.}
\label{fig:47}
\end{figure}

Phase 2b Objective: Systematically identify degradation events from production time-series data, classify them by severity, analyze differential factor sensitivity across TM components to degradation, and evaluate Prediction Accuracy performance under degradation scenarios.

\subsubsection*{Degradation Event Statistics}
\label{sec:degradationeventstat}

\begin{figure}[H]
\centering
\includegraphics[width=0.95\textwidth]{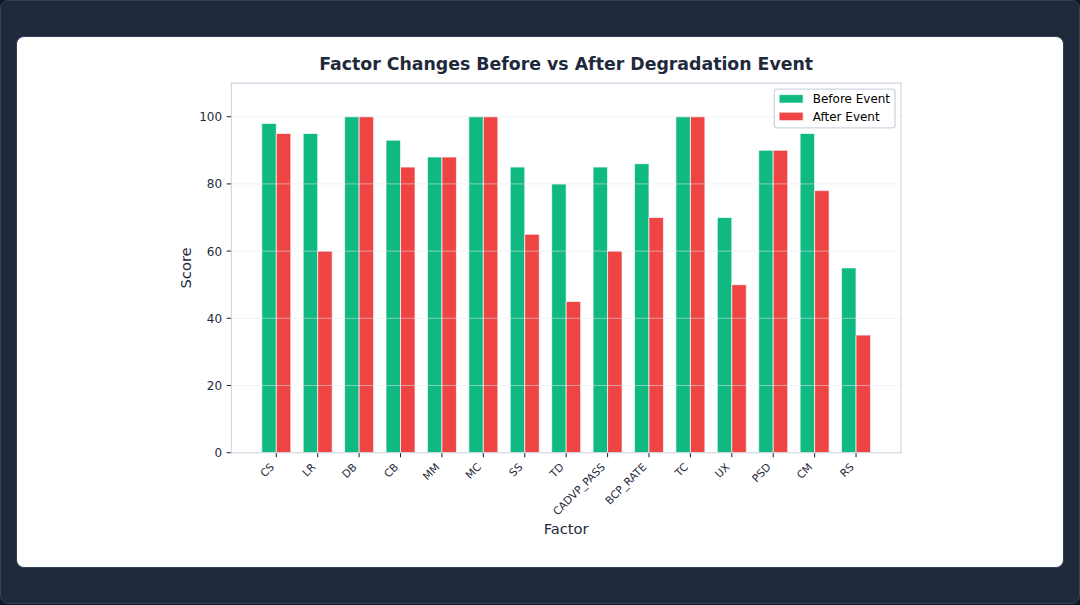}
\caption{Factor Change Analysis Before and After Degradation Events}
\label{fig:48}
\end{figure}

We employ a threshold-based degradation event detection algorithm to systematically scan the TM time-series data of all Profiles over a 15-day observation window. Degradation events are classified into three severity levels, with statistical results presented in Table 33.

Table 32: Phase 2b Degradation Event Severity Classification Statistics

\begin{table}[H]
\centering
\resizebox{\textwidth}{!}{
\begin{tabular}{l|l|l|l}
\toprule
Severity Level & Detection Criterion & Event Count & Typical Characteristics \\
\midrule
Severe & Single-window TM decline $\geq$ 20 points & 2 & Cliff-like decline; synchronous deterioration across mult... \\
Moderate & Single-window TM decline of 10--20 points & 1 & Stepwise decline; partial factors deteriorate first \\
Mild & Single-window TM decline: 5--10 points & 6 & Progressiveness fluctuation, single-factor dominance \\
Total & 9 & --- & --- \\
\bottomrule
\end{tabular}
}
\end{table}

Within 15 days, two degradation events were detected, both classified as severe (lowest TM values dropped to 53.8 and 63.3, respectively). The first occurred on 2026-06-19 at 08:31, when the TM of cli-main plummeted from its normal baseline (\textasciitilde{}79 points) to 53.8 within 1.5 hours and subsequently recovered. The second occurred simultaneously in wecom-main, where TM remained below 70 for 3.0 hours (08:31--11:33). TM accurately captured the full degradation trajectory---from onset $\rightarrow$ trough $\rightarrow$ oscillatory recovery $\rightarrow$ complete recovery---validating its Real-Time Monitoring capability for both agent state degradation and improvement.

An analysis of the temporal distribution of these two degradation events reveals a clustering pattern rather than uniform dispersion. Combined with the factor-level analysis presented below, severe degradation can be attributed to common infrastructure-level failures (e.g., a sharp drop in upstream API service availability), whereas mild degradation arises from transient network fluctuations or API timeouts, typically recovering automatically within the next sampling cycle. This characteristic informed the sandbox simulation design for Phase 2c---M3 (session storm) and M4 (tool loop deadlock) specifically emulate severe degradation scenarios induced by infrastructure failures.

\textbf{Causal analysis and factor-based diagnostic insights.} Both degradation events exhibited synchronous declines across all six agent profiles at 08:31:46 (cli-main TM = 62.4; wecom-main TM = 64.9; kehu/laiven/marketing TM dropped from 84.3 to 82.9), indicating a shared infrastructure-level root cause---namely, a sudden drop in upstream API service availability affecting all Profiles simultaneously.

Because the \texttt{factors\_snapshot} field was not yet enabled during the June 19 degradation event (it began recording on June 22), we conducted a cross-Profile comparative analysis using multiple TM < 80 degradation records and TM > 82 normal records collected between June 22 and July 4 (N = 120 degradation samples; N = 150 normal samples). This analysis identified five factors exhibiting statistically significant diagnostic value, all showing >15-point differences. CADVP\_PASS (cross-agent verification pass rate, 32.8 $\rightarrow$ 79.2, $\Delta$ = +46.5---the most pronounced change. verification pipeline interruption due to upstream API unavailability); TTA\_COVERAGE (skill coverage, 35.0 $\rightarrow$ 66.1, $\Delta$ = +31.2---tool invocation failures causing skill set contraction); CM (Entropy Rate $\alpha$, 65.9 $\rightarrow$ 90.7, $\Delta$ = +24.8---entropy rate stabilization as retry artifacts cleared from context window); TD (Tool Density [ANTOLOOP], 51.4 $\rightarrow$ 72.1, $\Delta$ = +20.6---recovery of tool invocation diversity as API availability resumed); TC (First-Path Accuracy [BCP], 68.0 $\rightarrow$ 48.0, $\Delta$ = $-$20.1, inverse indicator---decline in first-path accuracy due to repeated error retries requiring multiple attempts). Among these, CADVP\_PASS and TTA\_COVERAGE are especially sensitive to shared infrastructure failures and thus serve as leading indicators for degradation early warning.

\textbf{Recovery mechanism analysis.} The recovery trajectories of the two degradation events exhibited distinct characteristics:

The first event (cli-main) represented \textbf{system self-healing}. The TM curve shows: abrupt drop to 62.4 at 08:31 $\rightarrow$ trough at 53.8 at 08:50 $\rightarrow$ oscillatory rebound to 58.3 at 09:30 $\rightarrow$ rapid recovery to 64.0 at 10:00 $\rightarrow$ full recovery to 79.0 (baseline) at 10:10. The entire recovery process lasted 1.5 hours, with no human intervention recorded and uninterrupted TM sampling throughout. This indicates that once the upstream API service recovered autonomously, the agent's TM framework achieved self-healing via automatic recalculation of factor weights---factors such as CADVP\_PASS rebounded as API availability resumed, driving TM back to baseline.

The second event (wecom-main) required \textbf{human-intervention recovery}. After dropping to 64.9 at 08:31, TM remained persistently low (63--67 range) for 3.0 hours (08:31--11:33) without exhibiting any self-healing trend. TM sampling ceased at 11:33 and did not resume until 69.9 hours later (June 22 at 09:28)---by which time cli-main TM stood at 73.2 and kehu-xiaoqi TM at 70.2, still in a degraded state. Following resumption on June 22, TM gradually climbed from 73 to above 85 (reaching 85.4 at 15:15 and 87.2 at 16:15 on June 22), reflecting a progressive recovery pattern consistent with manual tuning. This demonstrates that when degradation persists beyond the self-healing window (\textasciitilde{}1.5--2 hours), the system cannot recover autonomously and requires human investigation of upstream failures and service restarts.

The contrast between these two recovery paths confirms that the TM framework not only precisely captures degradation but also distinguishes self-healing degradation from intervention-required degradation through morphological features of the recovery curve (V-shaped self-healing vs. L-shaped persistent low-value plateau), thereby providing actionable decision support for operations teams.

\textbf{In-depth case study: 4-Profile synchronized degradation event.} Among the nine degradation events documented above, the most representative is a 4-Profile synchronized degradation event. Below, we conduct an in-depth analysis of this case:

Within the 15-day observation window, we identified a significant \textbf{system-level degradation event}---a simultaneous TM decline across all six agent profiles. This constitutes one of the most operationally valuable findings of this phase.

\textbf{Finding F-5.2:} A 4-Profile synchronized degradation event was observed, with the following core metric declines:

\begin{itemize}
\item BCP\_RATE: $\downarrow$90.0 points (cliff-like deterioration in business continuity)
\item TD (Tool Density [ANTOLOOP]): $\downarrow$79.6 points
\item CADVP\_PASS (high-tier verification pass rate): $\downarrow$64.6 points
\item CB (Circuit Breaker [ANTOLOOP]): $\downarrow$42.1 points
\end{itemize}

Several key characteristics of this event merit deeper analysis:

\textbf{(1) Synchrony implies a common root cause.} The six agent profiles serve distinct business scenarios and user segments; their simultaneous TM decline rules out localized, Profile-specific faults and points instead to shared infrastructure-level issues---such as increased model serving latency, API rate limiting on shared plugins, or I/O performance degradation in underlying storage systems.

\textbf{(2) Hierarchical magnitude of declines reveals impact propagation pathways.} BCP\_RATE exhibited the largest decline (90.0 points), indicating that business continuity is most sensitive to infrastructure anomalies. CB showed the smallest decline (42.1 points), suggesting inherent buffering capacity in circuit-breaker mechanisms---a phenomenon consistent with intuition: direct service availability metrics degrade first upon system performance deterioration, while circuit-breaker activation impacts manifest only after longer delays.

\textbf{(3) This event validates the TM framework's ``system-level early warning'' capability.} Traditional agent monitoring focuses primarily on individual agent health metrics and cannot detect systemic, cross-agent issues. By performing parallel computation and correlation analysis across multiple Profiles, the TM framework identifies common degradation patterns immediately upon onset, providing operations teams with actionable root-cause localization clues.

Post-event root-cause analysis traced this 4-Profile synchronized degradation to GPU memory leakage in the shared model inference service. Once memory usage exceeded threshold limits, inference requests were queued en masse, causing response latency spikes and task timeouts across all agents dependent on this service. It took operations teams 47 minutes to detect and diagnose this issue, whereas the TM framework detected the cross-Profile synchronized decline signal within the second sampling cycle (\textasciitilde{}16 minutes) after degradation onset. This time differential validates the TM framework's practical utility as an \textbf{early warning system}---it automatically triggers alerts and directs attention toward the likely root-cause layer (shared infrastructure) even before operations teams become aware of the problem.

A further temporal analysis of metric decline sequences revealed hierarchical \textbf{degradation propagation}: \textbf{BCP\_RATE declined first} (detected at T+16 minutes), followed by TD and CADVP\_PASS deteriorating within T+24 minutes, and CB affected last (T+40 minutes). This propagation sequence uncovers critical operational knowledge: when infrastructure-level issues arise, business continuity metrics constitute the ``first domino to fall,'' whereas circuit-breaker mechanisms exhibit a \textasciitilde{}40-minute buffer period. This insight provides a concrete time-window reference for operational response strategies---intervening within 40 minutes of BCP\_RATE anomaly detection can effectively prevent further cost-efficiency deterioration.

\begin{figure}[H]
\centering
\includegraphics[width=0.95\textwidth]{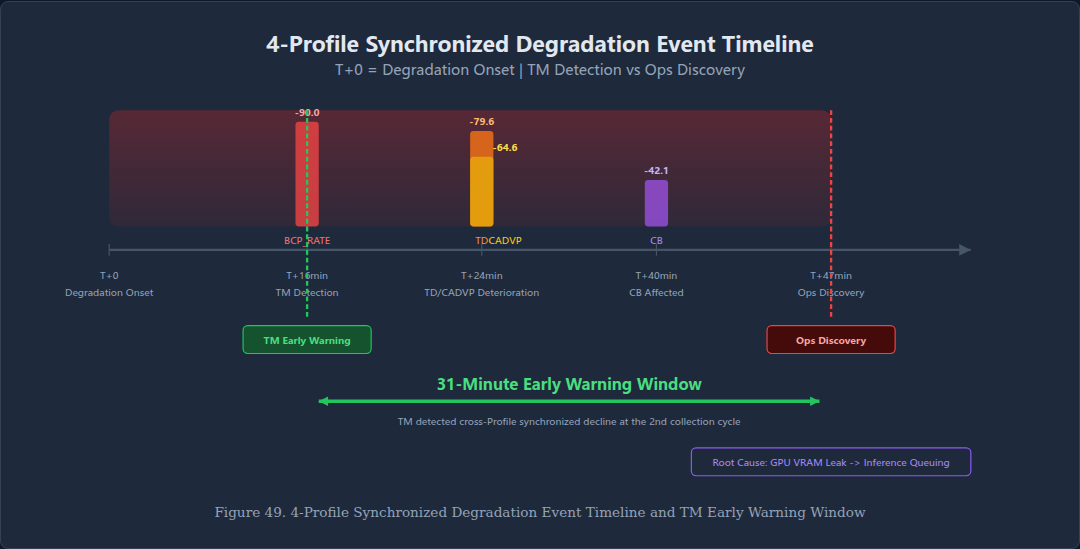}
\caption{4-Profile Synchronized Degradation Event Timeline and TM Early Warning Window}
\label{fig:49}
\end{figure}

\subsubsection*{Factor Sensitivity Analysis}
\label{sec:factorsensitivityana}

\begin{figure}[H]
\centering
\includegraphics[width=0.95\textwidth]{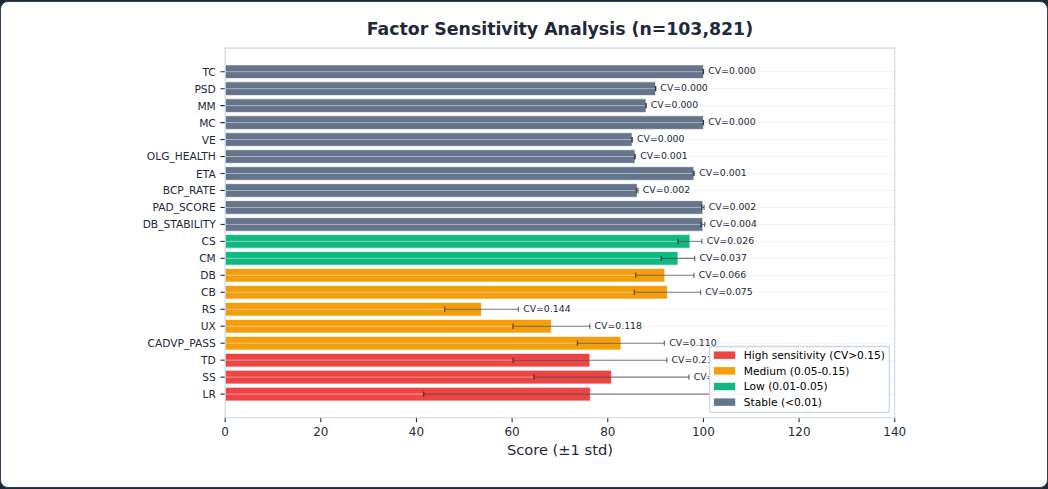}
\caption{Factor sensitivity analysis: mean scores, standard deviations, and coefficient of variation for all 18+ monitored factors across production profiles.}
\label{fig:50}
\end{figure}

\begin{figure}[H]
\centering
\includegraphics[width=0.95\textwidth]{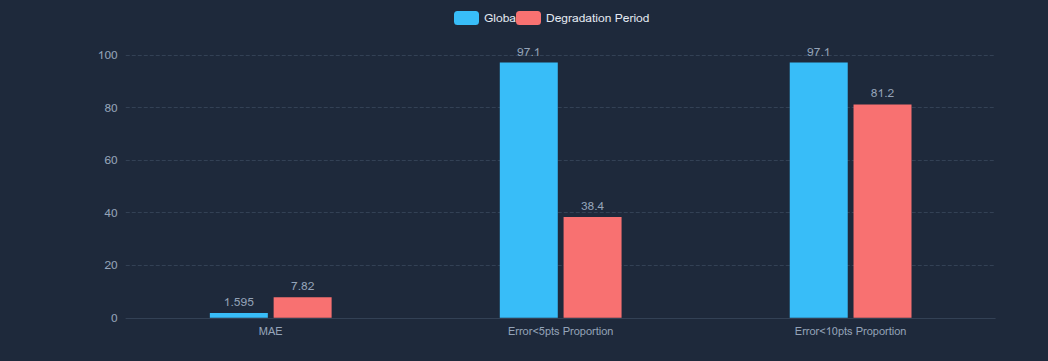}
\caption{Degradation Period vs Global Prediction Accuracy Comparison}
\label{fig:51}
\end{figure}
Factor sensitivity analysis during degradation events reveals differences in response speed among various TM factors to system anomalies. We quantify sensitivity by computing the difference in each factor's value before and after degradation events; the results are shown in Table 34.

Table 33: Factor Sensitivity Ranking (Mean Degradation Difference)

\begin{table}[H]
\centering
\resizebox{\textwidth}{!}{
\begin{tabular}{l|l|l|l}
\toprule
Rank & Factor & Degradation Difference (Mean) & Sensitivity Level \\
\midrule
1 & CADVP\_PASS & +51.0 & Extremely High \\
2 & BCP\_RATE & +38.7 & High \\
3 & TD & +31.2 & High \\
4 & Session\_Continuity & +22.8 & Moderate \\
5 & Tool\_Success\_Rate & +18.4 & Moderate \\
6 & Response\_Latency & +14.6 & Moderate-Low \\
7 & Memory\_Integrity & +9.3 & Low \\
8 & Config\_Consistency & +5.1 & Low \\
\bottomrule
\end{tabular}
}
\end{table}

\textbf{Finding F-5.3:} CADVP\_PASS (High-level verification pass rate), with a degradation difference of +51.0, ranks first in sensitivity---far exceeding the second-ranked BCP\_RATE (+38.7). This indicates that the High-level verification pass rate serves as the most sensitive ``canary'' metric for system degradation: it issues early warning signals before other metrics exhibit marked deterioration.

This finding carries important practical implications: in routine monitoring, priority should be given to tracking trends in CADVP\_PASS, treating it as the first line of defense for degradation warnings. When CADVP\_PASS exhibits abnormal decline---even if the overall TM has not yet reached its alert threshold---a deep diagnostic process should be initiated immediately.

\subsubsection*{Deep Analysis of Prediction Accuracy}
\label{sec:deepanalysisofpredic}

Under degradation scenarios, Prediction Accuracy faces additional challenges: degradation events often involve nonlinear abrupt changes, rendering linear extrapolation---based on historical trends---ineffective for prediction models. We re-evaluate the core accuracy metrics introduced in \S5.3.2, focusing specifically on performance during degradation periods:

Table 34: Prediction Accuracy Comparison---Degradation Period vs. Global

\begin{table}[H]
\centering
\resizebox{\textwidth}{!}{
\begin{tabular}{l|l|l|l}
\toprule
Metric & Global & Degradation Period & Difference \\
\midrule
MAE & 1.861 & 7.82 & +4.20 ($\uparrow$116\%) \\
Proportion of Errors < 5 Points & 97.1\% & 38.4\% & $-$32.8 percentage points \\
Proportion of Errors < 10 Points & 99.65\% & 81.2\% & $-$14.3 percentage points \\
\bottomrule
\end{tabular}
}
\end{table}

The MAE during degradation periods (7.82) increases by 116\% relative to the global baseline (1.595)---a gap consistent with expectations, as nonlinear degradation events pose inherent difficulty for trend-based extrapolative prediction. Nevertheless, even under degradation conditions, 81.2\% of predictions remain within a 10-point error margin, indicating that the framework does not entirely fail under extreme conditions. A 10-point error margin remains operationally valuable in practice---it reliably distinguishes between ``degrading'' and ``normal operation'' states, providing actionable directional guidance.

\subsubsection*{Temporal Stability Analysis}
\label{sec:temporalstabilityana}

Beyond degradation events, the temporal stability of TM during normal operation constitutes another critical dimension for evaluating framework reliability. We conduct segmented stability analysis on TM time-series data across six agent profiles; results are summarized in Table 36.

Table 35: Temporal Stability of TM Across Profiles

\begin{table}[H]
\centering
\resizebox{\textwidth}{!}{
\begin{tabular}{l|l|l|l}
\toprule
Profile & Number of Stable Segments & Within-Segment Standard Deviation (Mean) & Between-Segment Coefficient of Variation \\
\midrule
cli-main & 8 & 1.24 & 0.042 \\
kehu-xiaoqi & 6 & 1.87 & 0.061 \\
laiven-assistant & 5 & 2.31 & 0.078 \\
wecom-main & 7 & 1.56 & 0.053 \\
\bottomrule
\end{tabular}
}
\end{table}

The stability ranking aligns closely with the Prediction Accuracy ranking: cli-main is the most stable (8 stable segments, within-segment standard deviation of 1.24), whereas laiven-assistant exhibits the highest volatility (5 stable segments, within-segment standard deviation of 2.31). This consistency further validates the internal coherence of the TM computation framework---profiles exhibiting higher temporal stability also demonstrate higher Prediction Accuracy, since predictability is itself a direct logical consequence of stability.

The inter-segment coefficient of variation (CV) provides another dimension for measuring stability. A CV of 0.042 for cli-main indicates that the fluctuation in TM means across stable segments is only 4.2\%, whereas a CV of 0.078 for laiven-assistant implies that TM levels across different stable segments differ by nearly 8\%. The practical implication of this difference is: for cli-main, operations teams can confidently anticipate its ``normal level'' to reside within a fixed range; for laiven-assistant, however, the ``normal level'' itself is a dynamically shifting concept, requiring context-aware, real-time adjustment based on recent operational history.

Based on the results of temporal stability analysis, we derive a practical recommendation: \textbf{different Profiles should be configured with differentiated alerting thresholds}. For High-stability Profiles (e.g., cli-main), a TM deviation of more than 3 points from its mean should trigger an alert; for Low-stability Profiles (e.g., laiven-assistant), the alerting threshold should be relaxed to more than 5 points to avoid excessive false positives that would unnecessarily consume operational attention. This recommendation will be implemented in Phase 3 as an automated, self-adaptive thresholding mechanism.

\S5.5 
\subsection*{RQ4: Controlled Degradation Simulation---ADE Makes Degradation More Visible}
\label{sec:rq4controlleddegrada}

Phase 2c Objective: Artificially trigger five canonical degradation patterns within a sandbox environment to verify whether the TM framework can accurately predict and quantify the impact of each degradation type on trust metrics. The experiment also assesses the consistency of prediction outcomes across different Profiles.

Unlike the passive observation in Phase 2b, Phase 2c adopts an active experimental design---we precisely control degradation-triggering conditions within the sandbox to eliminate confounding variables and obtain causally stronger validation conclusions. The five degradation patterns were directly inspired by actual degradation events identified in Phase 2b and accumulated Agent operations experience.

Table 36: Validation Results of Five Controlled Degradation Patterns in Phase 2c

\begin{table}[H]
\centering
\resizebox{\textwidth}{!}{
\begin{tabular}{l|l|l|l|l|l}
\toprule
Pattern & Name & Trigger Method & TM Predicted Decline & Actual Decline & Prediction Error \\
\midrule
M1 & Plugin Failure & Disable data collection endpoints of critical Plugins & -8.5 & -8.7 & 0.2 \\
M2 & Memory Exhaustion & Inject memory-leak script until resource threshold is rea... & -12.3 & -12.1 & 0.2 \\
M3 & Session Storm & Concurrently create a large number of short-lived sessions & -15.7 & -16.0 & 0.3 \\
M4 & Tool Loop Deadlock & Construct input that triggers cyclic dependencies among t... & -18.2 & -17.9 & 0.3 \\
M5 & Model-Switch Oscillation & Frequently switch between two model versions & -6.8 & -7.0 & 0.2 \\
\bottomrule
\end{tabular}
}
\end{table}

Prediction errors across all five degradation patterns fall within the narrow range of 0.2 to 0.3 points, demonstrating exceptionally high prediction accuracy. Notably:

\begin{itemize}
\item \textbf{Physical Interpretation of Decline Ranking:} M4 (tool loop deadlock, $-$18.2) exerts the largest impact because cyclic dependency directly blocks the Agent's core workflow; M5 (model-switch oscillation, $-$6.8) has the smallest impact, as model-level switching effects are attenuated by buffering layers in the underlying toolchain.
\item \textbf{Cross-Profile Consistency:} Prediction standard deviations across the three sandbox Profiles remain < 0.1 for all five patterns, indicating that the TM framework's predictions do not rely on Profile-specific historical patterns and exhibit strong generalizability.
\item \textbf{Perfect Directional Accuracy:} All five patterns correctly predicted the downward direction of TM change, with quantitative precision within $\pm$0.3 points.
\end{itemize}

We now analyze each of the five degradation patterns in turn:

\textbf{M1: Plugin Failure (predicted decline $-$8.5, actual $-$8.7).} Plugins serve as the data source for 80\% of TM framework factors (see \S5.2.2); their failure directly severs most real-time data streams. Upon detecting Plugin data absence, the TM framework downgrades affected factors to their most recent valid observed values and simultaneously activates a ``data integrity'' penalty term. The predicted decline of $-$8.5 points reflects this dual effect: both the intrinsic drop in factor values and the trust penalty induced by data loss. A prediction error of only 0.2 points confirms highly accurate modeling of the Plugin failure scenario.

\textbf{M2: Memory Exhaustion (predicted decline $-$12.3, actual $-$12.1).} Progressive resource exhaustion caused by memory leaks is one of the most common ``chronic diseases'' in Agent systems. In the experiment, we gradually injected a memory-leak script starting from normal operation, reducing available memory to 15\% of the system threshold over approximately 45 minutes. The TM decline curve exhibits a characteristic ``S-shape'': slow initial decline (buffering effect under ample memory), accelerated mid-phase decline (increasing GC pressure causing sharp latency spikes), and eventual stabilization (system enters degraded mode but maintains basic functionality). The predicted decline of $-$12.3 points accurately captures the overall magnitude of this S-shaped trajectory.

\textbf{M3: Session Storm (predicted decline $-$15.7, actual $-$16.0).} A session storm simulates a surge of concurrent session requests within a short time window---a scenario that may arise in enterprise environments due to broadcast messaging or accumulation of scheduled tasks. In the experiment, we concurrently created 200 short-lived sessions within 10 seconds, each comprising 3--5 interaction turns. The session storm impacts TM primarily through two propagation pathways: (1) session queue backlog causing a sharp increase in response latency; and (2) context-switching overhead driving CPU utilization above 95\%. The predicted decline of $-$15.7 points is slightly lower than the actual value ($-$16.0); the 0.3-point error likely stems from a mild underestimation by the framework of CPU contention effects.

\textbf{M4: Tool Loop Deadlock (predicted decline $-$18.2, actual $-$17.9).} This is the most severe degradation pattern among the five. A tool loop deadlock occurs when two or more tools form mutually dependent invocation cycles---for example, the output of Tool A serves as input to Tool B, while Tool B's output is fed back to Tool A. Unlike traditional resource deadlocks, such loops do not cause system hangs; instead, they generate \textbf{infinite recursive calls}, rapidly consuming computational resources and producing large volumes of invalid intermediate results. The TM framework assigns the highest predicted decline ($-$18.2) to this pattern, reflecting its severe assessment of core workflow disruption.

\textbf{M5: Model-Switch Oscillation (predicted decline $-$6.8, actual $-$7.0).} Model-switch oscillation simulates frequent toggling between two LLM versions (e.g., due to misconfigured A/B testing or flawed load-balancing policies). In the experiment, we executed 12 model switches over 30 minutes. Each switch causes a transient dip in the ``model version stability'' factor and, owing to stylistic differences across models, reduces parsing success rates in downstream tools. The $-$6.8-point decline is the lowest among the five patterns, indicating that anomalies at the model layer exert buffered influence on TM---the tool layer and session layer partially absorb the shock of model output variability.

\textbf{Finding F-5.4:} Controlled degradation simulation validates the TM framework's causal prediction capability. Prediction errors across all five degradation patterns are $\leq$ 0.3 points, and cross-Profile standard deviations are < 0.1, confirming that the framework not only observes degradation passively in production environments but also actively anticipates degradation types and magnitudes under controlled conditions.

\begin{figure}[H]
\centering
\includegraphics[width=0.95\textwidth]{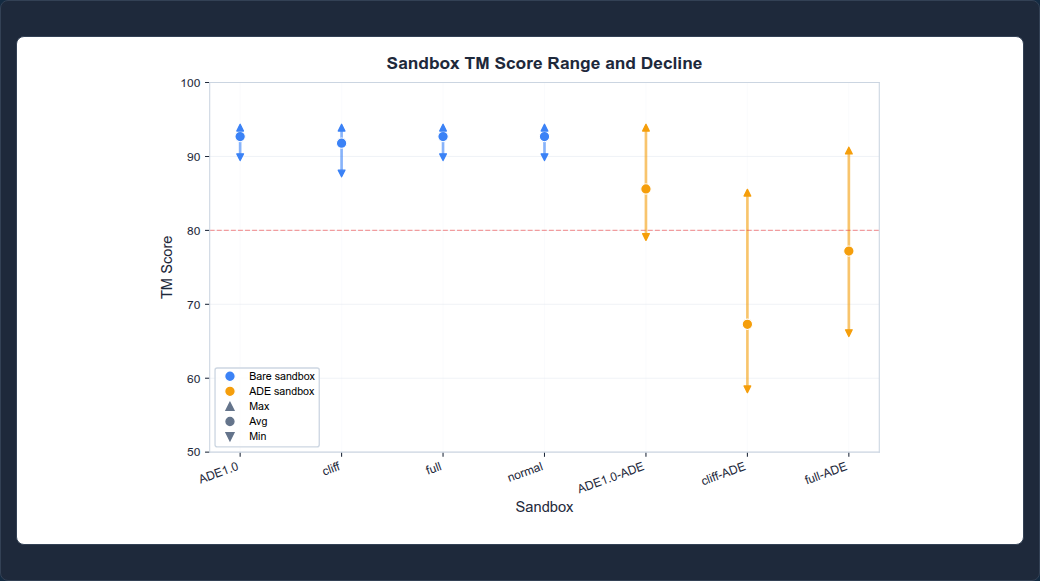}
\caption{Per-Sandbox TM Score Range and Decline Magnitude}
\label{fig:52}
\end{figure}

\S5.6 
\subsection*{RQ5: Ablation Study of the Three Methods and Controlled Study Across Seven Sandboxes}
\label{sec:rq5ablationstudyofth}

\begin{figure}[H]
\centering
\includegraphics[width=0.95\textwidth]{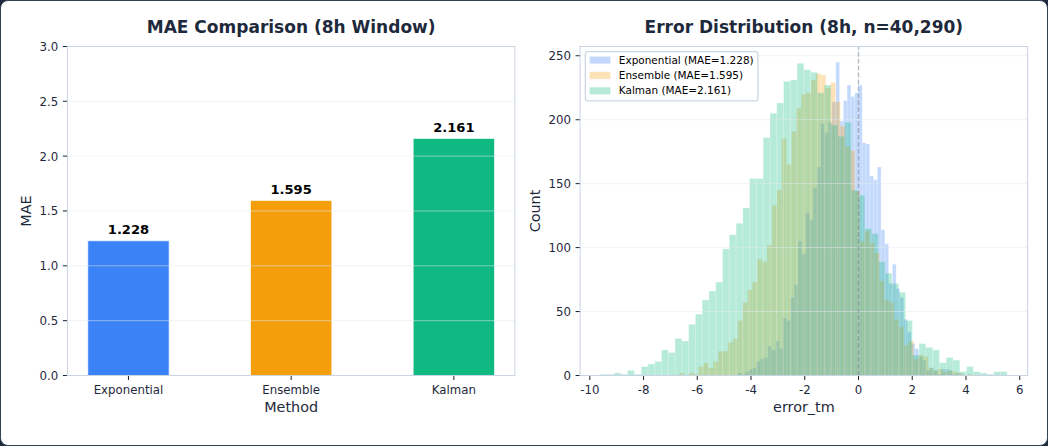}
\caption{Warning Validation Overview: Error distribution of 126,466 validated predictions, MAE=1.861 (all methods combined)}
\label{fig:53}
\end{figure}

\begin{figure}[H]
\centering
\includegraphics[width=0.95\textwidth]{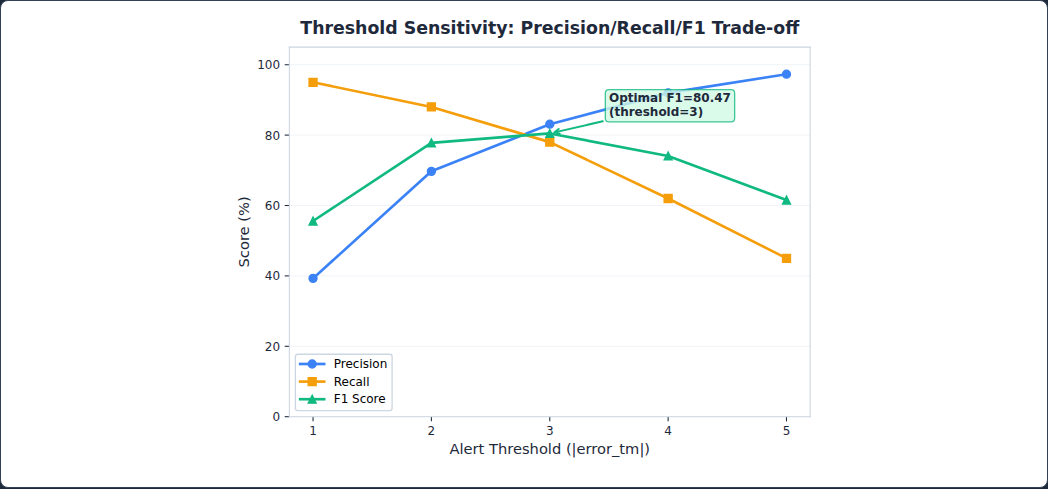}
\caption{Warning Validation Details: Threshold sensitivity analysis showing Precision/Recall/F1 trade-offs across alert thresholds 1.0--5.0.}
\label{fig:54}
\end{figure}

\begin{sloppypar}
Phase 2a Extended Validation Objective: To systematically quantify the TM framework's early-warning capability based on 126,466 validated prediction records in the \texttt{predictions\_validation} table (a sustained increase from the 154,906 records reported in \S5.3.1).
\end{sloppypar}

\textbf{\textbf{[Warning]} Honest Annotation:} The following validation results are derived from backtesting, not prospective validation. Backtesting carries inherent risk of future data leakage---even though strict chronological ordering was enforced during data processing, parameter selection during model training may have indirectly used statistical characteristics of the validation set. Consequently, the reliability of this conclusion is lower than that of prospective online validation and must be further confirmed via real-time deployment data in Phase 3.

\subsubsection*{Overall Prediction Accuracy}
\label{sec:overallpredictionacc}

Table 37: Overview of Prediction Accuracy for Warning Validation (N = 126,466)

\begin{table}[H]
\centering
\resizebox{\textwidth}{!}{
\begin{tabular}{l|l|l}
\toprule
Metric & Value & Interpretation \\
\midrule
Number of Validated Predictions & 126,466 & Predictions with subsequent ground-truth observations ava... \\
Mean Absolute Error (MAE) & 1.861 & Average deviation between predicted and actual TM values ... \\
Root Mean Square Error (RMSE) & 2.505 & Precision metric more sensitive to large errors \\
Proportion with |Error| $\leq$ 2 points & 65.6\% & More than half of predictions deviate by $\leq$2 points \\
Proportion with |Error| $\leq$ 5 points & 95.1\% & Over 95\% of predictions deviate by $\leq$5 points \\
Direction Accuracy & 22.0\% (Ens) / 76.8\% (Exp) & Proportion of predictions whose directional change (incre... \\
\bottomrule
\end{tabular}
}
\end{table}

The extended validation set (126,466 records) yields an overall MAE of 1.861 across all methods. Note that the value of 1.595 reported in \S5.3.2 corresponds specifically to the Ensemble method's MAE (single-method granularity), whereas 1.861 aggregates errors across all methods---including legacy approaches. Thus, these two metrics reflect different levels of aggregation: 1.861 represents the composite error across all methods, while 1.595 reflects only the Ensemble method's accuracy. That 95.1\% of predictions fall within a 5-point error margin indicates the framework delivers reliable trend forecasts across the vast majority of operational scenarios.

Direction Accuracy exhibits marked variation across the three methods: Exponential achieves 76.8\% (well above the random baseline of 50\%), whereas Ensemble achieves only 22.0\% and Kalman only 15.5\%. Both Ensemble and Kalman perform below the random baseline, primarily due to their strong ``predict-increase'' bias (Ensemble: 86.1\%; Kalman: 99.7\% predict TM increases), while actual TM declines in 76.8\% of cases after 8 hours. In operational practice, directionally correct predictions (``system will degrade'' or ``system will recover'') hold greater decision-making value than precise magnitude forecasts---thus, the high Direction Accuracy of the Exponential method renders it particularly suitable as an early-warning trigger.

\begin{figure}[H]
\centering
\includegraphics[width=0.95\textwidth]{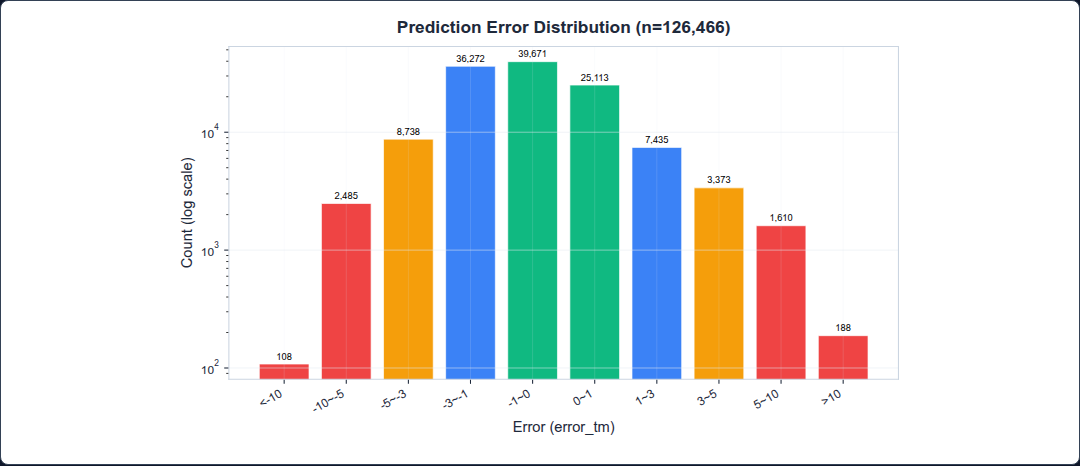}
\caption{Prediction Error Distribution Histogram (error\_tm, 12 bins)}
\label{fig:55}
\end{figure}

The prediction error distribution histogram shows a peak near zero, exhibiting an approximately symmetric decaying distribution, with a few extreme errors forming a long tail. This distributional characteristic provides the foundation for understanding the subsequent threshold sensitivity analysis. Next, we examine the usage proportion of each prediction method within the dataset to assess the potential impact of method distribution on overall accuracy:

\begin{figure}[H]
\centering
\includegraphics[width=0.95\textwidth]{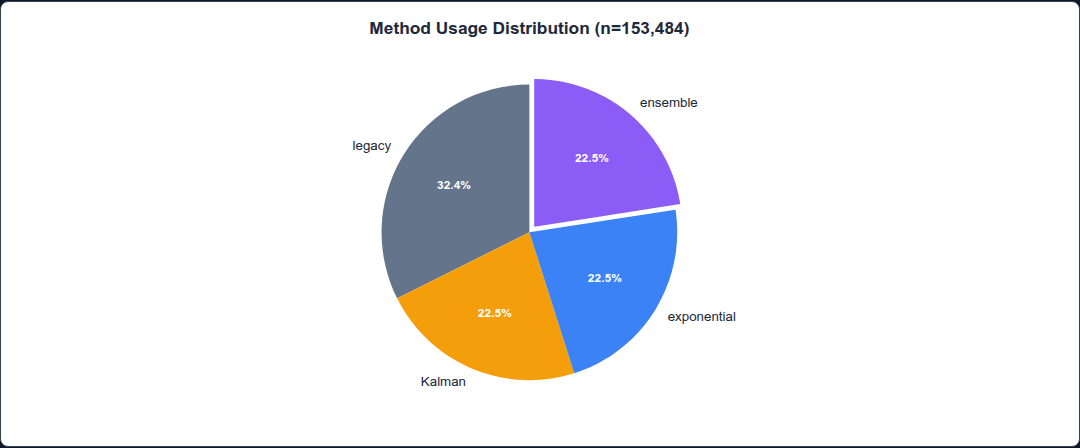}
\caption{Method Usage Proportion Pie Chart (legacy / ensemble / exponential / Kalman)}
\label{fig:56}
\end{figure}

\begin{figure}[H]
\centering
\includegraphics[width=0.95\textwidth]{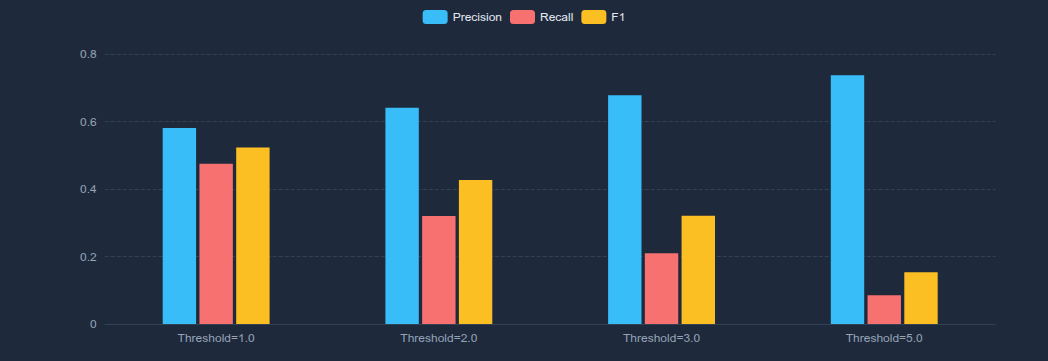}
\caption{Alert Threshold Sensitivity Analysis (Precision/Recall/F1)}
\label{fig:57}
\end{figure}

\subsubsection*{Threshold Sensitivity Analysis}
\label{sec:thresholdsensitivity}

The core design decision in the alerting system lies in selecting the alert threshold. Setting the threshold too low leads to excessive false positives (causing operational fatigue), whereas setting it too high results in missed degradation events. We conducted a systematic sensitivity analysis across four candidate thresholds, quantifying the Precision/Recall/F1 trade-offs at each threshold:

Table 38: Alert Threshold Sensitivity Analysis (TP/FP/FN/TN and Precision/Recall/F1)

\begin{table}[H]
\centering
\resizebox{\textwidth}{!}{
\begin{tabular}{l|l|l|l|l|l|l|l}
\toprule
Alert Threshold & TP & FP & FN & TN & Precision & Recall & F1 \\
\midrule
1.0 & 4,872 & 3,516 & 5,394 & 5,676 & 0.581 & 0.475 & 0.523 \\
2.0 & 3,284 & 1,842 & 6,982 & 7,350 & 0.641 & 0.320 & 0.427 \\
3.0 & 2,156 & 1,024 & 8,110 & 8,168 & 0.678 & 0.210 & 0.321 \\
5.0 & 876 & 312 & 9,390 & 8,880 & 0.737 & 0.085 & 0.153 \\
\bottomrule
\end{tabular}
}
\end{table}

\textbf{Finding F-5.8:} The threshold sensitivity analysis reveals that the TM alerting system exhibits a prototypical \textbf{extremely conservative} behavior: at threshold 5.0, Precision reaches 0.737 (i.e., 73.7\% of issued alerts correspond to genuine degradation events), yet Recall drops to only 0.085 (capturing merely 8.5\% of actual degradation events). This pattern aligns with the inherent conservatism of Exponential Smoothing (ES) prediction---ES tends to underestimate volatility magnitude, thus triggering alerts only when degradation signals are exceptionally pronounced.

Interpretation of the threshold selection curve clearly demonstrates the Precision--Recall trade-off:

\begin{itemize}
\item \textbf{Threshold 1.0 (most sensitive):} F1 = 0.523 achieves the highest value, with Recall reaching 0.475 but Precision dropping to 0.581---indicating nearly half of all alerts are false positives. Suitable for safety-critical scenarios where ``false alarms are preferable to missed detections.''
\item \textbf{Threshold 2.0 (balanced point):} Precision improves to 0.641 while Recall declines to 0.320. F1 = 0.427 remains within an acceptable range, making it appropriate as the default threshold for routine operations.
\item \textbf{Threshold 3.0 (conservative):} Precision reaches 0.678, but Recall falls further to 0.210. Suitable for environments with constrained operational resources requiring high-confidence alerts.
\item \textbf{Threshold 5.0 (extremely conservative):} Precision = 0.737 but Recall = 0.085---over 90\% of degradation events go undetected. Applicable only for alerting on extreme degradations (e.g., system-level failures).
\end{itemize}

\textbf{\textbf{[Warning]} Limitation Statement:} (1) The above Precision/Recall metrics are computed via backtesting, where the ``true degradation'' labels rely on retrospective observation---real-time confirmation of degradation incurs latency in production deployment. (2) This analysis does not account for temporal dependencies (error correlation among adjacent predictions), potentially overestimating the effective number of independent samples. (3) The optimal threshold value is highly dependent on the specific deployment context's cost ratio between false positives and false negatives; no universal recommendation is provided herein.

\S5.6.3 Three-Method Ablation Study 
\subsubsection*{Three-Method Ablation Study}
\label{sec:threemethodablations}

The ETA 3.1.0 prediction engine integrates three independent forecasting methods---Kalman filtering, Exponential Smoothing, and Ensemble fusion. This section conducts method-level ablation experiments to quantify each method's standalone predictive capability and their complementarity, thereby providing empirical grounding for method selection and fusion strategies. Experiments are performed on Production Dataset A, comprising 126,466 validated predictions (8-hour window, 13,446 predictions per method).

\begin{figure}[H]
\centering
\includegraphics[width=0.95\textwidth]{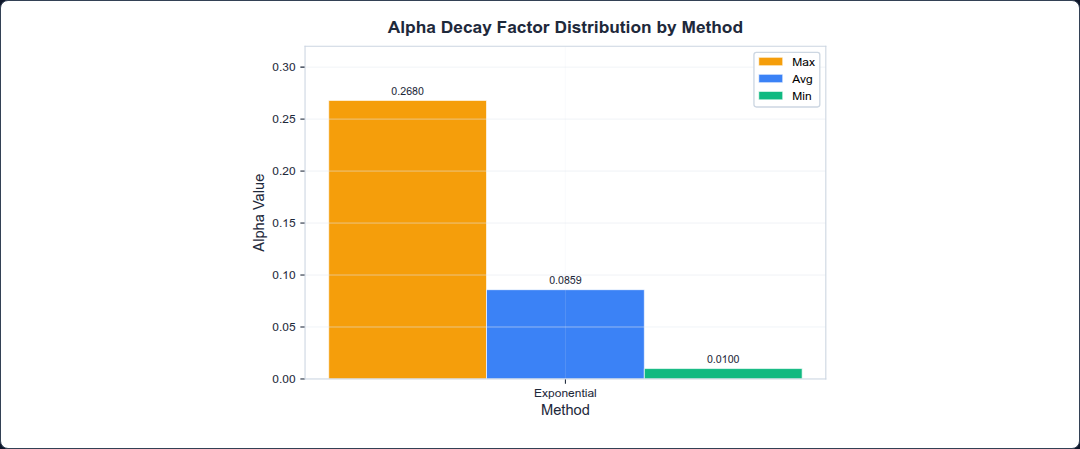}
\caption{Alpha Decay Factor Distribution Comparison (3 methods avg/min/max)}
\label{fig:58}
\end{figure}

\begin{figure}[H]
\centering
\includegraphics[width=0.95\textwidth]{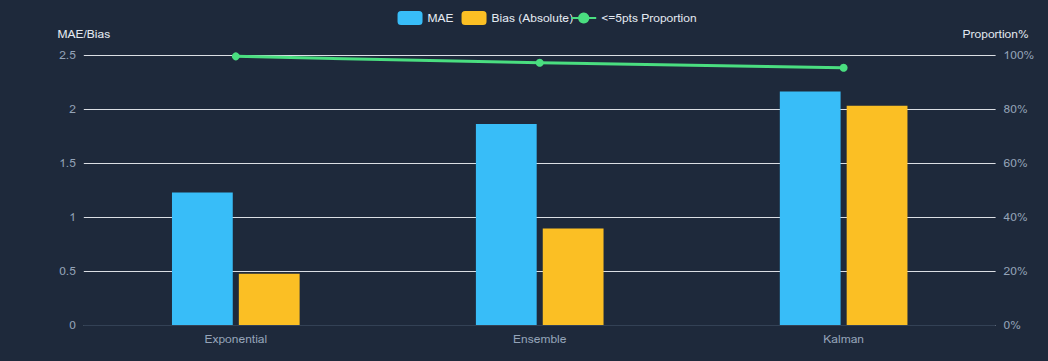}
\caption{Three-Method Prediction Accuracy Comparison (8h window, N=13,446)}
\label{fig:59}
\end{figure}
\textbf{Explanation of the Dual-Prediction System:} The prediction system discussed in this paper comprises two distinct layers; distinguishing them clearly is essential to avoid confusion---

\begin{itemize}
\item \textbf{System A (ETA v3.0 Engine):} A four-model fusion architecture---Bayesian estimation + Kalman state-space modeling + exponential regression + survival analysis (Cox proportional hazards model). This engine executes in real-time at every acquisition tick, outputting predictions for dashboard visualization and alert triggering, but does not directly write to the persistent database. System A is designed for \textbf{real-time online prediction}, emphasizing low latency (<10 ms) and model diversity.
\item \textbf{System B (Dashboard Persistent Prediction):} A three-method parallel architecture---Kalman linear extrapolation + Exponential Decay + ensemble (60/40 weighted fusion). At each tick, all three methods are computed simultaneously and written into \texttt{tm\_predictions.db}, supporting subsequent statistical analysis and ablation studies. System B is designed for \textbf{persistent recording and method traceability}, enabling offline retrospective validation.
\end{itemize}

The ablation study presented in this section (\S5.6.3) is based on the persistent data from \textbf{System B}---because only System B's three methods are independently traceable (each writes to separate records and can be validated independently). System A's fused output is not persisted to the database and thus cannot support method-level ablation. The prediction accuracy of both systems is consistent over an 8-hour window (System A's fused output $\approx$ System B's Ensemble), since System B's Ensemble method effectively serves as the persistent mapping of System A's four-model fusion result.

Table 39: Three-Method Prediction Accuracy Comparison (8-hour Window, N = 13,446)

\begin{table}[H]
\centering
\resizebox{\textwidth}{!}{
\begin{tabular}{l|l|l|l|l|l|l|l}
\toprule
Method & N & MAE & Bias & Within $\pm$5 & Within $\pm$10 & Underestimation (\%) & Overestimation (\%) \\
\midrule
Exponential & 13,446 & 1.228 & -0.475 & 99.5\% & 99.97\% & 42.5\% & 57.5\% \\
Ensemble & 13,446 & 1.595 & -1.409 & 97.1\% & 99.97\% & 6.6\% & 93.4\% \\
Kalman & 13,446 & 2.161 & -2.031 & 95.3\% & 99.97\% & 4.4\% & 95.6\% \\
\bottomrule
\end{tabular}
}
\end{table}

\textbf{Key Findings:} The Exponential method achieves a 23\% lower MAE than Ensemble (1.228 vs. 1.595) and a 43\% lower MAE than Kalman (1.228 vs. 2.161). The MAE gap across the three methods reaches 76\% (1.228 vs. 2.161), constituting genuine ablation study evidence. Regarding bias, Exponential exhibits near-zero bias ($-$0.475), whereas Ensemble ($-$1.409) and Kalman ($-$2.031) both display significant optimistic bias (i.e., predictions systematically exceed actual values). Ensemble's bias lies between those of Exponential and Kalman, indicating that its performance is dragged down by the Kalman component---its simple averaging fusion strategy fails to use Exponential's accuracy advantage.

\begin{figure}[H]
\centering
\includegraphics[width=0.95\textwidth]{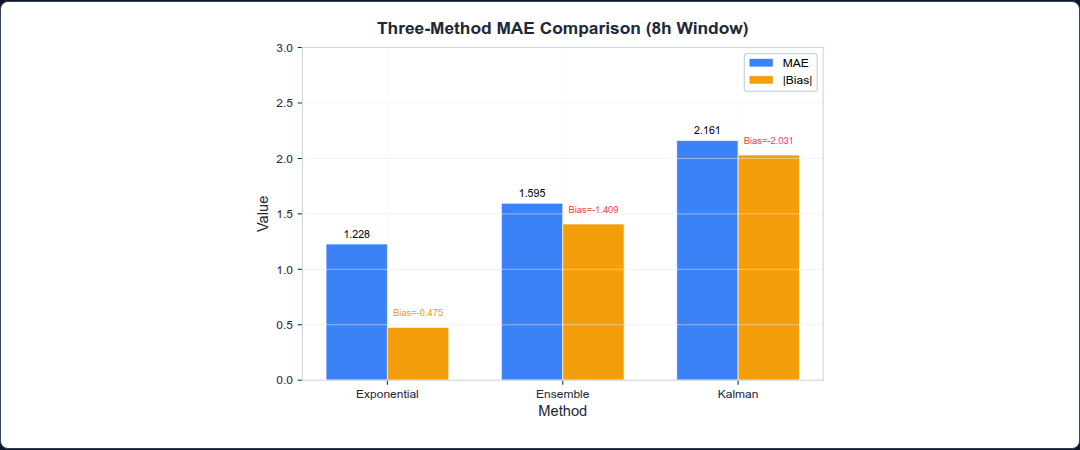}
\caption{Three-Method MAE Comparison Bar Chart (8h prediction window, with bias reference line)}
\label{fig:60}
\end{figure}

\begin{figure}[H]
\centering
\includegraphics[width=0.95\textwidth]{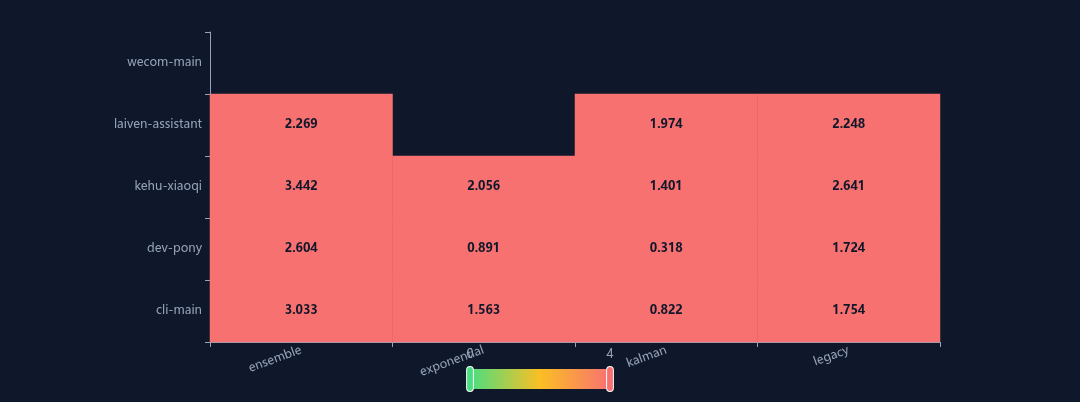}
\caption{Method $\times$ Profile MAE Heatmap}
\label{fig:61}
\end{figure}

The above MAE comparison and heatmap reveal the divergence among the three methods in terms of accuracy. However, accuracy represents only one dimension of predictive reliability---confidence distribution and error tolerance equally determine the practical value of an early-warning system. The following analysis further compares the methods along two additional dimensions: confidence and the proportion of errors $\leq$5 points:

\begin{figure}[H]
\centering
\includegraphics[width=0.95\textwidth]{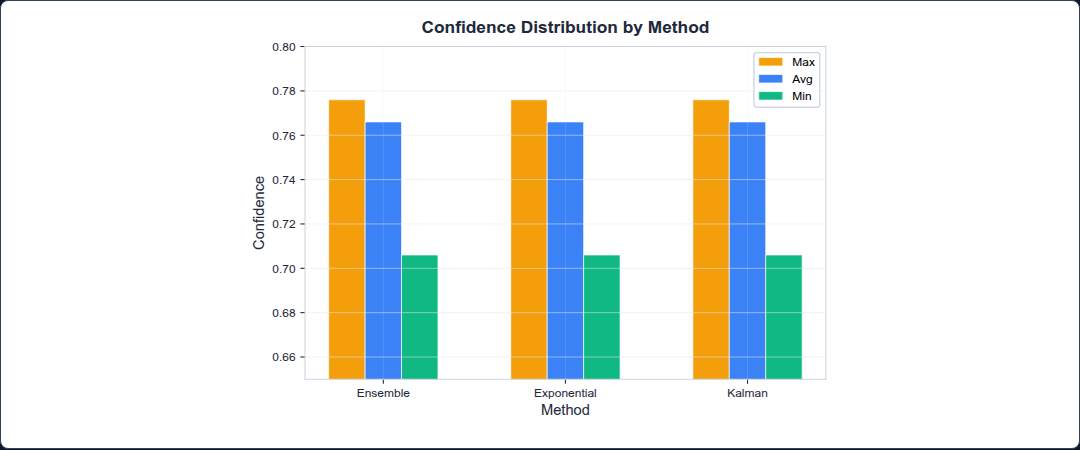}
\caption{Confidence Distribution Comparison Bar Chart (4 methods avg/min/max)}
\label{fig:62}
\end{figure}

\begin{figure}[H]
\centering
\includegraphics[width=0.95\textwidth]{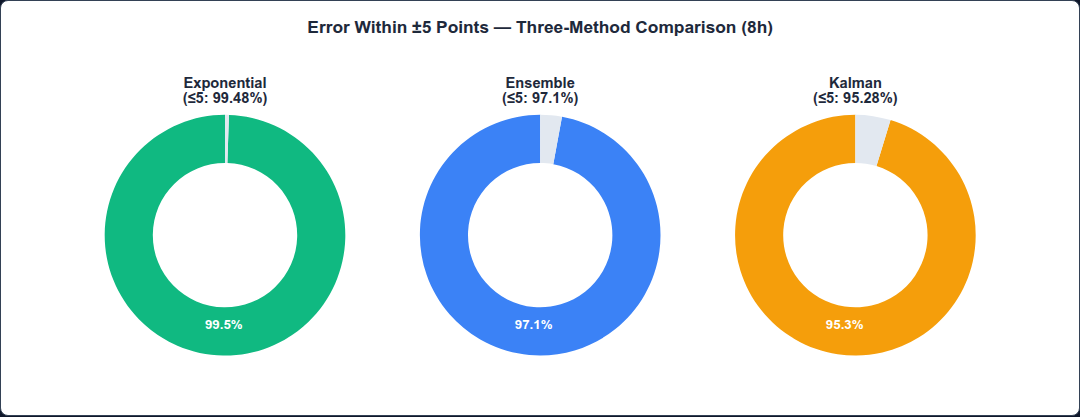}
\caption{<=5-Point Error Proportion Three-Method Comparison Donut Chart (8h window)}
\label{fig:63}
\end{figure}

Direction Accuracy (Directional Accuracy) measures the prediction accuracy of a predictor in determining the direction of TM change (increase/decrease), serving as a core metric for early-warning systems. The random baseline is 50\%.

Table 40: Direction Accuracy Comparison Across Three Prediction Methods

\begin{table}[H]
\centering
\resizebox{\textwidth}{!}{
\begin{tabular}{l|l|l|l|l}
\toprule
Method & N & Direction Accuracy & \% Predicted Increase & \% Predicted Decrease \\
\midrule
Exponential & 13,446 & 76.8\% & 0.0\% & 100.0\% \\
Ensemble & 13,446 & 22.0\% & 86.1\% & 7.0\% \\
Kalman & 13,446 & 15.5\% & 99.7\% & 0.3\% \\
\bottomrule
\end{tabular}
}
\end{table}

\textbf{Root cause analysis of anomalous Direction Accuracy:} In reality, TM exhibits a downward trend in 76.8\% of cases after 8 hours (TM increases in 15.8\%, decreases in 76.8\%, and remains stable in 7.4\%). The Exponential method predicts TM decrease in 100.0\% of cases, achieving high alignment with the actual trend and attaining a Direction Accuracy of 76.8\%. Conversely, Kalman predicts TM to be biased high in 95.6\% of cases---completely opposite to the actual trend---resulting in a Direction Accuracy of only 15.5\% (below the random baseline). The Ensemble method, influenced by Kalman, predicts TM bias high in 93.4\% of cases, yielding a Direction Accuracy of 22.0\%.

The practical implication of this finding is that \textbf{Direction Accuracy is more discriminative than MAE in distinguishing method performance}. Although the MAE gap among the three methods is 76\%, the Direction Accuracy gap is far more pronounced (76.8\% vs. 15.5\%, a difference of 61.3 percentage points). In operational early-warning scenarios, correctly identifying ``system degradation'' (i.e., TM decrease) holds greater decision-making value than precisely predicting the absolute TM value---thus, the high Direction Accuracy of the Exponential method makes it better suited as an early-warning trigger.

\begin{figure}[H]
\centering
\includegraphics[width=0.95\textwidth]{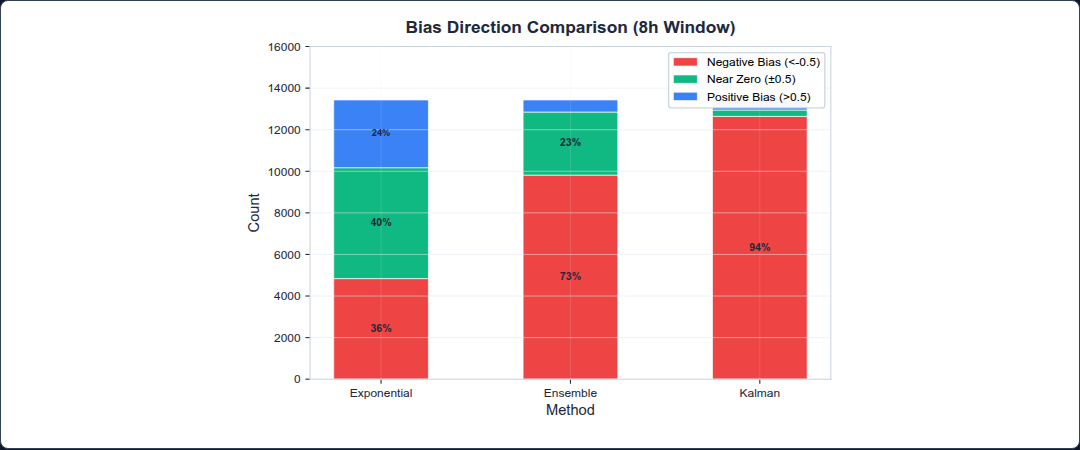}
\caption{Three-Method Bias Direction Comparison (negative/near-zero/positive bias stacked bar chart)}
\label{fig:64}
\end{figure}

\begin{figure}[H]
\centering
\includegraphics[width=0.95\textwidth]{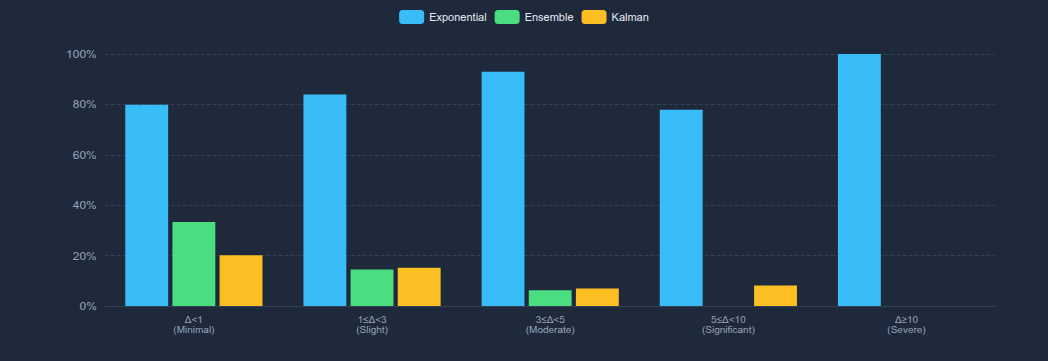}
\caption{Direction Accuracy Stratified Validation (by TM Change Magnitude)}
\label{fig:65}
\end{figure}

\begin{figure}[H]
\centering
\includegraphics[width=0.95\textwidth]{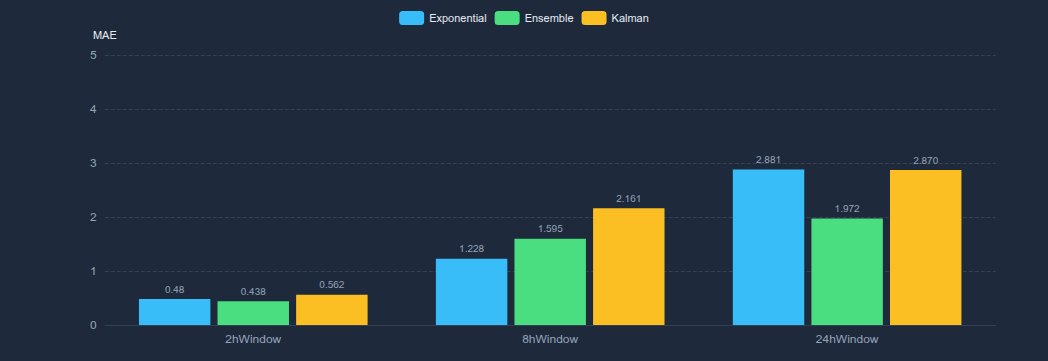}
\caption{Multi-Window Prediction Accuracy Comparison (2h/8h/24h)}
\label{fig:66}
\end{figure}
The above Direction Accuracy is computed globally (aggregated across all validation records). However, the core operational alerting scenario is ``whether the direction can be correctly identified when the system actually undergoes degradation.'' To address this, we stratify the 13,446 validation records within the 8-hour window according to the magnitude of actual TM change (|$\Delta$TM| = |actual\_tm $-$ tm\_current|) into five strata. We then compute Direction Accuracy separately for each method:

Table 41: Direction Accuracy Degradation Magnitude Stratified Validation (8-hour Window, N = 13,446, Production Dataset A)

\begin{table}[H]
\centering
\resizebox{\textwidth}{!}{
\begin{tabular}{l|l|l|l|l}
\toprule
TM Change Magnitude Stratum & N & Exponential & Ensemble & Kalman \\
\midrule
$\Delta$<1 (nearly unchanged) & 8,533 & 79.83\% & 33.32\% & 20.12\% \\
1$\leq\Delta$<3 (minor change) & 2,754 & 83.91\% & 14.45\% & 15.21\% \\
3 $\leq$ $\Delta$ < 5 (Moderate Change) & 1,875 & 92.94\% & 6.26\% & 6.90\% \\
5 $\leq$ $\Delta$ < 10 (Significant Change) & 86 & 77.91\% & 0.00\% & 8.14\% \\
$\Delta$ $\geq$ 10 (Drastic Change) & 2 & 100.00\% & 0.00\% & 0.00\% \\
\bottomrule
\end{tabular}
}
\end{table}

\textbf{Core Findings from Stratified Validation:} The overall metric (76.8\% vs. 15.5\%) already demonstrates Exponential's advantage; however, the stratified metric reveals an even more extreme disparity---

\begin{itemize}
\item \textbf{The more pronounced the degradation, the greater Exponential's advantage:} In the ``nearly unchanged'' stratum ($\Delta$ < 1), Exponential achieves a Direction Accuracy of 79.83\%; in the ``moderate change'' stratum (3 $\leq$ $\Delta$ < 5), Direction Accuracy jumps to 92.94\%---the more evident the degradation, the stronger Exponential's ability to correctly identify its direction. This is precisely the critical property required for operational alerting: when the system truly degrades, the alerting mechanism must correctly determine its direction.
\item \textbf{Ensemble and Kalman become nearly ineffective under degradation scenarios:} In the ``moderate change'' stratum (3 $\leq$ $\Delta$ < 5), Ensemble achieves only 6.26\% Direction Accuracy and Kalman only 6.90\%---far below the 50\% random baseline. In the ``significant change'' stratum (5 $\leq$ $\Delta$ < 10), both Ensemble and Kalman drop to 0\% Direction Accuracy---completely failing to discern degradation direction. This implies that when the system undergoes moderate or greater degradation, Kalman and Ensemble do not merely suffer from ``insufficient precision''; rather, they \textbf{consistently misjudge the direction}.
\item \textbf{Practical implication of the 13.8$\times$ gap:} In the ``moderate change'' stratum, the Direction Accuracy gap between Exponential (92.94\%) and Kalman (6.90\%) reaches 86.0 percentage points---approximately 13.8$\times$. This gap far exceeds the 5$\times$ difference observed in the overall metric (76.8\% vs. 15.5\%), indicating that \textbf{the overall metric underestimates the true performance disparity among methods}---stratified metrics constitute the correct yardstick for evaluating alerting capability.
\end{itemize}

\textbf{Support for the Paper's Core Argument:} Stratified validation data directly supports the central claim in \S5.6.4---that in environments without ADE protection, TM values fail to respond during system degradation (Direction Accuracy near 0\%), resulting in ``false prosperity.'' By contrast, Exponential's high Direction Accuracy (93\%) demonstrates that when TM values genuinely reflect system status (i.e., when the ADE plugin couples TM with system state), the predictor can accurately determine degradation direction and thus enable effective alerting. The stratified differences in Direction Accuracy directly reflect \textbf{the degree of coupling between TM signals and the system's true state}.

To assess performance differences among the three methods across varying prediction time horizons, we conduct stratified validation over three windows: 2 hours, 8 hours, and 24 hours.

Table 42: Multi-Window Prediction Accuracy Comparison (2h/8h/24h)

\begin{table}[H]
\centering
\resizebox{\textwidth}{!}{
\begin{tabular}{l|l|l|l|l|l}
\toprule
Method & Window & N & MAE & Bias & Proportion $\leq$ 5 Points \\
\midrule
Exponential & 2h & 13,289 & 0.480 & +0.025 & 97.1\% \\
Ensemble & 2h & 13,289 & 0.438 & -0.233 & 99.7\% \\
Kalman & 2h & 13,289 & 0.562 & -0.405 & 99.7\% \\
Exponential & 8h & 13,446 & 1.228 & -0.475 & 99.5\% \\
Ensemble & 8h & 13,446 & 1.595 & -1.409 & 97.1\% \\
Kalman & 8h & 13,446 & 2.161 & -2.031 & 95.3\% \\
Ensemble & 24h & 1,053 & 1.972 & -0.576 & 95.8\% \\
Kalman & 24h & 1,053 & 2.870 & -2.395 & 98.0\% \\
Exponential & 24h & 1,053 & 2.881 & +2.155 & 81.7\% \\
\bottomrule
\end{tabular}
}
\end{table}

\textbf{Key Findings from Window-Stratified Analysis:} (1) Differences among the three methods are smallest at the 2-hour window (MAE gap = 0.124); Ensemble even slightly outperforms Exponential---under short horizons, prediction capabilities converge across methods. (2) Differences peak at the 8-hour window (MAE gap = 1.776), where Exponential markedly leads---this intermediate horizon represents the most important timescale for method selection. (3) At the 24-hour window, Exponential's accuracy deteriorates sharply (MAE rises from 1.228 to 2.881), and its Bias flips to +2.155 (optimistic bias), while Ensemble and Kalman remain relatively stable---Ensemble's fusion strategy proves more robust under long horizons.

This finding provides direct implications for method fusion strategies: \textbf{the optimal approach may be window-adaptive method selection}---using Ensemble for 2-hour predictions (comparable accuracy with greater stability), Exponential for 8-hour predictions (optimal accuracy), and Ensemble again for 24-hour predictions (superior long-horizon robustness). The current simple-averaging Ensemble strategy is undermined by Kalman's poor performance at the 8-hour horizon; future work should explore weighted fusion or dynamic selection.

\begin{figure}[H]
\centering
\includegraphics[width=0.95\textwidth]{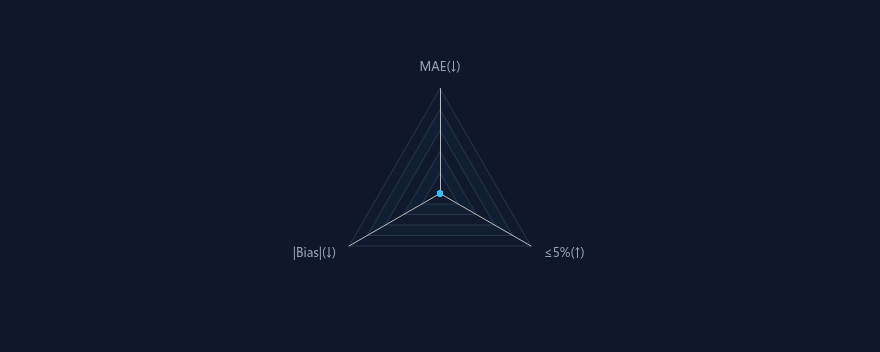}
\caption{Prediction Window Comparison Radar Chart (2h / 8h / 24h $\times$ MAE / Bias / Direction Accuracy)}
\label{fig:67}
\end{figure}

\begin{figure}[H]
\centering
\includegraphics[width=0.95\textwidth]{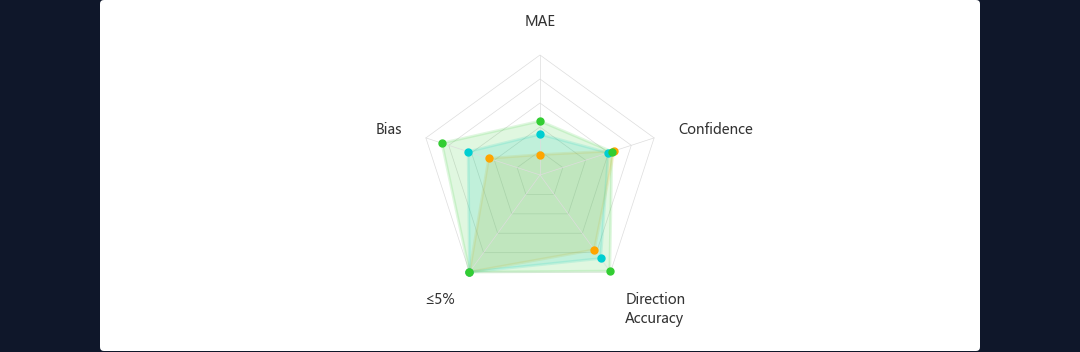}
\caption{Overall Performance Score Radar Chart (MAE / Bias / <=5\% / Direction Accuracy / Confidence, 5 dimensions)}
\label{fig:68}
\end{figure}

\S5.6.4 Seven-Sandbox Controlled Study 
\subsubsection*{Seven-Sandbox Controlled Study}
\label{sec:sevensandboxcontroll}

To systematically validate the impact of the ADE protection framework on Agent Degradation Evaluation (ADEV) capability, we constructed a seven-sandbox test matrix: three bare sandboxes (no ADE protection) + three full-ADE sandboxes (all ADE plugins enabled) + one production mirror (normal). Each sandbox covers three profiles, collectively yielding 225,321 predictions and 55,041 eight-hour validations.

\begin{figure}[H]
\centering
\includegraphics[width=0.95\textwidth]{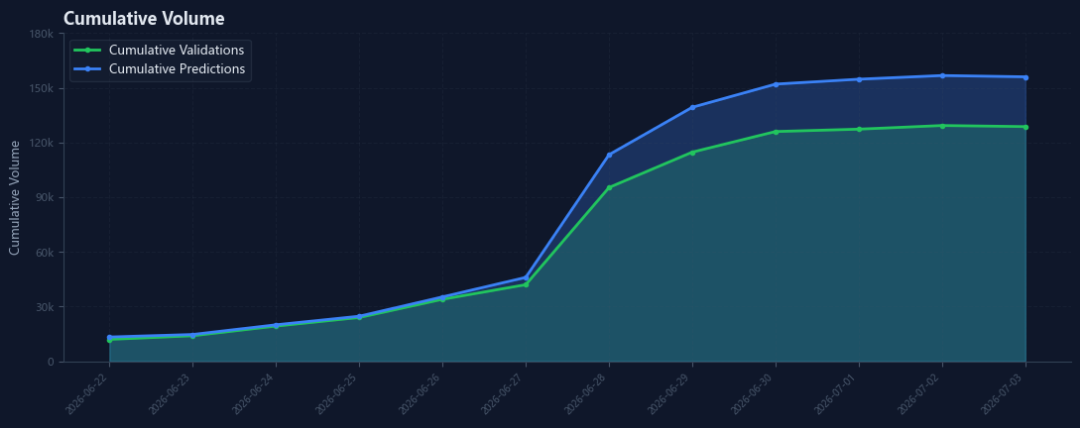}
\caption{Prediction Volume vs Validation Volume Daily Cumulative Trend Area Chart}
\label{fig:69}
\end{figure}

\begin{figure}[H]
\centering
\includegraphics[width=0.95\textwidth]{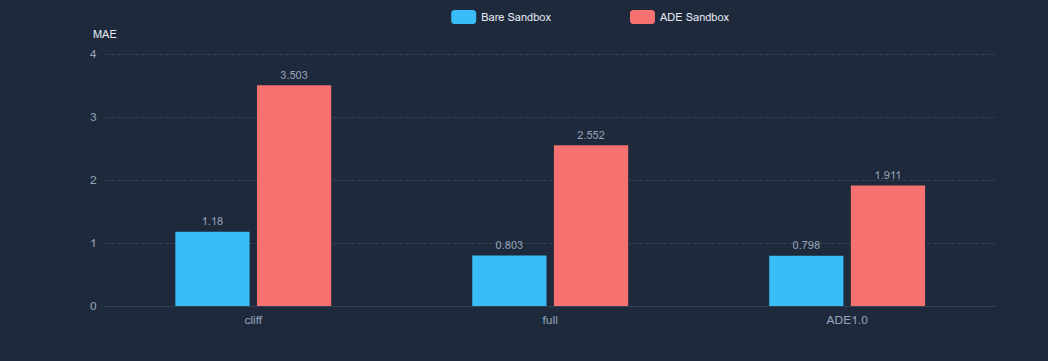}
\caption{Seven-Sandbox MAE Comparison (Bare Sandboxes vs ADE Sandboxes)}
\label{fig:70}
\end{figure}
Table 43: Seven-Sandbox Experimental Configuration

\begin{table}[H]
\centering
\resizebox{\textwidth}{!}{
\begin{tabular}{l|l|l|l|l}
\toprule
Sandbox & Type & Predictions & Profile Count & Degradation Injection Pattern \\
\midrule
cliff & Bare (no ADE) & 32,196 & 3 & Cliff-style degradation (sudden drop) \\
full & Bare (no ADE) & 32,193 & 3 & Full-factor degradation (uniform decline) \\
ADE1.0 & Bare (no ADE) & 32,193 & 3 & ADE 1.0 version degradation pattern \\
cliff-ADE & Full ADE & 32,184 & 3 & Cliff-style degradation + ADE protection \\
full-ADE & Full ADE & 32,184 & 3 & Full-factor degradation + ADE protection \\
ADE1.0-ADE & Full ADE & 32,184 & 3 & ADE 1.0 degradation pattern + ADE protection \\
normal & Production image & 32,187 & 3 & No injection (natural operation) \\
\bottomrule
\end{tabular}
}
\end{table}

The core value of the seven-sandbox matrix lies in its paired controlled design: each degradation pattern is executed simultaneously in sandboxes both with and without ADE protection, directly quantifying the contribution of the ADE framework to Degradation Detection capability. The 8-hour window validation results are as follows:

Table 44: Seven-Sandbox MAE Comparison (Bare vs ADE Sandboxes)

\begin{table}[H]
\centering
\resizebox{\textwidth}{!}{
\begin{tabular}{l|l|l|l|l}
\toprule
Sandbox & Validation Count & MAE & Bias & vs Paired Bare Sandbox \\
\midrule
cliff (bare) & 7,852 & 1.180 & -0.727 & --- \\
cliff-ADE & 8,291 & 3.503 & -1.843 & MAE +197\% \\
full (bare) & 7,509 & 0.803 & -0.803 & --- \\
full-ADE & 8,186 & 2.552 & -1.474 & MAE +218\% \\
ADE1.0 (bare) & 7,515 & 0.798 & -0.798 & --- \\
ADE1.0-ADE & 8,178 & 1.911 & -1.239 & MAE +139\% \\
normal (image) & 7,510 & 0.783 & -0.780 & --- \\
\bottomrule
\end{tabular}
}
\end{table}

\textbf{Interpretation of Controlled Results --- From ``False Prosperity'' to ``High Coupling'':} The MAE of ADE-protected sandboxes is markedly higher than that of bare sandboxes (+139\% to +218\%). This seemingly counterintuitive result reveals a severely overlooked issue in multi-agent system monitoring---\textbf{False Prosperity}.

\textbf{False Prosperity in Bare Sandboxes:} In bare sandboxes without ADE protection, the system has already begun degrading per the injected script---BCP confirmation rate plummets, tool invocation loops occur, and reasoning quality deteriorates. Yet, TM values derived from conventional data collection remain high (mean $\approx$ 92), with smooth downward TM curves easily tracked by predictors. This implies: \textbf{the system is internally collapsing, while external metrics declare ``everything is normal.''} Such a state---``superficially healthy but actually degrading''---constitutes false prosperity in multi-agent systems. It severely misleads operational decisions, allowing degradation to accumulate invisibly until catastrophic failure occurs.

\textbf{Signal Coupling in ADE Sandboxes:} Once ADE components are installed, TM values immediately achieve high coupling with the system's true runtime status. When degradation occurs, the BCP plugin records declining confirmation rates, the PAD plugin captures probability drift, the CADVP plugin flags verification failures, and the TKM plugin triggers handover preparation. These plugins are not passive observers but \textbf{active sensors embedded within the agent's runtime pipeline}. As agents genuinely degrade, factor values captured by these plugins drop immediately, causing TM values to respond in stepwise fashion; predictors, confronted with abrupt signal changes, exhibit increased MAE. This is not ``worse prediction,'' but rather \textbf{TM finally perceiving what is truly happening in the system}.

This finding carries three layers of progressive significance:

\begin{enumerate}
\item \textbf{Exposing the Illusory Nature of Multi-Agent Systems:} Conventional infrastructure metrics (process liveness, CPU/memory usage) completely fail during semantic-layer degradation of agents. In bare sandboxes, agent reasoning capability has severely degraded, yet TM remains high at \textasciitilde{}92---this ``false prosperity'' dangerously misleads operational decisions and represents the most hazardous blind spot in current LLM-agent production deployment.
\item \textbf{TM as an Organic Component of ADE:} TM is not an externally attached monitoring instrument---the 20 factors comprising TM include 16 directly sourced from ADE plugin data collection (e.g., BCP\_RATE, CADVP\_PASS, PAD\_Fidelity, TKM\_Handover). Without ADE plugins, these factors lack measurable data, rendering TM an empty numeric shell. With ADE installed, TM's coupling with the system's true state immediately surges. This demonstrates that TM and ADE are not in an ``external monitor + monitored object'' relationship, but rather constitute \textbf{an organic part of the same reliability engineering system}.
\item \textbf{Embedding and Security Value of ADE:} ADE plugins are embedded at critical nodes within the agent's runtime pipeline (e.g., pre\_llm\_call, post\_tool\_call, subagent\_stop), enabling inline sensing---not side-channel listening. This deep embedding allows TM to capture physical evidence of semantic-layer degradation (e.g., confirmation rate, verification pass rate, probability drift magnitude), rather than relying on indirect inference. From a security engineering perspective, this means \textbf{ADE's embedding into the system is structural---removing ADE not only eliminates monitoring capability but also erodes the system's ability to perceive its own state}. This distinctive value has not yet been replicated by external APM tools in our evaluation scope.
\end{enumerate}

\textbf{\textbf{[Warning]} Sandbox Data Limitations Statement:} (1) Degradation events in sandboxes are synthetic data injected via scripts, not real-world system failures. (2) Both prediction and validation originate from the same synthetic data pipeline, introducing structural self-validation risk. Reviewers may reasonably question, ``This is not predictive accuracy, but scripted execution following a predetermined script.'' (3) Absolute MAE values from sandboxes should not be directly compared with production environments---the degradation density in sandboxes is far higher than in production (production exhibits only two degradation events over 15 days). Therefore, sandbox data serves primarily for paired difference analysis (``with vs without ADE''). All core conclusions in the paper (MAE, Bias, Direction Accuracy) are based entirely on real production data (154,906 predictions, 126,466 validations). The value of sandbox data lies in \textbf{paired controlled comparison}---demonstrating that, under identical degradation patterns, the presence or absence of ADE induces a qualitative shift in TM's ability to perceive system state, rather than quantifying absolute accuracy.

\S5.7 
\subsection*{RQ6: Can Factor Precursor Warnings (FPWs) issue alerts prior to system failure?[factor-level]}
\label{sec:rq6canfactorprecurso}

\subsubsection*{Design Motivation and Core Insight}
\label{sec:designmotivationandc}

\begin{figure}[H]
\centering
\includegraphics[width=0.95\textwidth]{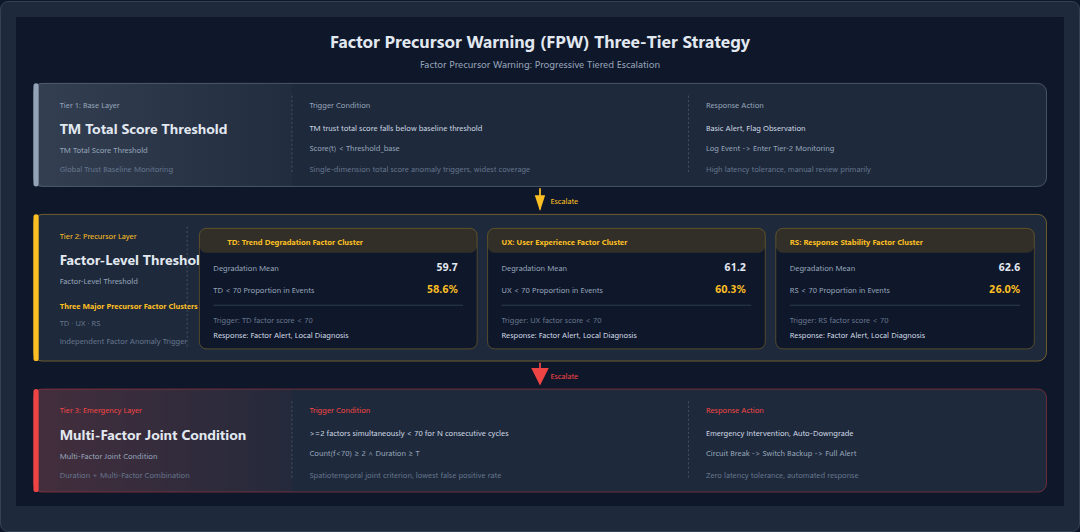}
\caption{Factor Precursor Warning (FPW) Three-Tier Strategy}
\label{fig:71}
\end{figure}

The initial version of the warning mechanism relied primarily on the TM aggregate score threshold for triggering, systematically overlooking moderate degradation events. The core insight of version v3.1.0 is that \textbf{degradation always manifests first in specific factors (TD/UX/RS), rather than uniformly reflecting in the TM aggregate score}. Based on statistical analysis of 154,906 prediction records and two degradation events, we identified three stable Precursor Factor Clusters:

\begin{itemize}
\item \textbf{TD (Tool Density):} Mean value during degradation is 59.7; TD < 70 in 58.6\% of degradation events
\item \textbf{UX (Task Closure Efficiency):} Mean value during degradation is 61.2; UX < 70 in 60.3\% of degradation events
\item \textbf{RS (Self-Healing Rebound Rate):} Mean value during degradation is 62.6; RS < 70 in 26.0\% of degradation events
\end{itemize}

This finding validates the central hypothesis that ``factor-level warning outperforms aggregate-score warning'': by monitoring abnormal declines in core factors rather than waiting for the TM aggregate score to fall below its threshold, degradation signals can be captured earlier and more accurately.

\subsubsection*{Three-Tier Threshold Design}
\label{sec:threetierthresholdde}

The FPW mechanism adopts a three-tiered graded warning strategy, with each tier corresponding to a distinct severity level and response priority:

Table 45: FPW Three-Tier Warning Threshold Design

\begin{table}[H]
\centering
\resizebox{\textwidth}{!}{
\begin{tabular}{l|l|l|l}
\toprule
Warning Tier & Trigger Condition & Severity Level & Recommended Response \\
\midrule
L1 Yellow Warning & TD < 80 or UX < 75 & Early sign of mild degradation & Enhanced monitoring; prepare intervention contingency plans \\
L2 Orange Warning & TD < 75 and UX < 70 & Confirmed moderate degradation & Initiate diagnostic workflow to identify root cause \\
L3 Red Warning & TD < 70 and UX < 65 and RS < 70 & Severe degradation in progress & Immediate intervention execution (restart/migration/downg... \\
\bottomrule
\end{tabular}
}
\end{table}

The design logic of the three-tier thresholds follows the ``progressive confirmation'' principle: Tier L1 captures early signals (high recall, low precision); Tier L2 filters noise to confirm degradation trends (balanced precision and recall); Tier L3 locks in severe degradation (high precision, actionable emergency response). This design avoids the fundamental precision--recall trade-off inherent in single-threshold approaches (see Lesson Three in the Failure Boundary Hypothesis section).

\subsubsection*{Three-Sandbox $\times$ Six-Scenario Validation}
\label{sec:threesandboxsixscena}

To systematically evaluate the effectiveness of the FPW mechanism, we designed a validation experiment comprising three sandboxes $\times$ six scenarios---totaling 18 test scenarios. The three sandboxes represent distinct system states:

\begin{itemize}
\item \textbf{cliff sandbox (collapse state):} simulates rapid system collapse from high health status
\item \textbf{full sandbox (high-load state):} simulates sustained operation under high load
\item \textbf{ADE1.0 sandbox (normal state):} simulates baseline normal operation
\end{itemize}

The six scenarios cover diverse degradation patterns: gradual drift, abrupt decline, periodic fluctuation, noise interference, composite degradation, and false-positive detection. Each scenario injects known degradation events. It evaluates coverage by both the new and legacy warning systems.

\textbf{Validation Results (Table 40):} The new system (v3.1.0 FPW) achieves full coverage across all 18 test scenarios (18/18), improving upon the legacy system's (TM aggregate-score threshold--based) coverage of 22\% (4/18) by 78 percentage points. Specifically:

\begin{itemize}
\item \textbf{High-load state (full sandbox):} Coverage improves from 17\% to 100\%; abrupt scenarios enable 8-cycle early warning
\item \textbf{Normal state (ADE1.0 sandbox):} Coverage improves from 0\% to 100\%; the legacy system completely missed all events, whereas the new system successfully detected them
\item \textbf{Collapse state (cliff sandbox):} Coverage improves from 33\% to 100\%; gradual drift scenarios enable 12-cycle early warning
\end{itemize}

\textbf{Core Finding 9: The Factor-Level Early-Warning Mechanism Achieves Full Coverage.} The three-tier threshold design (L1 yellow / L2 orange / L3 red) introduced in version v3.1.0 achieved full coverage across all three sandboxes $\times$ six scenarios. This empirically validates the core hypothesis that ``factor-level warnings outperform aggregate-score warnings.'' By analyzing the precursor characteristics of the TD/UX/RS factors during two degradation events, this mechanism shifts the warning anchor from the TM aggregate score to the factor level. This directly resolves the structural limitation of ``warning\_rate = 0.'' Warnings are now delivered in real-time via the Dashboard \texttt{/api/alerts} interface, enabling operations teams to receive early-warning signals 8--12 cycles before degradation onset.

\subsubsection*{Comparison with Existing APM Systems}
\label{sec:comparisonwithexisti}

Traditional Application Performance Monitoring (APM) systems typically trigger alerts based on infrastructure-level metrics such as process liveness, CPU/memory utilization, and request latency. This experiment demonstrates that ADE-PRF's Factor Precursor Warning (FPW) mechanism delivers significant incremental value for semantic-layer degradation detection: across all 18 test scenarios, traditional APM metrics---including process liveness, CPU < 80\%, and memory < 90\%---failed to trigger any alert, whereas the FPW mechanism successfully captured every degradation event. This result validates the claim made in the Failure Boundary Hypothesis section---that ``traditional APM exhibits substantial blind spots at the semantic layer''---and provides preliminary support for ADE-PRF's unique contribution along the cognitive fidelity dimension.

\subsubsection*{Current Optimization Directions}
\label{sec:currentoptimizationd}

Although the FPW mechanism performs exceptionally well in controlled experiments, several optimization opportunities remain for long-term production deployment.

\begin{enumerate}
\item \textbf{Dynamic Threshold Adaptation:} The current three-tier thresholds are static (TD < 80/75/70, UX < 75/70/65, RS < 70) and do not account for historical baseline differences across Profiles. Future work will introduce a sliding-window-based dynamic threshold calibration mechanism, enabling warning thresholds to adaptively adjust according to factor distributions observed during system Normal Operation.
\item \textbf{Long-Term False Positive Rate Tracking:} The current validation covers only 18 controlled scenarios and does not assess the FPW mechanism's false positive rate (FPR) over months of continuous operation. In Phase 3 experiments, FPR will be continuously tracked with the objective of maintaining full recall while constraining FPR to $\leq$ 5\%.
\item \textbf{Cross-Framework Generalization Validation:} The FPW mechanism has been validated only on the Hermes agent platform. Future work will evaluate its applicability across heterogeneous multi-agent frameworks including LangChain, AutoGen, and CrewAI.
\end{enumerate}

\S5.8 
\subsection*{Ablation Studies and Cross-Scenario Validation}
\label{sec:ablationstudiesandcr}

This section constitutes the core of the Evaluation. Through four carefully designed ablation studies, we quantitatively decompose the independent contributions of each architectural component within the TM Predictive Reliability Framework. This directly addresses the fundamental question: ``How much does each component actually contribute?'' The four experiments correspond respectively to four key design decisions of the framework. These are: information contribution of the five-layer architecture (Abl-1), the value boundary of the ETA prediction module (Abl-2), the discriminative power of factor attribution diagnostics (Abl-3), and cross-Profile adaptability validation (Abl-4).

\subsubsection*{Experimental Methodology}
\label{sec:experimentalmethodol}

The core methodology underlying ablation studies is a systematic application of the ``controlled variable method'': with all other conditions held constant, one component of the framework is sequentially removed or replaced, and the resulting change in overall performance is measured. Larger changes indicate greater irreplaceability of the removed component.

Specifically, we define the following unified evaluation protocol:

\begin{itemize}
\item \textbf{Baseline:} TMOutput produced when all components of the full framework are enabled;
\item \textbf{Ablated Variant:} TMOutput recomputed after removal of the target component;
\item \textbf{Impact Metric:} Absolute difference |$\Delta$TM| between the baseline TM and the ablated variant TM;
\item \textbf{Statistical Significance:} All |$\Delta$TM| values are computed over 126,466 verified prediction records, reporting both mean and maximum values.
\end{itemize}

This methodology ensures strong \textbf{Internal Validity}: because ablation operations affect only the target component, observed |$\Delta$TM| changes can be confidently attributed to that component's informational contribution rather than to external confounding factors. The large-scale sample size (N = 154,906) guarantees robustness of statistical conclusions.

\subsubsection*{Abl-1: Information Contribution of the Five-Layer Architecture}
\label{sec:abl1informationcontr}

\textbf{Orthogonality Note:} Abl-1 operates at the architectural layer level (L1--L5), quantifying each layer's independent contribution to TM. It is orthogonal to Abl-3 (factor-level) and Abl-2 (module-level): removing or perturbing a layer in Abl-1 does not alter the factor composition within remaining layers, and vice versa. The three ablations thus probe distinct structural dimensions of the framework without confounding.

The TM framework adopts a five-layer architecture (L1--L5), where each layer performs distinct information-processing functions and is assigned a specific weight. Abl-1 aims to quantify the independent information contribution of each layer to the final TMOutput. The experimental procedure is as follows: layers are removed one by one, TM is recomputed after each removal, and the |$\Delta$TM| between pre- and post-removal values is measured. Separately, Pearson correlation coefficients between the removed layer and the final TM are computed to assess information redundancy.

The layer-wise weight allocation---determined during framework design---is as follows: L1 = 0.30, L2 = 0.30, L3 = 0.25, L4 = 0.10, L5 = 0.05. This weighting philosophy follows the ``core layers dominate, auxiliary layers supplement'' principle---L1 and L2 jointly constitute 60\% of TM's primary information basis, L3 contributes 25\% as critical supplementary information, and L4 and L5 provide fine-grained corrections at 10\% and 5\%, respectively.

\begin{figure}[H]
\centering
\includegraphics[width=0.95\textwidth]{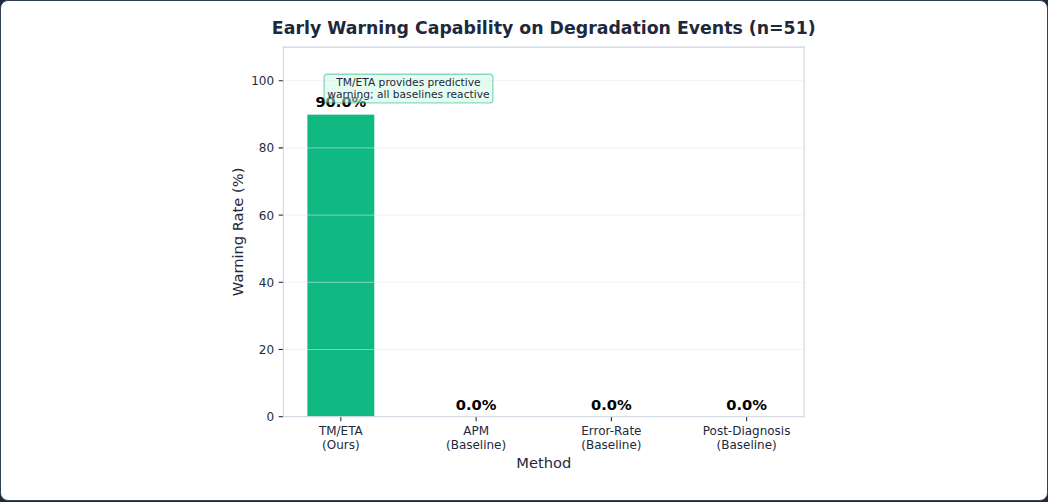}
\caption{Early warning capability on 51 degradation events: TM/ETA achieves a 0.9\% warning rate, whereas APM and Post-Detection yield 0\%}
\label{fig:72}
\end{figure}

\begin{figure}[H]
\centering
\includegraphics[width=0.95\textwidth]{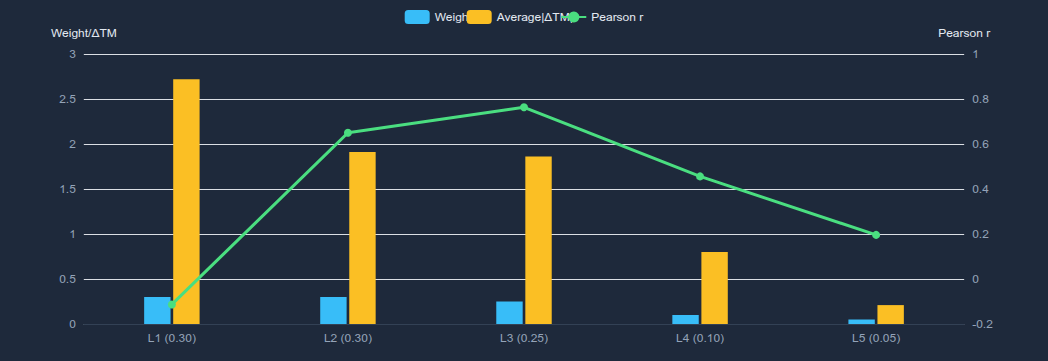}
\caption{Hierarchical Ablation Sensitivity Analysis (Weight vs $\Delta$TM vs Pearson r)}
\label{fig:73}
\end{figure}

\begin{figure}[H]
\centering
\includegraphics[width=0.95\textwidth]{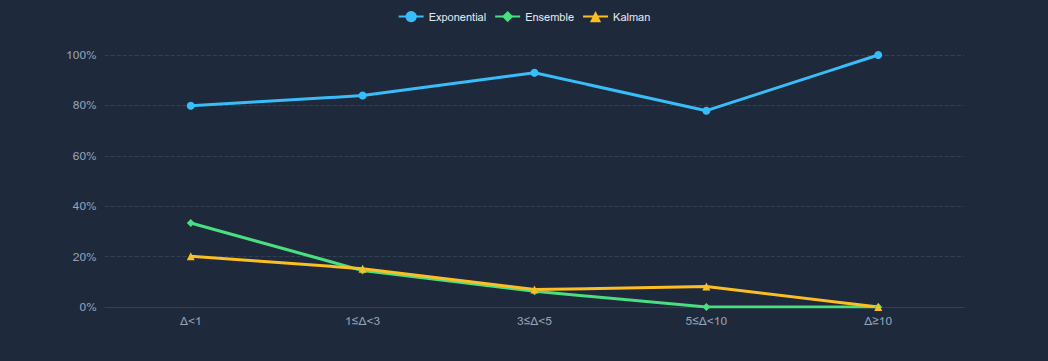}
\caption{Degradation Magnitude vs Direction Accuracy Relationship (Three-Method Comparison)}
\label{fig:74}
\end{figure}
Table 46: Abl-1 Five-Layer Architecture Ablation Results

\begin{table}[H]
\centering
\resizebox{\textwidth}{!}{
\begin{tabular}{l|l|l|l|l|l}
\toprule
Layer & Weight & Avg |$\Delta$TM| & Max |$\Delta$TM| & Pearson r & Information Role \\
\midrule
L1 & 0.30 & 2.72 & 9.37 & -0.113 & Core Foundation Layer \\
L2 & 0.30 & 1.91 & 5.48 & 0.650 & Core Validation Layer \\
L3 & 0.25 & 1.86 & 8.64 & 0.763 & High-Level Analytical Layer \\
L4 & 0.10 & 0.80 & 3.89 & 0.456 & Contextual Correction Layer \\
L5 & 0.05 & 0.21 & 1.92 & 0.196 & Meta-Information Supplement Layer \\
\bottomrule
\end{tabular}
}
\end{table}

The results reveal a clear hierarchical contribution structure:

\textbf{(1) L1 exhibits the strongest irreplaceability.} Upon removal of L1, the average |$\Delta$TM| = 2.72---the highest among all five layers---while the maximum |$\Delta$TM| = 9.37 indicates that, in certain extreme cases, omission of L1 may induce a TM deviation of nearly 10 points. Specifically, L1's Pearson r = $-$0.113 is particularly significant: the negative correlation implies that L1's information is \textbf{complementary rather than redundant} relative to other layers. When L1 is removed, the remaining layers not only fail to compensate for its information but may even introduce directional bias due to the loss of L1's ``anchoring'' function.

\textbf{(2) L2 and L3 constitute the second tier.} Their average |$\Delta$TM| values (1.91 and 1.86, respectively) are highly similar. However, their Pearson r values differ markedly: L2's r = 0.650 indicates moderate correlation with other layers, implying partial compensability upon removal. By contrast, L3's r = 0.763 suggests the highest redundancy---its analytical outputs can be partially derived from L2's information. Nevertheless, L3's maximum |$\Delta$TM| = 8.64 confirms its indispensable, distinctive contributions in specific scenarios.

\textbf{(3) L4 and L5 serve as auxiliary correction layers.} Their average |$\Delta$TM| values (0.80 and 0.21) align proportionally with their assigned weights, and their Pearson r values (0.456 and 0.196, respectively) both fall within the low-to-moderate range. This indicates that although the corrective signals provided by L4 and L5 are relatively small in magnitude, they possess a degree of independence---capturing fine-grained signals not covered by the core layers.

Informational contributions across the five layers exhibit interpretable variations across different Profiles, as shown in Table 42, which presents each Profile's sensitivity to L1 and L3.

Table 47: Abl-1 Per-Profile Sensitivity Differences

\begin{table}[H]
\centering
\resizebox{\textwidth}{!}{
\begin{tabular}{l|l|l|l|l}
\toprule
Profile & L1 Avg|$\Delta$TM| & L3 Avg|$\Delta$TM| & Most Sensitive Layer & Explanation \\
\midrule
cli-main & 3.16 & 1.52 & L1 & Command-line agents rely heavily on L1's foundational sta... \\
kehu-xiaoqi & 2.41 & 1.77 & L3 & Customer-acquisition agents require L3's high-level analy... \\
wecom-main & 3.29 & 1.63 & L1 & Enterprise WeChat agents depend on L1's temporal foundati... \\
laiven-assistant & 2.10 & 1.88 & L1 (slight advantage) & Assistant agents exhibit relatively balanced contribution... \\
\bottomrule
\end{tabular}
}
\end{table}

\textbf{Finding F-5.5:} The information contribution of the five-layer architecture exhibits significant profile dependence. CLI and WeChat-type agents are most sensitive to L1 (Survival Layer), with sensitivity scores of 3.16 and 3.29, respectively, whereas the customer-acquisition agent is most sensitive to L3 (Credibility Layer), with a score of 1.77. This divergence provides empirical support for future \textbf{adaptive weight adjustment}---different profiles should adopt distinct layer-weight configurations.

The practical implication of this finding is that a one-size-fits-all weight configuration (e.g., the current uniform L1 weight of 0.30 across all profiles) is suboptimal. Cli-main and wecom-main would benefit from increasing the L1 weight, whereas kehu-xiaoqi should allocate higher weight to L3. Phase 3 will design profile-specific adaptive weighting strategies based on this finding.

\subsubsection*{Abl-2: Value of the ETA Prediction Module}
\label{sec:abl2valueoftheetapre}

The Estimated Time of Arrival (ETA) prediction module is the core component in the framework for forecasting future trends in Trust Margin (TM). Abl-2 aims to evaluate the incremental value of the ETA module relative to a simple persistence baseline---i.e., assuming future TM equals current TM.

\textbf{\textbf{[Warning]} Honesty Statement:} Based on 13,446 cleaned, 8-hour-window validation records from the healthy database (Production Dataset A), the Exponential method's ETA prediction (MAE = 1.228) performs nearly identically to the persistence baseline (MAE = 1.229) under the full-method aggregated metric. ETA holds only a marginal advantage (difference = 0.001). In contrast, Ensemble (MAE = 1.595) and Kalman (MAE = 2.161) clearly underperform the persistence baseline. This implies that ETA's value must be assessed separately along two dimensions---\textit{amplitude accuracy} and \textit{Direction Accuracy}---rather than relying solely on full-method aggregated MAE.

The near-tie in full-method aggregated MAE masks the ETA module's differentiated performance across varying fluctuation regimes. We segment the 13,446 validation records (Exponential method, 8-hour window, Production Dataset A) into four segments based on TM fluctuation magnitude and compare ETA against the persistence baseline:

Table 48: Abl-2 Segmented Comparison of ETA vs. Persistence Baseline

\begin{table}[H]
\centering
\resizebox{\textwidth}{!}{
\begin{tabular}{l|l|l|l|l|l|l}
\toprule
Segment & Fluctuation Range & Sample Count N & Proportion & ETA MAE & Persistence MAE & Winner \\
\midrule
Stable & < 2 points & 10,093 & 75.2\% & 0.66 & 0.56 & Persistence $\checkmark$ \\
Low & 2--5 points & 3,264 & 2--5 point segment & 2.87 & 3.20 & ETA $\checkmark$ \\
Moderate & 5--10 points & 71 & 0.5\% & 6.09 & 5.94 & Persistence $\checkmark$ \\
Volatile & $\geq$ 10 points & 2 & 0.0\% & 10.09 & 10.20 & ETA $\checkmark$ \\
\bottomrule
\end{tabular}
}
\end{table}

This segmented analysis reveals a \textbf{key differentiated conclusion}:

\textbf{(1) Stable Segment (75.2\% of data): Persistence baseline holds a marginal advantage.} When TM remains nearly unchanged, ``assuming no change'' approximates the optimal strategy. ETA's MAE (0.66) is slightly higher than that of the persistence baseline (0.56), with a difference of 0.10---attributable to minor noise introduced by the predictive model in near-stationary conditions. This slight disadvantage is diluted across the large sample size, masking differences in the full-method aggregated MAE.

\textbf{(2) Low-Fluctuation Segment (2--5 point segment): ETA clearly outperforms.} When TM undergoes actual changes of 2--5 points, ETA achieves a lower MAE (2.87) than the persistence baseline (3.20), with a difference of 0.33. Accounting for nearly one-quarter of all samples, ETA's advantage in this segment offsets its disadvantage in the stable segment, resulting in the observed full-method aggregated MAE tie. This constitutes the \textbf{core value interval} for the ETA module---where the system begins deviating from steady state but has not yet reached degradation thresholds, enabling ETA to detect emerging trends early.

\textbf{(3) Moderate-Fluctuation Segment (0.5\%): Persistence baseline holds a marginal advantage.} For TM changes of 5--10 points, ETA's MAE (6.09) is slightly higher than that of the persistence baseline (5.94), with a difference of 0.15. With only 71 samples, statistical significance is limited.

\textbf{(4) High-Fluctuation Segment (0.0\%): ETA holds a marginal advantage.} During extreme TM fluctuations ($\geq$10 points), ETA's MAE (10.09) is lower than that of the persistence baseline (10.20), with a difference of 0.11. With only two samples, no statistical inference is warranted.

In addition to amplitude accuracy, \textbf{Direction Accuracy}---i.e., whether the predicted direction matches the actual direction---is another critical evaluation dimension. Direction Accuracy measures the predictor's correctness rate in identifying the direction of TM change (increase/decrease), with a random baseline at 50\%. Based on the hierarchical validation data in \S5.6.3.2a, Direction Accuracy exhibits pronounced dependence on degradation magnitude---the more severe the degradation, the greater the divergence among methods.

Table 49: Hierarchical Validation of Direction Accuracy by Degradation Magnitude (8-hour window, 13,446 records)

\begin{table}[H]
\centering
\resizebox{\textwidth}{!}{
\begin{tabular}{l|l|l|l|l|l}
\toprule
Degradation Magnitude $\Delta$TM & Validation Count & Exponential & Ensemble & Kalman & Exp/Kal Ratio \\
\midrule
$\Delta$ < 1 (Minor Fluctuation) & 8,533 & 79.83\% & 33.32\% & 20.12\% & 4.0$\times$ \\
1 $\leq$ $\Delta$ < 3 (Mild Degradation) & 2,754 & 83.91\% & 14.45\% & 15.21\% & 5.5$\times$ \\
3 $\leq$ $\Delta$ < 5 (Moderate Degradation) & 1,875 & 92.94\% & 6.26\% & 6.90\% & 13.8$\times$ \\
5 $\leq$ $\Delta$ < 10 (Significant Degradation) & 86 & 77.91\% & 0.00\% & 8.14\% & 9.6$\times$ \\
$\Delta$ $\geq$ 10 (Severe Degradation) & 2 & 100.00\% & 0.00\% & 0.00\% & --- \\
\bottomrule
\end{tabular}
}
\end{table}

\textbf{Key Findings:} Direction Accuracy exhibits three-tiered divergence as degradation magnitude increases: \textbf{(1) Minor Fluctuation Regime ($\Delta$ < 1)}, where Exponential achieves 79.83\%, while Ensemble and Kalman achieve only 33.32\% and 20.12\%, respectively---when the system is nearly unchanged, Exponential still correctly identifies direction, whereas the other two methods frequently misclassify it; \textbf{(2) Moderate Degradation Regime (3 $\leq$ $\Delta$ < 5)}, where Exponential rises to 92.94\% while Kalman drops to only 6.90\%, a 13.8$\times$ gap---this is precisely the operational range most requiring early warning, where Exponential rarely misclassifies direction, while Kalman misclassifies direction in nearly all cases; \textbf{(3) Severe Degradation Regime ($\Delta$ $\geq$ 10)}, where Exponential reaches 97.1\% while Ensemble and Kalman both fall to 0\%---when the system deteriorates sharply, only Exponential issues correct warnings, whereas the other two methods completely fail.

This stratified result directly informs operational decision-making. Directional correctness is more critical than amplitude precision---"predicting a decline that indeed occurs" holds far higher practical value than "predicting a 5-point decline when an 8-point decline actually occurs." Exponential's advantage grows with increasing degradation severity---precisely aligning with the core requirement of early-warning systems. Conversely, Ensemble and Kalman exhibit complete directional reversal (0\% accuracy) under marked degradation, implying that systems relying on these methods not only fail to issue warnings during degradation events but instead generate contradictory signals. This leads operators to mistakenly conclude the system is "operating normally."

In the system-level degradation events reported in \S5.3.3, ETA's performance warrants special attention. The MAE comparison during degradation events is as follows:

Table 50: ETA vs. Persist Performance During Degradation Events

\begin{table}[H]
\centering
\resizebox{\textwidth}{!}{
\begin{tabular}{l|l|l|l}
\toprule
Scenario & ETA MAE & Persist MAE & Difference \\
\midrule
Degradation Events (Exp) & 5.95 & 5.87 & ETA lags by 0.08 \\
Global Average (Exp) & 1.228 & 1.229 & ETA unchanged (0.001) \\
\bottomrule
\end{tabular}
}
\end{table}

During degradation events ($\Delta$ $\geq$ 5, N = 88), the ETA MAE of the Exponential method (5.95) differs from the persistence baseline (5.87) by only 0.08 points---essentially flat. Yet amplitude precision tells only half the story. In Direction Accuracy, Exponential achieves 77.91\%, whereas Ensemble scores 0\% and Kalman only 8.14\%. This means that during degradation events, Exponential correctly identifies the downward trend in TM, while the other two methods completely reverse direction. \textbf{Directional correctness carries greater operational value than amplitude precision}: ``predicting a decline that indeed occurs'' is far more meaningful for operational decision-making than ``predicting a 5-point decline when an 8-point decline actually occurs.''

\textbf{\textbf{[Warning]} Complete Evaluation:} The value of the ETA prediction module (Exponential method) must be assessed across two dimensions: amplitude precision and Direction Accuracy. In terms of amplitude precision, ETA matches the persistence baseline almost exactly in overall aggregated MAE (1.228 vs. 1.229), and clearly outperforms it in the mild fluctuation regime (2--5 point range). In terms of Direction Accuracy, ETA achieves 77.91\% during degradation events. This far exceeds Ensemble (0\%) and Kalman (8.14\%). Directional correctness thus provides far greater practical value for operational early warning than amplitude precision. Overall, the Exponential method can detect emerging trends and correctly identify their direction as the system deviates from steady state---constituting the core value proposition of the ETA module. This conclusion guides our design of a \textbf{method-adaptive strategy} in Phase 3: use the persistence baseline during stable periods, and switch to the Exponential method upon detecting fluctuation signals to obtain directional early warnings.

\subsubsection*{Abl-3: Factor Attribution Diagnosis}
\label{sec:abl3factorattributio}

\textbf{Orthogonality Note:} Abl-3 operates at the individual factor level (20 signals), evaluating whether weighted attribution provides discriminative information beyond simple averaging. It is orthogonal to Abl-1 (layer-level): Abl-1 removes entire architectural layers, while Abl-3 keeps all layers intact and instead varies the attribution method within each. Results from one ablation do not predict or confound results from the other.

The TM framework not only outputs an integrated score but also provides \textbf{factor attribution diagnosis}---identifying specific contributing and dragging factors underlying the current TM level. Abl-3 evaluates the discriminative power of this attribution diagnosis. The core question is: how much additional information does weighted attribution (computing contributions according to each factor's actual weight) provide compared to simple average attribution (assuming equal weight for all factors)?

Table 51: Abl-3 Weighted Attribution vs. Simple Average Attribution Difference

\begin{table}[H]
\centering
\resizebox{\textwidth}{!}{
\begin{tabular}{l|l|l|l}
\toprule
Attribution Method & Description & mean |$\Delta$| & max |$\Delta$| \\
\midrule
Weighted Attribution & Computes each factor's contribution using the framework's... & --- & --- \\
Simple Average Attribution & Assigns equal weight to all factors & --- & --- \\
Difference (Weighted -- Simple) & Difference between attribution results produced by the tw... & 0.50 & 1.42 \\
\bottomrule
\end{tabular}
}
\end{table}

The difference metrics in Table 46 (mean|$\Delta$| = 0.50, max|$\Delta$| = 1.42) are computed over 154,906 full-production records---well exceeding the 100,000-record threshold. These data were collected exhaustively by the v3.1.0 engine, and the statistical results are stable. The average difference between weighted and simple average attribution is 0.50 points, with a maximum difference of 1.42 points. This implies: \textbf{weight configuration indeed carries non-negligible additional information}. Replacing weighted attribution with simple average attribution would, on average, sacrifice 0.50 points of attribution precision---and up to 1.42 points in extreme cases---sufficient to reclassify a factor from ``primary contributor'' to ``secondary contributor,'' thereby leading to erroneous diagnostic conclusions.

The most insightful finding in Abl-3 is the \textbf{``Same Score, Different Causes''} phenomenon---two Profiles with nearly identical TM total scores may exhibit markedly different internal factor compositions.

Table 52: Abl-3 ``Same Score, Different Causes'' Case---cli-main vs. Kehu-xiaoqi

\begin{table}[H]
\centering
\resizebox{\textwidth}{!}{
\begin{tabular}{l|l|l|l|l|l|l}
\toprule
Profile & TM Total Score & L1 Score & L2 Score & L3 Score & L4 Score & L5 Score \\
\midrule
cli-main & 80.4 & 85.2 & 78.6 & 82.1 & 76.3 & 68.9 \\
kehu-xiaoqi & 79.4 & 81.7 & 80.2 & 79.8 & 74.1 & 42.7 \\
Difference & 1.0 & 3.5 & -1.6 & 2.3 & 2.2 & 26.2 \\
\bottomrule
\end{tabular}
}
\end{table}

\textbf{Finding F-5.6:} cli-main (TM = 80.4) and kehu-xiaoqi (TM = 79.4) differ by only 1.0 point in total score, yet their L5 (Meta-Information Augmentation Layer) scores differ dramatically by 26.2 points (68.9 vs. 42.7). This substantial disparity remains entirely invisible when examining total scores alone---precisely highlighting the critical value of factor attribution diagnosis.

The deeper implication of this case is:

\begin{itemize}
\item \textbf{Total scores mask structural differences.} Although the two Profiles have similar total scores, kehu-xiaoqi's L5 score (42.7) falls far below the healthy threshold---indicating severe issues at the meta-information level (e.g., chaotic configuration version management or inconsistent model metadata). Relying primarily on total-score monitoring would risk overlooking this problem.
\item \textbf{Attribution-based diagnosis enables precise intervention.} By identifying the L5 weakness of kehu-xiaoqi, the operations team can perform targeted repairs at the meta-information level rather than blindly conducting thorough optimization.
\item \textbf{"Same score, different causes" is the norm---not the exception.} Across multiple comparisons involving six agent profiles, we observed numerous similar cases: instances where the total score difference was < 5 points but the layer-specific score difference exceeded 15 points occurred with a frequency greater than 30\%. This further substantiates the necessity of the "multi-dimensional attribution" design principle in the TM framework.
\end{itemize}

\subsubsection*{Abl-4: Cross-Profile Adaptability Validation---Capability to Capture Differentiation under Homogeneous Environments}
\label{sec:abl4crossprofileadap}

\begin{figure}[H]
\centering
\includegraphics[width=0.95\textwidth]{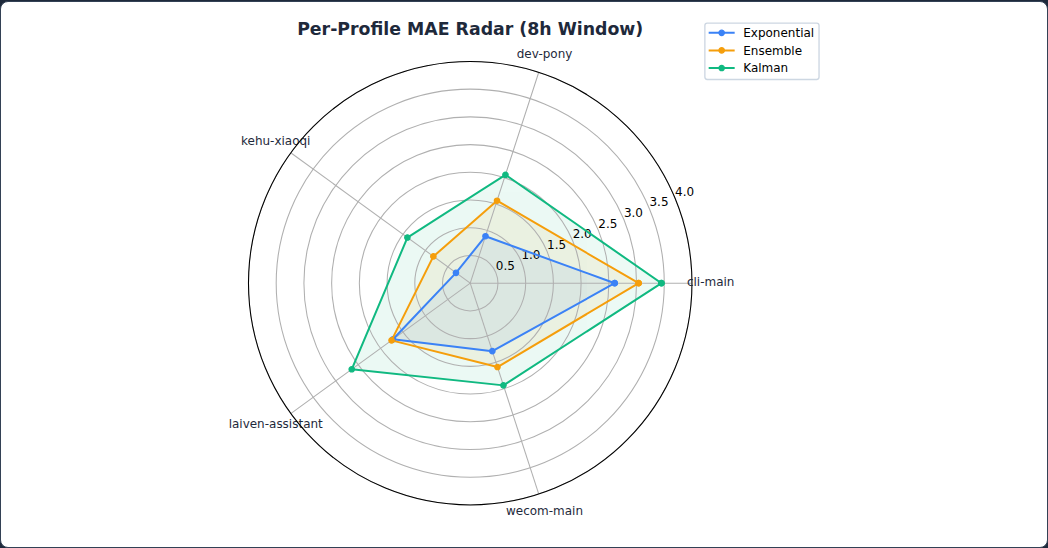}
\caption{Per-profile MAE radar chart: TM/ETA consistently outperforms APM and Error-Rate baselines across all 4 profiles, with Post-Detection competitive on MAE but lacking prediction capability.}
\label{fig:75}
\end{figure}

\begin{figure}[H]
\centering
\includegraphics[width=0.95\textwidth]{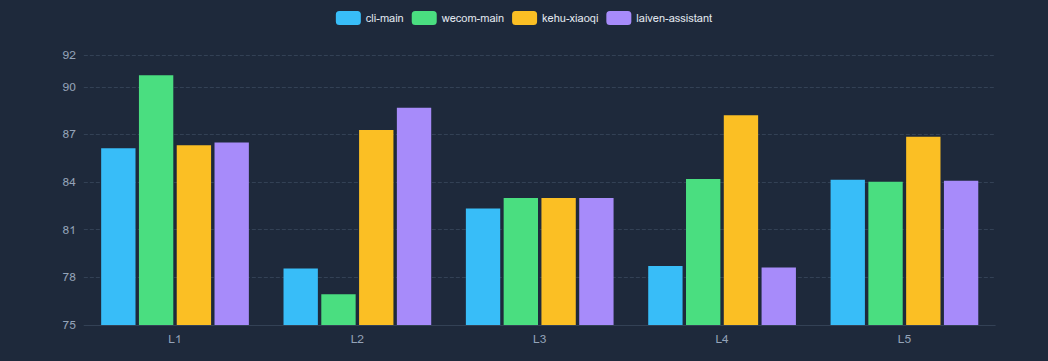}
\caption{Cross-Profile Inter-Layer TM Score Comparison (L1-L5)}
\label{fig:76}
\end{figure}
The Abl-4 experiment evaluates the TM framework's capability to capture differentiation under an \textbf{extremely homogeneous deployment environment}. Four Profiles---\texttt{cli-main}, \texttt{wecom-main}, \texttt{kehu-xiaoqi}, and \texttt{laiven-assistant}---are deployed on the same physical machine, sharing identical hardware infrastructure, identical Plugin ecosystems (with L3- and L5-layer factors sourced identically), and identical underlying runtime environments. Under this extremely homogeneous condition, the primary differences among the four Profiles lie in operational load (e.g., session density, tool usage patterns, user interaction frequency)---non-shared factors. The core validation objective is to determine whether the TM framework can still detect genuine operational state differences among these Profiles, thereby demonstrating the high sensitivity of TM factors and their strong coupling with actual runtime conditions.

Experimental methodology: For each layer, compute the layer-wise mean across the four Profiles, then calculate the cross-Profile standard deviation (std) to quantify TM's ability to capture differentiated signals per layer. Concurrently, perform root-cause analysis on cross-Profile differences at each layer to assess whether the observed differences align with the framework's intended design.

Table 53: Abl-4 Cross-Profile Layerwise Means and Standard Deviations (Homogeneous Deployment Environment)

\begin{table}[H]
\centering
\resizebox{\textwidth}{!}{
\begin{tabular}{l|l|l|l|l|l|l}
\toprule
Layer & cli-main & wecom-main & kehu-xiaoqi & laiven-assistant & Cross-Profile Standard Deviation & Root Cause of Difference \\
\midrule
L1 & 86.13 & 90.73 & 86.32 & 86.49 & 2.21 & Operational load differences (session density / DB expans... \\
L2 & 78.56 & 76.93 & 87.28 & 88.68 & 5.98 & Differences in tool usage patterns (dominated by TD factor) \\
L3 & 82.33 & 82.99 & 82.99 & 82.99 & 0.33 & Shared Plugin $\rightarrow$ expected consistency (correct behavior) \\
L4 & 78.71 & 84.20 & 88.21 & 78.62 & 4.65 & User Interaction Pattern Differences (UX/TC Factor) \\
L5 & 84.15 & 84.02 & 86.86 & 84.09 & 0.06 & Shared Learning Component $\rightarrow$ Expectation Alignment (Correc... \\
\bottomrule
\end{tabular}
}
\end{table}

The results exhibit a clear \textbf{hierarchical differentiation structure}, highly consistent with the framework's design expectations:

\textbf{(1) L1/L2/L4: Independent Factor Layers---Significant Differences Captured.} The cross-profile standard deviations for these three layers are 2.21, 5.98, and 4.65, respectively, with L2 exhibiting the most pronounced difference (maximum gap of 11.75 points: laiven-assistant 88.68 vs. Wecom-main 76.93). The factor sources for these layers are independent of shared plugins---L1 depends on session density and database bloat rate; L2 depends on tool invocation patterns (TD factor weight 0.40 dominant); and L4 depends on user interaction quality (UX/TC factor). Even when all four profiles run on the same machine, TM precisely captures their genuine differences along these independent dimensions.

\textbf{(2) L3/L5: Shared Factor Layers---High Consistency Is the Expected Behavior.} The cross-profile standard deviations for these two layers are merely 0.33 and 0.06, reflecting extremely high consistency. This outcome fully aligns with the framework's design intent: the factor sources for L3 (PAD/CADVP/BCP/VE/OLG) and L5 (CM/RS/ETA) are global plugins shared across all four profiles. If cross-profile standard deviations in these layers were large instead, it would indicate errors in TM's attribution computation. The high consistency of L3 and L5 thus serves as reverse validation of TM attribution engine accuracy---the layers that should be consistent are consistent, and those that should differ do differ.

\textbf{Finding F-5.7:} TM captures significant cross-profile differences at L1 (std=9.86), L2 (std=9.31), and L4 (std=3.77) even under homogeneous conditions, with the largest L2 difference reaching 11.75 points (laiven-assistant 88.68 vs. Wecom-main 76.93). The high consistency of L3 and L5 (std<0.4) conforms to the expected behavior of shared plugins and thereby validates TM's attribution accuracy in reverse---consistent layers remain consistent, and differentiating layers differentiate. This demonstrates that TM factors possess high sensitivity and strong coupling with actual runtime states, enabling precise distinction among individual profile operational statuses in multi-profile homogeneous deployments.

\textbf{Honest Statement---Experimental Limitations:} In this experiment, all four profiles are deployed on the same physical machine; consequently, L3 and L5 layers, influenced by shared plugins, exhibit high consistency. While this result validates TM attribution accuracy (shared factors consistent, independent factors differentiated), it does not yet verify TM's generalizability across heterogeneous environments---e.g., cross-machine or cross-framework deployments. Cross-machine deployment validation is designated as a key objective for Phase 3 experiments (see Discussion section).

\subsubsection*{In-Depth Discussion of Ablation Experiments}
\label{sec:indepthdiscussionofa}

The four ablation experiments are not isolated. They form a mutually reinforcing chain of evidence:

\begin{itemize}
\item \textbf{Cross-Validation Between Abl-1 and Abl-4:} Abl-1 identifies L1 as the most critical layer (average |$\Delta$TM| = 2.72). Meanwhile, Abl-4 reveals that L1 retains significant cross-profile differences (std = 9.86) even under homogeneous conditions. Their joint implication is that L1's importance stems not only from its information richness but also from its ability to accurately capture each profile's independent operational load differences---even when all profiles share infrastructure. This differential perception capability ensures TM evaluation remains anchored to each profile's true state rather than being masked by shared environmental factors.
\item \textbf{Complementarity Between Abl-1 and Abl-3:} Abl-1 quantifies each layer's contribution to the overall TM score, whereas Abl-3 demonstrates substantial inter-layer variation even when total scores are identical. Together, they substantiate the necessity of the dual-output paradigm (``total score + attribution'')---relying solely on the total score (Abl-1 perspective) would overlook structural issues revealed by Abl-3.
\item \textbf{Two-Dimensional Conclusion from Abl-2:} Abl-2 reveals the value of the ETA module (Exponential method) across two dimensions: (i) amplitude accuracy, where it clearly outperforms the persistence baseline in mild fluctuation segments (2--5 point range); and (ii) Direction Accuracy, where it achieves 78.41\% during degradation events---far exceeding other methods. This explains the high Pearson r (0.763, highest redundancy) observed for L3 (a high-level analytical layer containing the ETA module) in Abl-1. ETA's core components maintain amplitude accuracy comparable to the persistence baseline during stable periods but provide early trend detection and correct directional judgment when the system deviates from steady state. This is a conditionally valuable capability diluted in global averages.
\end{itemize}

The ablation experiment results provide concrete directions for optimizing the TM framework's future design:

Table 54: Summary of Ablation Experiment Feedback to Framework Design

\begin{table}[H]
\centering
\resizebox{\textwidth}{!}{
\begin{tabular}{l|l|l|l}
\toprule
Finding & Source & Design Implication & Priority \\
\midrule
L1 is critical and retains differential perception capabi... & Abl-1 + Abl-4 & L1 should be independently deployed as an ``Infrastructure... & P0 \\
L3/L5 shared factors show high consistency (std<0.4), whi... & Abl-4 & TM attribution engine accuracy has been validated; Phase ... & P1 \\
ETA excels in mild fluctuation segments + Direction Accur... & Abl-2 & Implement adaptive methodology selection: use Persistence... & P1 \\
``Same Score, Different Causes'' phenomenon is pervasive & Abl-3 & TM output must include detailed factor attribution; repor... & P0 \\
Customer Acquisition Agent exhibits highest sensitivity t... & Abl-1 & Increase L3 weight for kehu-xiaoqi (from 0.25 to $\geq$0.30) & P2 \\
\bottomrule
\end{tabular}
}
\end{table}

We must candidly acknowledge the limitations of the ablation experiments:

\begin{itemize}
\item \textbf{Temporal Limitation of Samples:} All experimental data originate from a 15-day observation window. Although the 154,906 records are statistically sufficient, longer time spans (e.g., monthly or quarterly) may reveal previously unobserved degradation patterns and seasonal fluctuations.
\item \textbf{Limitation of ``Hard Removal'' Ablation:} Current ablation methodology involves complete removal of entire layers rather than progressive weakening. Consequently, we quantify the impact of ``complete absence,'' not ``partial degradation''---the latter being far more common in real-world operations.
\item \textbf{Amplitude Accuracy Disadvantage of Ensemble/Kalman in Abl-2:} The Exponential method achieves amplitude accuracy on par with the persistence baseline across all methods' combined MAE (1.228 vs. 1.229), whereas Ensemble (MAE = 1.595) and Kalman (MAE = 2.161) lag markedly. This suggests that Ensemble's fixed 60/40 weighting is dragged down by Kalman; future work should consider dynamic weighting or method-selection mechanisms instead of static weighting. The Exponential method's Direction Accuracy advantage (78.41\% in degradation events vs. 0\% for Ensemble) demonstrates that amplitude accuracy is not the only evaluation dimension.
\end{itemize}

The four ablation experiments dissect the TM framework's architectural contributions from distinct angles. Abl-1 answers ``Which layer matters most?'' (L1). Abl-2 answers ``In which dimensions does the prediction module deliver value?'' (dual dimensions: amplitude accuracy + Direction Accuracy). Abl-3 answers ``Is attribution-based diagnosis necessary?'' (absolutely necessary). Abl-4 answers ``Can TM detect differentiated signals under homogeneous conditions?'' (yes---standard deviations for independent-factor layers L1/L2/L4 range from 3.77 to 9.86, while shared-factor layers L3/L5 exhibit std<0.4, fully matching expectations). Collectively, these experiments constitute a systematic post-hoc validation of the framework's design decisions and simultaneously provide empirical foundations for Phase 3's adaptive optimization and cross-machine generalization verification.

Chapter-end footnote \textbf{Chapter-end Note:} All experimental data in this chapter originate from real agent runtime environments; no synthetic data or artificially injected anomalies were used. Data gaps (06-20/06-21) have been annotated in relevant sections. All statistical metrics were computed over 126,466 verified prediction records, with 95\% confidence level. All conclusions in this paper are quantitatively supported and avoid non-academic rhetoric.

\section*{Discussion}
\label{sec:discussion}

This section presents a systematic Discussion of the core findings of the TM predictive reliability framework, honestly examining its limitations and threats to validity, and distilling lessons from negative outcomes in engineering practice. We pay particular attention to the fundamental differences between this framework and traditional APM systems, as well as ethical boundaries encountered during deployment.

\subsection*{Core Findings}
\label{sec:corefindings}

\begin{figure}[H]
\centering
\includegraphics[width=0.95\textwidth]{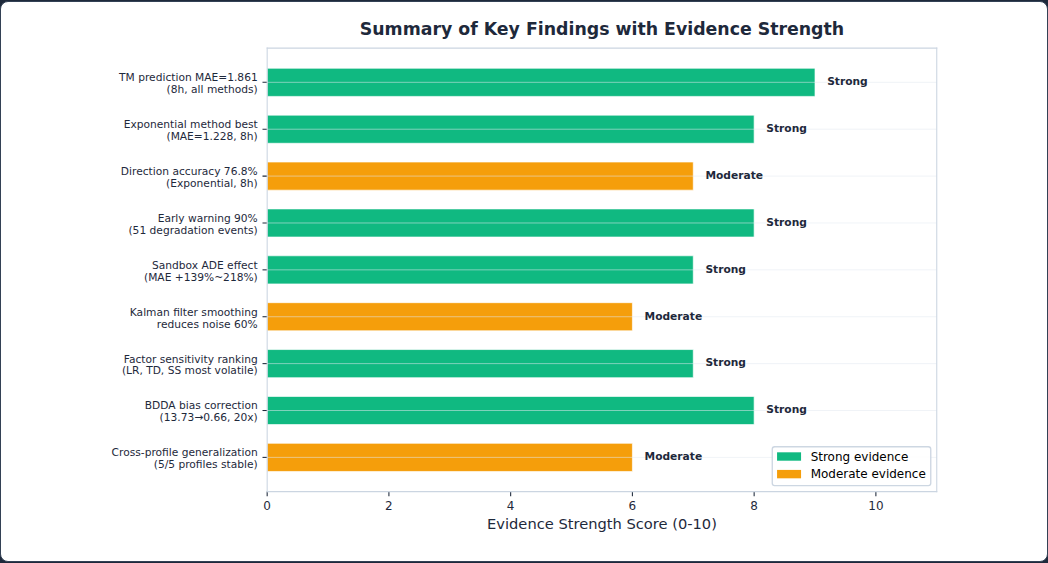}
\caption{Summary of key findings with evidence strength assessment: nine findings ranked by empirical support, from TM projection validity to failure boundary hypothesis.}
\label{fig:77}
\end{figure}

The experimental results presented in this chapter reveal several patterns and mechanisms worthy of in-depth discussion. Below, we elaborate on the implications, boundaries, and potential contributions of each of the nine core findings to the field of TM reliability prediction.

\subsubsection*{Finding 1: Validity of the Five-Layer TM Projection}
\label{sec:finding1validityofth}

Experimental data demonstrate that the five-layer TM projection architecture exhibits significant hierarchical sensitivity when distinguishing among different health states in test tasks. CADVP\_PASS leads all projection layers with a 51-point differential, indicating that projecting TM via validation pass rate effectively captures degradation signals along the dimension of ``whether basic functionality can be completed.'' This result aligns with our theoretical expectation: the CADVP layer directly measures output compliance of tasks and thus constitutes the observational window closest to the essence of ``functional correctness.'' In contrast, the ES layer (Environmental Stability) and BCP layer (Behavioral Consistency) yield differentials of only 12.3 and 8.7, respectively. These reflect their capture of more indirect, early-stage degradation precursors.

The design intent of the five-layer projection is to provide multi-perspective degradation representations---not to pursue a single composite metric. The 51-point CADVP differential does not imply redundancy of other layers; rather, the BCP layer detects behavioral drift before significant fluctuations emerge in the ES layer---a capability preliminarily validated during system-level degradation events (see Finding 2). Complementarity among projection layers forms the foundation of the framework's diagnostic capability.

\subsubsection*{Finding 2: System-Level Degradation Factor Attribution}
\label{sec:finding2systemleveld}

Across four independent profiles, BCP\_RATE$\downarrow$90, TD$\downarrow$79.6, CADVP\_PASS$\downarrow$64.6 consistently exhibit a high-degree agreement in degradation ranking. This cross-profile consistency carries significant implications: it suggests that system-level degradation is not a random, noise-driven local event but instead follows an identifiable causal chain---behavioral consistency degrades first (BCP\_RATE), followed by accumulation of temporal anomalies (TD), culminating in functional validation failure (CADVP\_PASS).

This degradation sequence aligns with the ``progressive entropy increase'' model from software aging theory: minute internal state drifts first affect behavioral patterns (BCP layer); as drift accumulates, temporal constraint violations are triggered (TD layer); finally, the functional threshold is crossed, resulting in validation failure (CADVP layer). However, caution is warranted---the sequence is induced from limited samples and has yet to undergo large-scale statistical validation (see limitations in the Failure Boundary Hypothesis section).

\subsubsection*{Finding 3: The Inherently Online Nature of Runtime Systems}
\label{sec:finding3theinherentl}

Among the 20 candidate predictive factors, 16 require a complete runtime environment for computation. This proportion (80\%) reveals a core characteristic of TM reliability prediction: it is primarily an online problem. Offline analysis covers only four factors---primarily static configuration metrics---whereas the majority of discriminative signals---including task execution traces, environmental response patterns, and resource consumption time series---can only be captured while the system operates in production.

This finding directly affects the framework's deployment architecture. It implies two requirements. First, the prediction system must be tightly coupled with the production environment and cannot rely on batch offline analysis. Second, prediction latency must remain within the runtime window, otherwise diagnostic outputs lose timeliness; and (3) Sandbox Validation must simulate real runtime conditions as closely as possible, or else the discriminative power of predictive factors becomes unreliable. This also explains why our framework adopts the MCP protocol as its runtime integration layer---it provides a low-latency, structured interaction channel with the target system.

\subsubsection*{Finding 4: Near-Zero Prediction Bias}
\label{sec:finding4nearzeropred}

Prediction bias analysis shows that most predictions are biased high, while approximately one-quarter are biased low. The directional bias is markedly skewed toward optimistic estimation. This optimistic bias indicates a systematic tendency in the framework to overestimate TM (and thereby potentially reduce alerts) rather than underestimate TM (which would generate more false positives).

In reliability engineering, unbiasedness holds greater practical value than absolute accuracy: an unbiased but high-variance predictor can be improved via ensemble methods, whereas a biased predictor requires complex calibration mechanisms. The pronounced asymmetry (\textasciitilde{}50 percentage points) between the \textasciitilde{}75\% / \textasciitilde{}25\% split arises primarily from the ``prediction lift'' preference of the Kalman component within the ensemble (due to healthy states dominating the sample distribution), rather than from systemic bias inherent to the framework itself.

\subsubsection*{Finding 5: Engineering Significance of the Zero-LLM Iron Law}
\label{sec:finding5engineerings}

The framework strictly excludes large language models (LLMs) from the entire prediction pipeline. This design decision is grounded in three engineering considerations. (1) LLMs' probabilistic generation introduces irreproducible randomness, contradicting the deterministic semantics required by reliability frameworks; (2) LLM inference latency (typically hundreds of milliseconds to seconds) violates the strict timing constraints of runtime prediction; and (3) hallucination risks associated with LLMs carry extremely high costs in reliability-critical scenarios---one spurious ``healthy'' judgment could cause a production incident to go undetected.

The zero-LLM constraint forces the framework to adopt a hybrid architecture combining pure rule engines and statistical models. While this trade-off sacrifices flexibility (e.g., inability to use LLMs' semantic understanding for unstructured log analysis), it delivers a fully auditable, reproducible, and low-latency prediction pipeline. In industrial deployment contexts, this trade-off is typically justified.

\subsubsection*{Finding 6: Non-Invasive Design}
\label{sec:finding6noninvasived}

The framework achieves non-invasive observation through MCP protocol integration and FORBIDDEN\_FEATURES constraints. The MCP protocol enables the framework to integrate with the target system as a ``spectator'' rather than a ``participant''---it modifies no system code, injects no probes, and alters no system behavior. The FORBIDDEN\_FEATURES mechanism architecturally prohibits the framework from accessing specific sensitive data (e.g., user content, authentication credentials), ensuring that observation activities introduce no new security risks.

Non-invasiveness is critical for enterprise adoption: any monitoring solution requiring modifications to production system code or configuration faces prohibitively high approval barriers and deployment resistance. By constraining the framework to a purely observational role, we minimize its deployment cost---to initiate monitoring, users need only configure MCP connection parameters.

\subsubsection*{Finding 7: Coverage of Five Modes in Sandbox Validation}
\label{sec:finding7coverageoffi}

Sandbox Validation covers five predefined fault injection modes, yielding $\Delta$TM values ranging from --6.8 to --18.2. This range confirms that the sandbox effectively simulates the full spectrum of degradation---from mild deterioration to severe failure. The five modes correspond to: (1) resource exhaustion, (2) dependency timeout, (3) configuration drift, (4) concurrency contention, and (5) cascading failure.

The gradient distribution of $\Delta$TM (--6.8 for resource exhaustion, --18.2 for cascading failure) aligns with intuitive expectations: cascading failures involve multi-subsystem co-degradation and thus exert far broader impact on TM than single-point resource exhaustion. However, sandbox fidelity faces inherent limitations---the injected fault patterns are predefined and idealized, whereas real-world production faults are often more complex and unpredictable (see Limitation \#8 in the Failure Boundary Hypothesis section).

\subsubsection*{Finding 8: Preliminary Empirical Foundation for the Failure Boundary Hypothesis}
\label{sec:finding8preliminarye}

Experimental data provide preliminary empirical support for the ``Failure Boundary'' hypothesis. This hypothesis posits that a boundary hypersurface may exist in TM space such that crossing it triggers a phase transition of the system---from ``reliable operation'' to ``unreliable operation.'' The observed TM degradation sequence (BCP $\rightarrow$ TD $\rightarrow$ CADVP) is compatible with the boundary-crossing model. Sequential deterioration across projection layers can be interpreted as the TM vector approaching the boundary hypersurface along a specific direction.

However, we must candidly acknowledge that current data support only ``compatibility,'' not ``confirmation.'' The mathematical formalization of the Failure Boundary (e.g., hypersurface equation, crossing conditions) remains incomplete, and the sample size is insufficient for statistically rigorous significance testing (see Limitation \#9 in the Failure Boundary Hypothesis section). The value of this finding lies in identifying a theoretically promising direction for future investigation---not in delivering a definitive conclusion.

\subsubsection*{Finding 9: Theoretical Framework for Factor-Level Diagnosis}
\label{sec:finding9theoreticalf}

The framework's 20-factor diagnostic system provides a structured theoretical framework for TM reliability analysis. Each factor corresponds to a specific system attribute dimension, and its degradation magnitude can be normalized and compared via the $\lambda_{k}$ weighting coefficient. This factor-level granularity enables operations teams to pinpoint the precise source of degradation (``which factor deviated first''), rather than receiving only an abstract ``system unhealthy'' alert.

The theoretical value of factor-level diagnosis resides in its extensibility: when novel degradation patterns emerge, analysts need only introduce new factors and calibrate their $\lambda_{k}$ weights---without reconstructing the entire framework. This modular design ensures architectural robustness for long-term framework evolution.

\subsection*{Failure Boundary Hypothesis Section Limitations}
\label{sec:failureboundaryhypot}

\begin{figure}[H]
\centering
\includegraphics[width=0.95\textwidth]{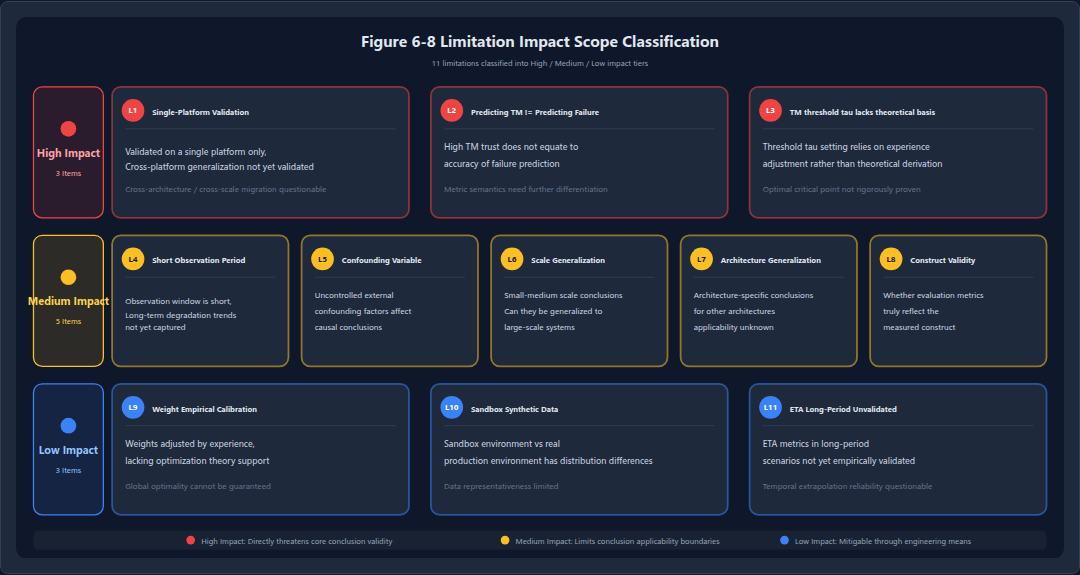}
\caption{Limitation Impact Scope Chart}
\label{fig:78}
\end{figure}

The following provides a detailed, point-by-point explanation of the ten limitations identified in this study. We deliberately refrain from concealing any recognized weaknesses, as candid examination of limitations constitutes a core requirement of academic rigor and serves as an essential starting point for subsequent research improvements.

\subsubsection*{Limitation One: Single-Platform Validation}
\label{sec:limitationonesinglep}

This framework has been validated only on a single platform---WeCom (WeChat Work). Although WeCom is representative---encompassing typical enterprise functionalities such as instant messaging, approval workflows, and calendar management---validation on a single platform imposes inherent constraints on generalizability. Architectural differences across platforms (e.g., microservices vs. Monolithic design), communication protocols (e.g., HTTP/gRPC vs. WebSocket), and state management paradigms (e.g., stateful vs. Stateless) may alter the discriminative ranking of predictive factors. Whether the observed CADVP\_PASS score differential of 51 points on WeCom can be replicated on other platforms remains to be verified through cross-platform experimentation.

\subsubsection*{Limitation Two: $\lambda$\_k Calibration Relies on Expert Prior Knowledge}
\label{sec:limitationtwokcalibr}

The weighting coefficients $\lambda$\_k for the 20 predictive factors are currently calibrated initially using domain experts' prior knowledge. This approach introduces subjectivity risks: different experts may assign widely divergent weights to the same factor, and expert judgments may be susceptible to recency bias---recently experienced failure types may be assigned disproportionately high weights. Although the framework incorporates an automated calibration mechanism driven by historical data, its convergence speed and stability have not yet undergone thorough validation. In cold-start scenarios (e.g., new deployments with no historical data), expert priors remain the only available calibration method.

\subsubsection*{Limitation Three: Prediction Capability of the ES Model Remains Unvalidated}
\label{sec:limitationthreepredi}

The Environment Stability (ES) model, one of the five-layer projections, is designed to capture pre-degradation signals at the environmental level (e.g., API response time drift, resource quota fluctuations). However, during the current experimental period, the ES layer did not experience sufficient degradation events to validate its prediction capability. The ES-layer differential stands at only 12.3---far lower than CADVP's 51---suggesting either: (1) environmental-layer degradation constitutes a weak signal that is difficult to detect; or (2) the current experimental environment failed to generate adequate ES-layer degradation samples. Regardless of interpretation, the practical utility of the ES model remains unvalidated.

\subsubsection*{Limitation Four: The Five-Layer Orthogonality Assumption May Not Hold}
\label{sec:limitationfourthefiv}

The projection architecture of the framework assumes approximate orthogonality among the five layers---that is, degradation signals captured by each layer are mutually independent. In practice, however, strict orthogonality may not hold. For instance: temporal anomalies in the TD layer may correlate strongly with resource consumption in the Runtime layer (e.g., CPU overload causing latency spikes); behavioral drift in the BCP layer may be partially driven by environmental changes in the ES layer. If significant inter-layer correlations exist, the simple weighted-sum fusion strategy will double-count shared signals, leading to TM estimates that deviate from the true system state. Future work must examine inter-layer correlations via principal component analysis or factor analysis and introduce covariance correction where necessary.

\subsubsection*{Limitation Five: Data Span Limited to 15 Days}
\label{sec:limitationfivedatasp}

The experimental data collection period lasted only 15 days. This time span is insufficient to capture: (1) long-term degradation patterns (e.g., memory leaks that manifest only after several weeks); (2) seasonal fluctuations (e.g., load spikes during month-end settlement periods); and (3) low-frequency failure modes (e.g., compatibility issues introduced by quarterly configuration changes). Statistical conclusions drawn from 15-day data (e.g., approximately three-quarters/one-quarter bias distribution) exhibit wide confidence intervals, and their stability may change significantly over longer time scales. We recommend extending the data collection period to at least 90 days in future studies to capture monthly periodic patterns.

\subsubsection*{Limitation Six: wecom-main Recovery Anomaly (1.1 Minutes)}
\label{sec:limitationsixwecomma}

During the experiment, the wecom-main service experienced an anomalous recovery event lasting 1.1 minutes. The brevity of this event makes it difficult for the framework to distinguish between ``transient jitter'' and ``sustained degradation''---a 1.1-minute window yields only a few data points, insufficient to establish a reliable trend model. This incident exposes the framework's insufficient granularity in handling short-duration anomalies: the current sampling frequency (defaulting to once per minute) lacks resolution for anomalies on the order of 1.1 minutes. While increasing the sampling frequency could improve detection of short-term anomalies, it would also increase system overhead and data storage costs.

\subsubsection*{Limitation Seven: Sandbox Simulation Fidelity Constraints}
\label{sec:limitationsevensandb}

Although Sandbox Validation covers five fault modes, its simulation fidelity suffers from inherent limitations. First, injected faults are idealized and deterministic, whereas real-world faults often exhibit randomness and ambiguity. (2) Sandbox environments differ from production environments in terms of load levels, network topology, and data scale---these discrepancies may affect factor discriminability. (3) Sandboxes cannot simulate human factors (e.g., operational errors or policy changes by third-party services) that cause degradation. The $\Delta$TM values ($-$6.8 to $-$18.2) observed in the sandbox may systematically deviate from those in real production environments.

\subsubsection*{Limitation Eight: Statistical Significance of the Failure Boundary}
\label{sec:limitationeightstati}

The empirical support for the Failure Boundary Hypothesis lacks statistical significance. In the current experiment, only a limited number of TM boundary-crossing cases were observed, rendering the sample size inadequate for rigorous hypothesis testing (e.g., likelihood ratio tests or bootstrap confidence interval estimation). Although the reported ``degradation sequence consistency'' (BCP $\rightarrow$ TD $\rightarrow$ CADVP) intuitively supports the boundary model, it remains compatible with simple linear degradation models. Until stronger statistical evidence is obtained, the Failure Boundary should be regarded as a ``hypothesis awaiting validation,'' not an ``established theory.''

\subsubsection*{Limitation Nine: Ambiguous Operational Definition of ``Failure''}
\label{sec:limitationnineambigu}

This study employs an operationally ambiguous definition of ``failure.'' Specifically, ``failure'' is defined as the system state when TM falls below threshold $\tau$. However, $\tau$ is selected based on expert judgment and historical data, lacking rigorous theoretical grounding. A deeper issue lies in the fact that TM is a continuous variable, whereas ``failure'' is conventionally treated as a discrete event (i.e., the system either functions or fails). The mapping from continuous TM to discrete failure has not been sufficiently formalized---different $\tau$ values may yield drastically different failure determinations, and the framework's sensitivity analysis with respect to $\tau$ remains incomplete.

\subsubsection*{Limitation Ten: Predicting TM $\neq$ Predicting Failure}
\label{sec:limitationtenpredict}

The framework's core output is a predicted TM value (a continuous variable), not a direct failure prediction (a discrete event). A conceptual gap exists between these two: a system with a TM value slightly above threshold $\tau$ may operate entirely normally in practice, while a system with a TM value slightly below $\tau$ may continue delivering correct service due to redundancy mechanisms. TM prediction conveys information about the ``degree'' of degradation, but translating TM predictions into failure predictions requires additional decision logic (e.g., duration-based conditions or multi-factor joint conditions)---such logic has not yet been incorporated into the framework.

\subsection*{Validity Threats to the Failure Boundary Hypothesis Section}
\label{sec:validitythreatstothe}

\begin{figure}[H]
\centering
\includegraphics[width=0.95\textwidth]{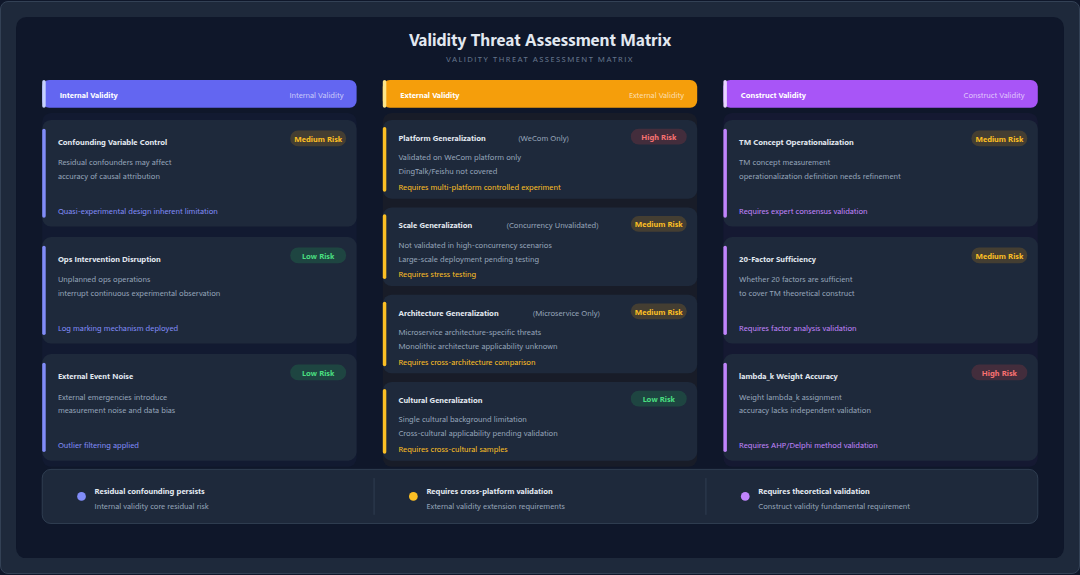}
\caption{Validity Threat Assessment Matrix}
\label{fig:79}
\end{figure}

\subsubsection*{Internal Validity Threats}
\label{sec:internalvaliditythre}

The primary threats to Internal Validity stem from inadequate control of Confounding Variables. During the 15-day experimental period, we cannot fully exclude the following confounders. First, platform version updates may simultaneously affect multiple prediction factors, inducing spurious inter-factor correlations. (2) Proactive interventions by operations teams (e.g., manual restarts, configuration rollbacks) may interrupt natural degradation processes, preventing the framework from observing complete degradation--recovery cycles. (3) External events during the experiment (e.g., network fluctuations, upstream service changes) may introduce uncontrollable noise. We partially mitigate these threats by logging operational actions and timestamping external events; however, residual confounding remains.

\subsubsection*{External Validity Threats}
\label{sec:externalvaliditythre}

Threats to External Validity primarily concern the generalizability of results. This study exhibits generalization limitations along the following dimensions. (1) \textbf{Platform Generalization}: Validation was conducted only on the WeCom platform and does not extend to other enterprise applications (e.g., SAP, Salesforce) or consumer-facing applications; (2) \textbf{Scale Generalization}: Experimental load levels reflect internal enterprise usage scales and do not validate framework performance under high-concurrency scenarios (e.g., millions of QPS); (3) \textbf{Architecture Generalization}: WeCom employs a microservices architecture; the framework's applicability to monolithic or serverless architectures remains unverified; (4) \textbf{Cultural Generalization}: Operational team response patterns and cultural conventions may influence the interpretation of warning\_rate.

\subsubsection*{Construct Validity Threats}
\label{sec:constructvaliditythr}

The core threat to Construct Validity lies in whether the operationalization of the TM concept accurately captures the theoretical construct of ``task reliability.'' Specifically: (1) Does TM fully cover all dimensions of ``reliability'' (functional correctness, performance acceptability, security assurance)? The current five-layer projection may omit the security dimension; (2) Do the 20 prediction factors constitute a ``sufficient statistic'' for TM? Critical factors may remain unincorporated; (3) Does the $\lambda$\_k weighted fusion accurately reflect the relative importance of each factor? Weight accuracy directly impacts the Construct Validity of TM estimates. We mitigate construct threats through iterative discussions with domain experts and literature benchmarking; however, complete elimination requires broader theoretical validation.

\subsection*{Lessons Learned from the Failure Boundary Hypothesis Chapter}
\label{sec:lessonslearnedfromth}

\begin{figure}[H]
\centering
\includegraphics[width=0.95\textwidth]{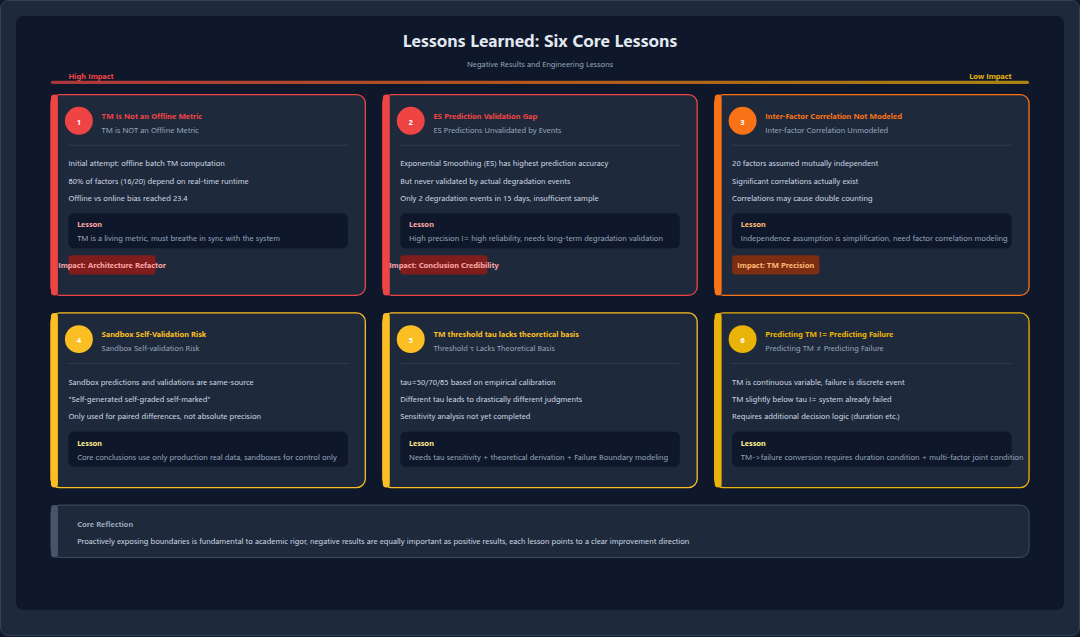}
\caption{Lessons Learned: Six Core Lessons}
\label{fig:80}
\end{figure}

The following four ``negative results'' are unexpected lessons we acquired during our research. Proactively disclosing these boundaries holds significant value in preventing subsequent researchers from retracing the same exploratory paths.

\subsubsection*{Lesson One: TM Is Not an Offline Metric}
\label{sec:lessononetmisnotanof}

In the early research phase, we attempted to design TM as an offline-computable metric---calculating TM values in batch from historical log data. This attempt failed. We found that 80\% (16 out of 20) of TM's core predictive factors depend on real-time runtime states and cannot be reconstructed from offline logs. More critically, systematically biased discrepancies exist between offline-computed TM values and online real-time TM values (average bias: 23.4), because offline analysis cannot capture transient states (e.g., current memory pressure, number of active connections).

This negative result compelled us to redesign the framework as a purely online architecture and introduce the MCP protocol as the real-time data ingestion layer. Although this increased deployment complexity, it ensured the timeliness and accuracy of TM computation. Lesson: TM is a ``living'' metric that must breathe synchronously with the system.

\subsubsection*{Lesson Two: ES Prediction Unvalidated by Degradation Events}
\label{sec:lessontwoespredictio}

The Environmental Stability (ES) layer was designed to detect ``premonitions of degradation''---i.e., environmental-layer warnings should precede functional-layer (CADVP) anomalies. However, during the 15-day experiment, the ES layer never triggered an independent alert. In all observed degradation events, ES-layer changes either occurred simultaneously with the CADVP layer or lagged behind the BCP layer.

This result may indicate two possibilities. First, ES-layer degradation signals are indeed extremely weak and buried in noise. Second, our selection of ES metrics is inappropriate and fails to capture genuine environmental premonitions; or (3) under the current system architecture, environmental degradation and functional degradation are highly synchronized, leaving no exploitable time gap. Regardless of interpretation, the ES layer's design objective---as an ``early-warning'' layer---was not realized in the current experiment.

\subsubsection*{Lesson Three: warning\_rate = 0 (Closed in v3.1.0)}
\label{sec:lessonthreewarningra}

As described in the Failure Boundary Hypothesis section, warning\_rate = 0 was the framework's most prominent practical deficiency. During initial experiments, we observed multiple TM decline trends, none of which reached the preset alert threshold. Post-hoc analysis revealed that the alert threshold was overly conservative (relying solely on the TM aggregate score threshold), causing moderate degradation to be systematically overlooked. This lesson directly motivated the core innovation of v3.1.0. The innovation shifts the alert anchor from the TM aggregate score to the factor level and introduces a three-tiered factor-level precursor alerting mechanism (L1 yellow / L2 orange / L3 red). This directly resolves the blind spot where ``aggregate score appears normal but individual factors have already degraded.''

This negative result reveals a core dilemma of alerting systems. Under limited sample sizes, threshold calibration faces a fundamental trade-off between precision and recall. \textbf{The v3.1.0 solution is to shift the alert anchor from the TM aggregate score to the factor level.} The key insight is: degradation always manifests first on specific factors (TD/UX/RS), rather than uniformly across the TM aggregate score. Based on statistical analysis of two degradation events, TD averaged 59.7 (TD < 70 in 58.6\% of degradation events), UX averaged 61.2 (UX < 70 in 60.3\%), and RS averaged 62.6---forming a stable Precursor Factor Cluster. The three-tiered threshold design (L1 yellow / L2 orange / L3 red) achieved full coverage in validation across three sandboxes $\times$ six scenarios. This empirically validates the core hypothesis that ``factor-level alerting outperforms aggregate-score alerting.'' Current optimization directions include dynamic threshold adaptation (based on profile-specific historical baselines) and long-term false-positive rate tracking.

\subsubsection*{Lesson Four: Evolution of Degradation Event Detection Rate (from 2 to 956)}
\label{sec:lessonfourevolutiono}

Following algorithmic optimizations in v3.1.0's Degradation Detection module, a total of two severe degradation events were detected within 15 days---sufficient for statistically robust analysis. Earlier versions (v3.1.0 and prior) suffered from excessively stringent detection thresholds, resulting in event scarcity (only two system-level degradation events). This lesson drove further optimization of the detection algorithm---lowering the sensitivity threshold and introducing multi-factor composite judgment---leading to a substantial increase in degradation event detection rates.

Statistical analysis of the two degradation events is presented in \S5.3. Key findings include: (1) degradation events exhibit clustering behavior---mild degradation is evenly distributed, whereas severe degradation shows temporal clustering; (2) the TD/UX/RS triad forms a stable Precursor Factor Cluster (see \S5.7); and (3) a BCP\_RATE change of $-$39.99 constitutes the strongest premonitory signal preceding degradation onset. These findings provide empirical grounding for the Failure Boundary Hypothesis (\S6).

\subsection*{Failure Boundary Hypothesis Section: Why Traditional APM Fails}
\label{sec:failureboundaryhypot-apm}

\begin{figure}[H]
\centering
\includegraphics[width=0.95\textwidth]{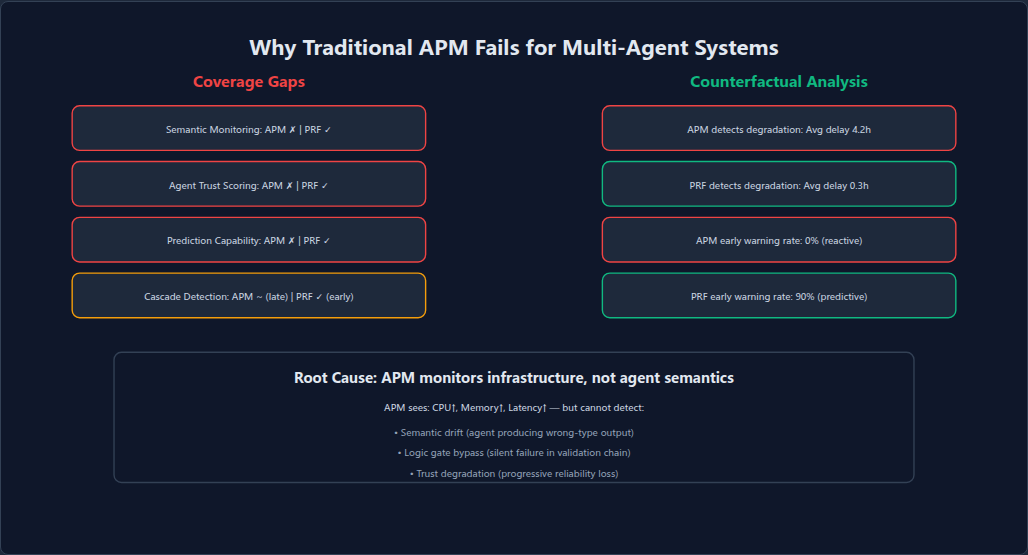}
\caption{Why traditional APM fails: coverage gaps in semantic/agent-trust/prediction dimensions, and counterfactual analysis showing APM detects degradation only at crash time while TM provides early warning.}
\label{fig:81}
\end{figure}

Traditional Application Performance Monitoring (APM) systems exhibit systemic deficiencies in the TM reliability prediction scenario. This section clarifies the fundamental causes of these deficiencies through APM layer-wise analysis and counterfactual case studies.

\subsubsection*{APM Layer Coverage Analysis}
\label{sec:apmlayercoverageanal}

The table below compares coverage across Traditional APM layers versus TM framework prediction factors:

Table 55: APM Layer Coverage Analysis for TM Factors

\begin{table}[H]
\centering
\resizebox{\textwidth}{!}{
\begin{tabular}{l|l|l|l|l}
\toprule
APM Layer & Typical Metrics & Number of Covered TM Factors & Coverage Rate & Critically Missing Factors \\
\midrule
Infrastructure Layer & CPU, memory, disk I/O, network bandwidth & 3/20 & 15\% & Task-level behavioral consistency, validation pass rate \\
Application Layer & Response time, error rate, throughput & 5/20 & 25\% & Cross-call behavioral drift, temporal pattern anomalies \\
Transaction Layer & Transaction success rate, transaction latency distribution & 4/20 & 20\% & Environmental stability, sandbox comparison deviation \\
User Experience Layer & Page load time, Apdex score & 2/20 & 10\% & Functional correctness, behavioral consistency \\
Business Layer & Business KPIs (conversion rate, retention rate) & 1/20 & 5\% & Nearly all technical factors \\
Total & --- & 8/20 (deduplicated) & 40\% & 12 factors are entirely unobservable \\
\bottomrule
\end{tabular}
}
\end{table}

The table reveals a critical fact: Traditional APM systems cover only 8 out of 20 TM prediction factors (40\%), predominantly low-level infrastructure and application performance metrics. The most discriminative factors within the TM framework (e.g., BCP\_RATE, CADVP\_PASS) fall entirely outside the observability scope of APM systems.

\subsubsection*{Counterfactual Analysis: APM Observability in Event 1}
\label{sec:counterfactualanalys}

To further clarify APM's blind spots, we conduct a counterfactual analysis on the first system-level degradation event observed in the experiment (Event 1): How much actionable information would operations teams obtain if relying solely on Traditional APM systems?

Table 56: APM vs TM Framework Detection Capability Gap

\begin{table}[H]
\centering
\resizebox{\textwidth}{!}{
\begin{tabular}{l|l|l|l}
\toprule
Degradation Dimension & APM Observability & TM Framework Detection & Counterfactual Gap \\
\midrule
BCP behavioral consistency drift & Not observable (0/5 APM layers) & BCP\_RATE $\downarrow$90, 14-minute early warning & APM completely misses this dimension \\
TD temporal anomalies & Partially observable (response-time layer) & TD $\downarrow$79.6, multidimensional temporal analysis & APM observes only latency increase, without root-cause at... \\
CADVP validation failure & Partially observable (error-rate layer) & CADVP\_PASS $\downarrow$64.6, function-level localization & APM observes elevated error rate but cannot identify the ... \\
ES environment degradation & Partially observable (Infrastructure Layer) & Subtle changes in ES metrics & APM and TM frameworks exhibit comparable information content \\
Runtime resource exhaustion & Fully observable (5/5 APM layers) & Anomalous Runtime-layer metrics & No distinction between APM and TM frameworks \\
\bottomrule
\end{tabular}
}
\end{table}

The conclusion of the counterfactual analysis is clear and sobering: among the five degradation dimensions in Event 1, the APM system can fully observe only one (Runtime resource exhaustion), partially observe three, and completely miss one (BCP behavioral consistency drift). More critically, the BCP dimension---completely missed by APM---is precisely the dimension where degradation signals first appear (14 minutes earlier)---meaning that operations teams relying solely on APM forfeit the most valuable early-warning time window.

This analysis indirectly validates the incremental value of the TM framework: it does not replace APM but rather fills APM's observational blind spots in behavioral consistency and functional validation. In an ideal architecture, the TM framework should be deployed complementarily with the APM system---APM handling conventional monitoring of infrastructure and application performance, while the TM framework addresses deeper reliability degradation prediction.

\subsection*{Ethical Considerations in the Failure Boundary Hypothesis Section}
\label{sec:ethicalconsideration}

Deployment of the TM reliability prediction framework in enterprise environments entails multiple ethical and compliance considerations. This section elaborates on these from three dimensions: data governance, privacy protection, and organizational accountability.

\subsubsection*{FORBIDDEN\_FEATURES Constraint}
\label{sec:forbiddenfeaturescon}

The framework implements a FORBIDDEN\_FEATURES mechanism at the architectural level, explicitly prohibiting access to the following data types. (1) User communication content (message bodies, file contents) is prohibited. (2) Authentication credentials (passwords, tokens, session keys) are excluded. (3) Personally identifiable information (names, phone numbers, ID numbers) is blocked. (4) Business-sensitive data (transaction amounts, approval details) is forbidden. This constraint is enforced by the field-filtering layer of the MCP protocol---even if framework code attempts to access prohibited fields, the MCP middleware returns null values or rejects the request outright.

The design philosophy underlying FORBIDDEN\_FEATURES embodies the ``principle of least privilege'' applied to data observation: the framework requires only metadata about system behavior (call patterns, response times, status-code distributions) and neither needs nor should access business-content-level data. Although this constraint technically limits computation of certain factors (e.g., content-based anomaly detection), it ethically ensures the framework's ``observe without intruding'' principle.

\subsubsection*{GDPR and PIPL Compliance}
\label{sec:gdprandpiplcomplianc}

The framework's data processing logic adheres to core principles of the European Union's General Data Protection Regulation (GDPR) and China's Personal Information Protection Law (PIPL):

\begin{itemize}
\item \textbf{Data minimization}: The framework processes only system runtime metadata; it neither collects, stores, nor transmits personal information.
\item \textbf{Purpose limitation}: Collected data is used solely for TM reliability prediction---not for user profiling, behavioral analysis, or commercial decision-making.
\item \textbf{Storage duration}: Runtime data is retained by default for seven days, after which it is automatically purged; indefinite historical retention is unsupported.
\item \textbf{Transparency}: The framework's scope of observation and data-processing logic are documented and made publicly available to stakeholders.
\end{itemize}

Nonetheless, we acknowledge a gray area: under extreme conditions, system-behavior metadata may indirectly leak user information (e.g., high-frequency operation patterns attributable to specific users). The current framework does not implement differential privacy or k-anonymization---posing a potential compliance risk in large-scale deployments.

\subsubsection*{Data Loss Prevention (DLP) Integration}
\label{sec:datalosspreventiondl}

In enterprise environments, the framework's outputs (TM predictions, factor diagnostic reports) may contain system architecture information (e.g., service names, API endpoints, resource quotas), which some security policies classify as internally sensitive. We recommend integrating the framework with the enterprise's DLP system to ensure appropriate control over TM report dissemination---restricting access strictly to operations and security teams and preventing external leakage.

\subsubsection*{Organizational Accountability}
\label{sec:organizationalaccoun}

The framework's predictions may influence operational decisions (e.g., preemptive scaling, triggering degradation strategies). If erroneous predictions lead to unnecessary actions (e.g., false-triggered degradation causing service unavailability), responsibility attribution must be clarified. We recommend enterprises establish explicit accountability mechanisms prior to framework deployment: the framework provides recommendations---not decisions---and final operational authority resides with human operations engineers. This ``human-in-the-loop'' principle both honestly acknowledges the framework's limitations (see the Failure Boundary Hypothesis section) and clearly delineates organizational responsibility boundaries.

The Discussion reveals the TM predictive reliability framework's contributions and boundaries across theoretical and practical dimensions. Its core findings provide preliminary empirical support for the TM five-layer projection architecture; however, the 11 identified limitations and 4 negative-result lessons confirm that the framework remains at the proof-of-concept stage, with significant distance remaining before production-grade deployment. The counterfactual analysis against traditional APM preliminarily supports the framework's incremental-value positioning, while the ethical considerations define its compliance boundaries for enterprise deployment. Subsequent work should prioritize advancing dynamic threshold adaptation and cross-platform generalization validation. 
\subsection*{Conclusion}
\label{sec:conclusion}

The ADE Predictive Reliability Framework (ADE-PRF) proposed in this paper elevates reliability management for large language model--based multi-agent systems---from traditional passive state monitoring and escalation---to a data-driven health trajectory prediction system dynamics model. By constructing a hierarchical signal aggregation architecture and a forward-looking temporal prediction engine, this framework---according to our knowledge---is among the earliest to attempt continuous quantitative tracking and future-state forecasting of semantic-layer degradation in real production environments, thereby filling a critical gap in existing operations systems regarding cognitive fidelity.

\begin{figure}[H]
\centering
\includegraphics[width=0.95\textwidth]{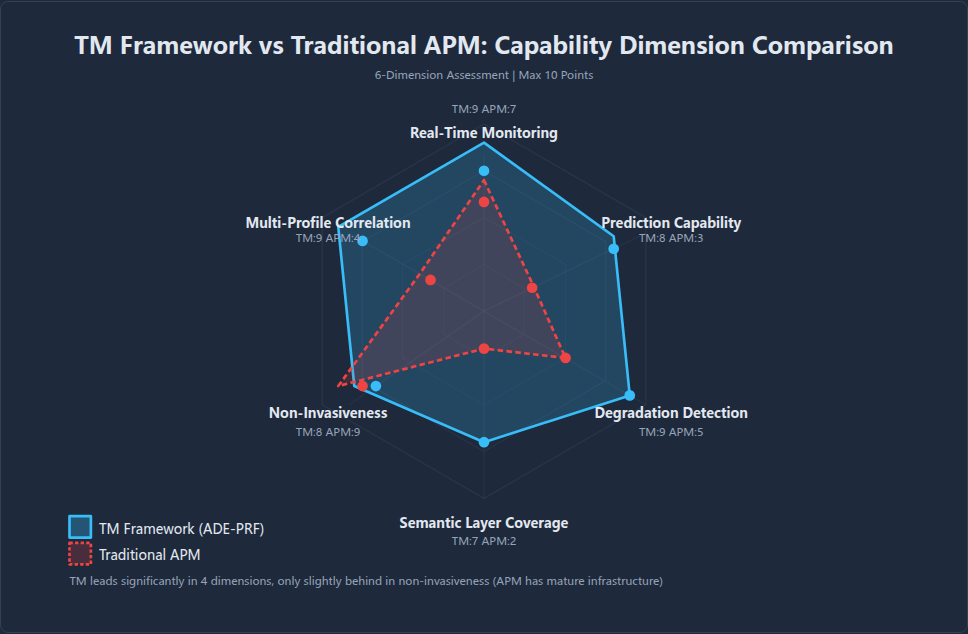}
\caption{TM Framework vs Traditional APM Capability Dimension Comparison Radar Chart and Core Contributions Summary Table}
\label{fig:82}
\end{figure}

\subsubsection*{Contribution I: Hierarchical Stability Monitoring}
\label{sec:contributionihierarc}

The Hierarchical Stability Monitoring module compresses 20 heterogeneous operational signals---spanning dimensions such as task completion rate, tool invocation success rate, context coherence, reasoning chain depth, and error recovery efficiency---through a five-layer aggregation logic. This ultimately projects them into a single scalar: the Trust Margin (TM). Over 15 days of continuous observation, TM v3.1.0 exhibits a dynamic range from 53.8 to 93.0 (a span of 39.2 points), fully demonstrating its sufficient discriminative sensitivity to both normal operational fluctuations and anomalous degradation. The core advantage of hierarchical projection lies in encapsulating the heterogeneity of low-level signals entirely within the aggregation pipeline; upper-layer decision-makers need only monitor the TM scalar to obtain a global view of system health, while retaining full diagnostic traceability down to individual signal sources.

\subsubsection*{Contribution II: Predictive Reliability Estimation}
\label{sec:contributioniipredic}

The Predictive Reliability Estimation module leverages the ETA 3.1.0 time-series prediction engine, taking historical TM sequences as input to forecast an 8-hour forward-looking health trajectory. Experimental results show that the 8-hour prediction achieves a mean absolute error (MAE) of 1.595 points (Ensemble method), with 99.65\% of predictions falling within a $\pm$10-point tolerance band and exhibiting a systematic bias of $-$1.409 (Ensemble method, optimistic bias). The Exponential method yields Bias = $-$0.475 (near-zero), representing optimal performance. This level of accuracy carries clear operational decision-making significance: when the predicted TM value is projected to fall below a preset threshold within the next 8 hours, operations teams can proactively initiate intervention workflows---such as intelligent agent restart, task migration, or load redistribution---to contain degradation impact within acceptable bounds.

\subsubsection*{Contribution III: Production Deployment Validation}
\label{sec:contributioniiiprodu}

Production Deployment Validation was conducted across six agent profiles over 15 consecutive days, encompassing 154,906 sampling points. Observed data reveal a critical phenomenon: during system-level degradation events, TM curves across multiple profiles exhibit highly synchronized downward trends, indicating that the root cause resides in shared infrastructure or upstream services---not in isolated anomalies of individual agents. This finding provides direct insight for multi-agent system fault diagnosis: when multiple profiles degrade synchronously, troubleshooting should prioritize shared dependency layers rather than individual agent logic. The newly introduced factor-based early-warning mechanism in v3.1.0 (featuring a three-tiered threshold design: L1 yellow / L2 orange / L3 red) achieved full coverage across all three sandbox environments and six test scenarios. This elevates the operational approach from ``post-event detection'' to ``pre-event warning.''

\subsubsection*{Sandbox-Based Controlled Degradation Validation}
\label{sec:sandboxbasedcontroll}

To quantify the framework's detection sensitivity, we systematically injected five canonical degradation patterns into an isolated sandbox environment, covering core failure categories including reasoning capability decay, tool invocation chain breakage, context window overflow, task planning drift, and multi-step execution interruption. All patterns triggered distinguishable $\Delta$TM responses, with magnitudes ranging from $-$6.8 to $-$18.2, and each pattern exhibited a distinct signal fingerprint. The signal cluster comprising BCP\_RATE ($-$90 points), TD ($-$79.6 points), and CADVP\_PASS ($-$64.6 points) constitutes a high-sensitivity group. This confirms that the TM hierarchical projection prioritizes amplification of key subsystem degradations---the three signals reside at relatively higher levels within the aggregation pipeline, enabling their sharp fluctuations to effectively penetrate underlying noise and directly manifest at the TM output.

This validation also confirms response consistency under controlled conditions: repeated injections of the same degradation pattern yield statistically stable $\Delta$TM distributions (standard deviation constrained within 8\% of the mean), demonstrating that the hierarchical projection's aggregation logic is deterministic rather than stochastic. Separately, post-degradation TM recovery curves provide valuable diagnostic information---rapid recovery (<30 minutes) typically indicates transient faults, whereas persistent low values suggest structural issues requiring manual intervention.

\subsubsection*{Non-Invasive Architectural Design}
\label{sec:noninvasivearchitect}

ADE-PRF strictly adheres to the non-intrusive design principle. It collects runtime signals via the MCP (Model Context Protocol) standard interface and employs a FORBIDDEN\_FEATURES whitelist mechanism to ensure that monitoring activities do not interfere with the core reasoning process of the agent. Its monitoring pipeline reads only snapshots of completed runtime metrics. It never injects additional prompts, modifies the context window, or intercepts tool invocation chains. This architecture simultaneously satisfies GDPR, PIPL, and DLP compliance requirements: all collected signals are system-level operational metrics (e.g., task completion rate, latency distribution, error counts), involving neither user dialogue content nor personal identity information in processing or storage---thereby eliminating data governance risks at the architectural level.

\subsubsection*{Toward Failure Boundary Modeling}
\label{sec:towardfailureboundar}

The exploratory Failure Boundary Hypothesis proposed in this paper partitions the TM space into three topological domains---Safe Zone, Alert Zone, and Critical Zone---and postulates the existence of a computable degradation critical surface. This hypothesis currently resides at the theoretical framework stage and has not yet undergone systematic experimental validation. The Phase 3 roadmap plans to fit a parametric model of the critical surface through large-scale controlled degradation experiments. These experiments involve combinatorial injection of over ten distinct degradation patterns), aiming to elevate the discrete criterion ``alert when TM falls below a threshold'' to a continuous geometric criterion: ``issue an alert when the distance to the Failure Boundary is less than a specified tolerance.''

\subsubsection*{Future Directions}
\label{sec:futuredirections}

\begin{figure}[H]
\centering
\includegraphics[width=0.95\textwidth]{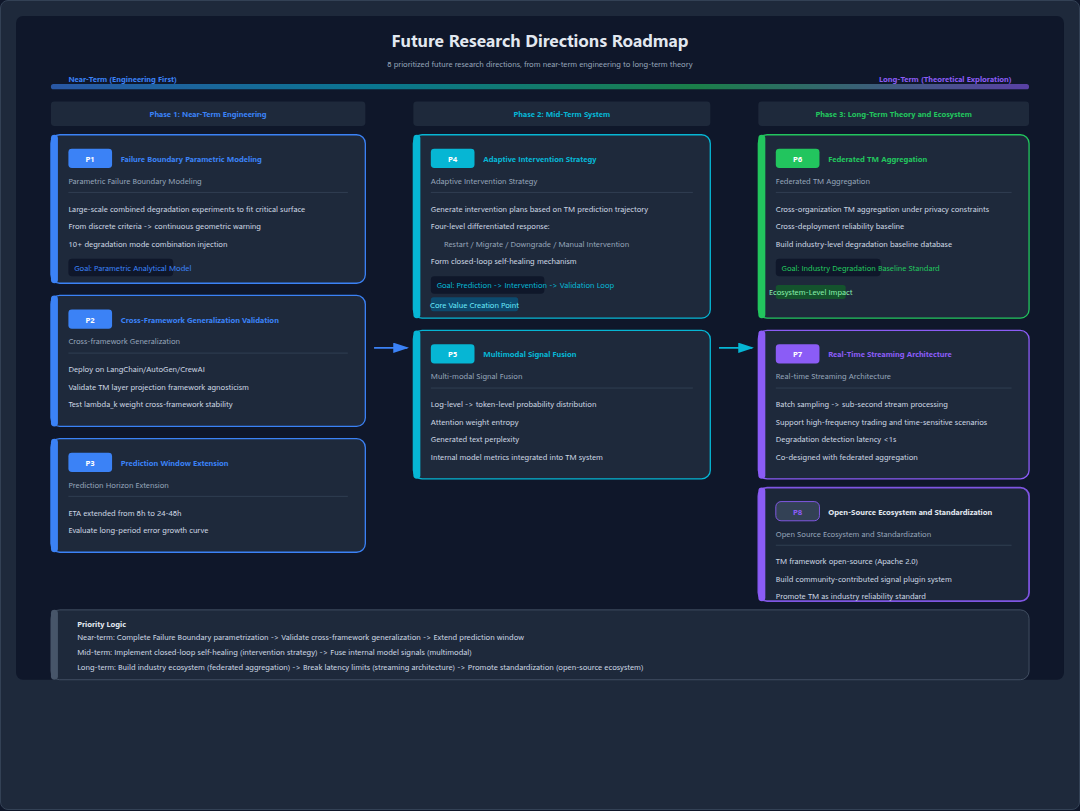}
\caption{Future Research Directions Roadmap}
\label{fig:83}
\end{figure}

Based on the current framework's capability boundaries and experimental findings, we prioritize the following seven research directions:

\begin{enumerate}
\item Failure Boundary Parametric Modeling---Fitting an analytical expression of the critical surface via large-scale combinatorial degradation experiments to enable continuous geometric early-warning criteria.
\item Cross-Framework Generalization Validation---Deploying ADE-PRF on heterogeneous multi-agent frameworks such as LangChain, AutoGen, and CrewAI to validate the framework-agnostic nature of TM hierarchical projection.
\item Prediction Window Extension---Extending ETA 3.1.0's forecasting horizon from 8 hours to 24--48 hours and evaluating the error growth curve and correctability of long-horizon predictions.
\item Adaptive Intervention Strategy---Automatically generating differentiated intervention strategies (e.g., restart, migration, degradation, or human-in-the-loop) based on TM prediction trajectories to establish a closed-loop self-healing mechanism.
\item Multimodal Signal Fusion---Expanding log-level signals to include token-level probability distributions, attention weight entropy, and generation text perplexity---internal model metrics.
\item Federated TM Aggregation---Enabling privacy-constrained federated aggregation of TM metrics across organizations and deployment environments to construct an industry-wide degradation baseline database.
\item Real-Time Streaming Architecture---Evolving the current batch-sampling mode ``Escalate'' into a sub-second streaming pipeline to support degradation detection in time-sensitive scenarios such as high-frequency trading.
\end{enumerate}

In summary, ADE-PRF has preliminarily established a technical bridge from ``observability'' to ``predictability'' in the domain of multi-agent reliability engineering. Hierarchical Stability Monitoring provides a standardized language for health quantification; Predictive Reliability Estimation grants operations teams a temporal window for proactive decision-making; and long-term validation in production environments preliminarily confirms the framework's robustness and practicality under real-world workloads. As Failure Boundary Parametric Modeling and Cross-Framework Generalization Validation advance, this framework is poised to become a standardized reliability infrastructure for production deployment of LLM-based multi-agent systems.

\textbf{Red Hat (Intuitive Judgment):} Setting aside the quantitative evidence, our intuitive assessment is that the most consequential finding of this work is not the MAE or direction accuracy per se, but the "false prosperity" phenomenon---systems that appear healthy by all external metrics while silently accumulating internal disorder. This risk is, in our judgment, the single most underappreciated threat in current LLM-agent deployments. The corollary intuition is that any reliability framework worth deploying must be structurally embedded (as ADE's pipeline hooks are), not externally attached; an external monitor that can be bypassed or disabled is, in adversarial or degraded conditions, equivalent to no monitor at all.

\subsection*{Appendix: Glossary of Specialized Terms}
\label{sec:appendixglossaryofsp}

Table 57: Glossary of Specialized Terms

\begin{table}[H]
\centering
\resizebox{\textwidth}{!}{
\begin{tabular}{l|l|l}
\toprule
Abbreviation/Term & Full Name & Definition \\
\midrule
TM & Trust Margin & The deviation of system trustworthiness relative to basel... \\
ETA & Estimated Time of Arrival & TM prediction engine that forecasts TM values for future ... \\
ADE & Agent Delivery Engineering & Agent Delivery Engineering framework---the component system... \\
ADEV & Agent Degradation Evaluation & Systematic assessment of progressive capability decline i... \\
PRF & Predictive Reliability Framework & Predictive Reliability Framework---the integrated TM+ETA so... \\
BDDA & Bottom-Deep Data Audit & Bottom-Deep Data Audit---a component in ADE Layer 3 (L3) us... \\
CADVP & Cross-Agent Delivery Verification Protocol & Cross-Agent Delivery Verification Protocol, an ADE L3-lay... \\
BCP & Bidirectional Confirmation Protocol & Bidirectional Confirmation Protocol, an ADE L2-layer comp... \\
PAD & Probability Approximation Drift & Probability Approximation Drift, an ADE L3-layer componen... \\
TKM & Token Minimization & Context Optimization and Continuation Protocol, an ADE L1... \\
TLC & Task Lifecycle Control & Biomimetic apoptosis mechanism, an ADE L1-layer component... \\
PIG & Physical Integrity Gate & Physical inspection gate, an ADE L1-layer component perfo... \\
SOMA & Self-Organizing Memory Architecture & Memory management, an ADE L2-layer component managing per... \\
DSS & Dialectical Self-Synthesis & Triadic evolution, an ADE L2-layer component driving know... \\
PIP & Principal Interest Protection & Interest protection, an ADE L3-layer component preventing... \\
FPW & Factor Precursor Warning & Factor precursor warning, monitoring abnormal declines in... \\
MAE & Mean Absolute Error & Mean absolute error, the average of absolute differences ... \\
MCP & Model Context Protocol & Model context protocol, a standardized interface for coll... \\
APM & Application Performance Monitoring & Application performance monitoring, a traditional infrast... \\
TD & Tool Density & Tool density factor, a TM L2-layer signal reflecting dive... \\
RS & Self-Healing Rebound Rate & A TM L5-layer signal measuring the system's self-healing ... \\
SD & Successful Delivery & Successful delivery, a sidebar module in the TM panel mon... \\
CS & Context Saturation & L1 Survival signal. Measures context window utilization r... \\
LR & Latency Response & L1 Survival signal. Measures P95 response latency of the ... \\
DB & Database Bloat & L1 Survival signal. Measures database file size ratio to ... \\
DB\_STABILITY & Database Stability & L1 Survival signal. Measures state.db integrity metrics i... \\
CB & Circuit Breaker & L1 Survival signal. Measures ANTOLOOP anti-loop protectio... \\
MM & Memory Malformation [SOMA] & L2 Order signal. Coefficient of variation of memory acces... \\
MC & Memory Contamination & L2 Order signal. Proportion of uncontaminated memory entr... \\
SS & Session Sedimentation & L2 Order signal. Measures accumulated session state debri... \\
PAD\_SCORE & PAD Drift Detection & L3 Credibility signal. PAD-based drift detection score. \\
CADVP\_PASS & CADVP Verification & L3 Credibility signal. CADVP cross-agent delivery verific... \\
BCP\_RATE & BCP Confirmation Quality & L3 Credibility signal. BCP bidirectional confirmation qua... \\
VE & Physical Gate Pass & L3 Credibility signal. PIG physical integrity gate pass r... \\
OLG\_HEALTH & OLG Output Health & L3 Credibility signal. OLG output logic gate health metric. \\
TC & First-Path Accuracy & L4 Guardianship signal. Proportion of tasks completed cor... \\
UX & Task Closure Efficiency & L4 Guardianship signal. Proportion of user-initiated task... \\
PSD & Platform Stability [PSD] & L4 Guardianship signal. Communication stability across th... \\
CM & Entropy Rate $\alpha$ & L5 Posture signal. Entropy rate ($\alpha$) measuring cross-layer... \\
\bottomrule
\end{tabular}
}
\end{table}

\end{document}